\documentclass[11pt]{article}
\usepackage[english]{babel}
\usepackage{amsmath}
\usepackage{amsthm}
\usepackage{amsfonts}
\usepackage{amssymb}
\usepackage{hyperref}
\usepackage{enumitem}
\usepackage{titlesec}
\usepackage{secdot}
\sectiondot{subsection}
\sectiondot{subsubsection}
\usepackage{fancyhdr}
\usepackage{color}
\usepackage{tcolorbox}
\usepackage{subcaption}
\usepackage{titlesec}
\usepackage{relsize}
\usepackage{mathtools}

\usepackage{pgf,tikz}
\usepackage{graphicx} % Required for inserting images

\usepackage{xcolor}

\newcommand{\ola}{\overleftarrow}
\newcommand{\f}{\frac}

\title{A Score-based Diffusion Model Approach for Adaptive Learning of Stochastic Partial Differential Equation Solutions}

\author{Toan Huynh, Ruth Lopez Fajardo, Guannan Zhang, Lili Ju, Feng Bao}
\date{}

\begin{document}
% \today
\maketitle

\begin{abstract}
We propose a novel framework for adaptively learning the time-evolving solutions of stochastic partial differential equations (SPDEs) using score-based diffusion models within a recursive Bayesian inference setting. SPDEs play a central role in modeling complex physical systems under uncertainty, but their numerical solutions often suffer from model errors and reduced accuracy due to incomplete physical knowledge and environmental variability. To address these challenges, we encode the governing physics into the score function of a diffusion model using simulation data and incorporate observational information via a likelihood-based correction in a reverse-time stochastic differential equation. This enables adaptive learning through iterative refinement of the solution as new data becomes available. To improve computational efficiency in high-dimensional settings, we introduce the ensemble score filter, a training-free approximation of the score function designed for real-time inference. Numerical experiments on benchmark SPDEs demonstrate the accuracy and robustness of the proposed method under sparse and noisy observations. 
\end{abstract}

\section{Introduction}

In this paper, we introduce a score-based diffusion model approach for adaptively learning the time-evolving solutions of stochastic partial differential equations (SPDEs) through recursive Bayesian inference.

Partial differential equations (PDEs) are fundamental tools for modeling the dynamic behavior of complex physical systems. While they have been widely successful in scientific and engineering applications, many practical scenarios involve inherent uncertainties due to limited physical knowledge and environmental variability.   For example, in climate and meteorological modeling, uncertainties in initial conditions, boundary data, and subgrid-scale physical processes can significantly affect the accuracy of predictions governed by PDEs such as the Navier–Stokes or advection–diffusion equations. Similarly, in porous media flow problems, spatial heterogeneity and limited characterization of subsurface properties — such as permeability or porosity — introduce substantial uncertainty into models governed by Darcy’s law and related PDEs, making accurate prediction particularly challenging. To capture these uncertainty effects and support reliable predictive analysis, it is essential to incorporate SPDEs into mathematical modeling framework. The numerical solution of SPDEs has thus become a central focus of the uncertainty quantification (UQ) community, where significant efforts have been dedicated to developing efficient solvers that can accurately characterize and propagate uncertainty in high-dimensional, nonlinear dynamical systems  (see, e.g.,~\cite{Abrall2013, Abgrall2017, Barth2012, Geraci2016, KNIO2006, Petrella2019, Walton2025} and the reference therein). 

Despite advances in SPDE solvers capable of quantifying uncertainty, significant challenges remain. In particular, model errors arising from limited physical knowledge and uncertainties due to unknown environmental variability can accumulate over time, leading to a progressive degradation in solution accuracy — especially when the underlying physical laws are incomplete or only approximately known \cite{Bao2025MWR, cheng2013comparison, evensen2009data, Toan_JCP_2025, Toan2025, kennedy2001bayesian, oliver2008inverse,  reich2015probabilistic}. 

A strategy to address these issues is to leverage measurement data to ``learn'' an estimated solution that reflects the underlying physical reality.  However, fully data-driven approaches are often impractical due to the fact that real-world observations are typically sparse, indirect, and contaminated with noise.  As a result, there is a critical need for adaptive learning frameworks that can effectively \textit{integrate partial and noisy measurement data into existing physical models}. Such frameworks are essential for correcting model discrepancies, reducing uncertainty, and ultimately improving the fidelity of SPDE solvers.

The standard framework for integrating data and models is known as ``data assimilation'', which systematically combines observational data with numerical model simulations to estimate the evolving state of a dynamical system. The mathematical formulation of this problem is referred to as ``optimal filtering'', which seeks to represent the ``optimal'' estimate of the target model state in the form the \textit{conditional expectation} -- conditioning on the observational information. The state-of-the-art approach for solving the optimal filtering problem is the recursive Bayesian filter, including the Kalman type filters \cite{LETKF_2020, evensen2009data, kalman1960new} and the particle filters \cite{MCMC-PF, Bao2019a, particle-filter, Kang-PF, MTAC2012, pitt1999filtering}. In addition to the Bayesian approach, there exists a class of methods based on stochastic partial differential equations that directly approximate the conditional distribution of the target state \cite{Bao2022, Bao2014, Bao_zakai, bao2011numerical, zakai, Bao_BSDEF_2022}. 

\vspace{0.5em}

In this work, we develop a novel framework that applies score-based diffusion models to the adaptive learning of SPDE solutions. Following the recursive Bayesian framework, our method used numerical SPDE solvers to encode first-principle and model-based dynamics for predictive simulation. When observational data becomes available, we apply Bayesian inference to integrate this data into the simulated model, enabling calibration of the model-based estimates. The adaptive learning in this context refers to the iterative refinement of the solution: as new data is assimilated,  the model dynamically adjusts its estimates, progressively capturing hidden physics and reducing epistemic uncertainty. This leads to improved predictive accuracy and robustness over time, even in the presence of incomplete or imperfect knowledge of the underlying system.

The central contribution of this work is the introduction of a score-based diffusion model technique~\cite{ho2020denoising, Bao2025a, YSong2021} that enables this adaptive learning process through data assimilation. While conventional optimal filtering methods have achieved considerable success, these methods face fundamental limitations. Kalman-type filters are efficient in high-dimensional settings but rely on linear or weakly nonlinear assumptions \cite{LETKF_2007, evensen2009data}. Particle filters, on the other hand, are well-suited for nonlinear systems but suffer from the ``curse of dimensionality'' issue in high-dimensional spaces due to its sequential Monte Carlo mechanism \cite{hu_vanLeeuwen_2021, sny}. As a result, no existing method can effectively address the joint challenges of high dimensionality (e.g., arising from spatial discretization in SPDEs) and strong nonlinearity — both of which are intrinsic to adaptive learning tasks for SPDEs. Our framework is designed to overcome this gap by taking the advantage of the flexibility and scalability of score-based generative models \cite{song2019generative}.

In a score-based diffusion model, a forward stochastic differential equation (SDE) is used to progressively transform an arbitrary data distribution into a standard Gaussian distribution. During this process, the information from the original data is encoded in the so-called ``score function'', defined as the gradient of the log-probability density of the forward SDE. Then, a reverse-time SDE, driven by the score function, is solved to effectively transform samples drawn from the standard Gaussian distribution back into samples that follow the original data distribution. When applying score-based diffusion models to the adaptive learning of SPDE solutions, we encode the underlying physics described by the SPDE into the score function using simulation samples. When observational data becomes available, a likelihood score is introduced to incorporate data information into the reverse-time SDE, enabling the model to iteratively adjust its estimates and improve solution accuracy ~\cite{Bao2024a}. To enhance the computational efficiency of score-based diffusion models in high-dimensional settings and enable real-time calibration of estimated solutions, we propose an ensemble-based approximation of the score function—rather than relying on neural network-based deep learning methods~\cite{Bao2024c}. We refer to this approach as the ensemble score filter.

We validate the proposed method through a series of numerical experiments involving prototypical SPDEs, including the Burgers' equations, incompressible Navier–Stokes equations, and Allen-Cahn equations -- all subjected to uncertainty perturbations.  These tests demonstrate the accuracy, robustness, and adaptability of the proposed framework, particularly in scenarios with sparse and noisy observations.

The rest of the paper is organized as follows: Section~\ref{Sec2_General} provides a detailed overview of the data assimilation framework for adaptively solving the SPDEs. We shall introduce the the score filter methodology and its extension, the ensemble score filter in Section~\ref{Sec3_EnSF}. In Section~\ref{Sec4_Numerics}, we present extensive numerical results to illustrate the effectiveness of the proposed approach. Some concluding remarks will be given in Section~\ref{Conclusion}.

\section{Adaptive learning for stochastic partial differential equations via data assimilation} \label{Sec2_General}

\subsection{Partial differential equations with uncertainty}
We consider the following partial differential equations:
\begin{equation}\label{PDE:deterministic}
d u_t(x) = \mathcal{L}(x, t) dt + h(u_t, x, t) dt,
\end{equation}
where $u_t$, is the unknown function (either scalar or vector-valued) defined over the spatial domain $x \in \Omega \subset \mathbb{R}^{dim}$, and $t \geq 0$ is the time variable. The operator $\mathcal{L}$ is a second-order (typically elliptic) spatial differential operator, and $h(u_t, x, t)$ represents a given source term. 

In practical applications, various uncertainties arise due to limited physical knowledge and incomplete environmental information. As a result, the deterministic PDE \eqref{PDE:deterministic} is often subject to random perturbations, leading to the following stochastic partial differential equation (SPDE): 
\begin{equation}\label{SPDE}
d u_t(x) = \mathcal{L}(\tilde{\omega}, x, t) dt + h(u_t, \tilde{\omega}, x, t) dt + \sigma_t d\tilde{W}(x, t),
\end{equation}
where $\tilde{\omega}$ encapsulates \textit{model uncertainties due to incomplete environmental information}, and we formulate this kind of uncertainties by unknown coefficients in the spatial differential operator and the source term. The term $\tilde{W}(x, t)$ denotes a space-time Wiener process that captures \textit{model errors arising from unresolved or unknown physical processes} during the formulation of the PDE model, and $\sigma_t$ is the noise coefficient.

% (see, e.g.,~\cite{Solano2016,  Harmon2025, Jung2024, Mo2019a, Mo2019b,  Cho2022, Tokareva2024, Zhao2024, Zhu2018, Zhu2019} and the reference therein)

Extensive research in the uncertainty quantification (UQ) community has focused on developing efficient numerical algorithms for characterizing and propagating the uncertainties inherent in stochastic partial differential equations (SPDEs) (see, e.g.,~\cite{Abrall2013, Abgrall2017, bao2011numerical, Barth2012, Geraci2016, KNIO2006, Petrella2019, Walton2025} and the reference therein). To simplify our presentation, we use the following generic form to represent the numerical solver that propagates the approximate solution of the SPDE \eqref{SPDE}:
\begin{equation}\label{SPDE:scheme}
u_{n+1} = \mathbb{F}(u_n, t_n, \tilde{\omega}, \Delta \tilde{W}_{t_n}),
\end{equation}
where $\Delta \tilde{W}_{t_n}: = \tilde{W}_{t_{n+1}} - \tilde{W}_{t_n}$, and $\mathbb{F}$ denotes a forward operator that incorporates the random variables $\tilde{\omega}$ and $\Delta \tilde{W}_{t_n}$ within a UQ-enabled PDE solver. This operator advances the solution $u$ in Eq.~\eqref{SPDE} from time $t_{n}$ to $t_{n+1}$, with $u_{n+1}$ being a numerical approximation to $u({t_{n+1}})$.  In the numerical experiments section (Section \ref{Sec4_Numerics}), we provide a more detailed discussion of the various forward solvers used for SPDE solution propagation, depending on the specific types of PDEs being solved.

\vspace{1em}

While most conventional UQ techniques aim to characterize the uncertainties inherent in SPDE systems~\cite{Solano2016,  Harmon2025, Jung2024, Mo2019a, Cho2022, Tokareva2024, Zhao2024}, the model errors and uncertainties stemming from limited knowledge and incomplete information remain unavoidable. \textit{The primary goal of this work is to introduce a robust and accurate data assimilation framework based on score-based diffusion models, designed to adaptively reduce model uncertainties and mitigate system errors by leveraging partial and noisy observations of the target system.}

In what follows, we briefly outline the data assimilation framework and explain how it can be applied to adaptively enhance the accuracy of SPDE solutions.

\subsection{A data assimilation framework for adaptive learning of SPDE solutions}

The data assimilation problem combines a spatial-temporal model typically derived from a forward numerical solver with an observational system.  Specifically, we consider the following formulation: 
\begin{equation}\label{DA:formulation}
\begin{array}{rll}
X_{n+1} &=  f(X_n, \omega_n), \qquad & \text{(State)} \\
Y_{n+1} &=   g(Y_n) + \xi_n. & \text{(Measurement)}
\end{array}
\end{equation}
Here, $X_{n} := [u_n(x^1), u_n(x^2), \cdots, u_n(x^d)]$ denotes the discretized state vector of the quantity of interest.  In this work, the state corresponds to to the solution of the SPDE,  evaluated at a set of spatial grid points $\{x^i\}_{i=1}^{d}$ over the domain $\Omega \in \mathbb{R}^{dim}$. 

In the state equation, the function $f$ represents the forward model, which incorporates the random variable $\omega_n$ to account for uncertainties.  When this forward model is implemented through a numerical solver for the SPDE~\eqref{SPDE:scheme}, the random input includes both model uncertainty $\tilde{\omega}$ and model error $\tilde{W}_{t_n}$. 

As the forward model propagates in time using UQ-enabled SPDE solvers, uncertainties and errors naturally accumulate over time, leading to progressively less accurate predictions with reduced confidence. In practice, these uncertainties often arises from stochastic components with unknown distributions, making precise solution prediction infeasible.  To mitigate this degradation in accuracy, observational data -- typically collected via measurement instruments -- are incorporated to calibrate the model and enhance solver accuracy. However, such observations are generally sparse, noisy, and incomplete, and therefore insufficient by themselves to fully reconstruct the underlying state.

The measurement equation in the data assimilation framework directly reflects the need for observational data to calibrate and correct the forward model predictions. As described in Eq.~\eqref{DA:formulation}, the observed data $Y$ are related to the true target state $X$ through an observation operator $g$, with additive measurement noise $\xi_n$. These observations help constrain the uncertainty in the model and improve the accuracy of state estimates. Throughout this work, we assume that the observational noise $\xi_n$ is Gaussian with covariance matrix $R$.

\vspace{0.5em}

The goal of data assimilation is to recover the best possible estimate of the target state given the available observational data $Y_{1:n+1}$. In the context of adaptive learning for SPDEs considered in this work, data assimilation plays a critical role in reducing uncertainties arising from incomplete information and model errors. The mathematical framework that solves the data assimilation problem is the \textit{optimal filter}, which seeks the ``best estimate'' in the form of the conditional expectation of $X_{n+1}$ given $Y_{1:n+1}$. Specifically, the optimal filter $\tilde{X}_{n+1}$ is defined as
$$\tilde{X}_{n+1} := \mathbb{E}\left[X_{n+1} | Y_{1:n+1}\right].$$
In practical implementations, rather than computing the conditional expectation directly, one typically aims to derive an approximation of the conditional distribution $P(X_{n+1} | Y_{1:n+1})$, referred to as the \textit{``filtering distribution''}. 

The standard mathematical framework for computing the filtering distribution is recursive Bayesian filter. Suppose that at time step $n$, the filtering distribution $P(X_{t_{n}}|Y_{1:t_n})$ is available, representing the current estimate of the state given all observations up to time $t_n$. The Bayesian filter then proceeds in two steps: a prediction step and an update step.

In the prediction step, we utilize the Chapman-Kolmogorov formula to obtain \begin{equation}\label{Prediction}
    \quad P(X_{t_{n+1}}|Y_{1:t_n}) = \int P(X_{t_{n+1}}|X_{t_{n}})P(X_{t_{n}}|Y_{1:t_n})dX_{t_n},  \qquad \textit{Prior filtering distribution} 
\end{equation}
where $P(X_{t_{n+1}}|X_{t_{n}})$ is the transition probability derived from the state dynamics \eqref{DA:formulation} governed by the forward solver $f$. The prior filtering distribution $P(X_{t_{n+1}}|Y_{1:t_n})$ contains the data information up to time $t_{n}$ and the model information captured by the transition probability.

In the update step, we apply the following Bayesian inference scheme to compute the posterior filtering distribution of the state variable:
\begin{equation}\label{Update}
    \quad P(X_{t_{n+1}} \mid Y_{1:t_{n+1}}) \propto P(X_{t_{n+1}} \mid Y_{1:t_{n}}) P(Y_{t_{n+1}} \mid X_{t_{n+1}}),    \qquad \textit{Posterior filtering distribution} 
\end{equation}
which allows us to adaptively refine the estimate of the SPDE solution. The likelihood function $P(Y_{t_{n+1}}|X_{t_{n+1}})$ quantifies the probability of the observed data given the current state:
\begin{equation}\label{Likelihood}
    P(Y_{t_{n+1}}|X_{t_{n+1}}) \propto \text{exp}\big[-\frac{1}{2}(g(X_{t_{n+1}})-Y_{t_{n+1}})^\top R^{-1}(g(X_{t_{n+1}})-Y_{t_{n+1}})\big], 
\end{equation}
where $R$ is the covariance matrix associated with the observational noise introduced in the measurement process described in Eq.~\eqref{DA:formulation}.

\vspace{0.25em}

Together, the prediction and update steps form a recursive model-data integration framework that enables adaptive learning of the system state -- specifically, the solution of the SPDE -- over time.  In the prediction stage, the filtering distribution is propagated forward using the transition probability $P(X_{t_{n+1}}|X_{t_{n}})$, which encodes the model dynamics and uncertainty. However, without incorporating observational data, model errors and uncertainty would accumulate, leading to increasingly inaccurate predictions. The update step plays a critical role in adaptively correcting these predictions by assimilating new observations through Bayesian inference. At the same time, the model information embedded in the prior filtering distribution helps reconstruct the unobserved components of the solution that are not captured by the indirect and potentially sparse measurement process. This recursive framework allows the system to adaptively refine its state estimates by incorporating new data, dynamically adjusting both the solution and its associated uncertainty in response to discrepancies between model predictions and observations. 

\vspace{0.25em}

In the following section, we shall introduce an accurate and robust ensemble score filter approach for solving the data assimilation problem.

\section{The ensemble score filter approach for data assimilation}\label{Sec3_EnSF}

The score-based filter methodology characterizes the filtering distribution in data assimilation problems using the score function within the diffusion model framework. To begin, we briefly outline the design of the score filter for data assimilation in Section \ref{Score_Filter}. We then introduce an ensemble-based Monte Carlo approximation scheme for the score function and derive the resulting ensemble score filter method in Section \ref{EnSF}.

\subsection{Score filter for data assimilation}\label{Score_Filter}
In the score-based diffusion model framework, two stochastic processes are defined as follows:
\begin{equation}\label{Diffusion_Model}
\begin{array}{llll}
    d Z_{\tau} = &  b_{\tau} Z_{\tau} d_{\tau} + \sigma_{\tau} dW_{\tau}, &Z_0 = Q_0(X), \qquad &\text{(Forward SDE)}\\
    d Z_{\tau} = &  \big( b_{\tau} Z_{\tau} - \sigma_{\tau}^2 S(Z_{\tau}, \tau) \big)d_{\tau} + \sigma_{\tau} d\ola{W}_{\tau}, &Z_T = N(0, \pmb{I}_d). \qquad &\text{(Reverse SDE)}
\end{array}
\end{equation}
The first equation is a forward stochastic differential equation (SDE) that propagates a given target data distribution $Z_0= Q_0(X)$, such as the filtering distribution in data assimilation, toward the standard Gaussian distribution over the pseudo time interval $[0, 1]$. The second equation is the corresponding time-reversed SDE, which transports the standard Gaussian random variable $Z_{1} = N(0, \pmb{I}_d)$ back to the original data distribution $Z_0$. The drift coefficient $b$ and diffusion coefficient $\sigma$ in the forward SDE are pre-defined deterministic functions. In this work, we let $b_{\tau} = \f{d \log \alpha_{\tau}}{ d \tau }$ and $\sigma_{\tau} = \sqrt{\f{d \beta_{\tau}^2}{d \tau} - 2 \f{d \log \alpha_{\tau} }{d \tau} \beta_{\tau}^2 }$ with $\alpha_{\tau} = 1 - \tau$ and $\beta_{\tau} = \tau$ (see \cite{Bao2024a}).
\vspace{0.5em}

The key component that enables the reverse SDE in Eq.~\eqref{Diffusion_Model} to transport samples from the standard Gaussian distribution to the target data distribution is the \textit{score function} $S(Z_{\tau}, \tau)$, defined by 
\begin{equation}\label{Def:Score}
S(Z_{\tau}, \tau) = \nabla_z \log Q_{\tau}(Z_{\tau}),
\end{equation}
where $Q_{\tau}$ is the distribution of the forward SDE solution $Z_{\tau}$. In other words, the score function $S$ acts as a guide that steers the reverse-time sampling process in the diffusion model framework back toward the original data distribution.

The methodology of the score filter is to connect the target filtering distribution to a score function and generate state samples that follow the filtering distribution through the reverse SDE process.

Specifically, suppose that we have access to the posterior score $S_{n|n}$ associated with the posterior filtering distribution $P(X_{t_{n}} \mid Y_{1:t_n})$. By simulating the reverse SDE in Eq.~\eqref{Diffusion_Model} using this score, we can generate samples that follow the filtering distribution. Let $\{x_{t_n}^{(j)}\}_{j=1}^{J}$ denote the set of generated samples, where $J$ is any positive integer. To incorporate model dynamics, we propagate these state samples forward using the state evolution model in Eq.\eqref{DA:formulation} and get 
$$x_{t_{n+1}}^{(j)} = f(x_{t_n}^{(j)}, \omega_n^{(j)}), \hspace{2em} j = 1, 2, \cdots, J.$$
Recall that in the context of adaptive learning for SPDEs, the state evolution function $f$ represents the numerical forward solver of the SPDE under consideration. 

The state samples $\{x_{t_{n+1}}^{(j)}\}_{j=1}^{J}$ carry information from both the previous posterior filtering distribution and the state dynamics (i.e., the PDE model). These samples are used to approximate the prior filtering score, denoted by $S_{n+1 \mid n}$, which plays a central role in the prediction step of the recursive Bayesian filtering procedure. This score is then used to generate samples from the prior filtering distribution $P(X_{t_{n+1}} \mid Y_{1:t_n})$. The approximation of $S_{n+1 \mid n}$ can be accomplished using deep learning techniques~\cite{Bao2024c}. However, a more computationally efficient approach in data assimilation is to use a Monte Carlo–based ensemble approximation \cite{Bao2024a, Bao2025a}, which will be detailed in the following subsection.

Once the prior score $S_{n+1 \mid n}$ is available, we incorporate the observational data to update the filtering distribution. To this end, we define the posterior score $S_{n+1 \mid n+1}(Z_{\tau}, \tau)$ through the following scheme:
\begin{equation}\label{Posterior-Score}
S_{n+1 \mid n+1}(Z_{\tau}, \tau) := S_{n+1 \mid n}(Z_{\tau}, \tau) + h(\tau) S_{LH}(Z_{\tau}, \tau), \qquad \tau \in [0, 1],
\end{equation}
where $S_{LH}(Z_{\tau}, \tau):= \nabla_z \log P(Y_{t_{n+1}} \mid X_{t_{n+1}})$ is referred to as the \textit{``likelihood score''}. This term represents the gradient of the log-likelihood function, which quantifies the discrepancy between the model prediction and the observation, and has the form of a score function. The function $h(\cdot)$ is a damping function satisfying $h(0) = 1$ and $h(1) = 0$. 

Accordingly, the terminal condition for Eq.~\eqref{Posterior-Score} at $\tau = 0$ is given by
$$S_{n+1 \mid n+1}(Z_{0}, 0) = S_{n+1 \mid n}(Z_{0}, 0) + \nabla_z \log P(Y_{t_{n+1}} \mid X_{t_{n+1}}), $$
which corresponds to the Bayesian update expressed in Eq.~\eqref{Update}.  In contrast, the terminal condition at $\tau = 1$ is
$$S_{n+1 \mid n+1}(Z_{1}, 1) = S_{n+1 \mid n}(Z_{1}, 1),$$
reflecting the fact that $Z_1 \sim N(0, \pmb{I}_d)$. This completes the update step of the filtering procedure. 

Note that the choice of the damping function $h(\tau)$ governs how the observational data (encoded in the likelihood score) is integrated with the model dynamics (encoded in the prior score). In this work, we adopt the simple linear form $h(\tau) = 1 - \tau$, for $\tau \in [0, 1]$. More advanced choices for $h(\tau)$ will be explored in future studies.

The remaining challenge in the score-based filtering framework lies in the approximation of the (prior) score function itself, which we now address in the following subsection.

\subsection{Monte Carlo-base ensemble approximation for the score function}\label{EnSF}

The methodology of the ensemble score filter (EnSF)  builds on the fact that the distribution of the forward SDE solution can be expressed as follows \cite{Bao2024c}:
$$Q_{\tau}(Z_{\tau}) = \int_{\mathbb{R}^d} Q_{\tau}(Z_{\tau} \mid Z_0) Q_0(Z_0) dZ_0,$$
where the transition probability $Q_{\tau}(Z_{\tau} \mid Z_0)$ over the pseudo-temporal interval $\tau \in [0,1]$ is given by 
$$Q_{\tau}(Z_{\tau} \mid Z_0) = N(\alpha_{\tau} Z_0, \beta_{\tau}^2 \pmb{I}_d)$$
due to the linearity of the diffusion process described in Eq.~\eqref{Diffusion_Model}. 

With the above expressions, the score function defined in Eq.~\eqref{Def:Score} can be rewritten as follows~\cite{Bao2024c, rafid24scalable}:
\begin{equation}\label{Integral:Score}
\begin{aligned}
S(Z_{\tau}, \tau) = & \nabla_z\log \left(\int_{\mathbb{R}^d} Q_{\tau}(Z_{\tau} \mid Z_0) Q_0(Z_0) dZ_0 \right) \\
= & \int_{\mathbb{R}^d} - \frac{Z_{\tau} - \alpha_{\tau} Z_0}{\beta_{\tau}^2} w(Z_{\tau}, Z_0) Q_0(Z_0) dZ_0,
\end{aligned}
\end{equation}
where the weight function $w(Z_{\tau}, Z_0)$ is defined by
$$w(Z_{\tau}, Z_0) := \frac{Q_{\tau}(Z_{\tau} \mid Z_0)}{\int_{\mathbb{R}^d} Q_{\tau}(Z_{\tau} \mid \tilde{Z}_0) Q_0(\tilde{Z}_0) d \tilde{Z}_0},$$
and satisfies the normalization condition $\int_{\mathbb{R}^d} w(Z_{\tau}, Z_0) Q_0(Z_0) dZ_0 = 1$. 

The integral representation of the score function described in Eq.~\eqref{Integral:Score} naturally leads to a Monte Carlo–based ensemble approximation: 
\begin{equation}\label{MC:Score}
S(Z_{\tau}, \tau) \approx \sum_{j=1}^J \frac{Z_{\tau} - \bar{\alpha}_{\tau} z^{(j)}_0}{\bar{\beta}_{\tau}^2} \bar{w}(Z_{\tau}, z_0^{(j)}),
\end{equation}
where $\{z_0^{(j)}\}_{j=1}^{J}$ are samples that serve as ``data samples'' in the diffusion model to characterize the target distribution $Q_0$ \cite{scalesc24}. The modified diffusion parameters are defined as $\bar{\alpha}{\tau} := 1 - \tau (1 - \epsilon_{\alpha})$ and $\bar{\beta}{\tau} := \sqrt{\epsilon{\beta} + \tau (1 - \epsilon_{\beta})}$, where $\epsilon_{\alpha}$ and $\epsilon_{\beta}$ are regularization hyper-parameters introduced to stabilize the reverse SDE dynamics.

\vspace{0.5em}

In the context of data assimilation for adaptive learning of SPDE solutions, the data samples are used to represent the predicted state variable through simulating the SPDE model via the UQ-enabled PDE solver $f$. Specifically, the prior score $S_{n+1 \mid n}$ associated with the prior filtering distribution $P(X_{t_{n+1}} \mid Y_{1:t_n})$ can be approximated as 
$$S_{n+1 \mid n}(Z_{\tau}, \tau) \approx \sum_{j=1}^J \frac{Z_{\tau} - \bar{\alpha}_{\tau} x_{t_{n+1}}^{(j)}}{\bar{\beta}_{\tau}^2} \bar{w}(Z_{\tau}, x_{t_{n+1}}^{(j)}),$$
where each sample $x_{t_{n+1}}^{(j)}$ is generated via $x_{t_{n+1}}^{(j)} = f(x_{t_n}^{(j)}, \omega_n^{(j)})$, $j = 1, 2, \cdots, J$.  The approximated prior score is then used to compute the posterior score according to the formulation in Eq.~\eqref{Posterior-Score}.

\vspace{1em}

In the next section, we present detailed numerical implementations of the EnSF method to demonstrate its adaptive learning mechanism across a range of SPDE models.

\section{Numerical experiments}\label{Sec4_Numerics}
We conduct numerical experiments to evaluate the effectiveness of our proposed adaptive learning approach in improving the accuracy of SPDE solvers across four test problems. In the first example, we study the two-dimensional (2D) Burgers' equations and examine the performance of our method in capturing sharp features that develop in the solution. The second example involves 2D Navier–Stokes equations with various boundary conditions and external forces. We also account for different levels of model error introduced in the forward solver. Finally, in the third example we study the 2D Allen–Cahn equations with general mobility, exploring multiple ways to incorporate model uncertainties.

For all cases, we simplify our demonstration by consider \eqref{SPDE:scheme} in the form
\begin{equation}\label{SPDE:scheme_simplifed}
u_{n+1} = \mathbb{F}(u_n, t_n, \tilde{\omega})+\sigma_n\tilde{W}_{t_n},
\end{equation}
where $\mathbb{F}$ denotes the forward operator that propagates the solution of the SPDE in time, $\tilde{W}_{t_n} \sim N(0,\pmb{I}_d)$ is a standard Gaussian random variable, representing a generic model uncertainty, and the noise magnitude is scaled by the coefficient $\sigma_n$. The random variable $\tilde{\omega}$ captures various uncertainties due to unknown physics or simulation errors.   We shall present a brief explanation of the implementation of the operator $\mathbb{F}$ for each numerical experiment.

\subsection{Burgers' equation}

Next, we consider the following 2D Burgers' equation:
\begin{equation}
\label{Burgers2D}
    \begin{array}{l}
        u_t + \left(\dfrac{u^2}{2}\right)_x+\left(\dfrac{u^2}{2}\right)_y = 0, \hspace{3em} (x, y, t) \in (-1, 1)^2 \times (0, T), \vspace{0.1cm} \\
        u_0(x, y) = \dfrac{1}{4}+\dfrac{1}{4}\sin\left(\pi(x+y)\right), \hspace{3em} (x, y) \in (-1, 1)^2,
    \end{array}
\end{equation} 
subject to periodic boundary conditions. It is well-known for this type of equations that even though the initial condition is smooth, the solution can develop sharp corners over time and eventually becomes discontinuous (see, e.g.,~\cite{Shu1988a}). Therefore, we aim to demonstrate the performance of our method for two terminal times: $T=0.2$ when the solution is smooth and $T=0.45$ when the solution starts to exhibit sharp corners. Moreover, since the mass of the solution to \eqref{Burgers2D} is preserved over time (see, e.g.,~\cite{Doan2025a}), we also focus on investigating whether the estimated solutions achieve similar behavior. 

To obtain the forward solver that propagates the PDE model, we employ the so-called Runge-Kutta discontinuous Galerkin method to solve \eqref{Burgers2D} numerically, where $\eqref{Burgers2D}$ is discretized in space by discontinuous Galerkin~\cite{Cockburn1991a} (DG) methods and in time using Strong Stability Preserving Runge-Kutta~\cite{Shu1998a} (SSP-RK) schemes. To handle moving shocks of the numerical solution after a finite time, the numerical solver is combined with the standard TVB (Total Variation Bounded) corrected slope
limiter~\cite{Cockburn2001}. 

Regarding the discretization of the model problem, we keep the mesh size fixed at $h=1/40$ for spatial discretization in the numerical solvers, while using different time-step sizes for the reference solution and for data assimilation. However, as the SSP-RK is an explicit time-stepping scheme, those time steps must be chosen carefully to satisfy the CFL conditions. Specifically, our synthetic ``true state'' is obtained by solving the fully discrete version of \eqref{Burgers2D} on a very fine temporal mesh with mesh size $\Delta{t} = T/800$, while the filtering process is performed on a coarser temporal mesh with $\Delta{t}_{Filter} = T/80$.  For simplicity, we only consider the second-order DG and SSP-RK methods. The discretized version of \eqref{Burgers2D} is also used in the forecast model. To ensure the stability of our estimates, we apply the TVB scheme to the analysis at each filtering step. More details of the implementation can be found on Github ~\url{https://github.com/Toanhuynh997/StateEst_PDEs/tree/main/Burgers2D}.  

For this Burgers' equation example, the uncertainty in the model is incorporated by adding the term $\sigma_n \tilde{W}_{t_n}$ to the numerical solver $\mathbb{F}$, with $\sigma_n = 0.01$. Additionally, uncertainty in the initial condition is introduced by sampling the initial state as $u_0 \sim 2 \cdot N(0, \bf{I}_d)$. We point out that at each filtering time $t_n$,  we add the perturbation term  $\sigma_n \tilde{W}_{t_n}$ to the current forecast and then feed this perturbed state into the solver to advance to the next time step. Consequently, each realization evolves along its own stochastic trajectory rather than the deterministic path of the reference solution. Thus, in our data assimilation procedure we solve an SPDE instead of the deterministic PDE used to generate the reference.

Observations are collected under three scenarios: (i) full observation of the state ($100\%$), (ii) partial observation covering $60\%$ of the state, and (iii) sparse observation with only $10\%$ of the state is observed. In all cases, the observational data of the reference (true) solutions are obtained through the \textbf{\textit{arctangent operator}}, i.e.,
\begin{align*}
    \begin{array}{l}
  Y_t = \arctan(X_t) + \varepsilon_t, \,\, t=1,2,..., N,\,\, \text{with}\, \, \varepsilon_t\sim \mathcal{N}(\vec{0}, 0.1^2I_d).
    \end{array}
    \vspace{-0.5cm}
\end{align*}
Note that the arctangent observation operator is highly nonlinear and is used here as a simplified proxy to reflect key challenges present in real-world measurement processes, such as those arising in radar or satellite observations.
For the $10\%$ observation case, we compare the performance of our method with the local ensemble transform Kalman filter (LETKF), a state-of-the-art approach for data assimilation. Regarding the setting of reversed SDE in the EnSF, the number of ensembles and the time steps are chosen to be $80$ and $300$ for the cases with $100\%$ and $60\%$ observations, while for the case with $10\%$, they are designated to be $100$ and $400$, respectively. 

\vspace{0.5em}

\begin{figure}[h!]
\centering
\begin{minipage}{0.3\textwidth}
\includegraphics[scale=0.25]{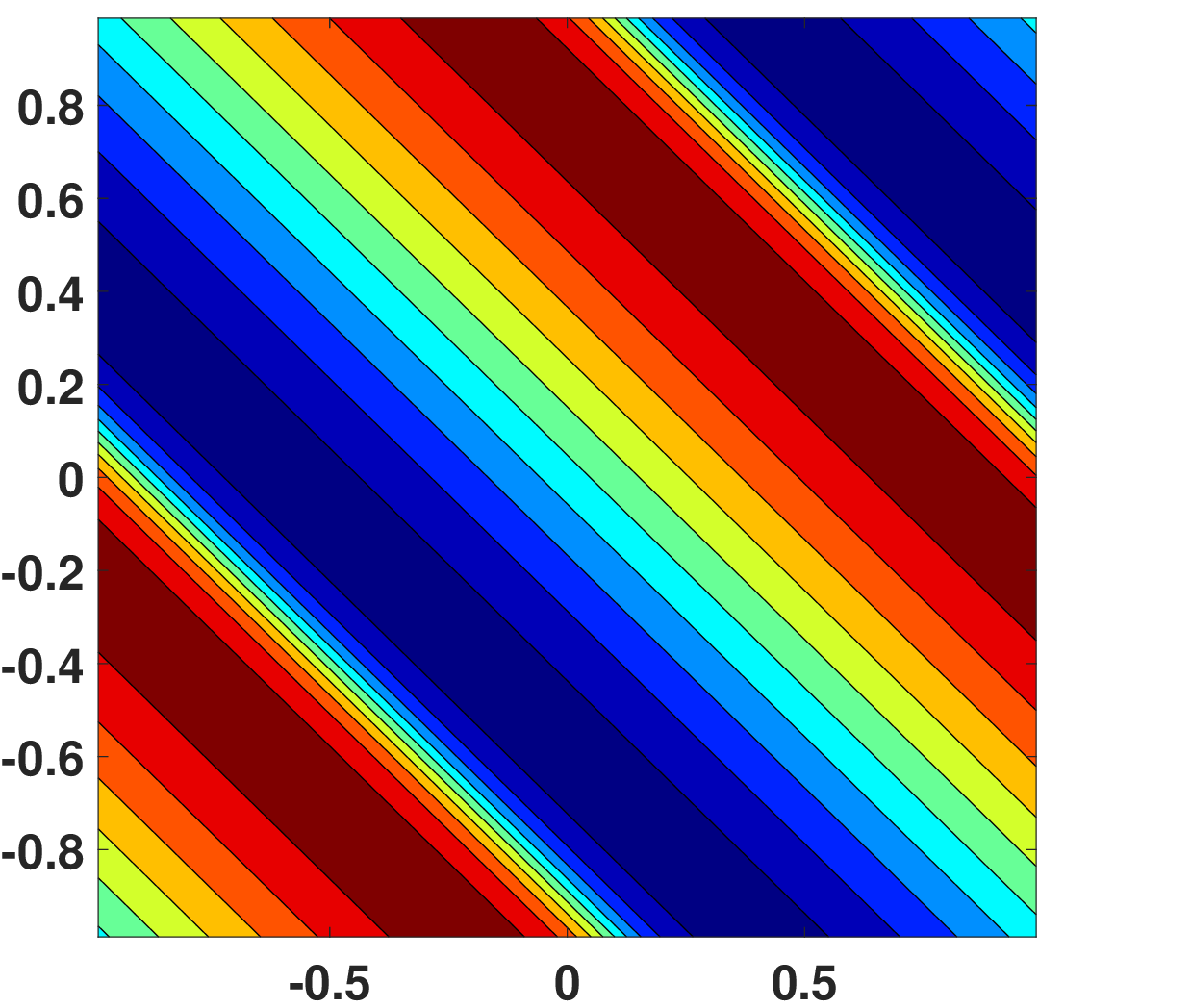}
\end{minipage}%
\begin{minipage}{0.3\textwidth}
\includegraphics[scale=0.25]{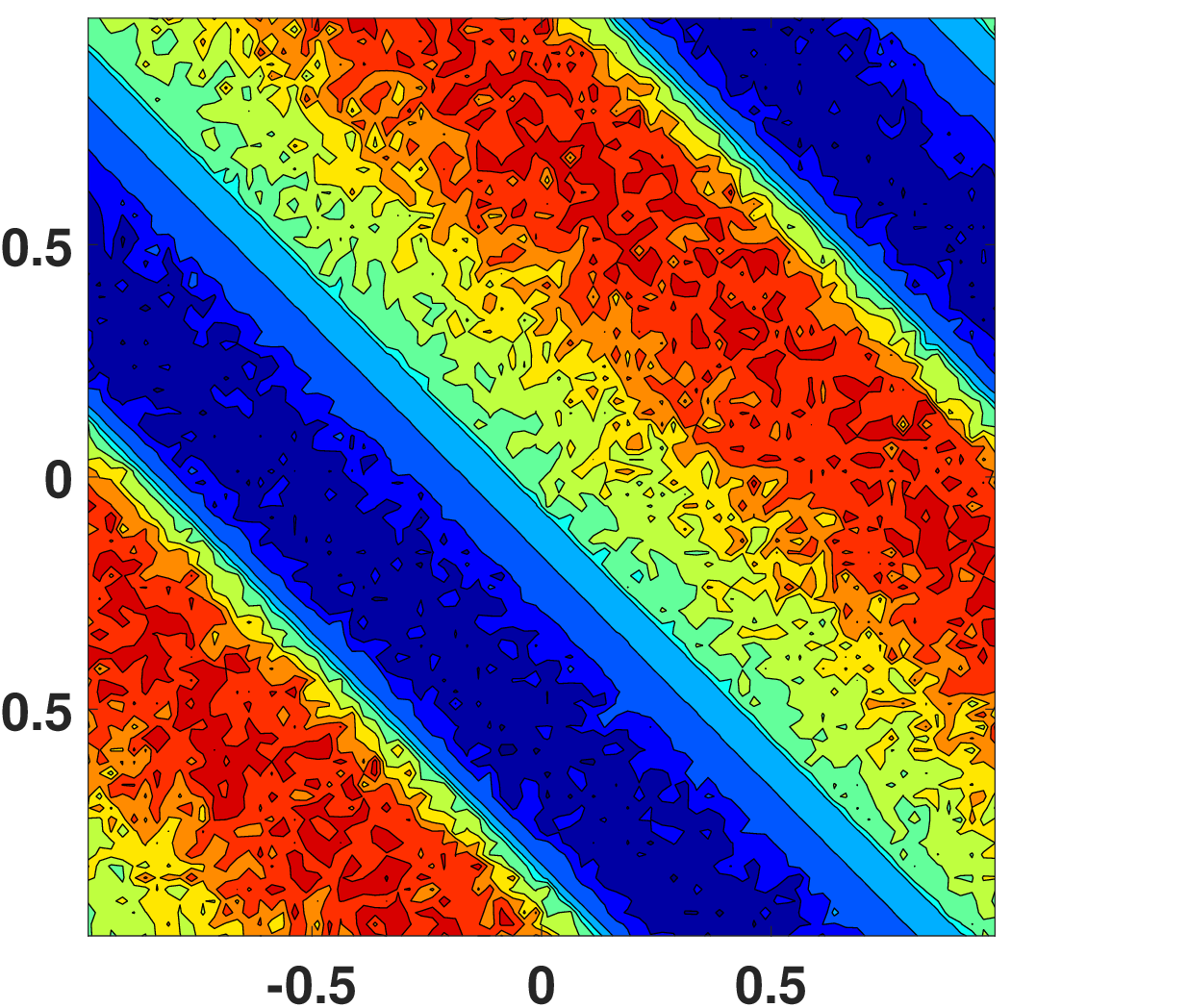} 
\end{minipage}%
\begin{minipage}{0.3\textwidth}
\includegraphics[scale=0.25]{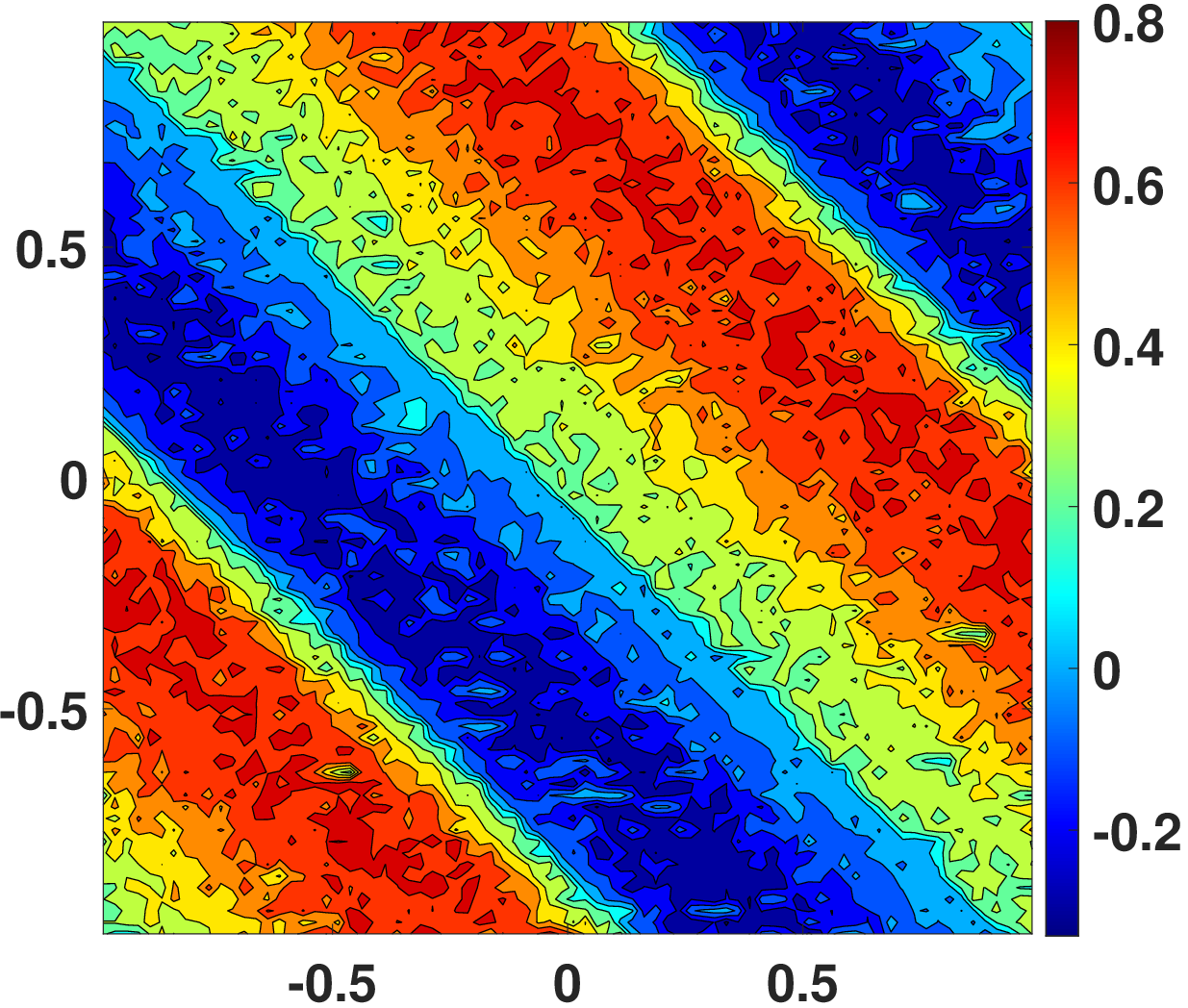} 
\end{minipage}
    \caption{\small Contour maps of the estimated solution to the Burgers' equation at $T = 0.2$. (Left) Reference solution. (Center) EnSF estimate with $100\%$ observations. (Right) EnSF estimate with $60\%$ observations.}
    \label{Contour_T02}
    \vspace{-0.2cm}
\end{figure}

\begin{figure}[h!]
    \centering
     \begin{minipage}{0.42\textwidth}   \hspace{1cm}\includegraphics[scale=0.28]{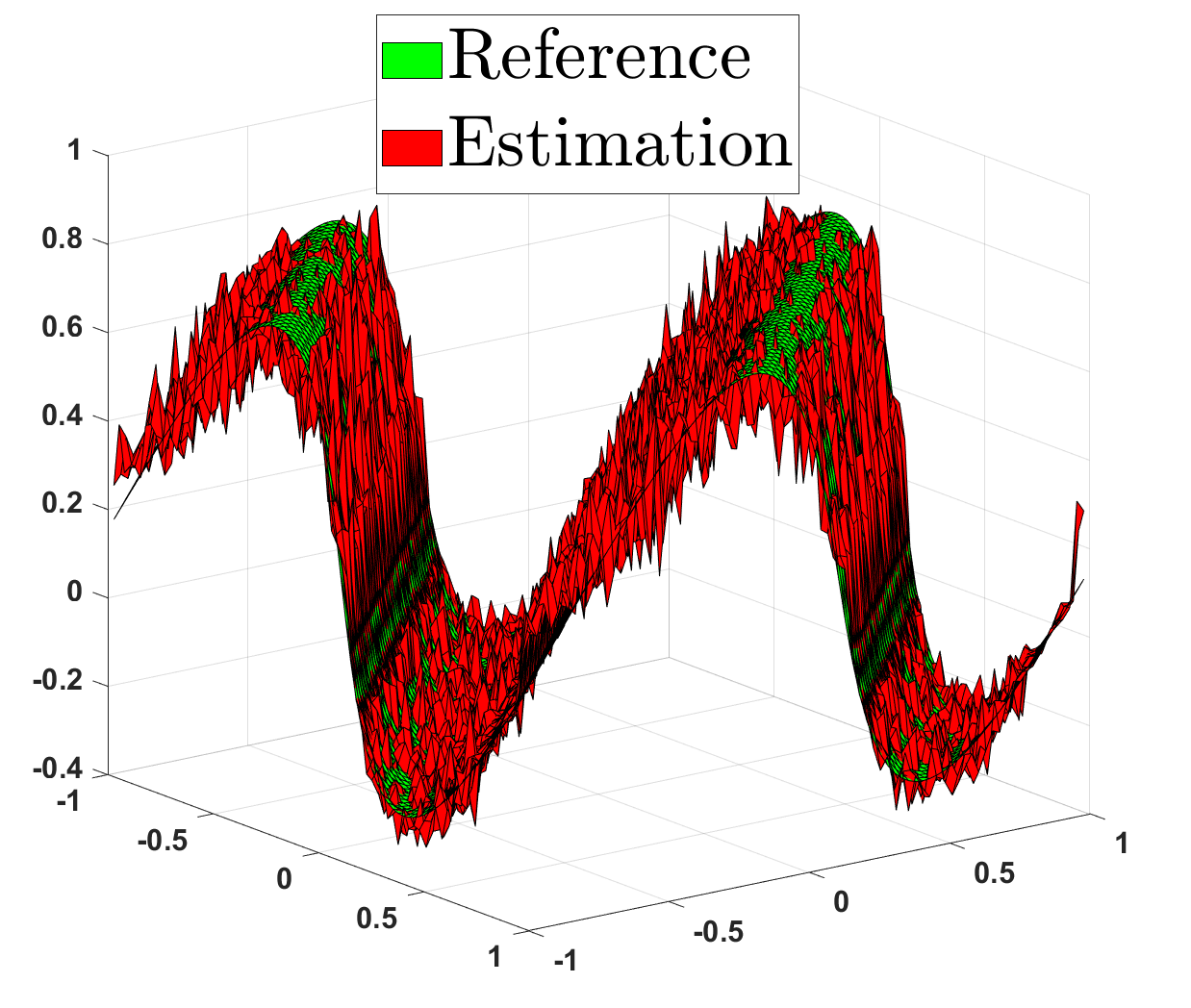}
     \end{minipage}  %
 \begin{minipage}{0.42\textwidth}
      \includegraphics[scale=0.28]{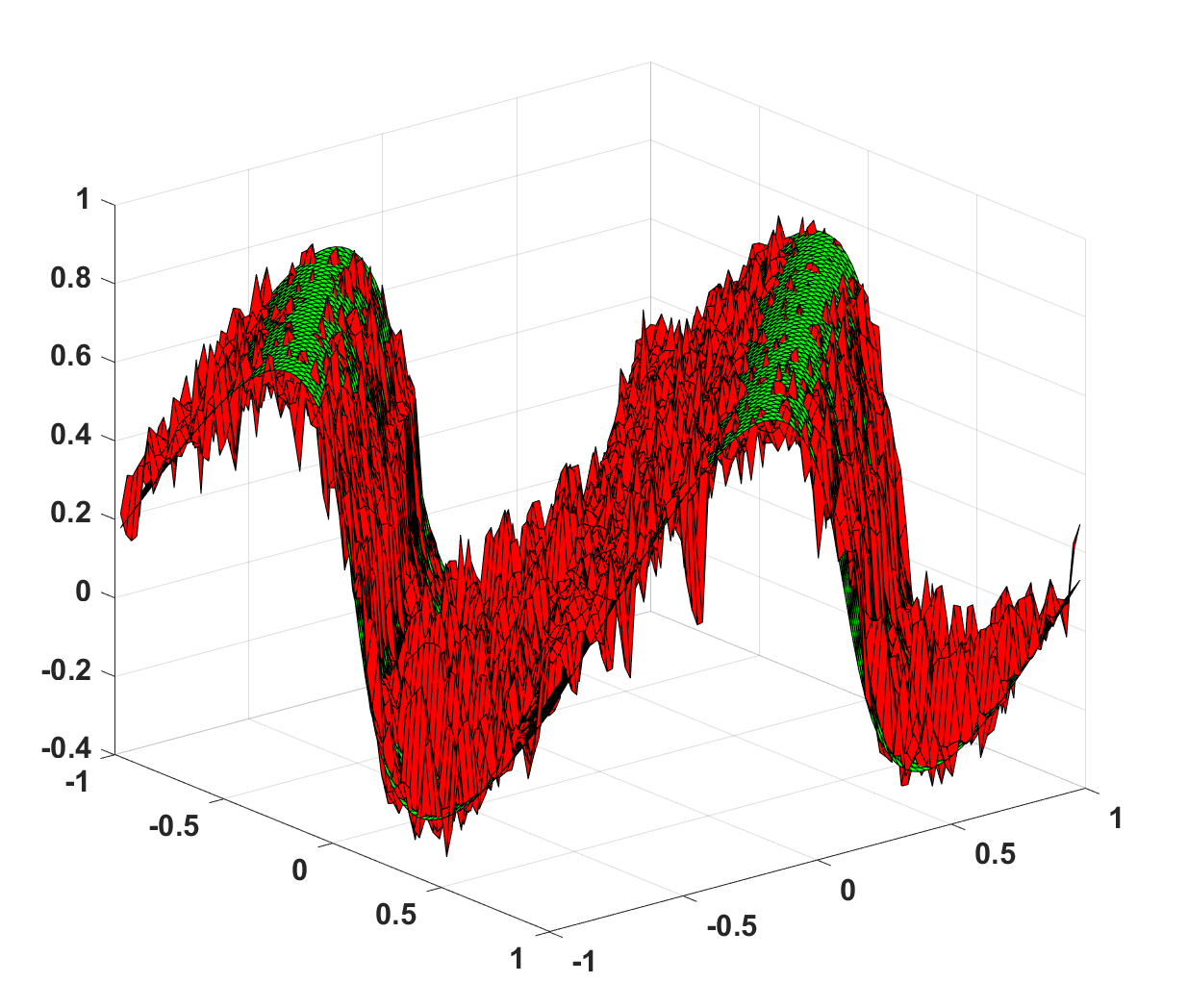}
 \end{minipage}
 \caption{\small 3D view of the estimated solution for the Burgers' equation at $T = 0.2$.  (Left) Estimate with $100\%$ observations. (Right) Estimate with $60\%$ observations.}
 \label{3DView_T02}
 \vspace{-0.2cm}
\end{figure}

\begin{figure}[h!]
\vspace{-0.25cm}
\centering
\begin{minipage}{0.3\textwidth}
\includegraphics[scale=0.25]{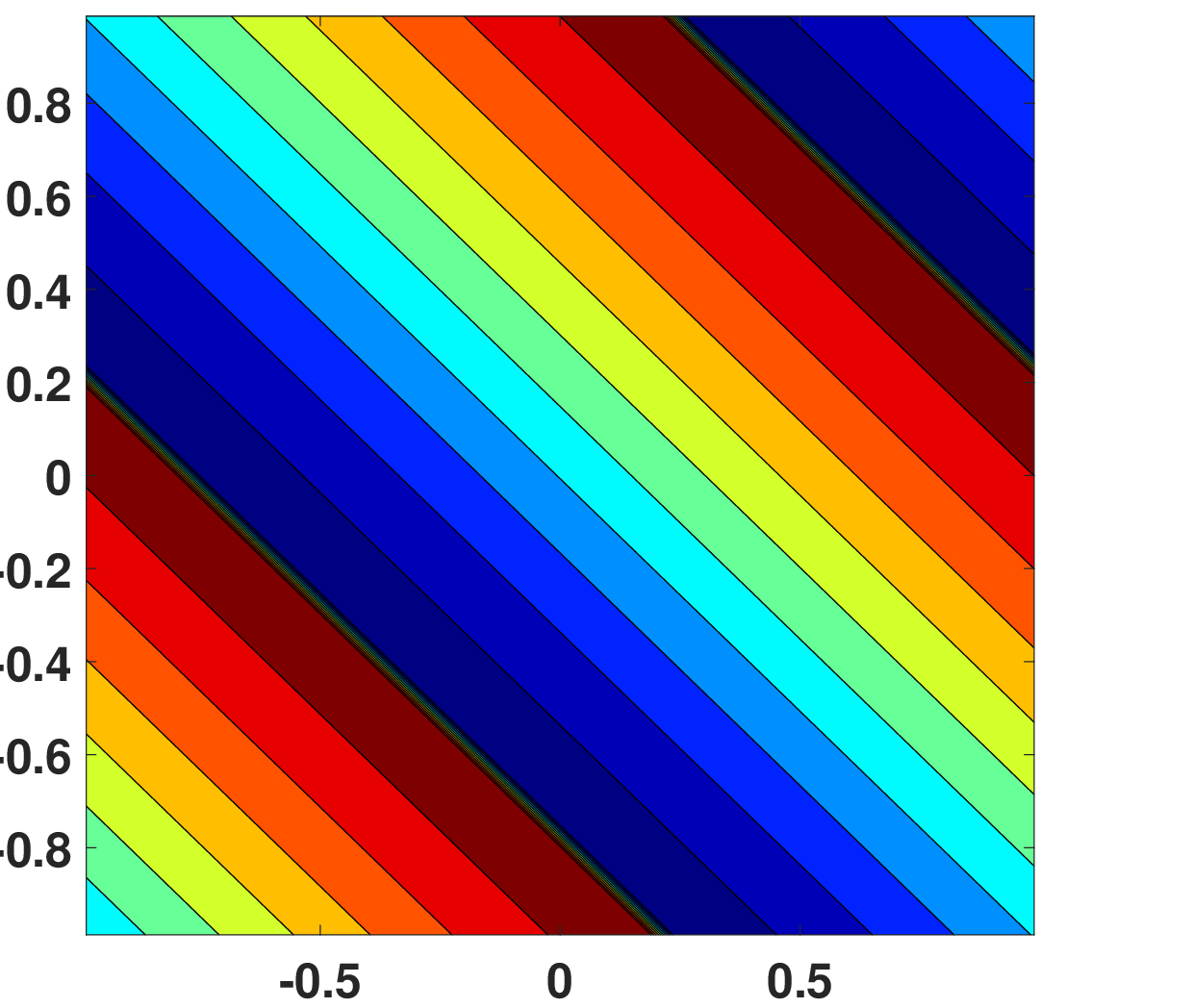}
\end{minipage}%
\begin{minipage}{0.3\textwidth}
    \includegraphics[scale=0.25]{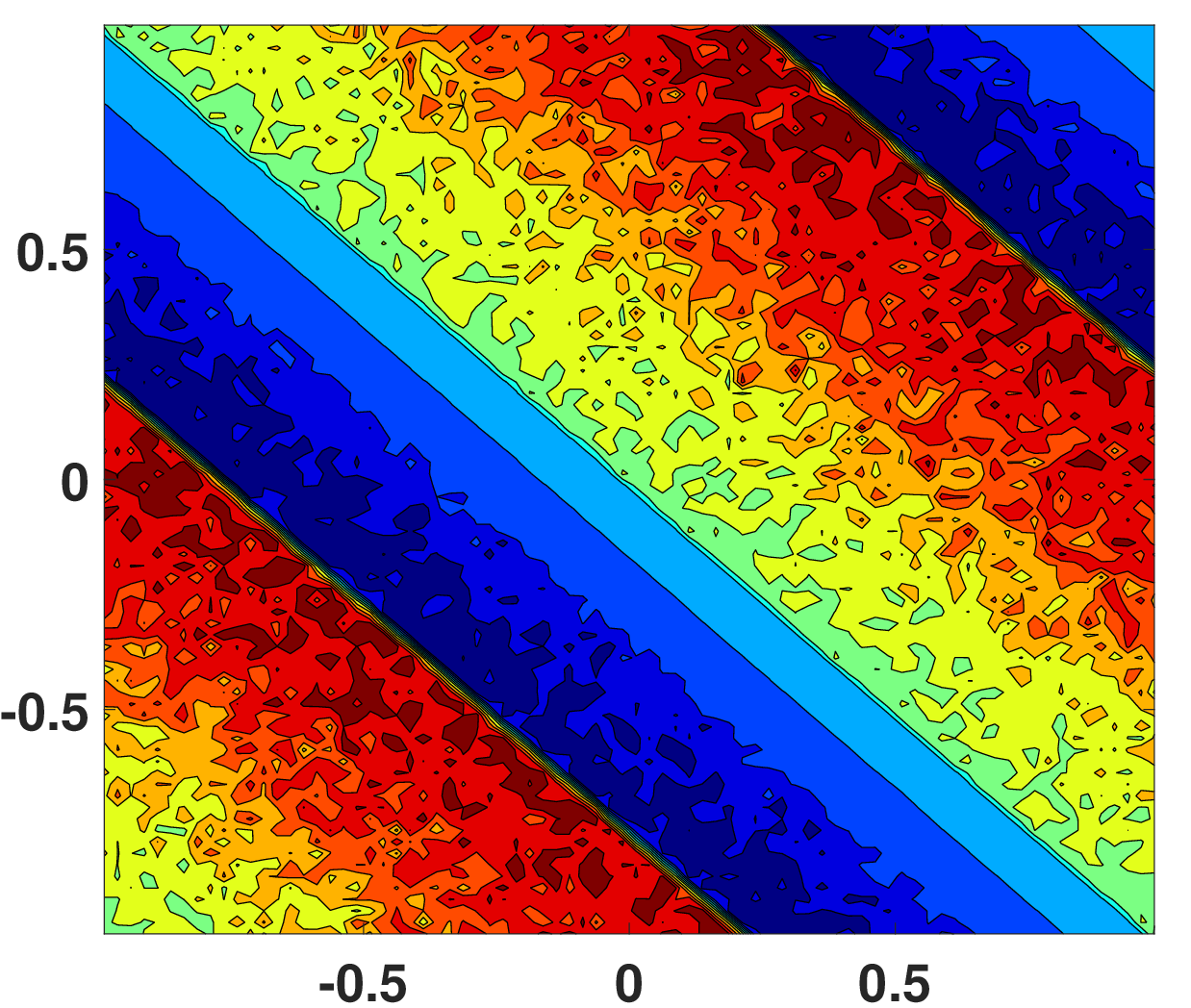} 
\end{minipage}%
\begin{minipage}{0.3\textwidth}
\includegraphics[scale=0.25]{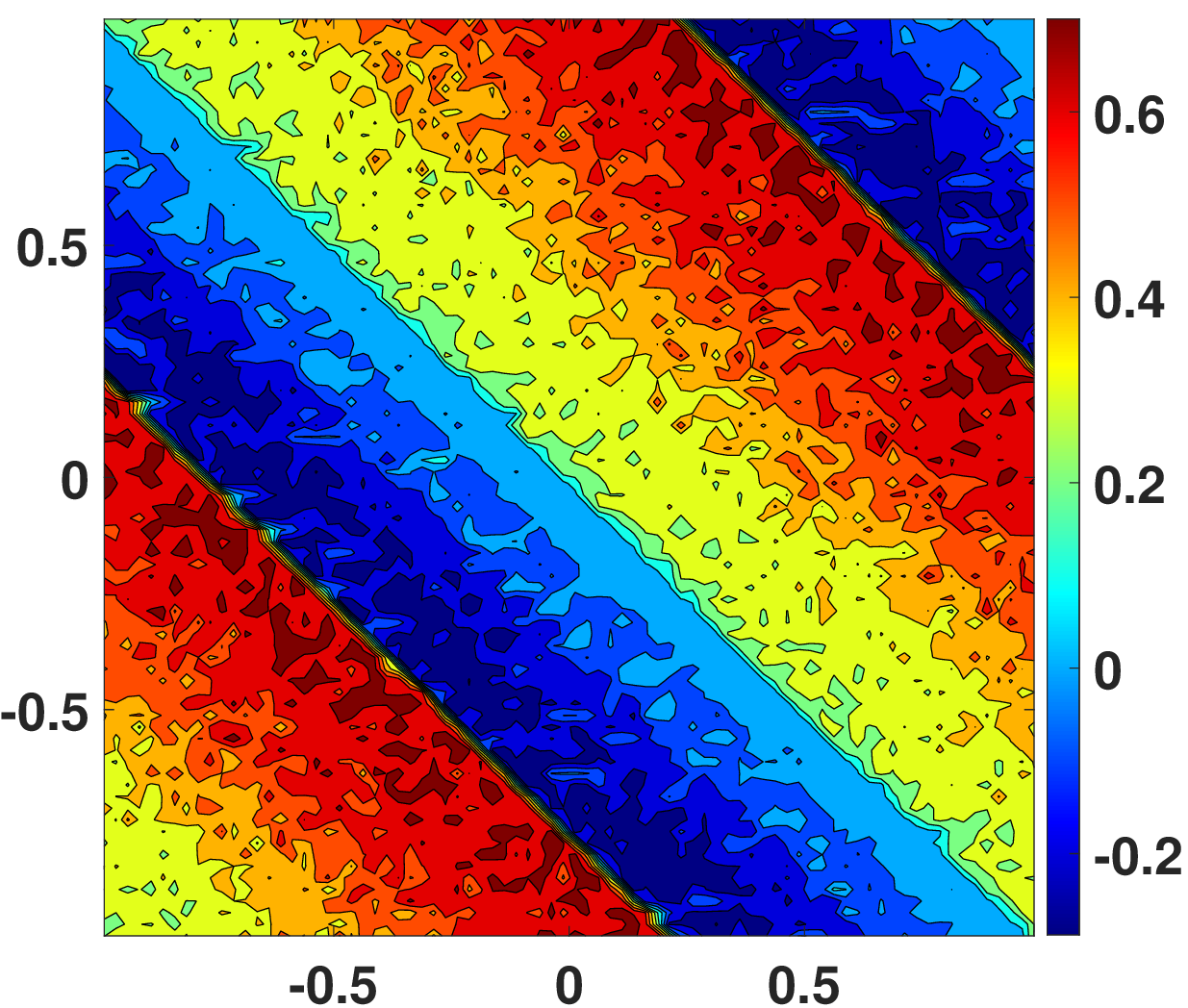} 
\end{minipage}
    \caption{\small Contour maps of the estimated solution to the Burgers' equation at $T = 0.45$. (Left) Reference solution. (Center) EnSF estimate with $100\%$ observations. (Right) EnSF estimate with $60\%$ observations.}
     \label{Contour_T045}
     \vspace{-0.2cm}
\end{figure}

\begin{figure}[h!]
 \centering
 \begin{minipage}{0.42\textwidth}    \hspace{0.5cm}\includegraphics[scale=0.28]{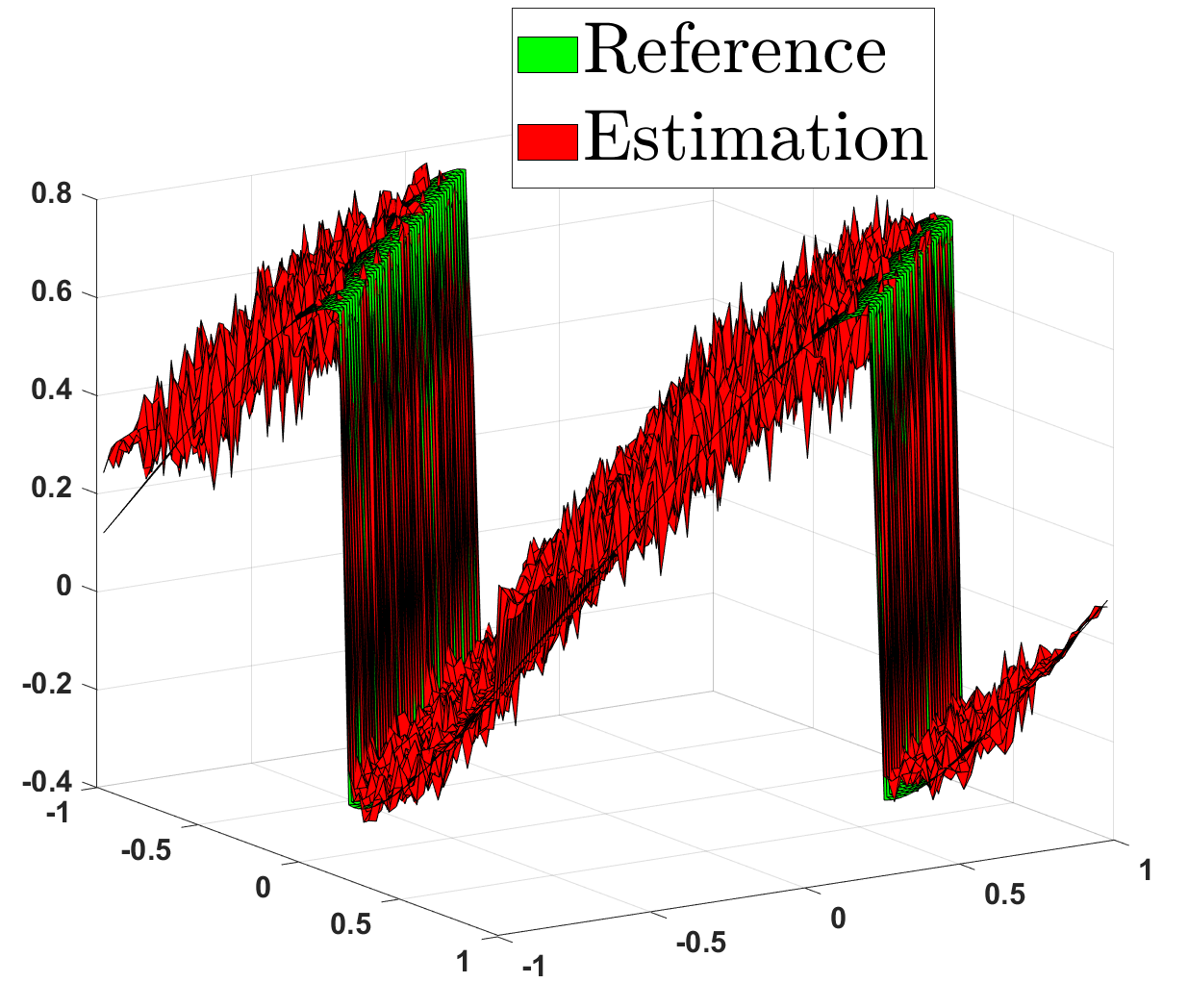}
 \end{minipage}  %
 \begin{minipage}{0.42\textwidth}
      \includegraphics[scale=0.28]{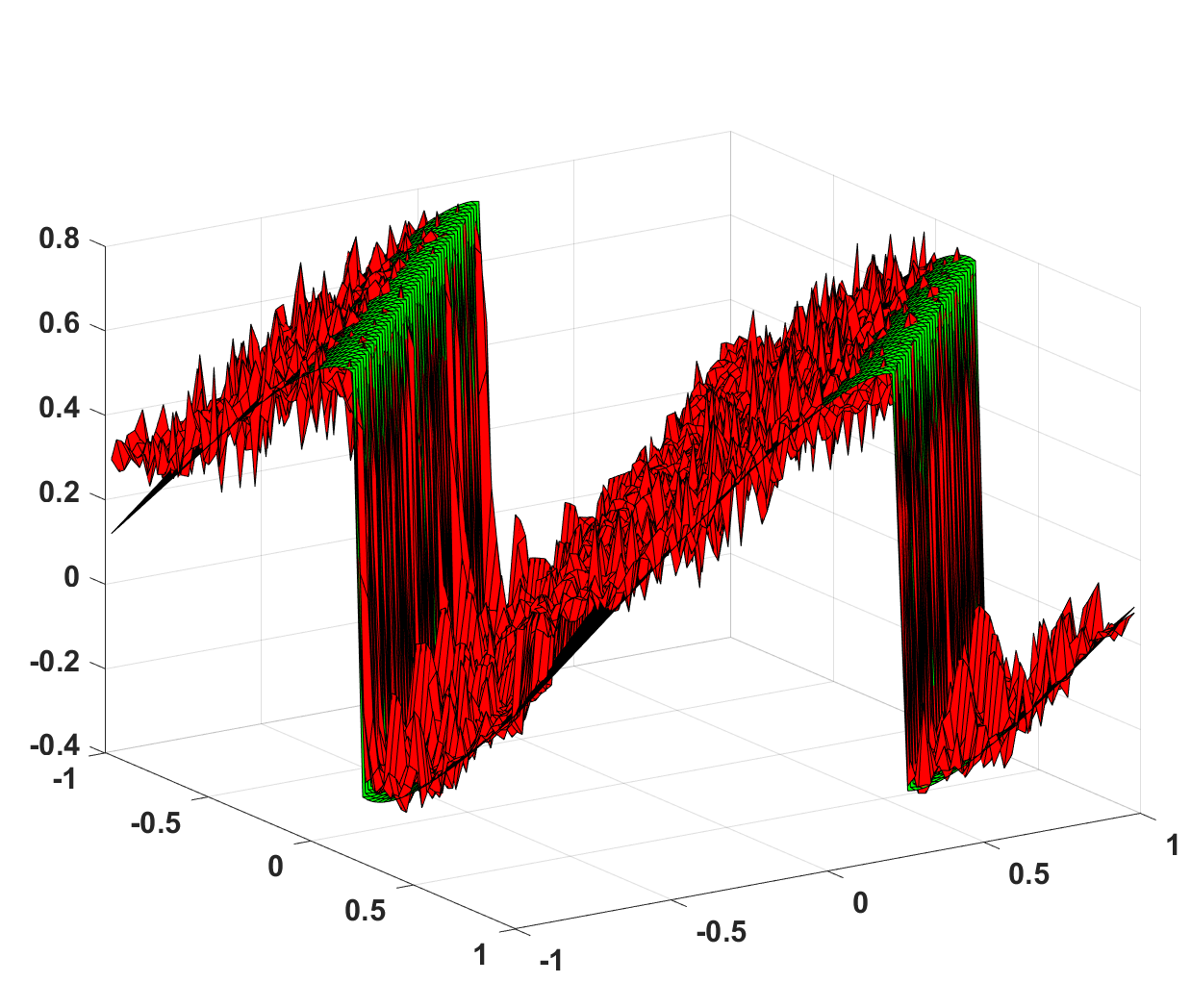}
 \end{minipage}
 \caption{\small 3D view of the estimated solution for the Burgers' equation at $T = 0.45$.  (Left) Estimate with $100\%$ observations. (Right) Estimate with $60\%$ observations.}
 \label{3DView_T045}
 \vspace{-0.2cm}
\end{figure}

We first evaluate the performance of the EnSF in cases where the solution of the Burgers' equation is fully observed ($100\%$) and partially observed ($60\%$). The contour plots of the EnSF-estimated solutions at $T=0.2$ and $T=0.45$ are presented in Figure~\ref{Contour_T02} and Figure~\ref{Contour_T045}, respectively. To provide a cleared view of the performance of the solution estimation, in Figure~\ref{3DView_T02} and Figure~\ref{3DView_T045} we display the 3D visualizations of the reference (true) solution alongside the corresponding EnSF estimates.

% To illustrate the temporal evolution of the solution estimates, we present contour plots of the EnSF-estimated solution at time $T = 0.45$ in Figure~\ref{Contour_T045}. Corresponding 3D visualizations are provided in Figure~\ref{3DView_T045} to offer a clearer view of the estimation performance. 

%It can be seen that the estimated solutions by the EnSF capture well the shapes of the reference solutions. However, at time $T=0.45$, it is more challenging for the estimated solution to recover the region near the sharp corners.

Figures~\ref{Contour_T02} - \ref{3DView_T045} demonstrate that the EnSF is capable of providing highly accurate solution estimates. When only $60\%$ of the solution is observed, the estimation becomes less accurate due to the reduced amount of observational data.

\begin{figure}[h!]
    \centering
     \begin{minipage}{0.42\textwidth}  \hspace{1cm}\includegraphics[scale=0.28]{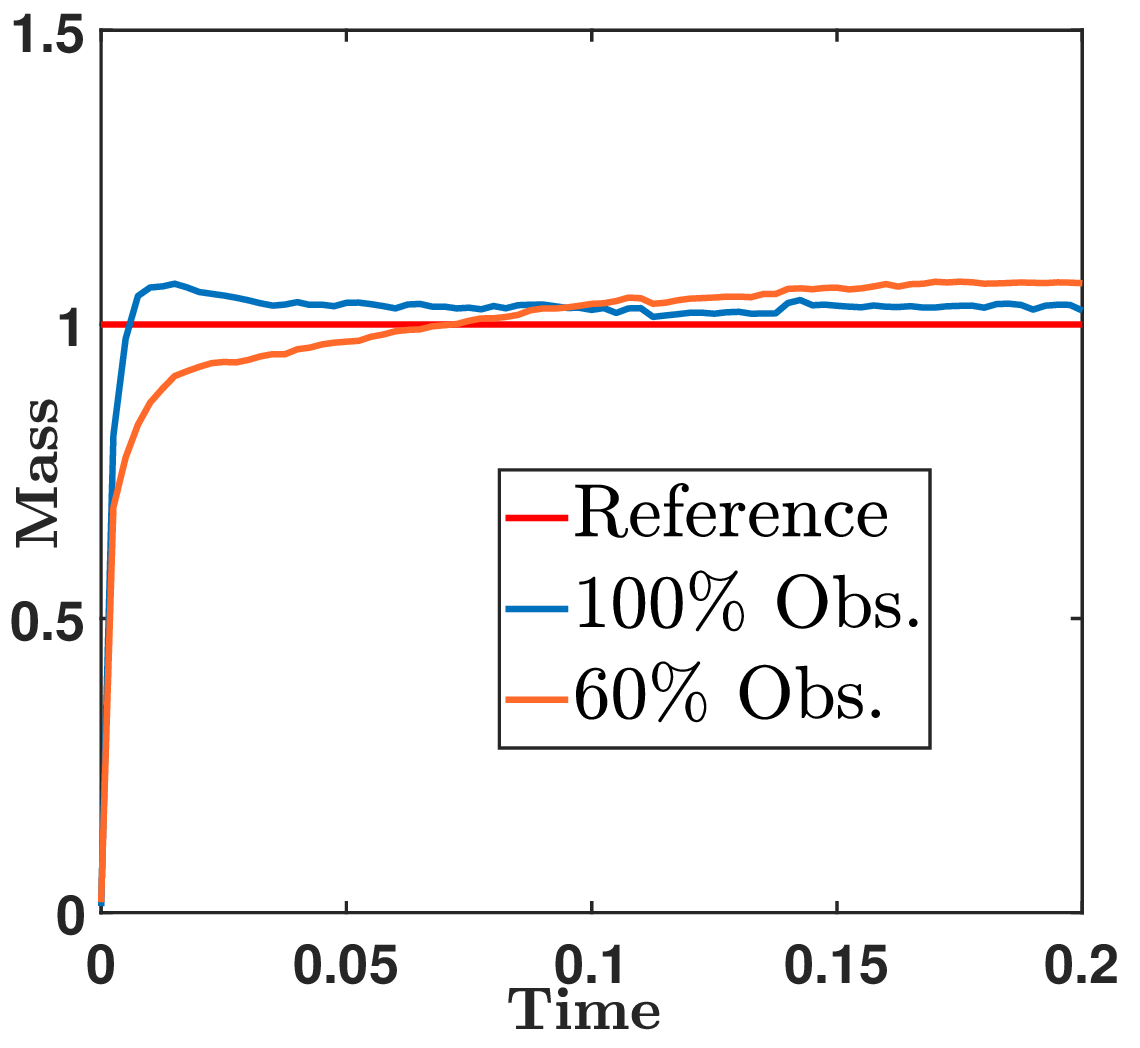}
     \end{minipage}  %
 \begin{minipage}{0.42\textwidth}
      \includegraphics[scale=0.28]{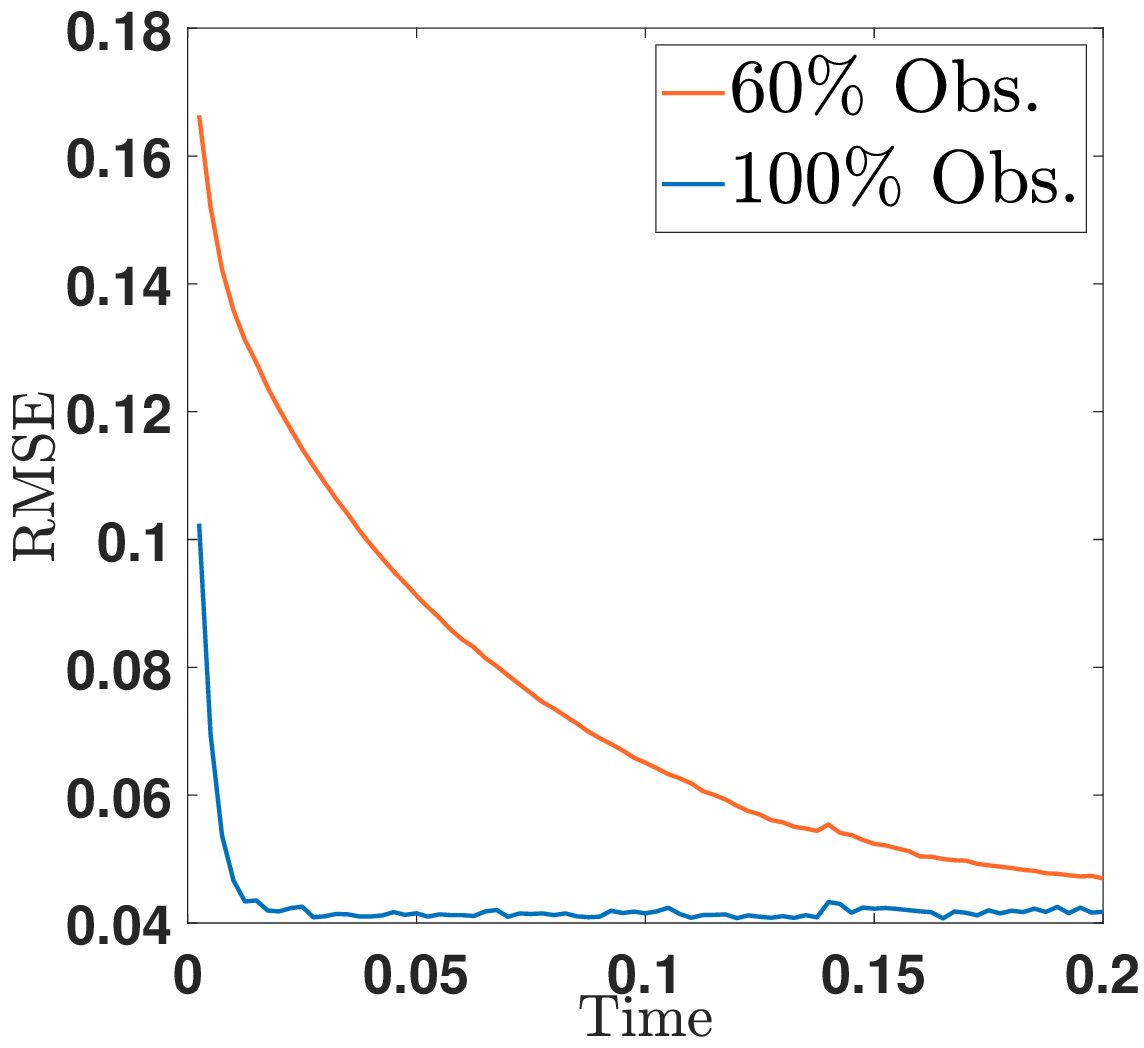}
 \end{minipage}
 \caption{\small (Left) Evolution of Mass with $T=0.2$. (Right) RMSE for state estimations with $T=0.2$.
 }
 \label{Mass_RMSE_T02}
 \vspace{-0.2cm}
\end{figure}

\begin{figure}[h!]
% \vspace{-0.8cm}
    \centering
 \begin{minipage}{0.42\textwidth}    \hspace{1cm}\includegraphics[scale=0.28]{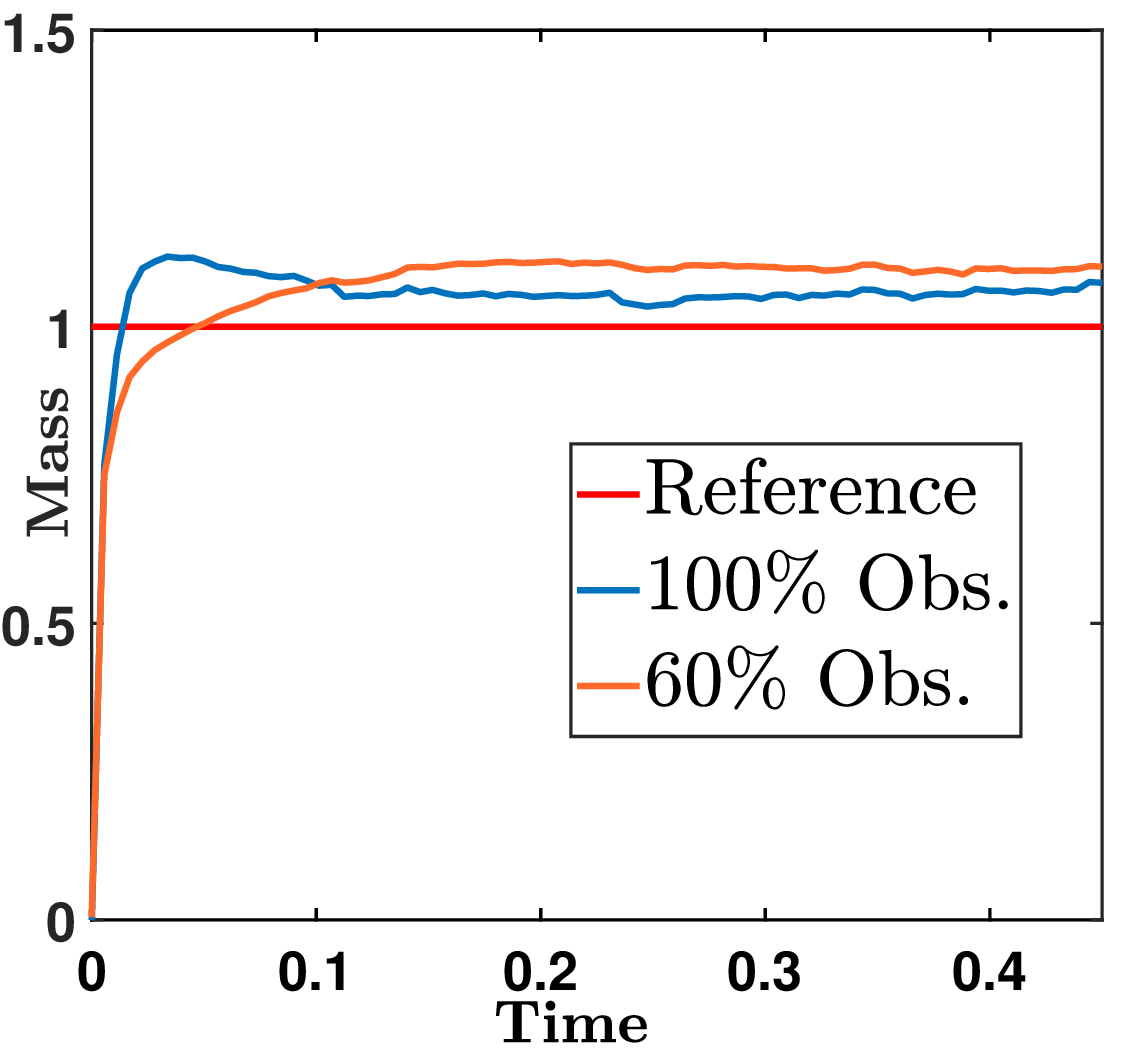}
 \end{minipage}  %
 \begin{minipage}{0.42\textwidth}
      \includegraphics[scale=0.28]{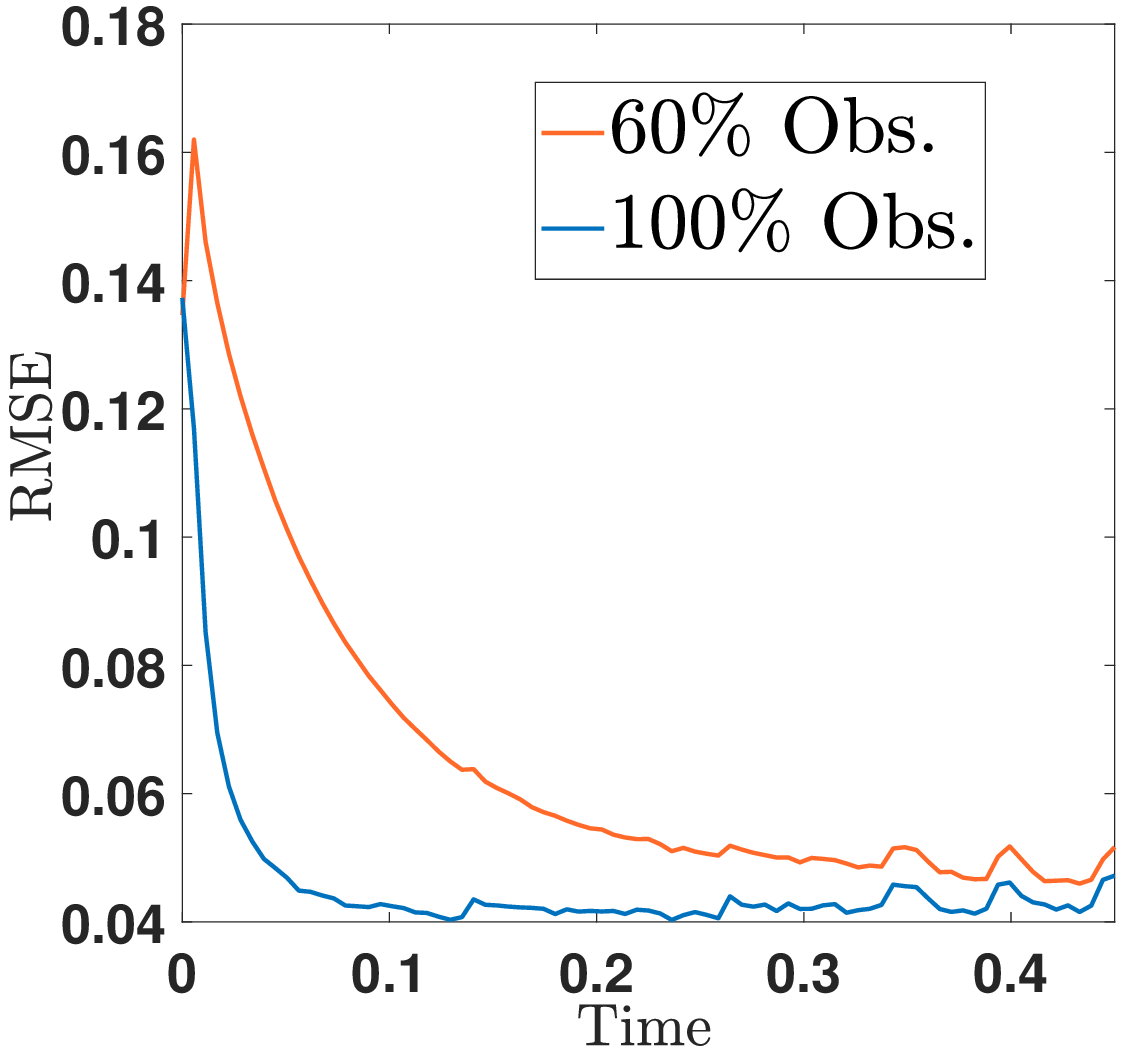}
 \end{minipage}
 \caption{\small (Left) Evolution of Mass with $T=0.45$. (Right) RMSE for state estimations with $T=0.45$.}
 \label{Mass_RMSE_T045}
\end{figure}

Finally, the evolution of the corresponding masses and the Root Mean Square Errors (RMSEs) for solution estimation over time are displayed in Figure~\ref{Mass_RMSE_T02} (for a total time of $T = 0.2$) and Figure~\ref{Mass_RMSE_T045} (for $T = 0.45$). Although the initial mass does not match the reference solution due to discrepancies in the guessed initial condition, the mass conservation property is quickly recovered. As for the RMSEs, both cases show a quick decrease over time. However, minor fluctuations appear in the RMSE curves after $t = 0.3$ in the $T=0.45$ case, which is expected, as this is when sharp corners begin to develop in the solution.

\vspace{0.5em}
To make the problem more challenging, we consider a scenario in which only $10\%$ of the reference solution is observed. In this case, we apply the EnSF-based data assimilation method to estimate the solution under uncertainty and sparse observational coverage. To address the issue of sparse observation, we incorporate a PDE-based inpainting technique into the EnSF framework. Specifically, we employ the biharmonic (Bih) inpainting method to reconstruct unobserved regions of the state \cite{Bao2025a}.

\begin{figure}[h!]
\vspace{-0.2cm}
\centering
\begin{minipage}{0.3\textwidth}
\includegraphics[scale=0.25]{Toanfigures/2DBurger/Burger2D_Ref_T02_contour.eps}
\end{minipage}%
\begin{minipage}{0.3\textwidth}
\includegraphics[scale=0.25]{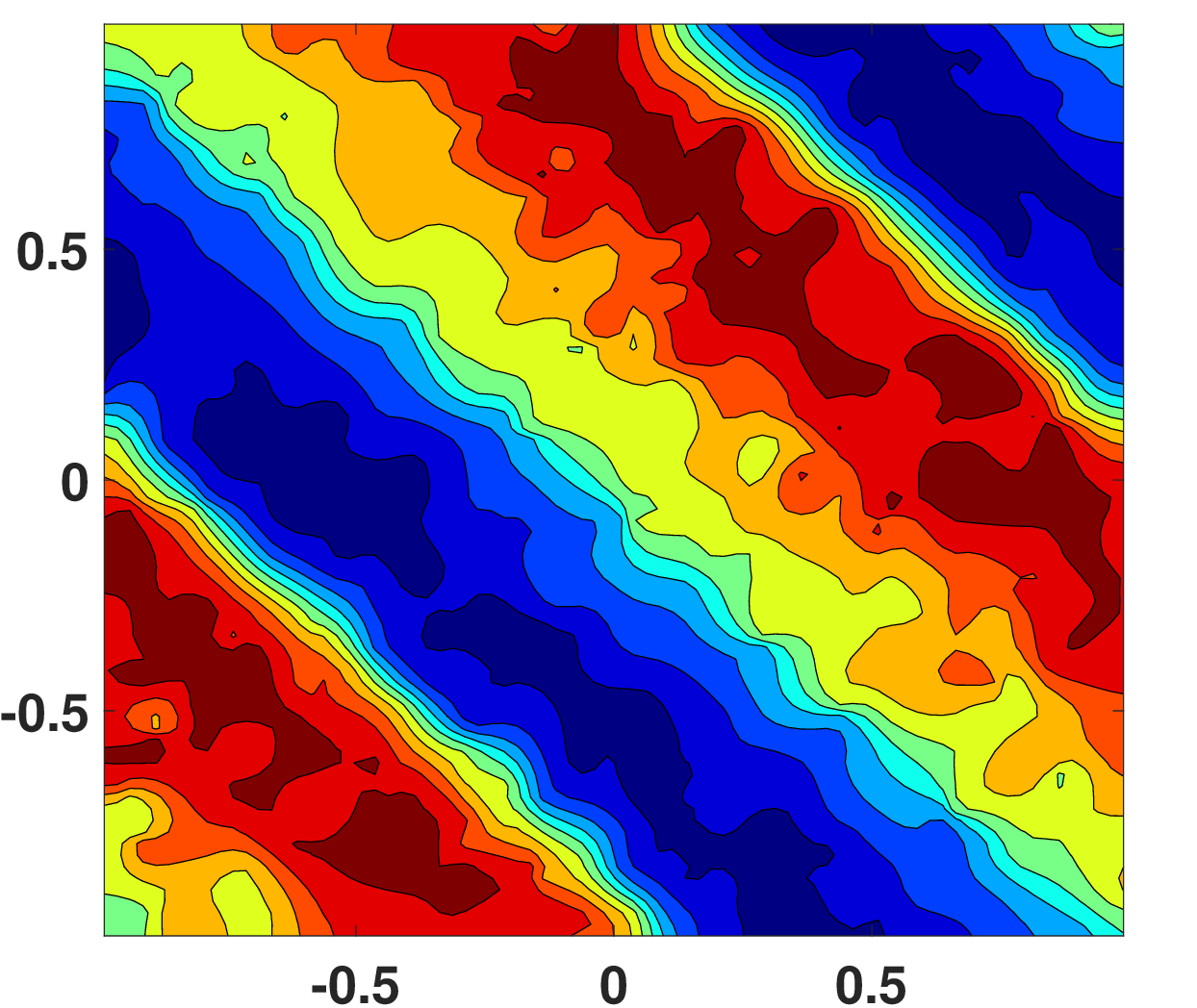} \end{minipage}%
\begin{minipage}{0.3\textwidth}
\includegraphics[scale=0.25]{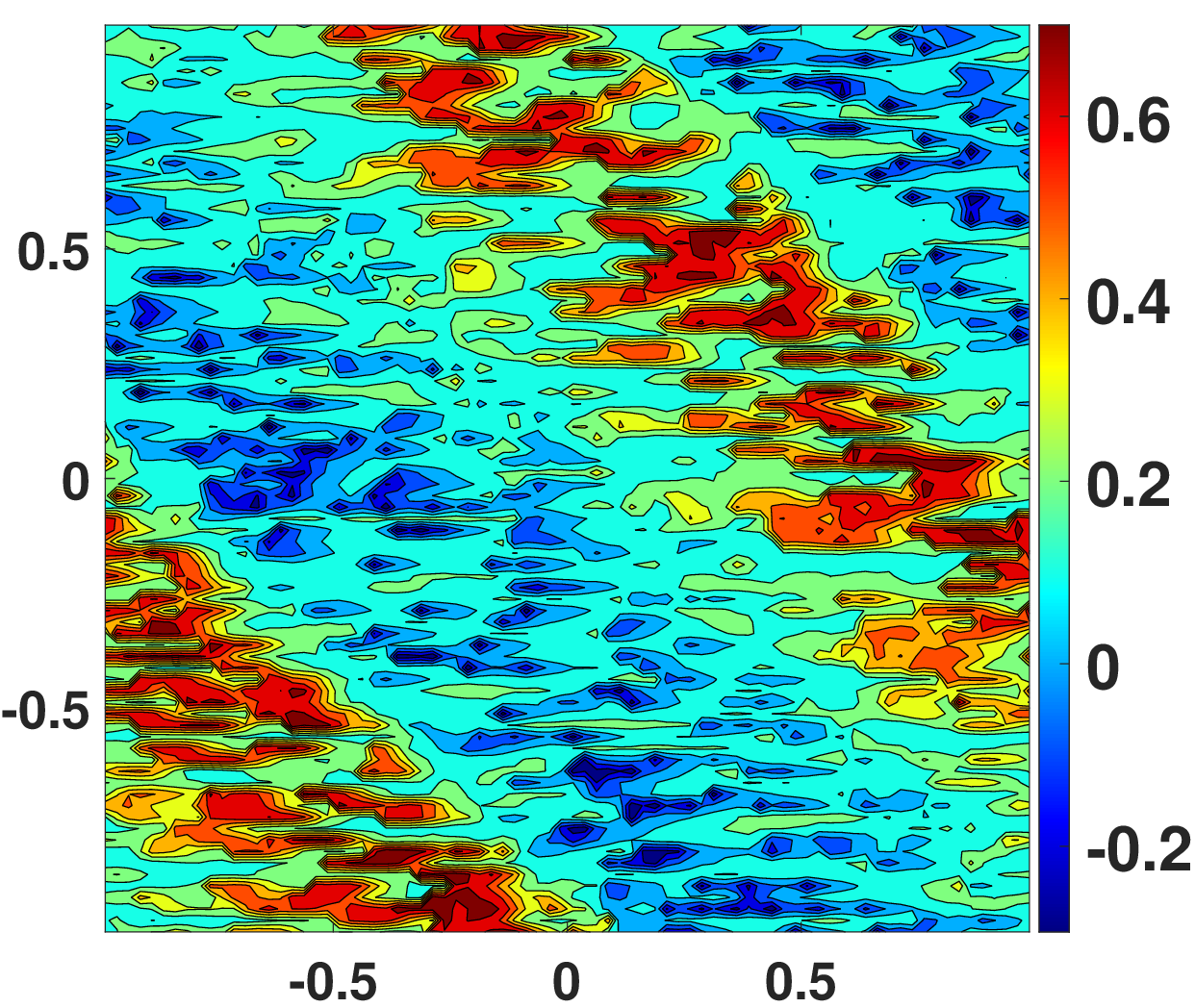}     
\end{minipage}
    \caption{\small Contour maps of the estimated solution to the Burgers' equation at $T = 0.2$ with $10\%$ observations. (First) Reference solution. (Second) EnSF estimate with biharmonic inpainting. (Third) LETKF estimate.}
    \label{Contour_T02_Inpainting}
    \vspace{-0.2cm}
\end{figure}

\begin{figure}[h!]
% \vspace{-0.8cm}
    \centering
 \begin{minipage}{0.42\textwidth}
 \includegraphics[scale=0.28]{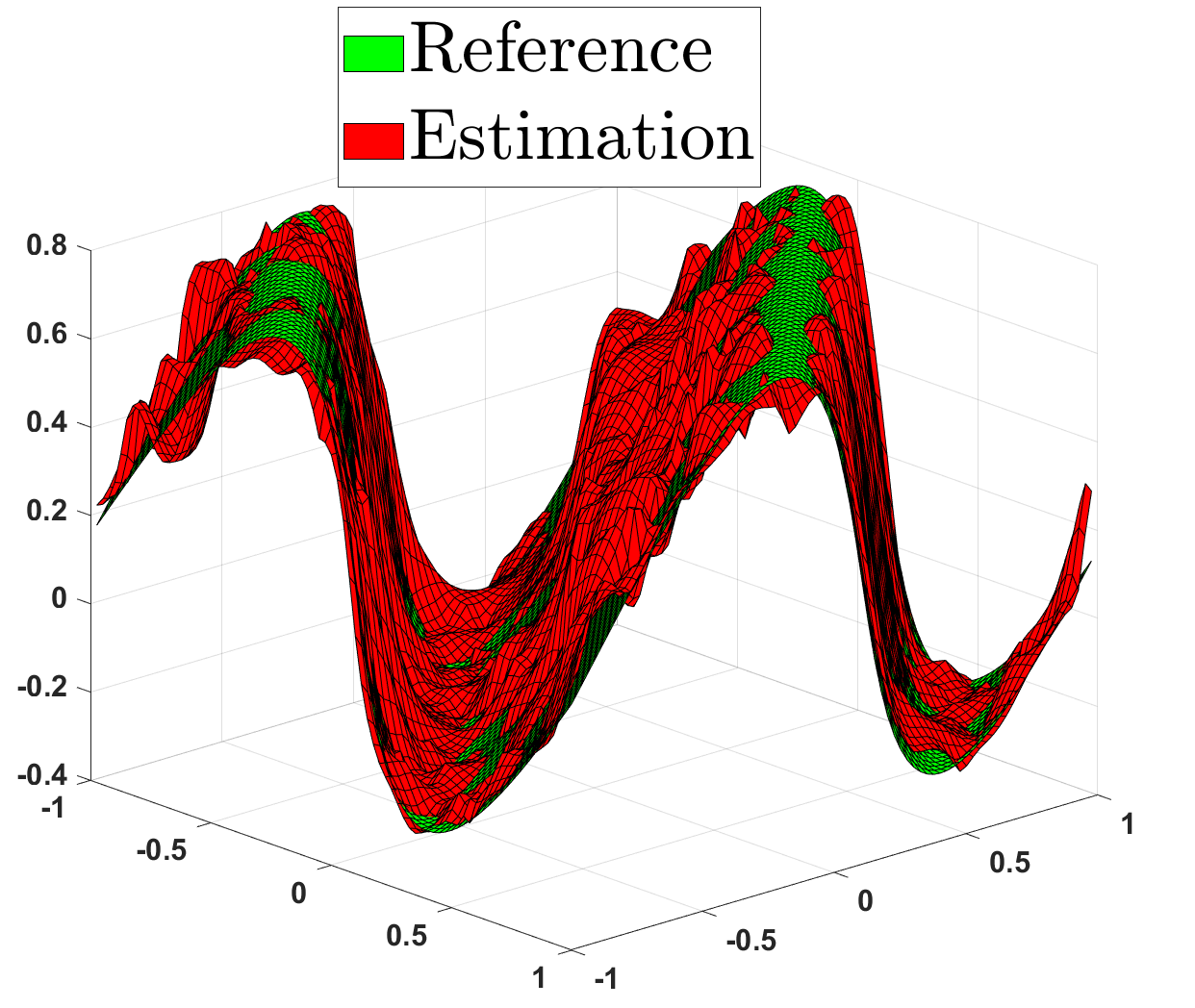}
 \end{minipage} %
 \begin{minipage}{0.42\textwidth}
      \includegraphics[scale=0.28]{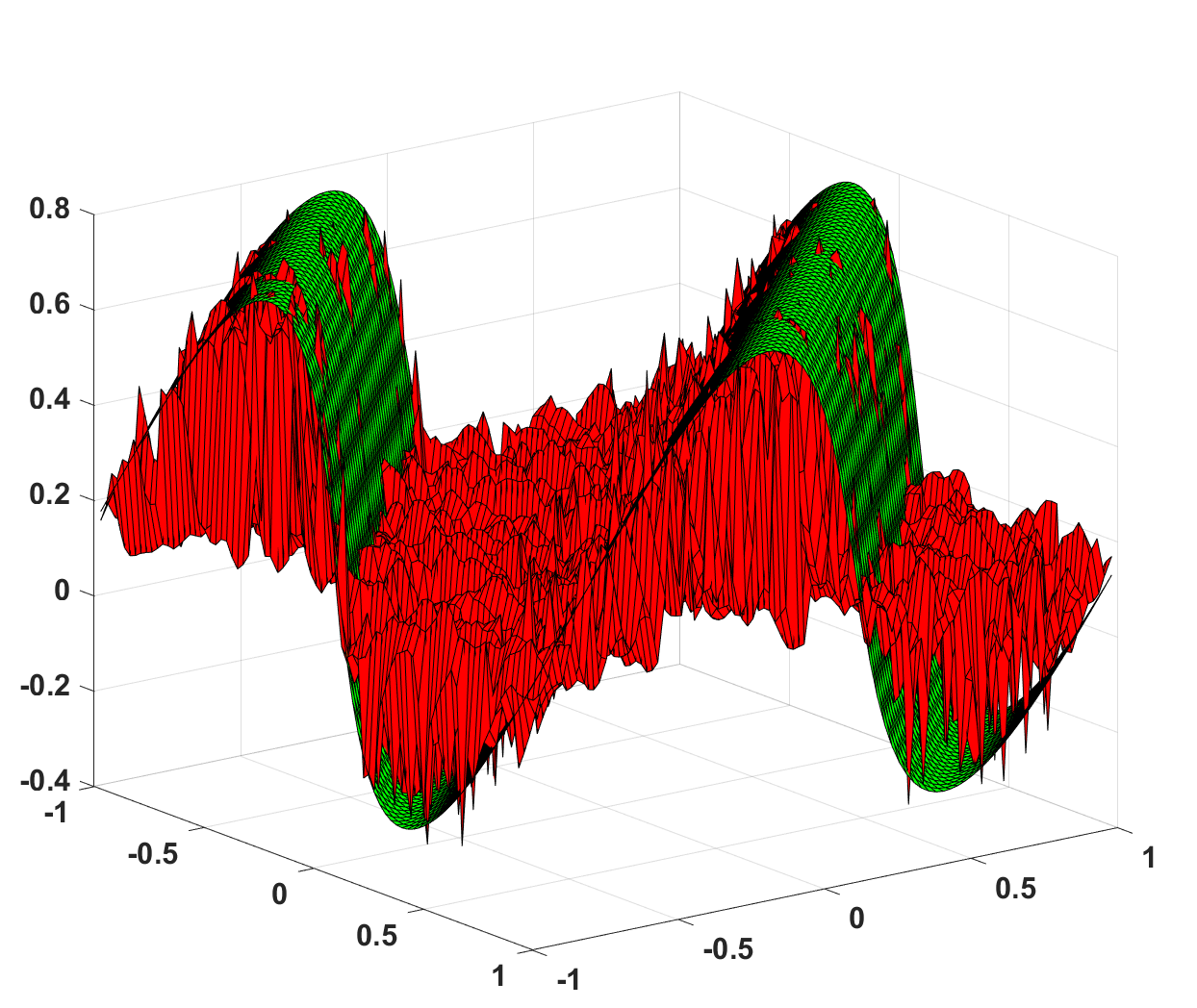}
 \end{minipage}%
 \caption{\small 3D view of the estimated solution for the Burgers' equation at $T = 0.2$ with $10\%$ observations.  (Left) EnSF estimate with biharmonic inpainting. (Right) LETKF estimate.}
\label{3DContour_T02_Inpainting}
\vspace{-0.2cm}
\end{figure}
\begin{figure}[h!]
    \centering
     \begin{minipage}{0.42\textwidth}   \hspace{1cm}\includegraphics[scale=0.28]{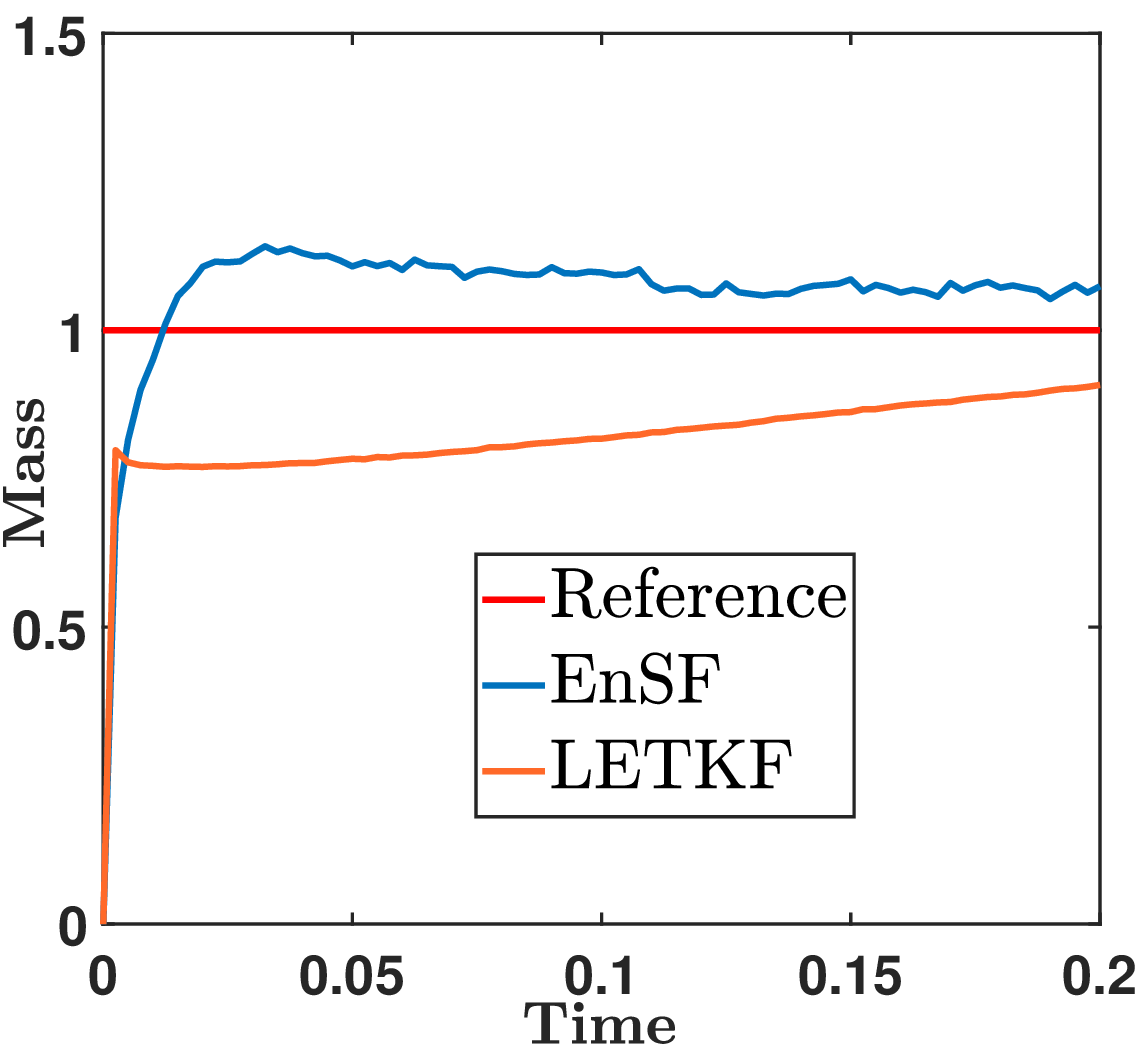}
     \end{minipage}  %
 \begin{minipage}{0.42\textwidth}
      \includegraphics[scale=0.28]{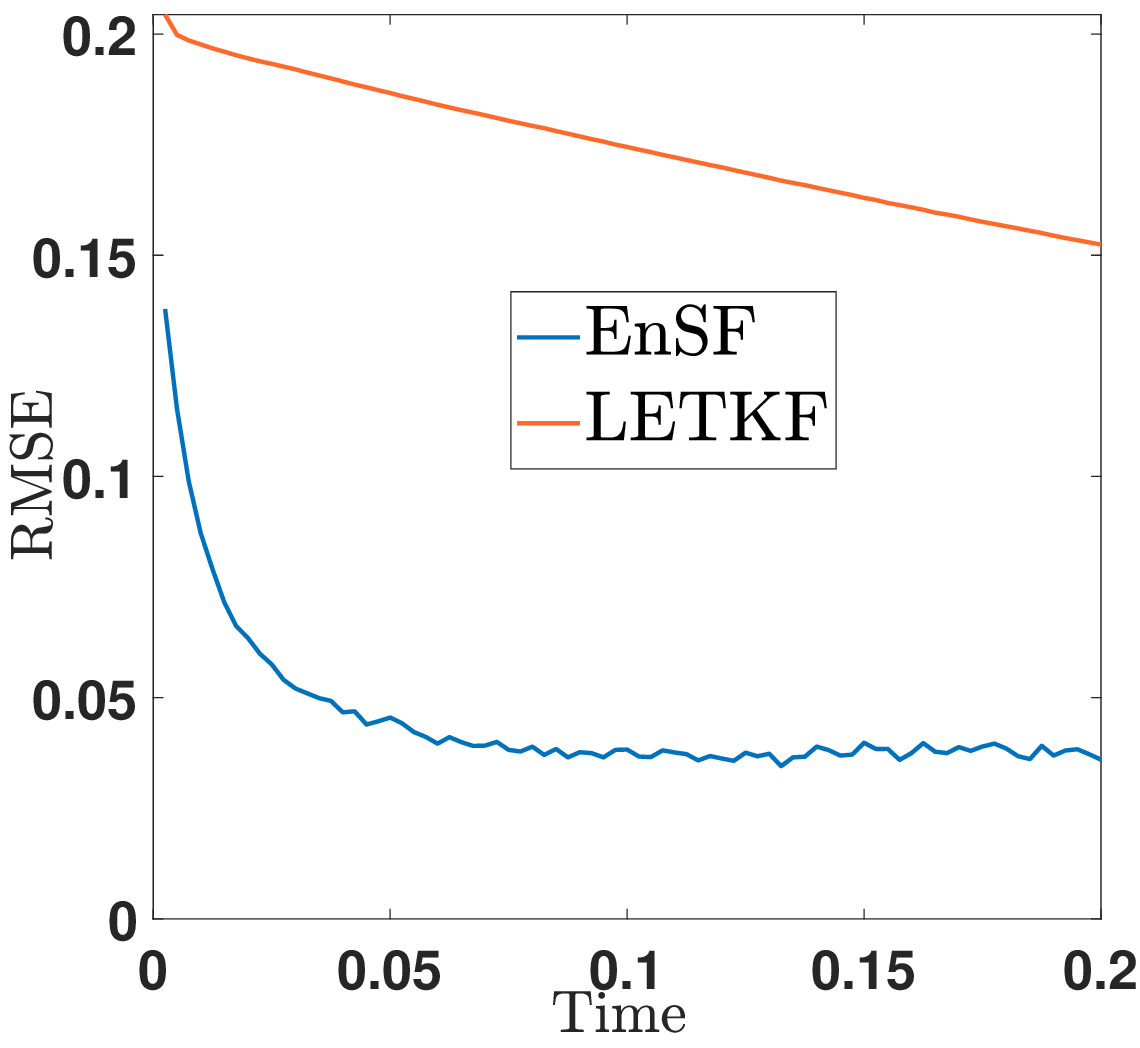}
 \end{minipage}
 \caption{\small (Left) Evolution of Mass with $T=0.2$ with $10\%$ observations. (Right) RMSE for state estimations with $T=0.2$ with $10\%$ observations.}
\label{Mass_RMSE_T02_Inpainting}
 \vspace{-0.2cm}
\end{figure}   
\begin{figure}[h!]
\centering
\begin{minipage}{0.3\textwidth}
\includegraphics[scale=0.25]{Toanfigures/2DBurger/Burger2D_Ref_T045_contour.eps}
\end{minipage}%
\begin{minipage}{0.3\textwidth}
\includegraphics[scale=0.25]{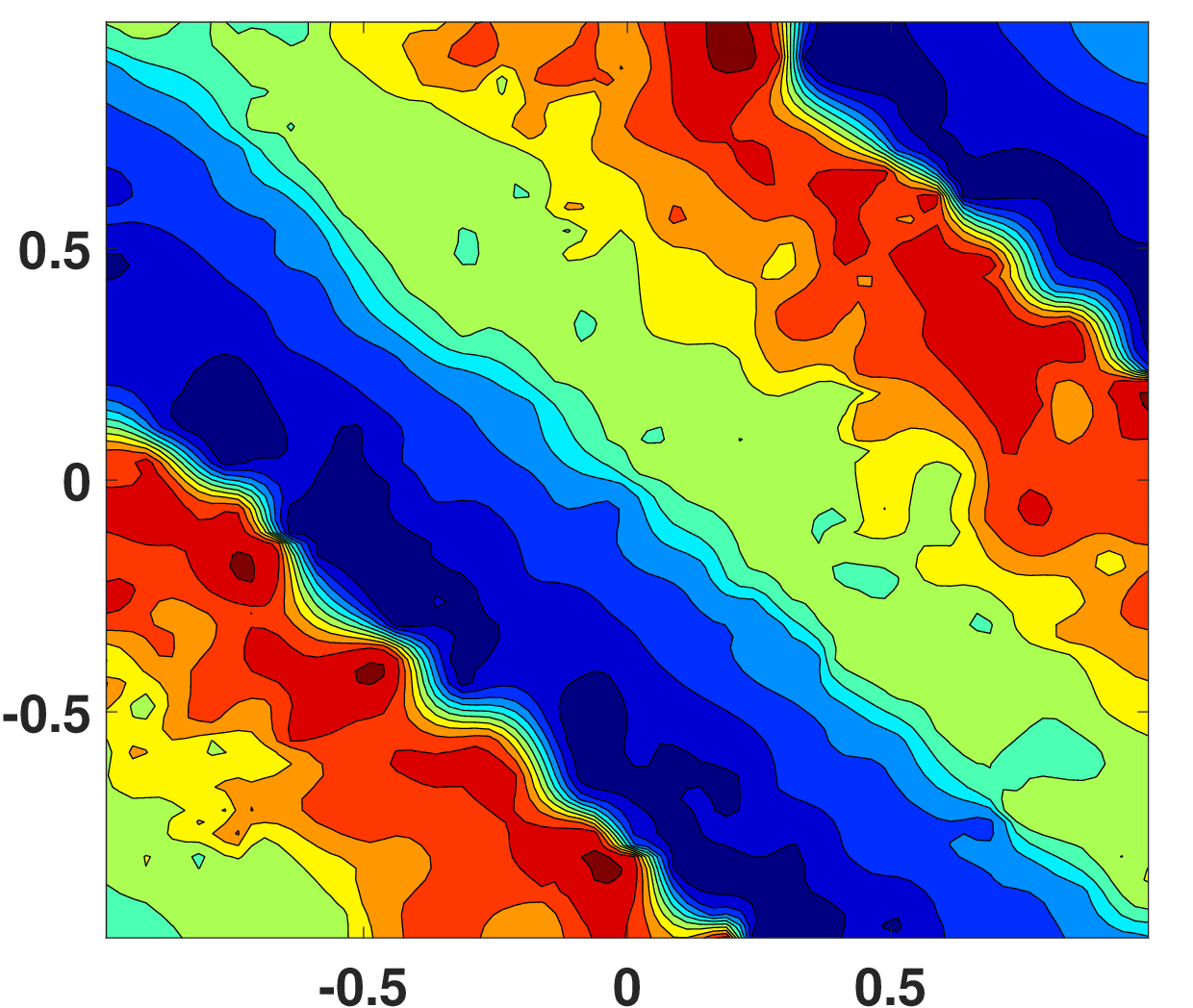} \end{minipage}%
\begin{minipage}{0.3\textwidth}
\includegraphics[scale=0.25]{Toanfigures/2DBurger/T02/Burger2D_EstState_10Obs_LETKF.eps}     
\end{minipage}
    \caption{\small Contour maps of the estimated solution to the Burgers' equation at $T = 0.45$ with $10\%$ observations. (First) Reference Solution. (Second) EnSF estimate with bi-harmonic inpainting. (Third) LETKF estimate.}
    \label{Contour_T045_Inpainting}
    \vspace{-0.2cm}
\end{figure}
\begin{figure}[h!]
% \vspace{-0.8cm}
    \centering
 \begin{minipage}{0.42\textwidth}
 \includegraphics[scale=0.28]{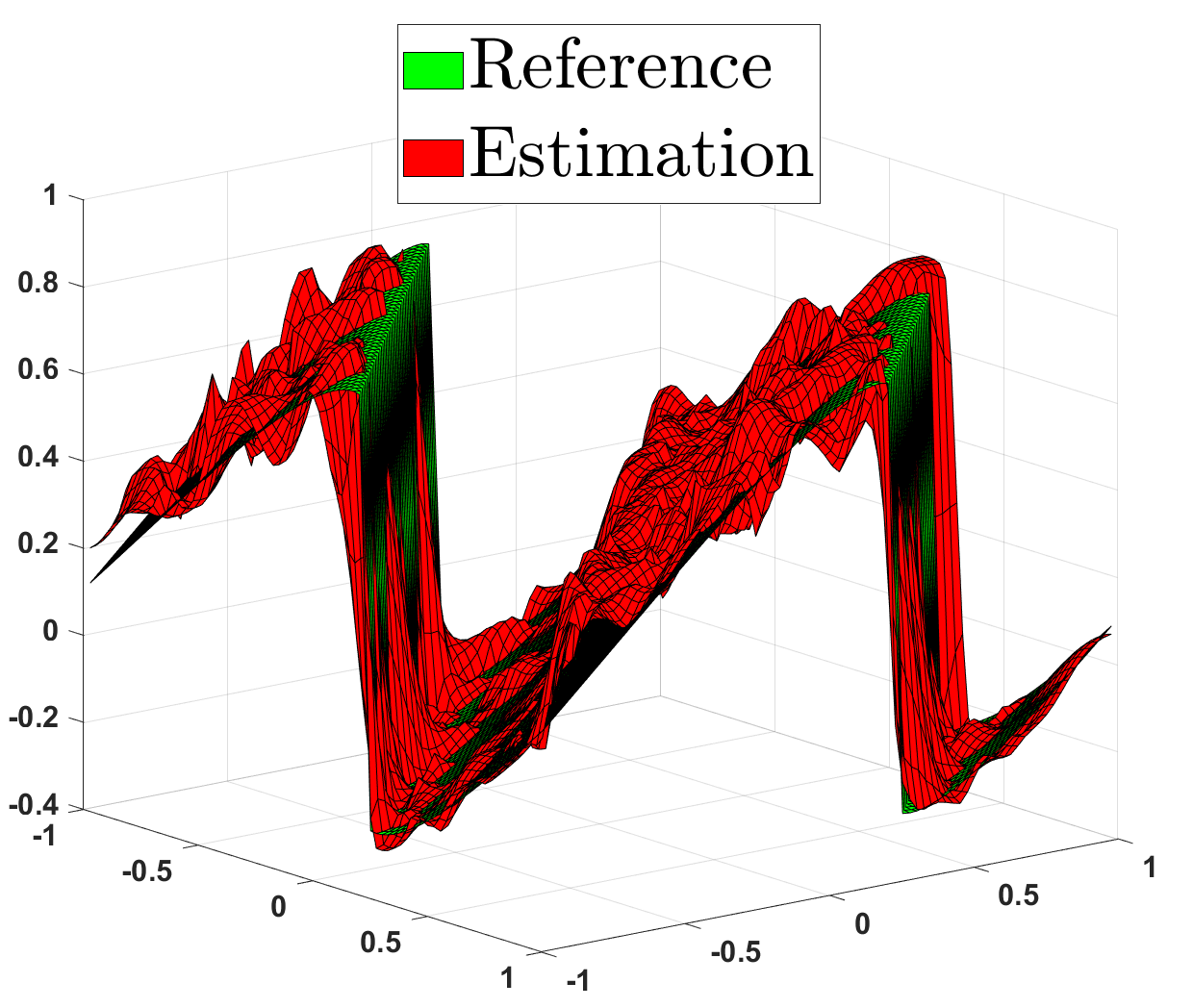}
 \end{minipage} %
 \begin{minipage}{0.42\textwidth}
      \includegraphics[scale=0.28]{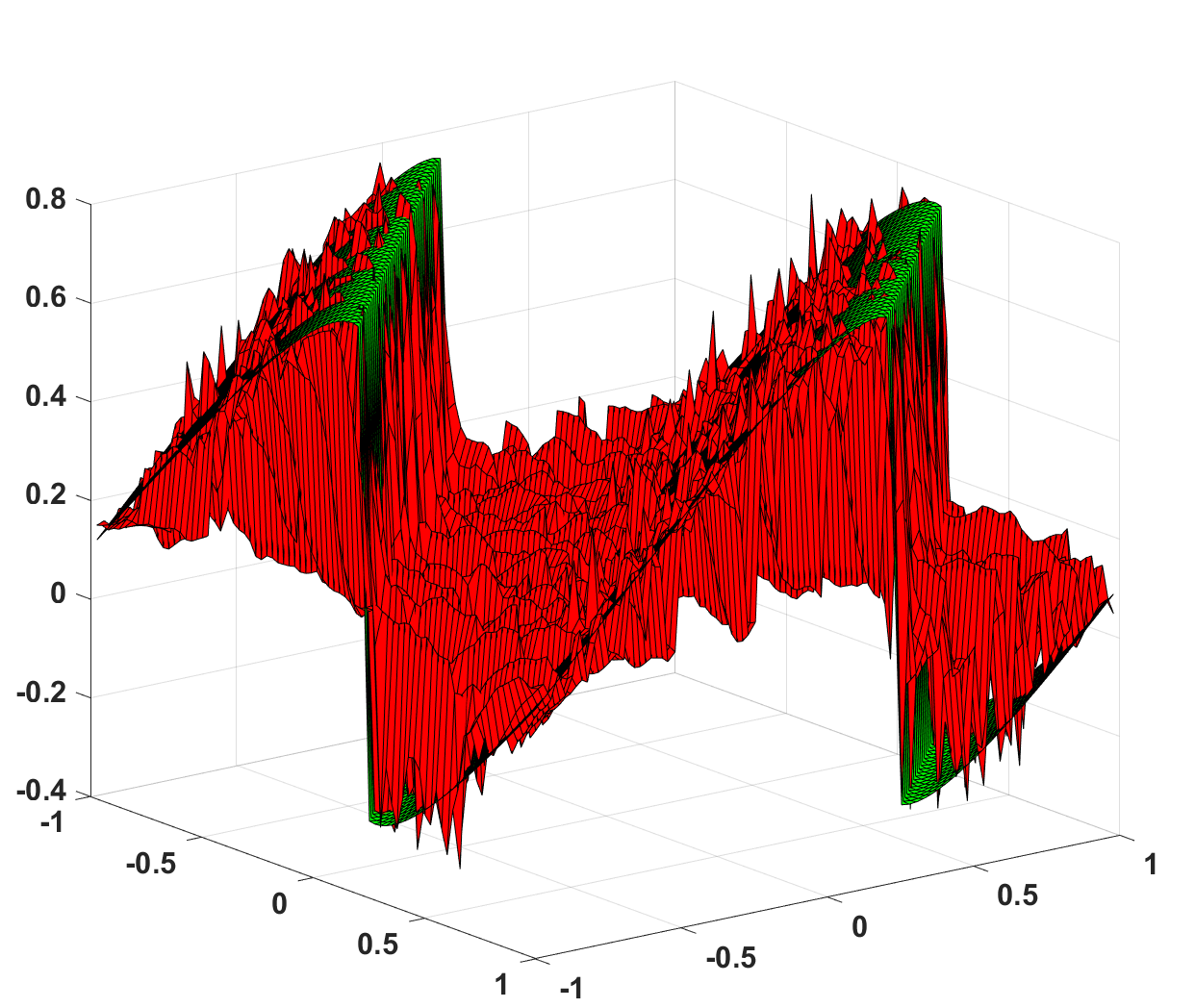}
 \end{minipage}
 \caption{\small 3D view of the estimated solution for the Burgers' equation at $T = 0.45$ with $10\%$ observations.  (Left) EnSF estimate with biharmonic inpainting. (Right) LETKF estimate. }    \label{3DContour_T045_Inpainting}
 \vspace{-0.2cm}
\end{figure}

\begin{figure}[h!]
    \centering
 \begin{minipage}{0.42\textwidth}   \hspace{1.5cm}\includegraphics[scale=0.28]{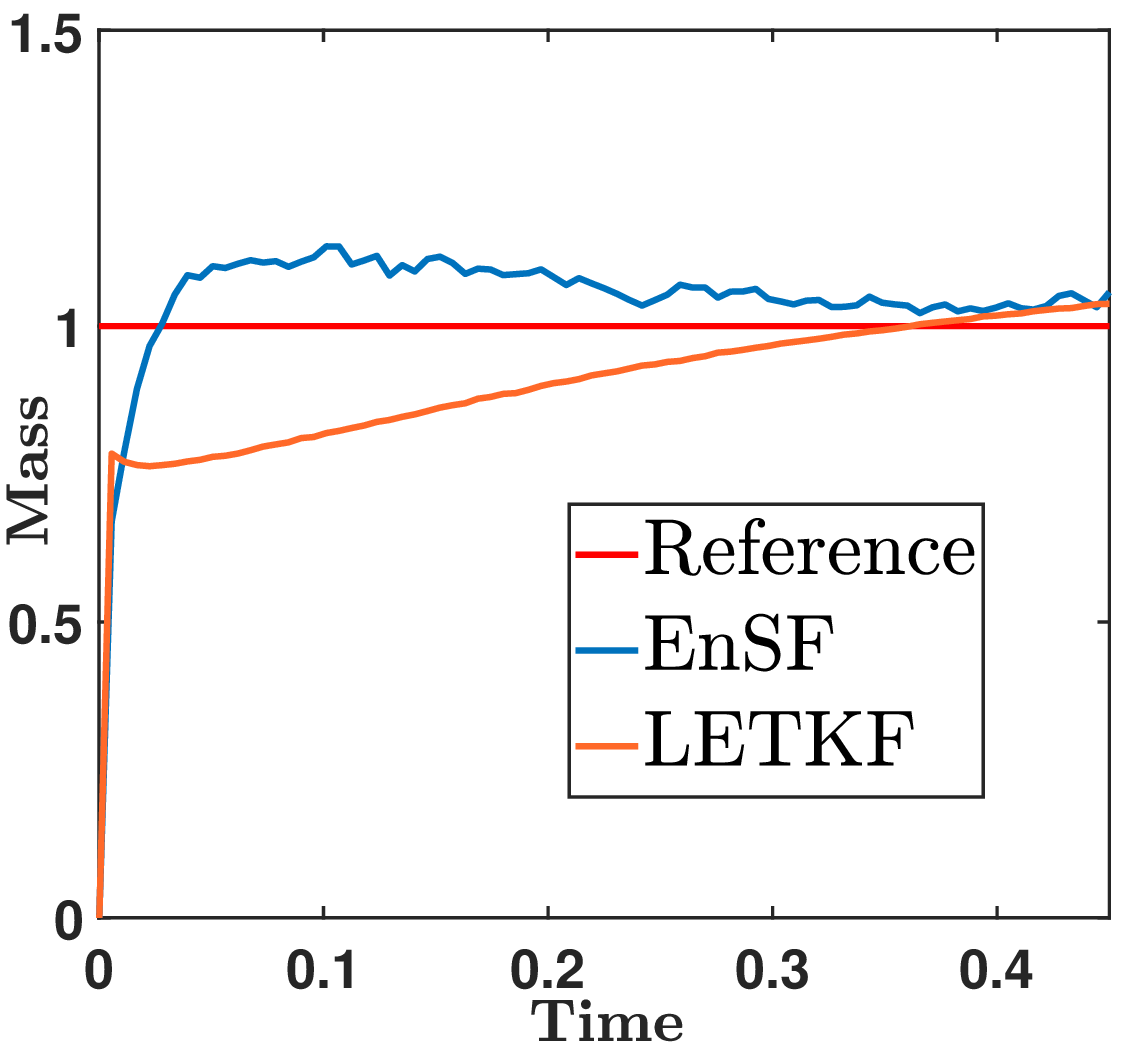}
 \end{minipage}  %
 \begin{minipage}{0.42\textwidth}    \includegraphics[scale=0.28]{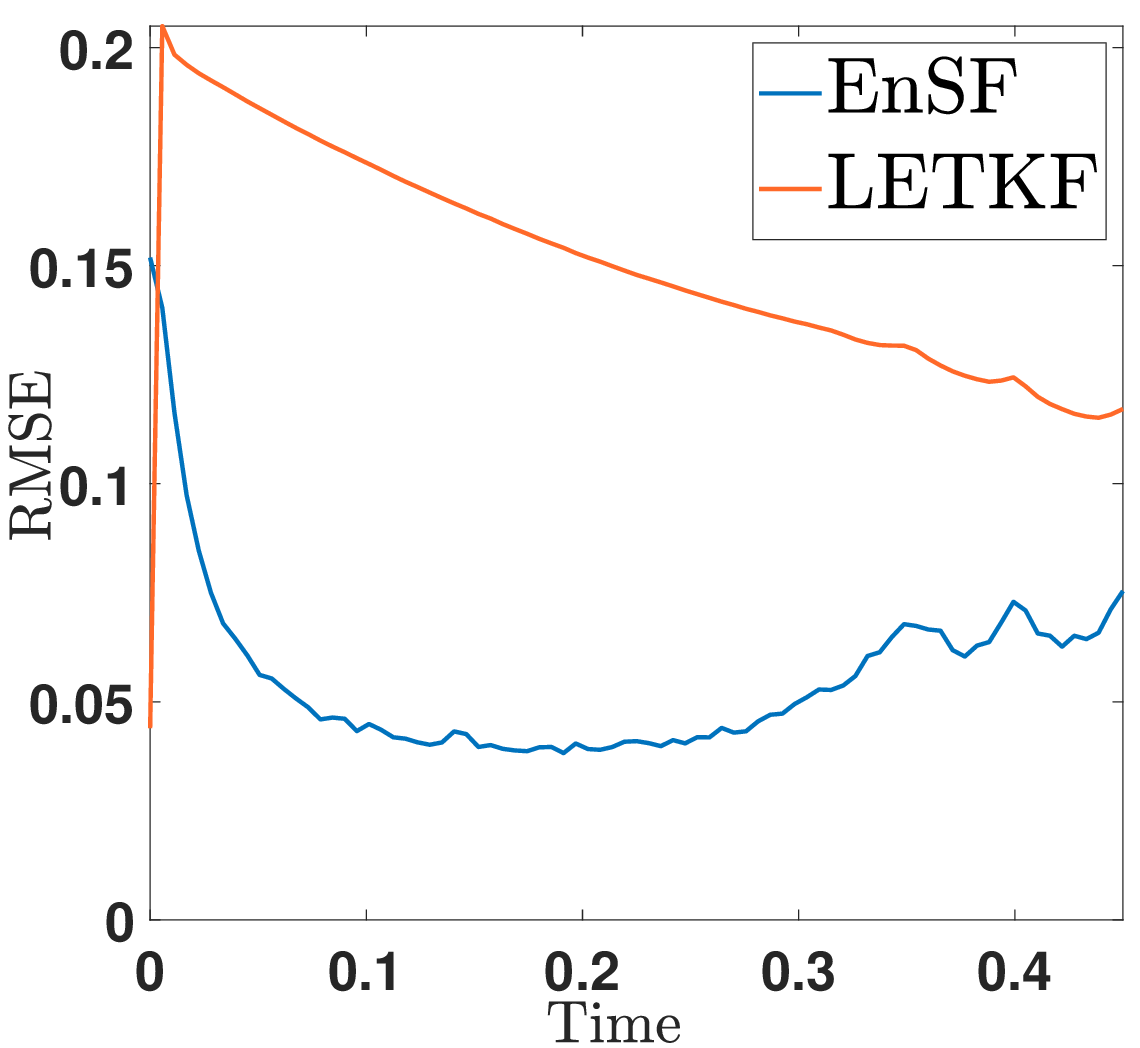}
 \end{minipage}
 \caption{\small (Left) Evolution of Mass with $T=0.45$ with $10\%$ observations. (Right) RMSE for state estimations with $T=0.45$ with $10\%$ observations.}
 \label{Mass_RMSE_T045_Inpainting}
 \vspace{-0.3cm}
\end{figure}

We compare the estimation performance obtained from two approaches: by using EnSF with the Bih inpainting and by LETKF. The estimated solutions for $T=0.2$ and $T=0.45$ are shown in Figure~\ref{Contour_T02_Inpainting} and Figure~\ref{Contour_T045_Inpainting}, respectively. The three-dimensional views are depicted in Figures~\ref{3DContour_T02_Inpainting}and~\ref{3DContour_T045_Inpainting}. The corresponding mass evolution and RMSEs are displayed in Figure~\ref{Mass_RMSE_T02_Inpainting} and Figure~\ref{Mass_RMSE_T045_Inpainting}, respectively. The displayed figures show that the EnSF approach yields more accurate solution estimates and outperforms the LETKF. Moreover,  by incorporating the inpainting technique, the estimated solutions appear significantly less noisy, even though only $10\%$ of the spatial domain was observed.

\subsection{Navier-Stokes equation}

In this subsection, we focus on the incompressible Navier-Stokes (NS) equations given as follows:
\begin{equation}
    \left\{\begin{array}{ll}
    \pmb{u}_t - \nu\Delta{\pmb{u}}+\pmb{F}(\pmb{u})+\nabla{p} = \pmb{f}, & \; \text{in} \; \Omega \times (0, T], \vspace{0.1cm} \\
    \nabla \cdot \pmb{u} = 0, & \; \text{in} \; \Omega \times (0, T],
    \end{array}\right.
    \label{NavierStokes}
\end{equation}
subject to the initial condition $\pmb{u}(\pmb{x}, 0) = \pmb{u}_0(\pmb{x})$ and the Dirichlet or the periodic boundary conditions. $\Omega$ in \eqref{NavierStokes} is an open bounded domain in $\mathbb{R}^2$, $\pmb{u} =(u, v)$ and $p$ are the unknown velocity field and pressure, respectively, $\pmb{F}(\pmb{u}) = (\pmb{u} \cdot \Delta) \pmb{u}$ denotes the nonlinear convection, $\pmb{f}$ is an external force, and $\nu$ represents the kinematic viscosity which is inversely proportional to the Reynolds number $\textit{Re}$. To formulate the forward solver, we employ a newly developed dynamically regularized Lagrange multiplier (DRLM) method proposed in~\cite{Doan2025b} to discretize the equations~\eqref{NavierStokes}. More specifically, an artificial Lagrange multiplier $q$ is introduced to reformulate~\eqref{NavierStokes} into
\begin{equation}
    \left\{\begin{array}{ll}
    \pmb{u}_t - \nu\Delta{\pmb{u}}+\pmb{F}(\pmb{u})+\nabla{p} = \pmb{f}, & \; \text{in} \; \Omega \times (0, T], \vspace{0.1cm} \\
    \nabla \cdot \pmb{u} = 0, & \; \text{in} \; \Omega \times (0, T], \vspace{0.1cm} \\
    \dfrac{d\pmb{\mathcal{K}}(\pmb{u})}{dt} +\theta\dfrac{dq^2}{dt}= \left(\pmb{u}_t+q\pmb{F}(\pmb{u}), \pmb{u}\right), & \; \text{in} \; \Omega \times (0, T],
    \end{array}\right.
    \label{NavierStokes_reform}
\end{equation}
where $\pmb{\mathcal{K}}(\pmb{u}) = \frac{1}{2}\int_{\Omega}\vert \pmb{u} \vert^2 d\pmb{x}$ is the kinetic energy associated with~\eqref{NavierStokes}, $q(0) = 1$ and $\theta>0$ is a regularization parameter. It is straightforward to verify that at the continuous level, we have $q(t) = 1$ for all $t>0$, and that the new system~\ref{NavierStokes_reform} is equivalent to the original one~\cite{Doan2025b}. We discretize the reformulated Navier–Stokes system \eqref{NavierStokes_reform} in space using a finite-difference scheme on a MAC staggered grid \cite{Li2020} and in time with backward-differentiation-formula (BDF) methods. Alternatively, one may substitute the finite-difference spatial discretization with a finite-element or finite-volume approach and adopt the Crank–Nicolson scheme for time stepping. We emphasize that the DRLM schemes employed in this work require solving two generalized Stokes systems at each time step. One can instead replace the Stokes solver with two successive Poisson solvers via a pressure-correction approach \cite{Guermond2006, Yang2022}. For a complete derivation of the fully discrete formulations under both strategies, see \cite{Doan2025b}. Accordingly, we augment the filter state with a scalar variable $\tilde{q}$ to represent the estimate of  $q$. Since $q$ is an artificial parameter with a known exact value, we simply use that true value as the observation for $\tilde{q}$.

Throughout the experiments presented in this example, different time-discretization schemes are used for generating the ``true state'' and for prediction during the filtering process. Specifically, the synthetic reference solution representing the ``true state'' is computed using the second-order DRLM solver, while the first-order DRLM scheme is used during prediction to reduce computational cost. The detailed numerical implementation can be found on \url{https://github.com/Toanhuynh997/StateEst_PDEs/tree/main/NavierStokes} .

\subsubsection{Taylor-Green vortex problem}

We begin with the Taylor-Green vortex problem on the domain $\Omega =(0, 1)^2$ where Eq.~\eqref{NavierStokes} is considered with zero external forcing, i.e., $\pmb{f} = [0, 0]^T$, and periodic boundary conditions are imposed. The Reynolds number is set at $\textit{Re} = 1000$, the parameter $\theta$ is fixed at $\theta=5$, and the final time is $T = 1$, and the mesh size is fixed at $h = 1/40$. To generate the reference solution as the ``true state'', we solve Eq.~\eqref{NavierStokes} numerically using the second-order DLMR method on a fine temporal grid with time step $\Delta t = T/600$. Observations are then obtained by applying the \textbf{\textit{tangential operator}} to this reference solution.

\begin{figure}[h!]
\centering
\begin{minipage}{0.333\textwidth}
\includegraphics[scale=0.245]{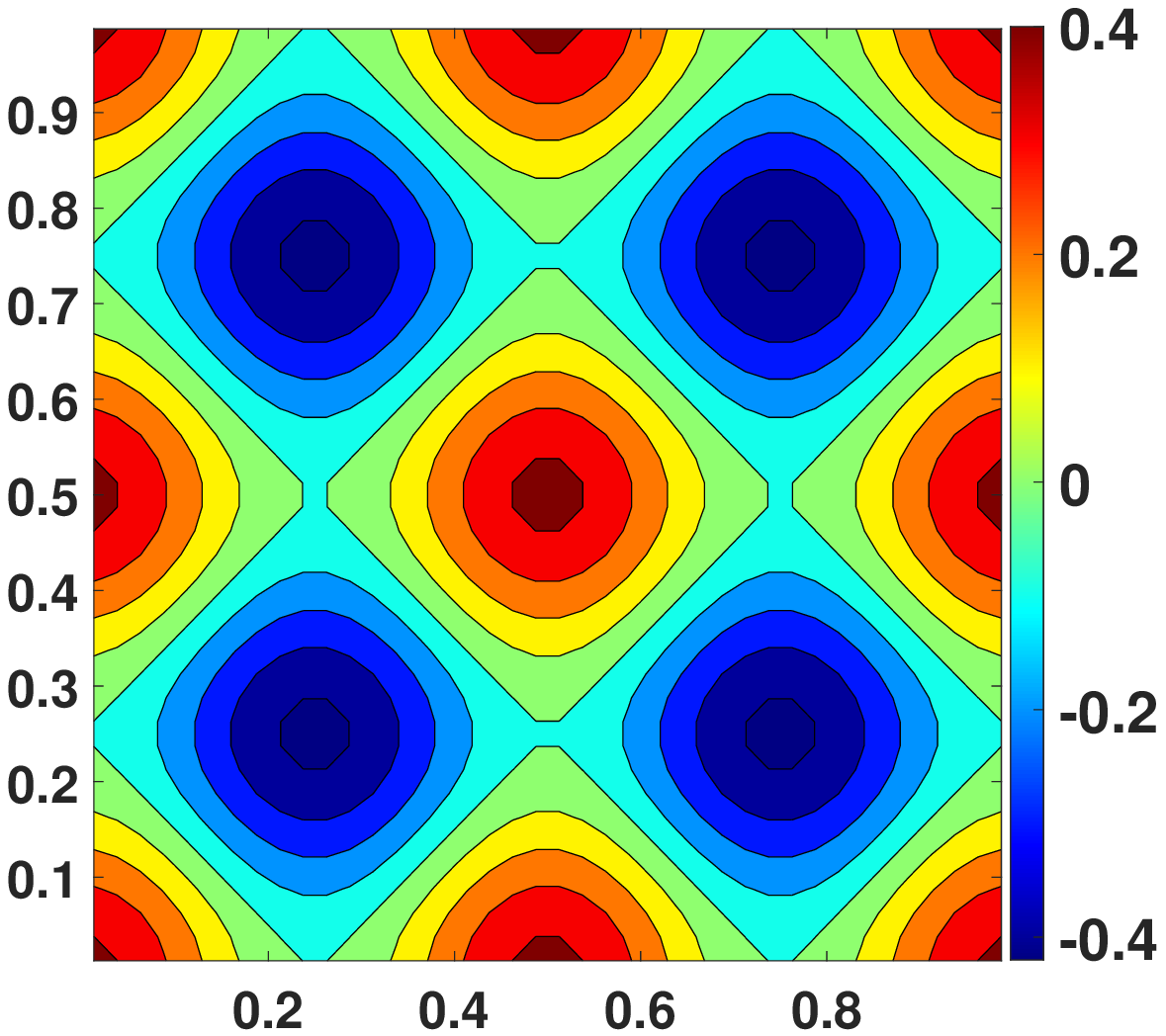}
\end{minipage}%
\begin{minipage}{0.333\textwidth}
\hspace{0.1cm}\includegraphics[scale=0.245]{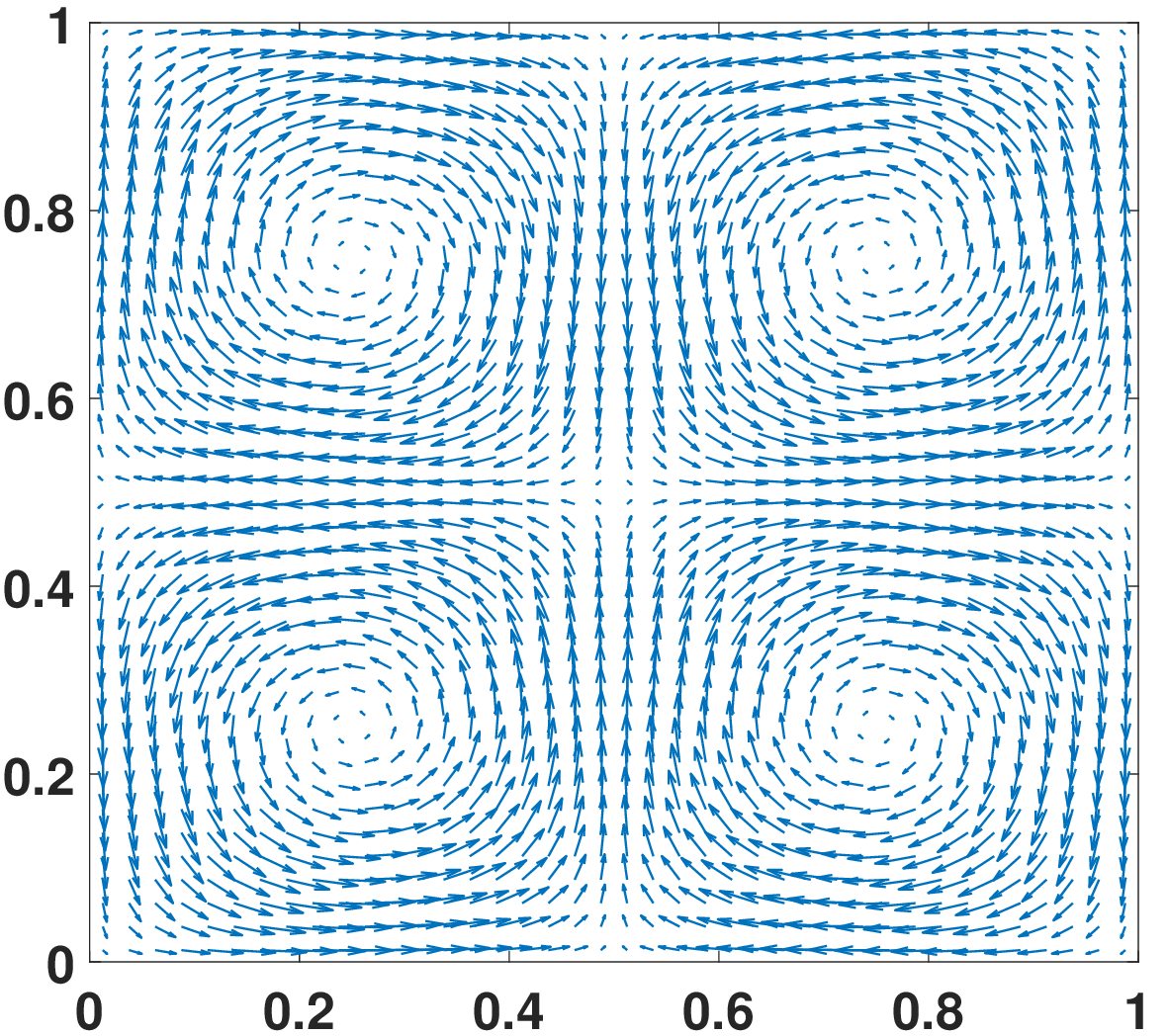}
\end{minipage}%
\begin{minipage}{0.333\textwidth}
\hspace{0.1cm}\includegraphics[scale=0.245]{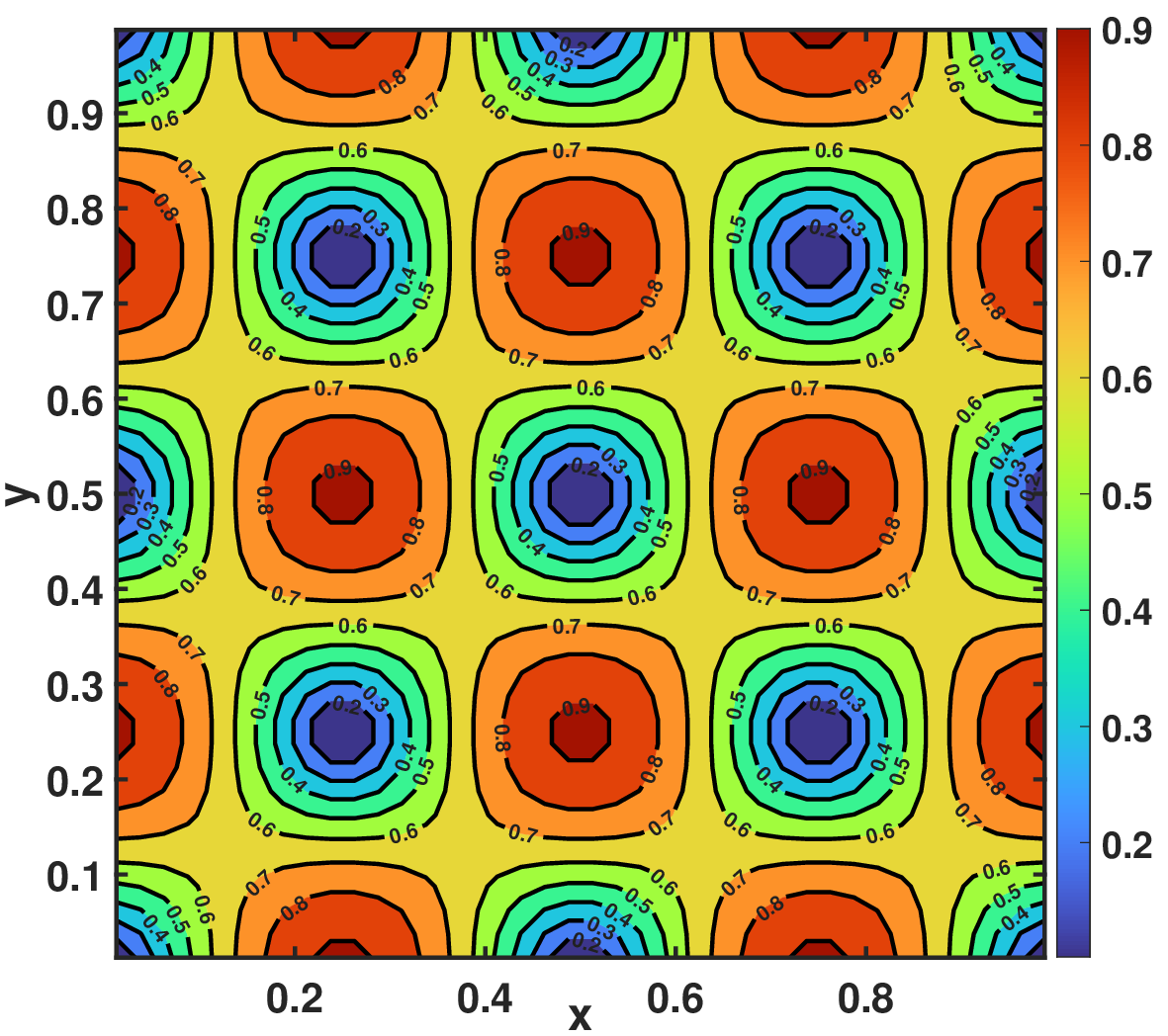}
\end{minipage}
\caption{\small Reference solution for the Taylor-Green vortex problem. (Left) Pressure field. (Center) Velocity field. (Right) Velocity magnitude.}
\label{TL_RefSol}
\vspace{-0.3cm}
\end{figure}

In this experiment, we let the initial conditions for the reference solution be
\begin{align*}
    \left\{\begin{array}{l}
    u(x, y) = \sin(2\pi{x})\cos(2\pi{y}), \vspace{0.1cm} \\
    v(x, y) = -\cos(2\pi{x})\sin(2\pi{y}).
    \end{array}\right.
\end{align*}
The reference pressure field and the reference velocity field are illustrated in Figure~\ref{TL_RefSol}. To implement the prediction step in data assimilation, we solve the PDE on a coarser temporal grid with a time step of $\Delta t_{Filter} = T/100$. The initial conditions for $u$ and $v$ are assumed to follow a Gaussian distribution: $2\cdot N(0, \pmb{I}_d)$. We explore two levels of additive model uncertainties by setting $\tilde{W}^{i}_{t_n} \sim \sigma_n^i N(0, \pmb{I}_{d}), i=1, 2$  where $\sigma_n^1 =0.001$ and $\sigma_n^2 =0.1$. Similarly to the Burgers' equation, the perturbed term $\sigma_n\tilde{W}^i_{t_n}$ is also added to the predicted solution at each filtering step, which results in solving an SPDE and forces the forecast to evolve in a different trajectory compared to the reference one.
\begin{figure}[h!]
\vspace{-0.3cm}
\centering
\begin{minipage}{0.333\textwidth}
\includegraphics[scale=0.24]{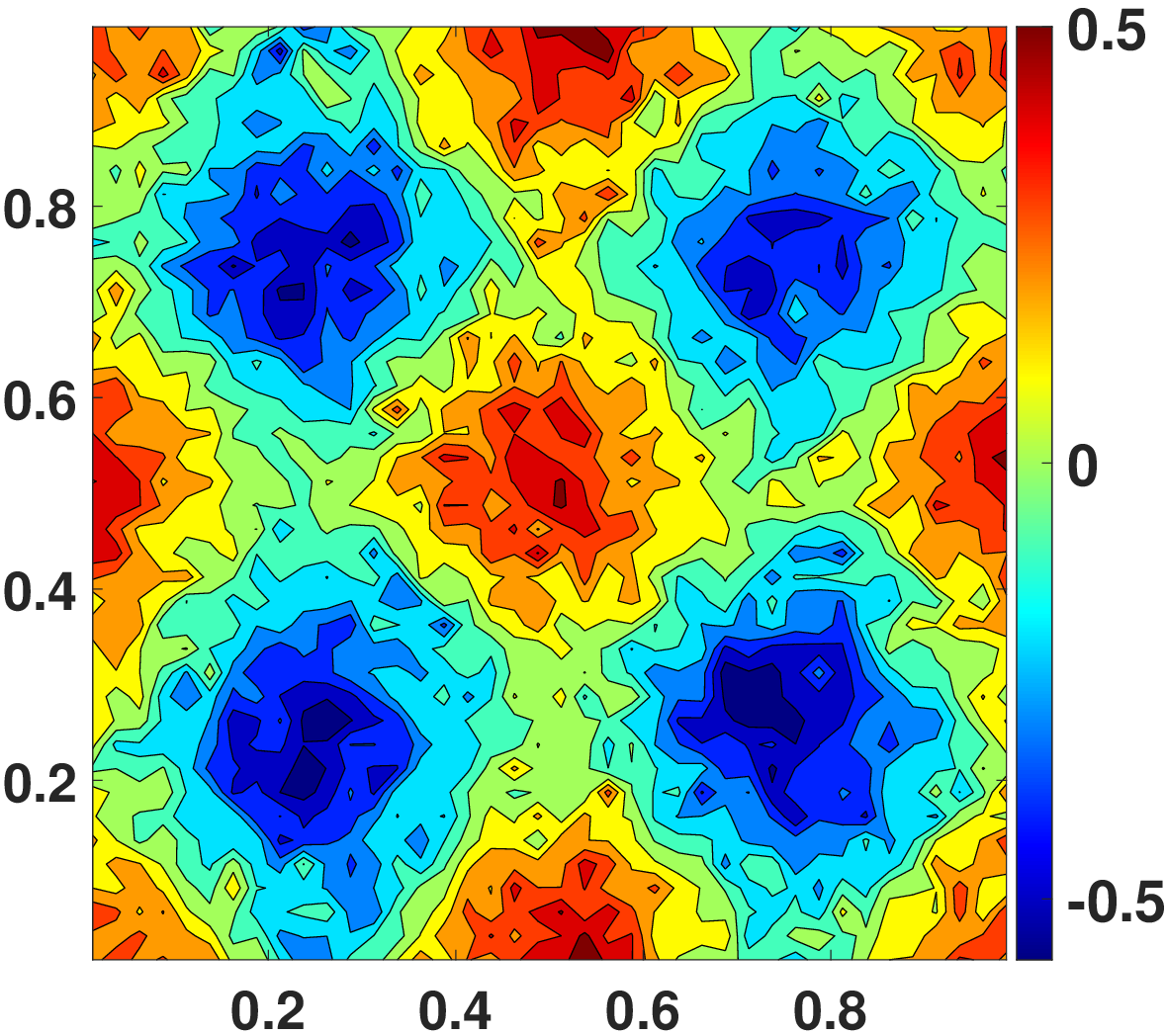}
\end{minipage}%
\begin{minipage}{0.333\textwidth}
\hspace{0.1cm}\includegraphics[scale=0.24]{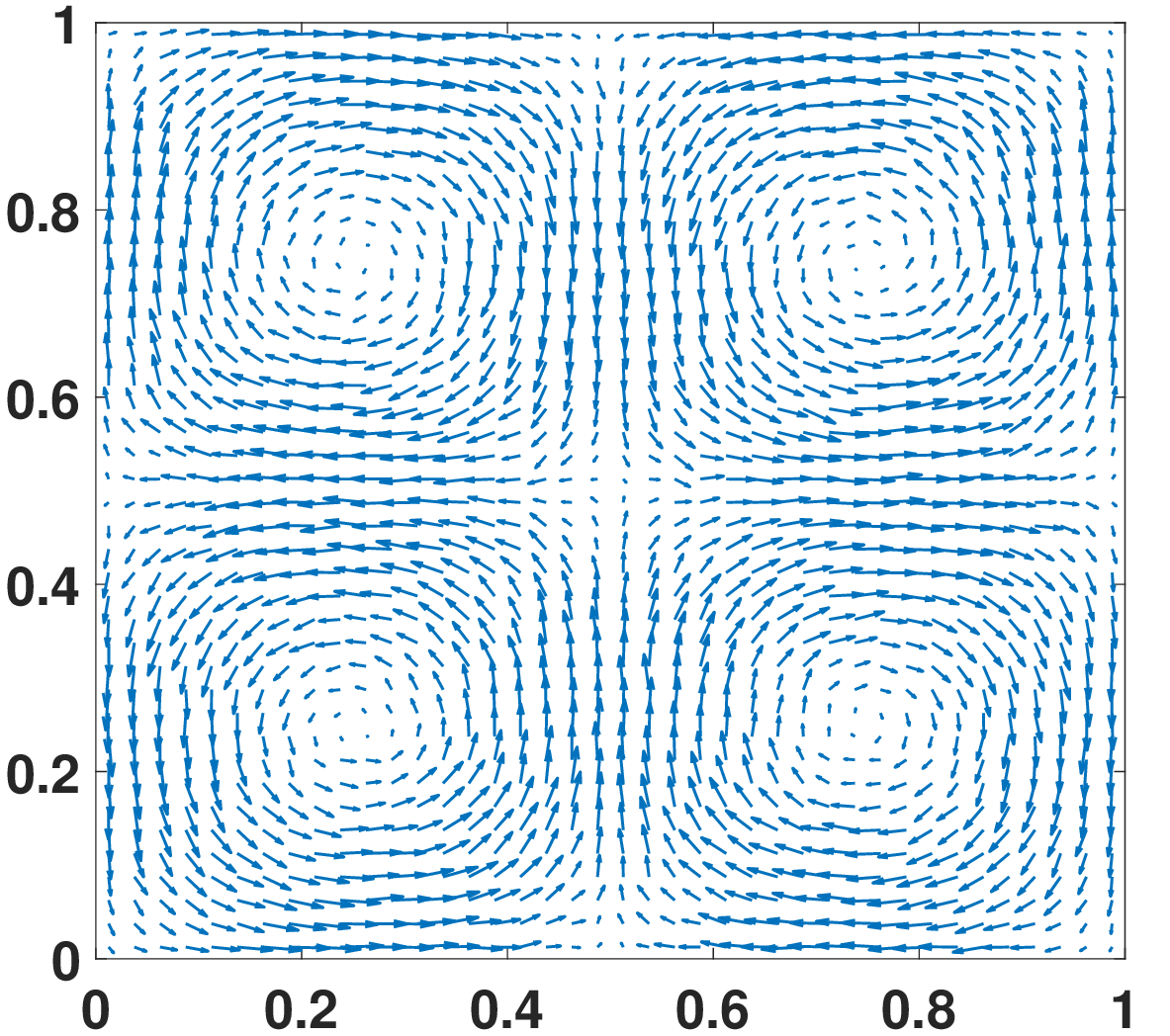}
\end{minipage}%
\begin{minipage}{0.333\textwidth}
\hspace{0.1cm}\includegraphics[scale=0.24]{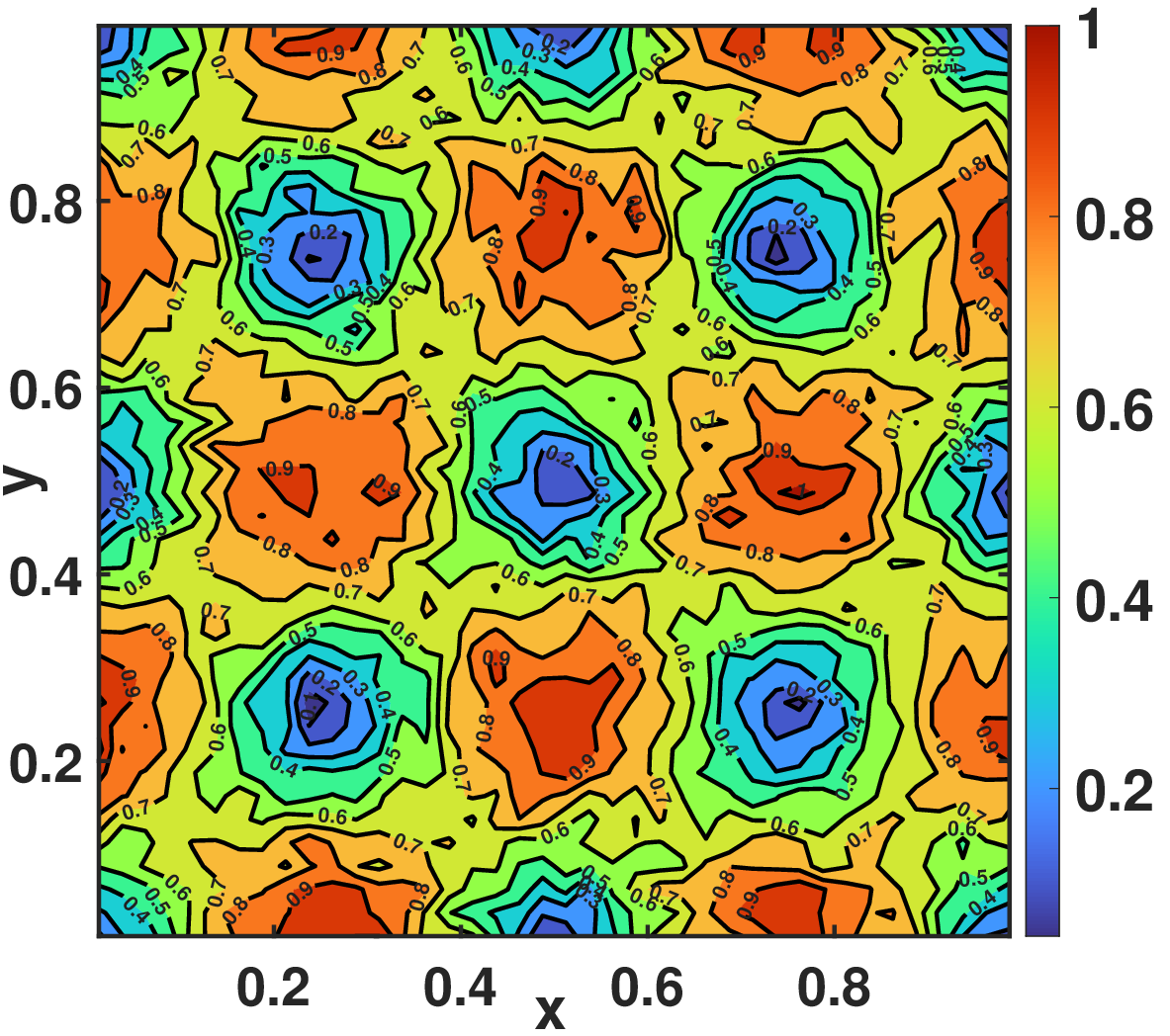}
\end{minipage}
\caption{\small Estimated state for the Taylor-Green vortex problem with $100\%$ observations and model uncertainty level $\sigma_n^1$. (Left) Pressure field. (Center) Velocity field. (Right) Velocity magnitude.}
\label{TL_EstSol_100Obs_omega1}
\vspace{-0.4cm}
\end{figure}
\begin{figure}[h!]
\centering
\begin{minipage}{0.333\textwidth}
\includegraphics[scale=0.24]{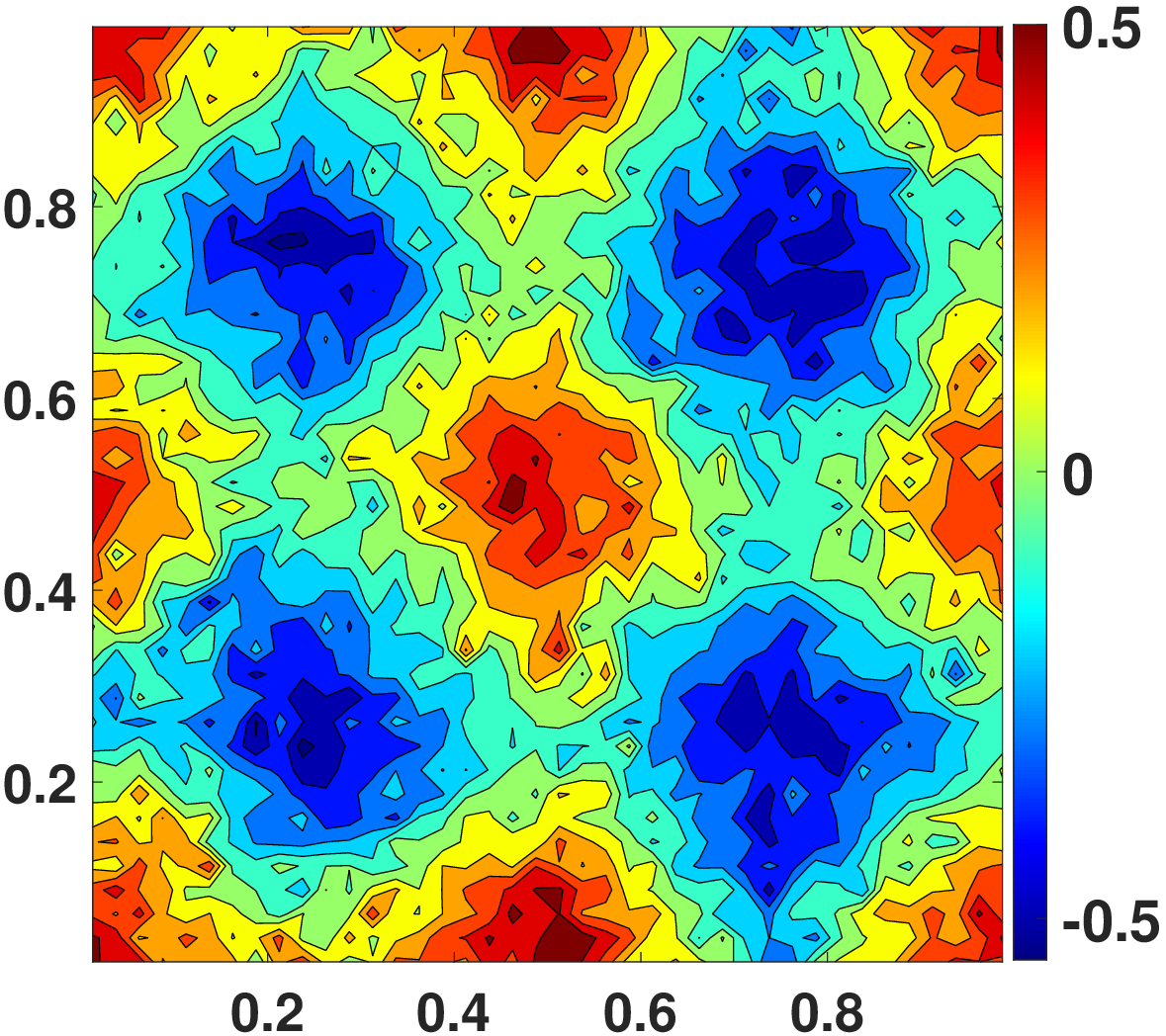}
\end{minipage}%
\begin{minipage}{0.333\textwidth}
\hspace{0.1cm}\includegraphics[scale=0.24]{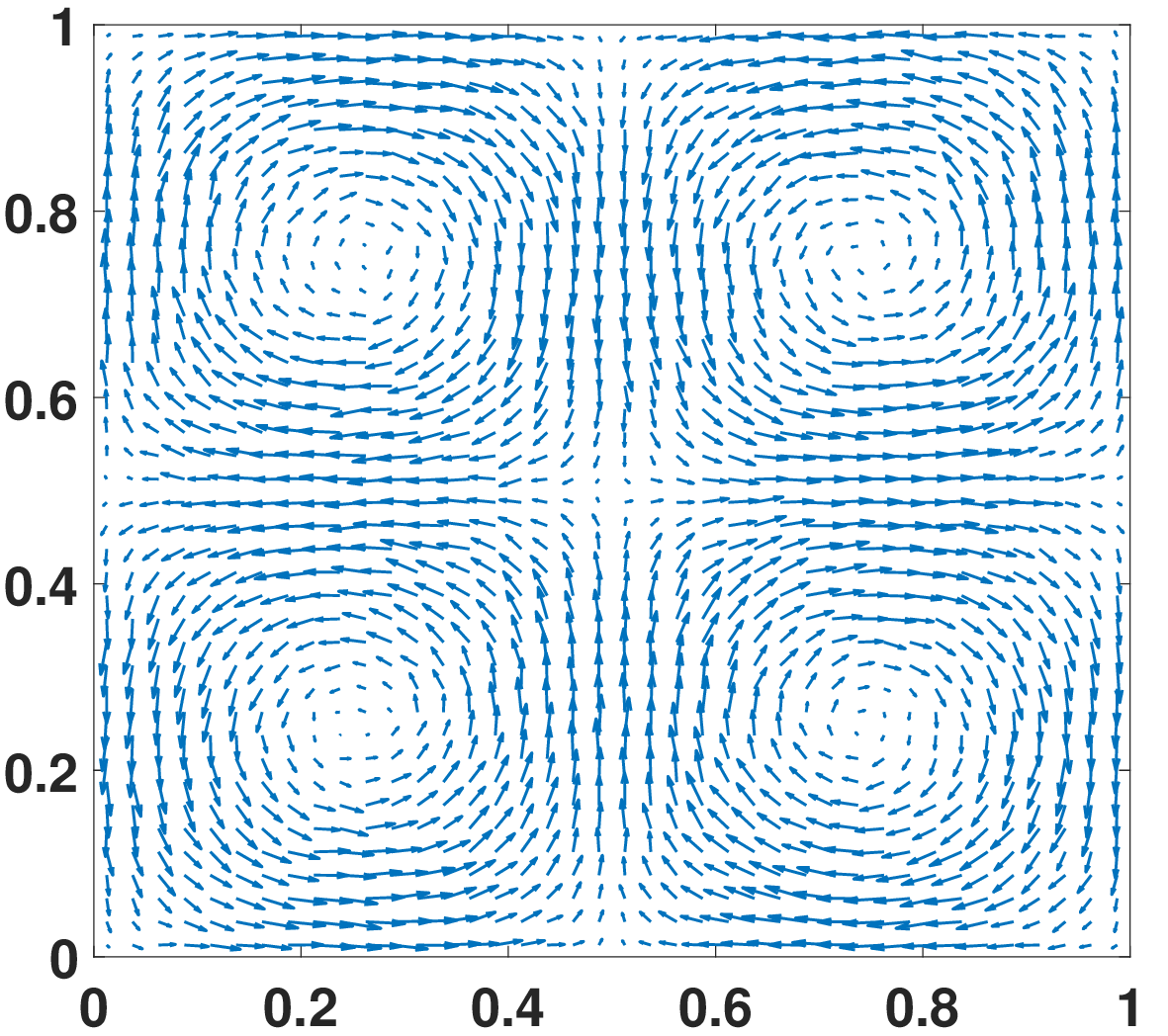}
\end{minipage}%
\begin{minipage}{0.333\textwidth}
\hspace{0.1cm}\includegraphics[scale=0.24]{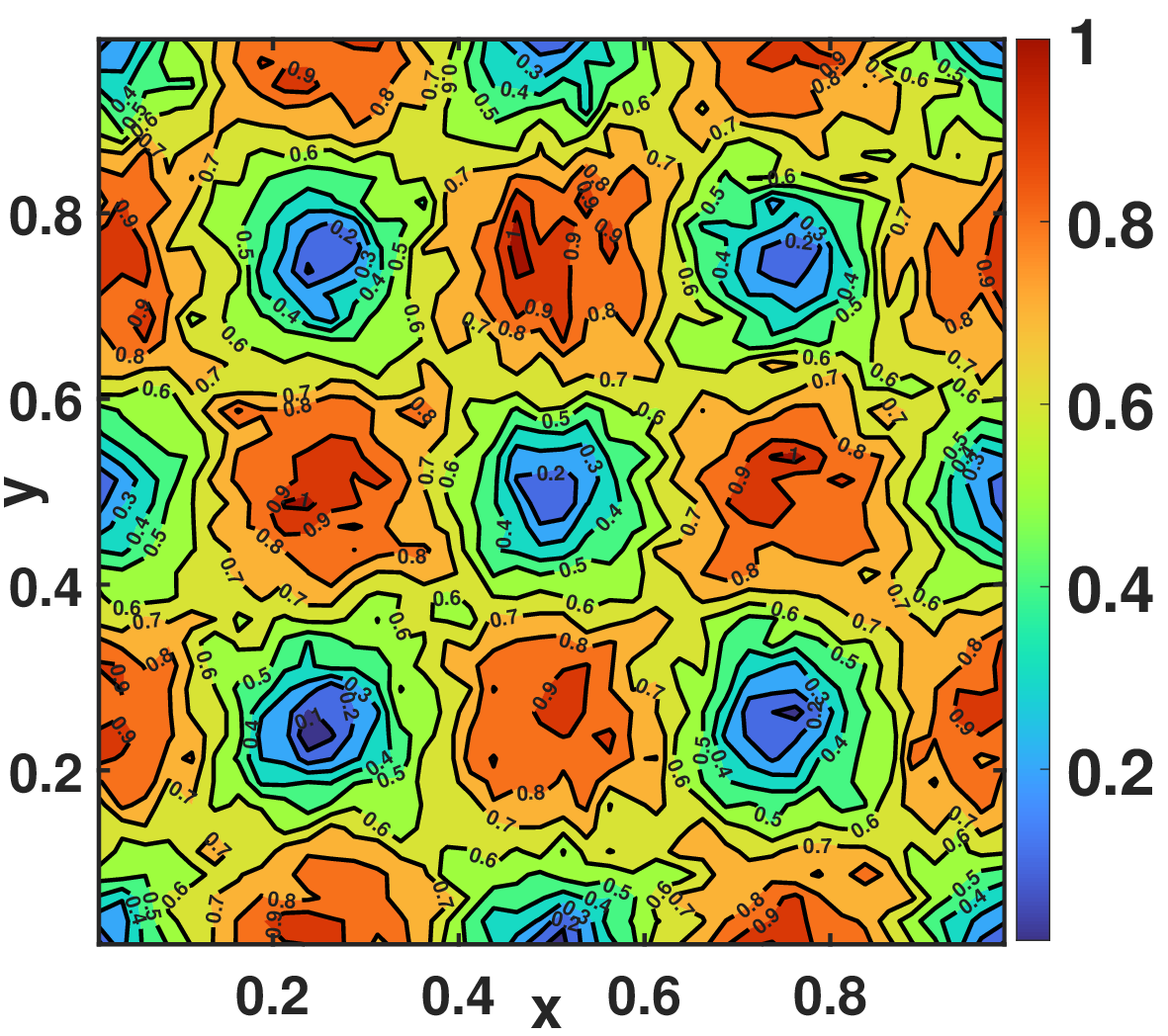}
\end{minipage}
\caption{\small Estimated state for the Taylor-Green vortex problem with $100\%$ observations and model uncertainty level $\sigma_n^2$. (Left) Pressure field. (Center) Velocity field. (Right) Velocity magnitude.}
\label{TL_EstSol_100Obs_omega2}
\vspace{-0.4cm}
\end{figure}
\begin{figure}[h!]
\centering
\begin{minipage}{0.333\textwidth}
\includegraphics[scale=0.24]{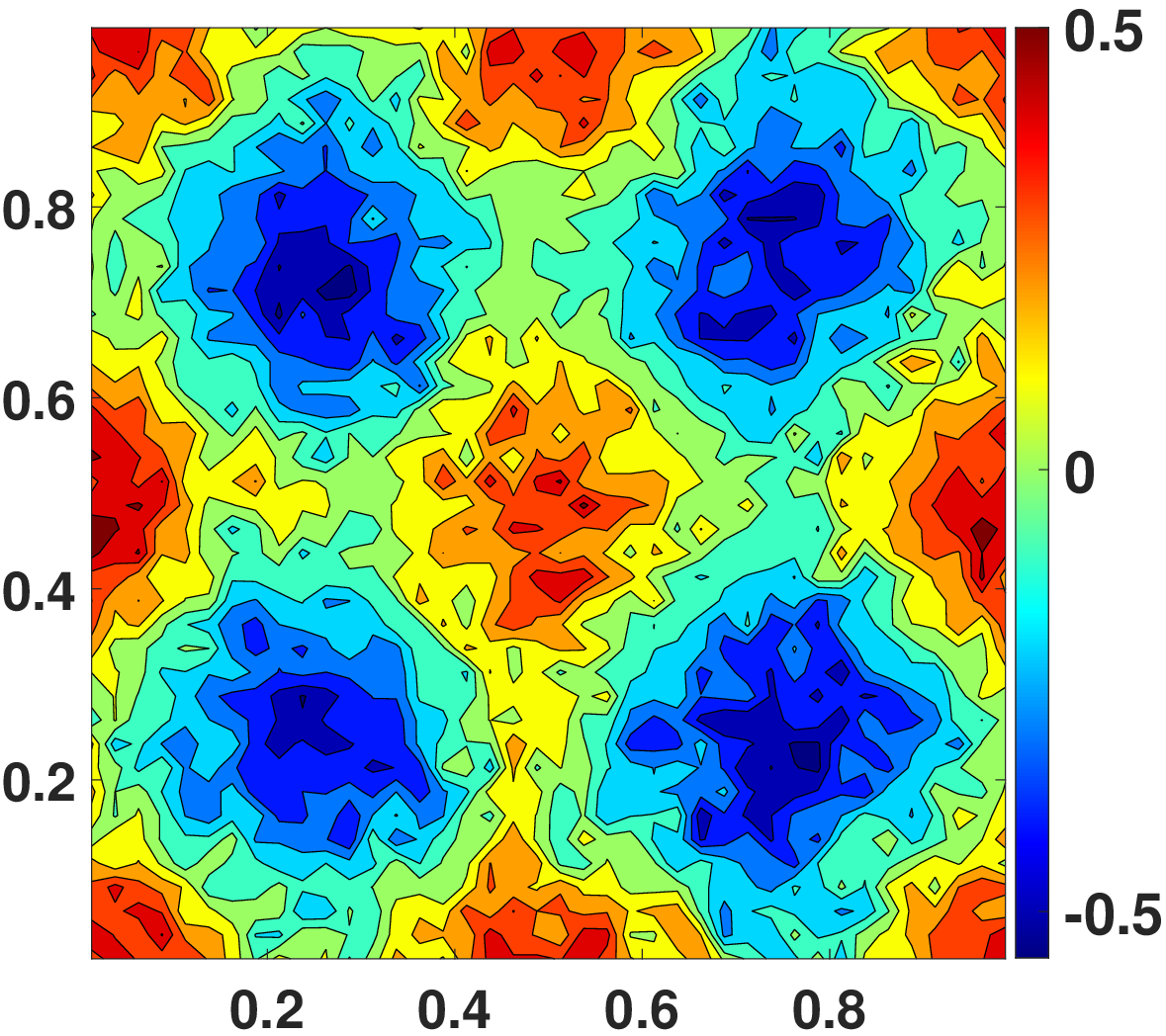}
\end{minipage}%
\begin{minipage}{0.333\textwidth}
\hspace{0.1cm}\includegraphics[scale=0.24]{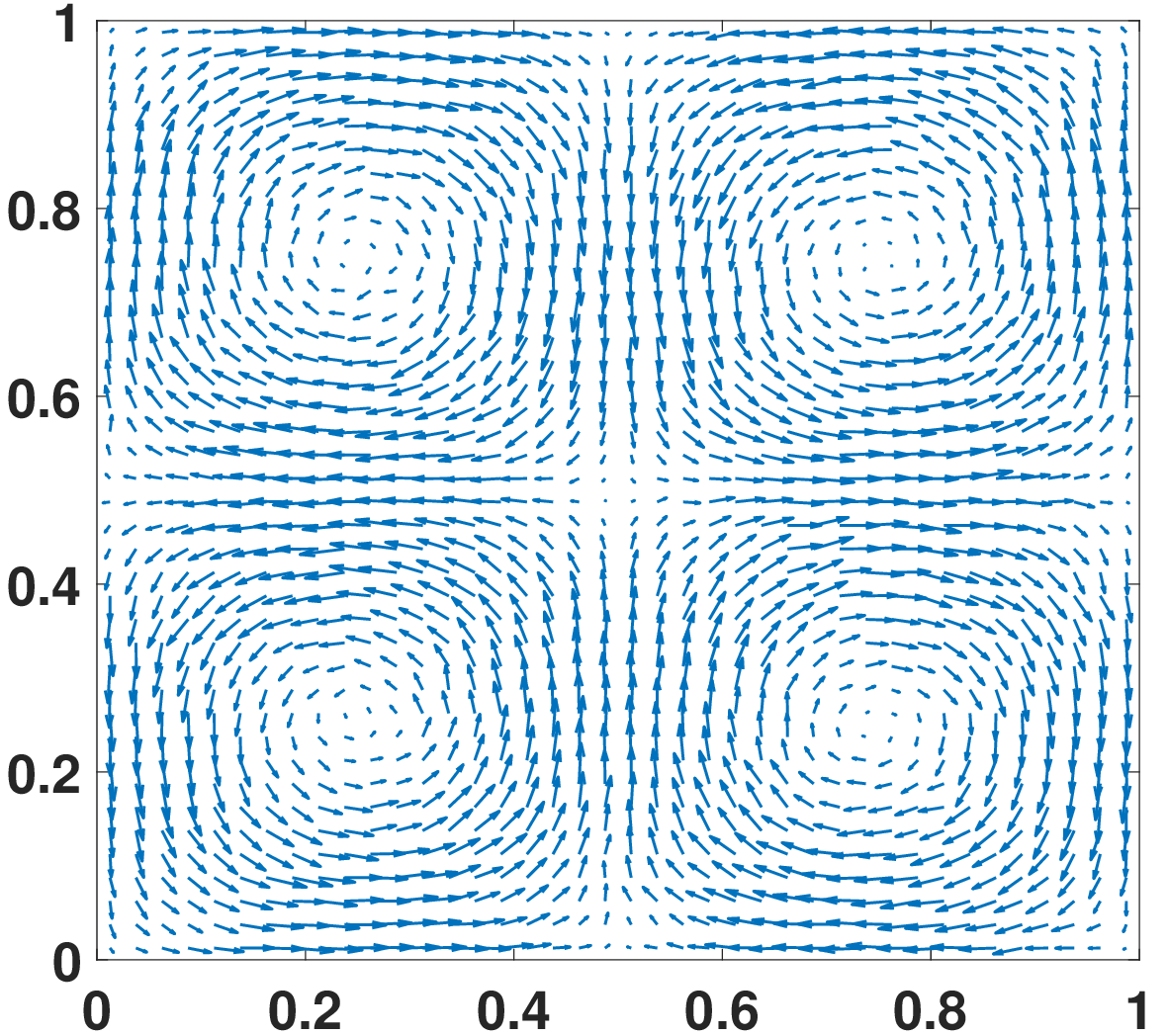}
\end{minipage}%
\begin{minipage}{0.333\textwidth}
\hspace{0.1cm}\includegraphics[scale=0.24]{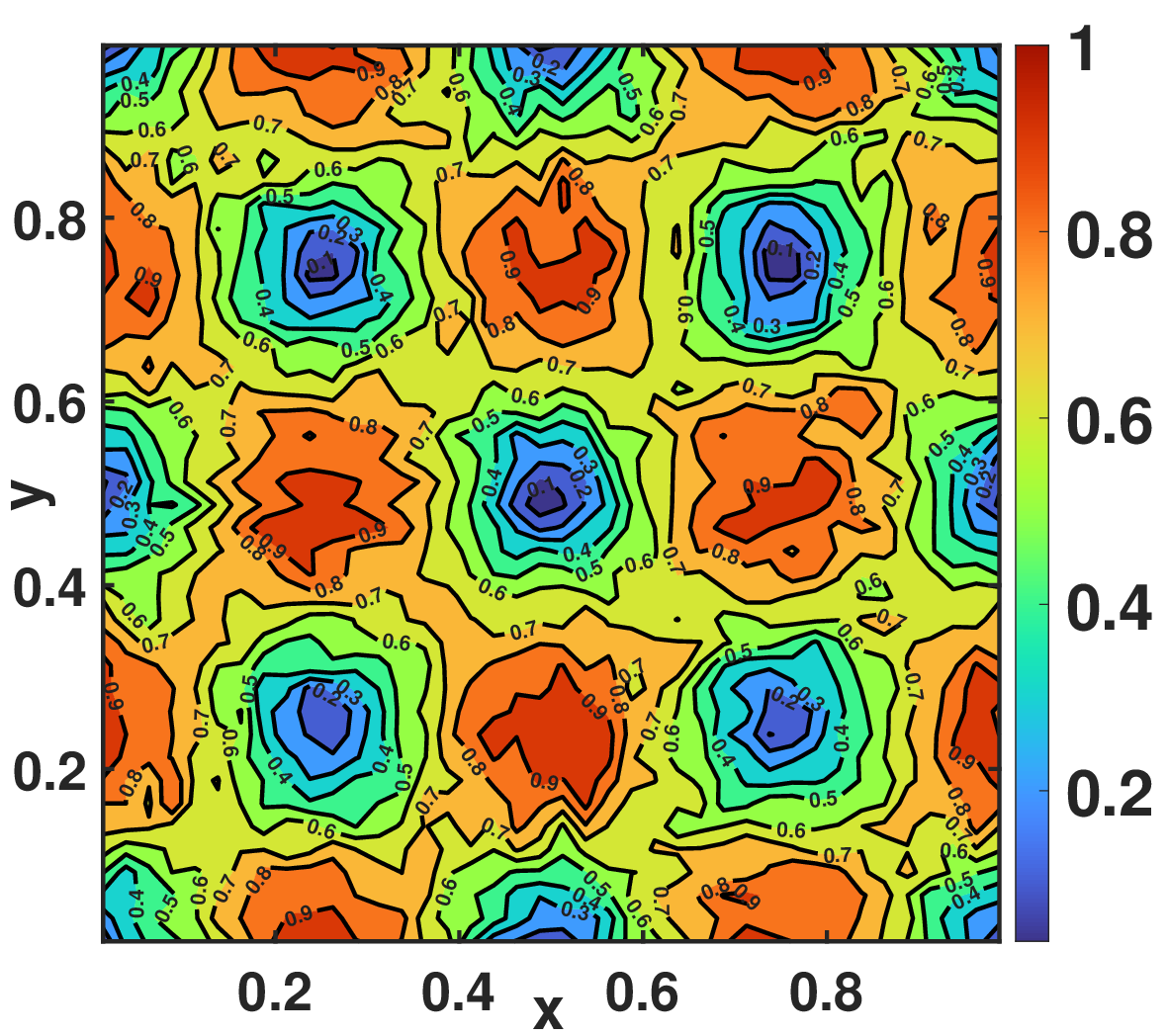}
\end{minipage}
\caption{\small Estimated state for the Taylor-Green vortex problem with $70\%$ observations and model uncertainty level $\sigma_n^1$. (Left) Pressure field. (Center) Velocity field. (Right) Velocity magnitude.}
\label{TL_EstSol_70Obs_omega1}
\vspace{-0.2cm}
\end{figure}

It is well known that, in the absence of the external forces, the solution to Eq.~\eqref{NavierStokes} under the prescribed boundary conditions satisfies the energy dissipation~\cite{Doan2025b, Yang2022}. Therefore, in addition to state estimation, we also investigate whether the estimated solution fulfills the discrete energy dissipation law.

To proceed, we consider three levels of observational coverage: $100\%$, $70\%$, and $7\%$. As in the previous example, the estimated solutions produced by our method for the $7\%$ observation case will be compared with those obtained using the LETKF.

\begin{figure}[h!]
\centering
\begin{minipage}{0.333\textwidth}
\includegraphics[scale=0.24]{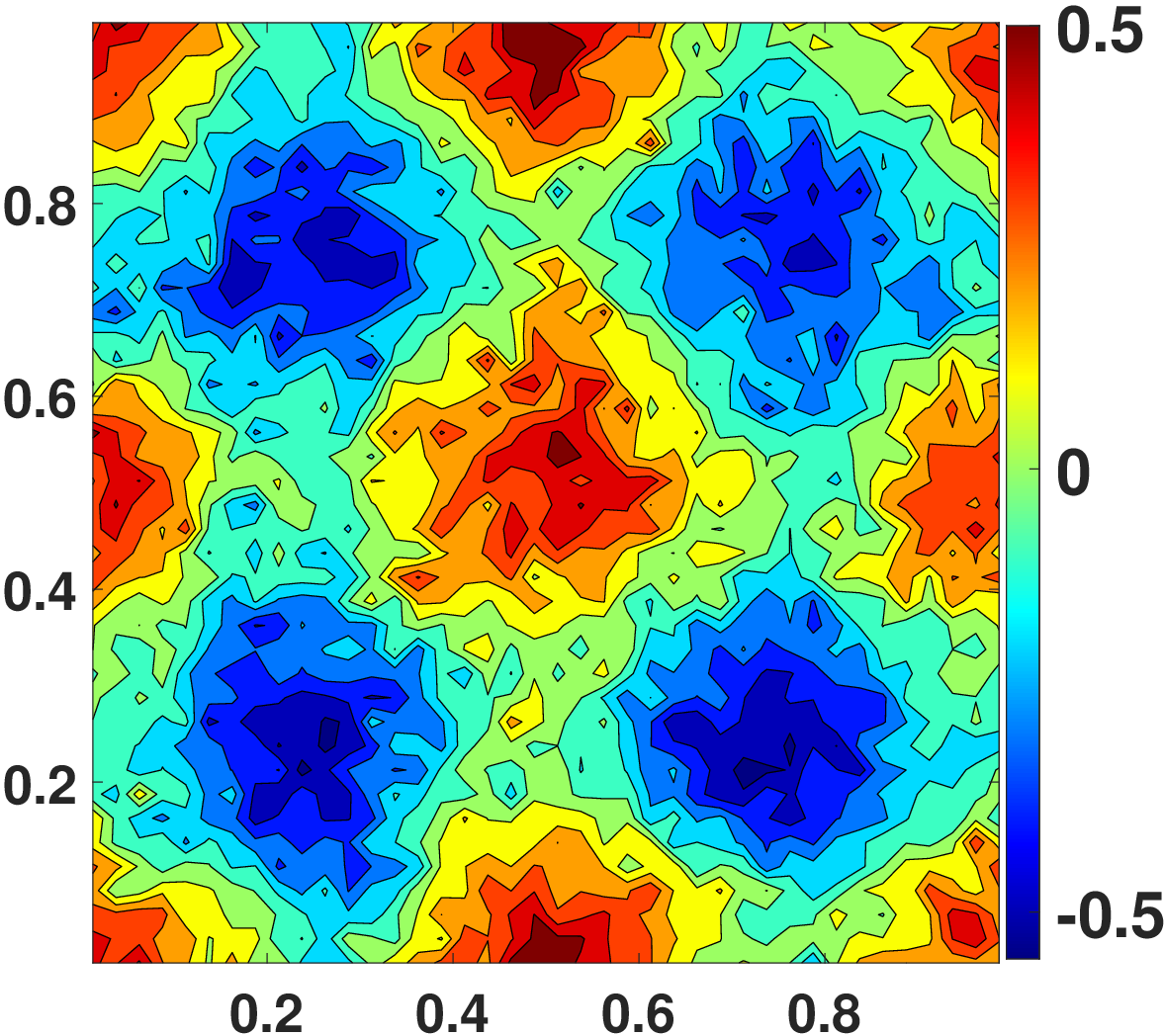}
\end{minipage}%
\begin{minipage}{0.333\textwidth}
\hspace{0.1cm}\includegraphics[scale=0.24]{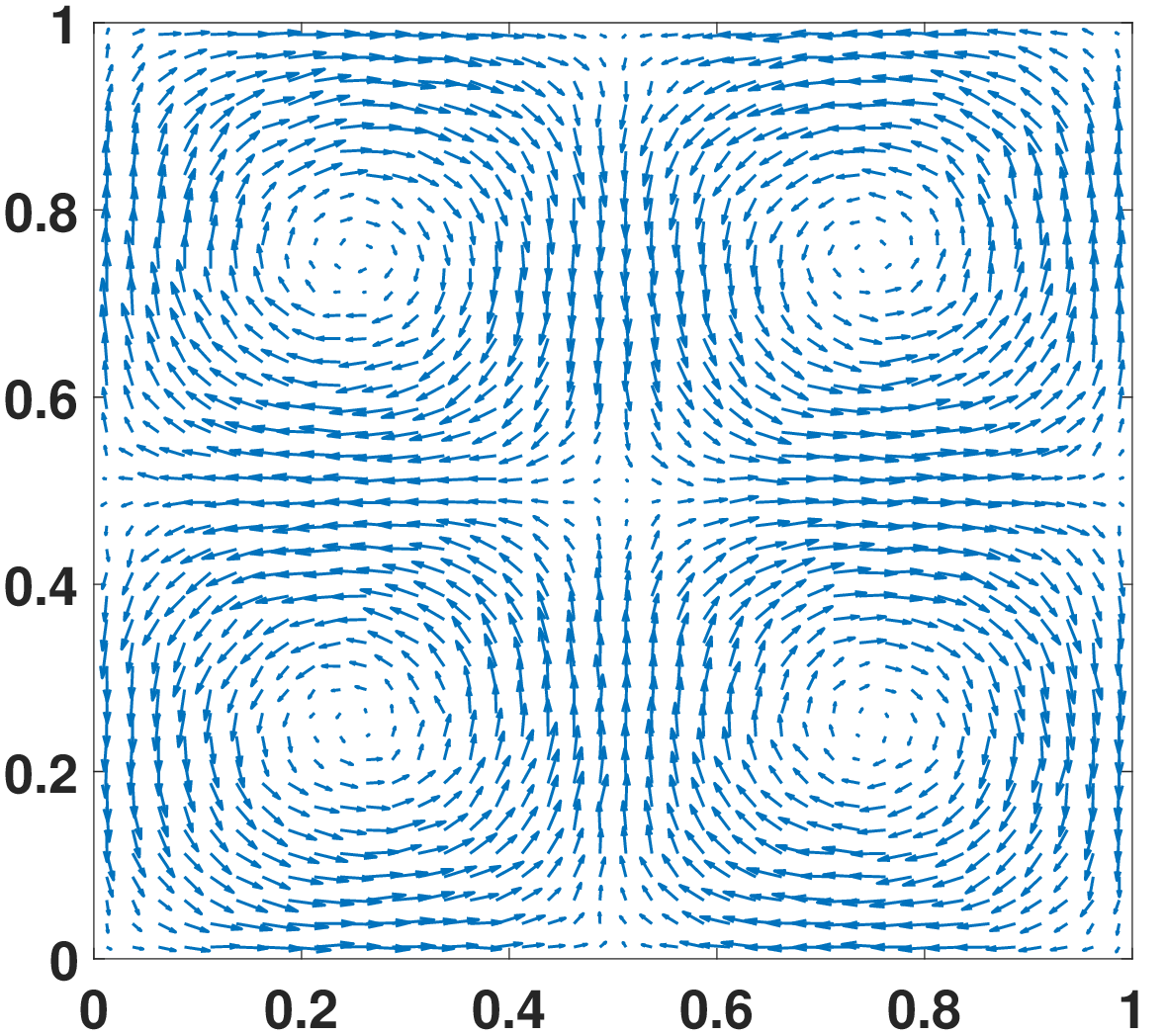}
\end{minipage}%
\begin{minipage}{0.333\textwidth}
\hspace{0.1cm}\includegraphics[scale=0.24]{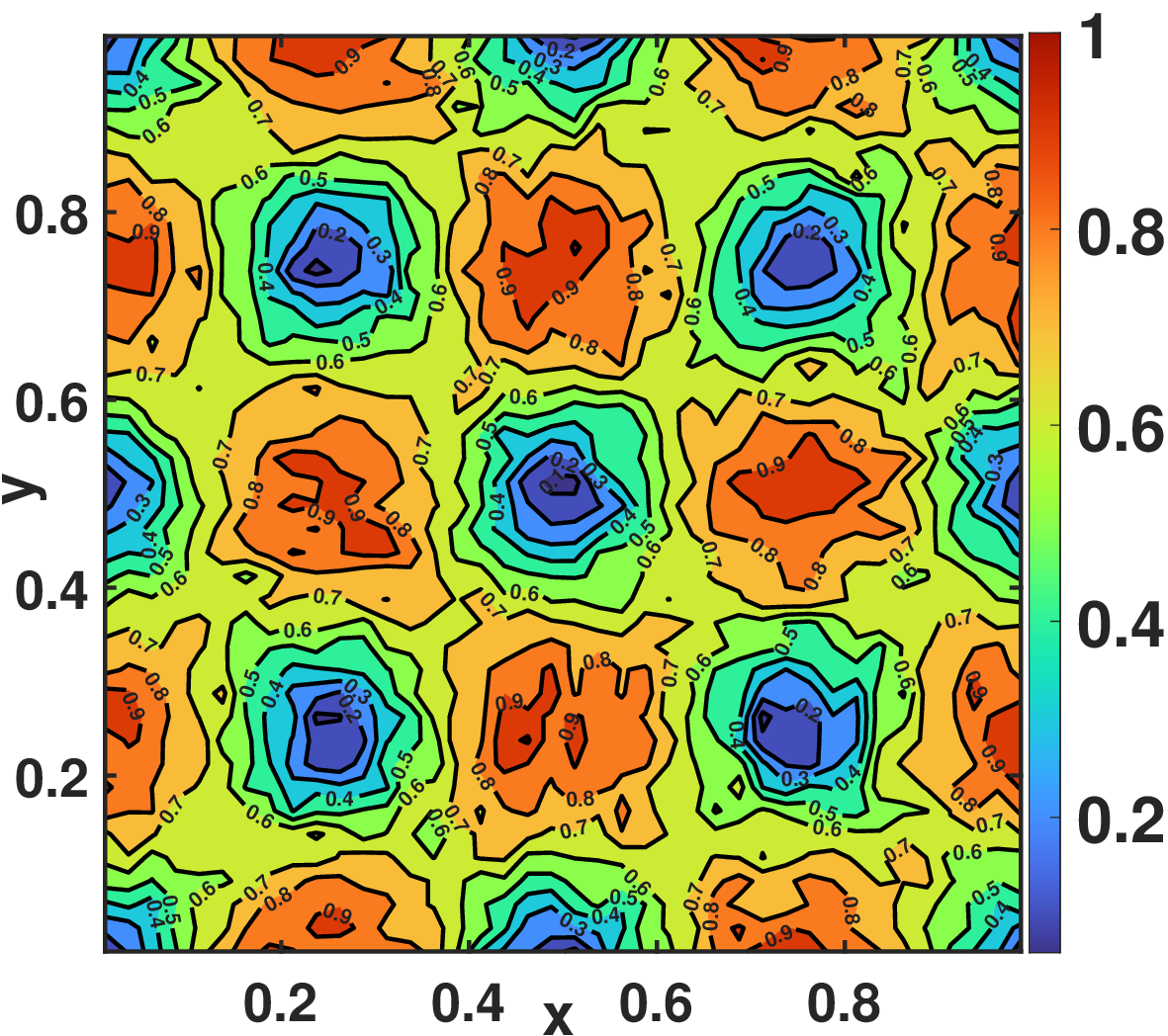}
\end{minipage}
\caption{\small Estimated state for the Taylor-Green vortex problem with with $70\%$ observations and model uncertainty level $\sigma_n^2$. (Left) Pressure field. (Center) Velocity field. (Right) Velocity magnitude.}
\label{TL_EstSol_70Obs_omega2}
\vspace{-0.2cm}
\end{figure}

\begin{figure}[h!]
\centering
\begin{minipage}{0.48\textwidth}
\hspace{1.5cm}\includegraphics[scale=0.25]{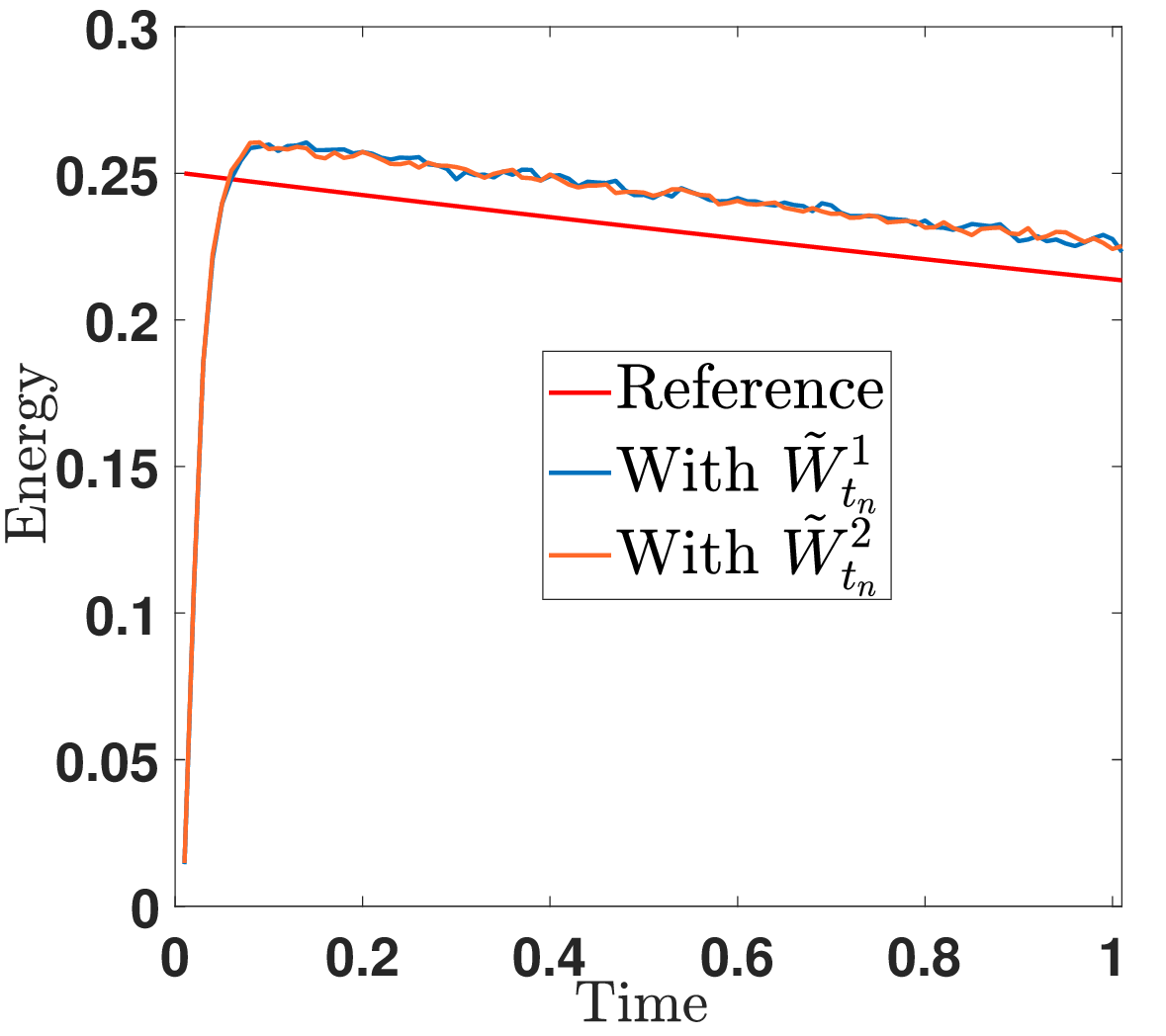}
\end{minipage}%
\begin{minipage}{0.48\textwidth}
\includegraphics[scale=0.26]{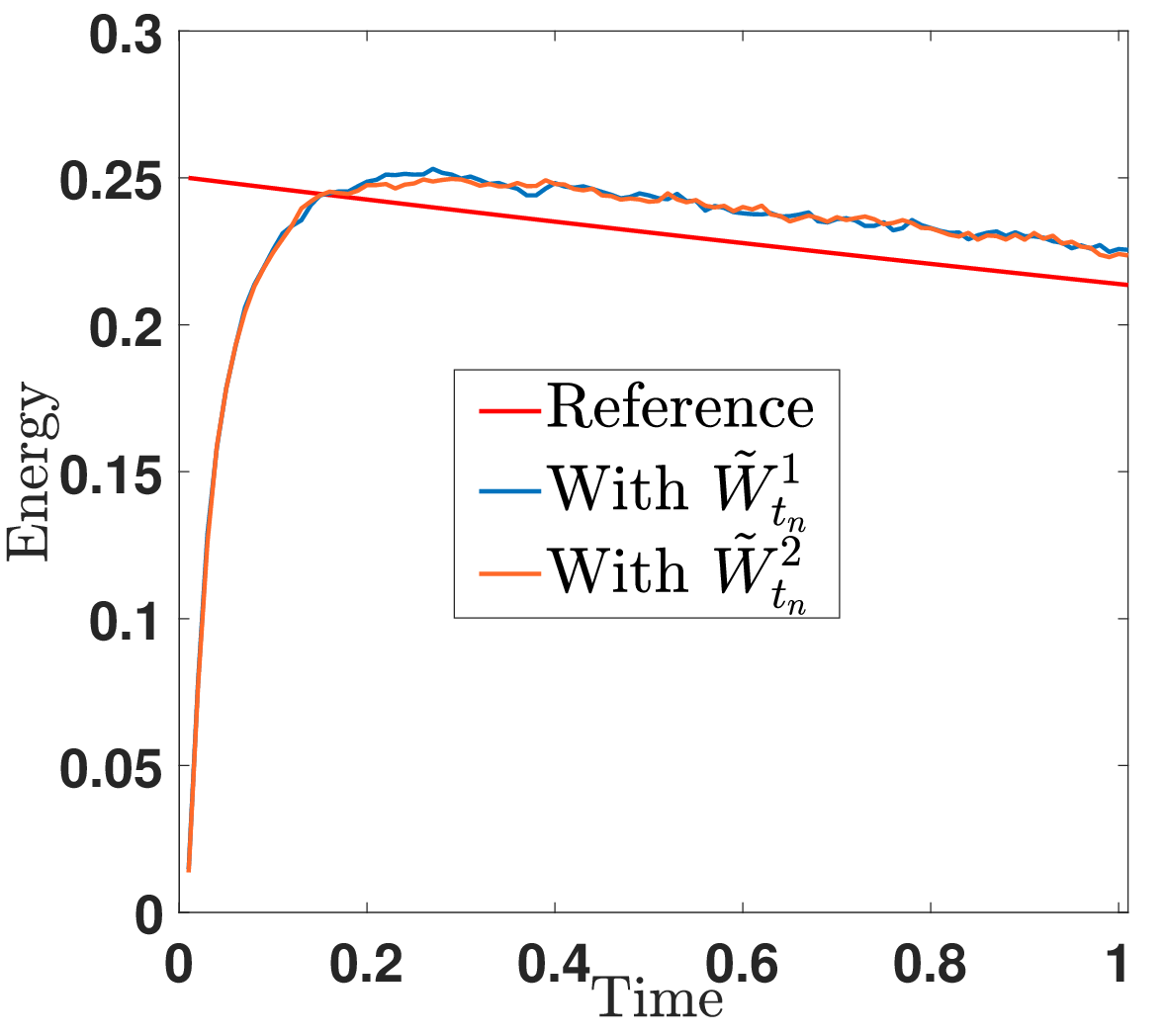}
\end{minipage}
\caption{\small Estimated energies: (Left) $100\%$ Observations. (Right) $70\%$ Observations.}
\label{TL_Engery}
\vspace{-0.2cm}
\end{figure}

% \subsubsection{$100\%$ and $70\%$ arctangent observations}
We begin with presenting the results with $100\%$ and $70\%$ observations. The estimated solutions with $100\%$ amount of observations are displayed in Figure~\ref{TL_EstSol_100Obs_omega1} and Figure~\ref{TL_EstSol_100Obs_omega2}, while those obtained with $70\%$ are illustrated in Figures~\ref{TL_EstSol_70Obs_omega1} and~\ref{TL_EstSol_70Obs_omega2}. The corresponding estimated energies are shown in Figure~\ref{TL_Engery}.

It can be observed that our proposed method recovers accurately the behavior of the reference solution across two types of noise. Moreover, as shown in Figure~\ref{TL_Engery}, although the estimated energy initially deviates from the reference, it quickly converges and begins to decrease after roughly 10 filtering iterations for the $100\%$ observation case and about 20 iterations for the $70\%$ observation case.  

% \subsubsection{$7\%$ arctangent observations}
\begin{figure}[h!]
\centering
\begin{minipage}{0.333\textwidth}
\includegraphics[scale=0.24]{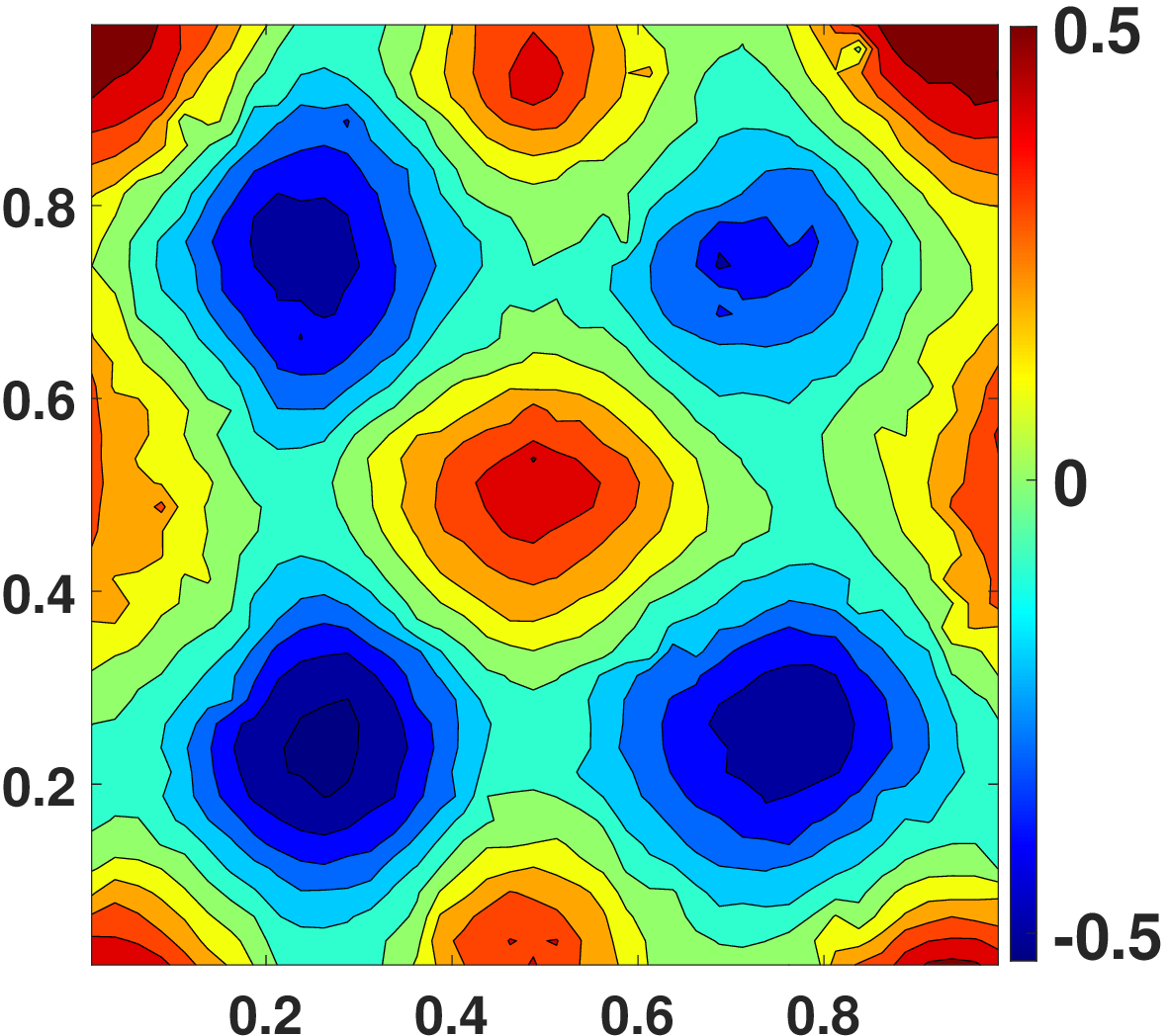}
\end{minipage}%
\begin{minipage}{0.333\textwidth}
\hspace{0.1cm}\includegraphics[scale=0.24]{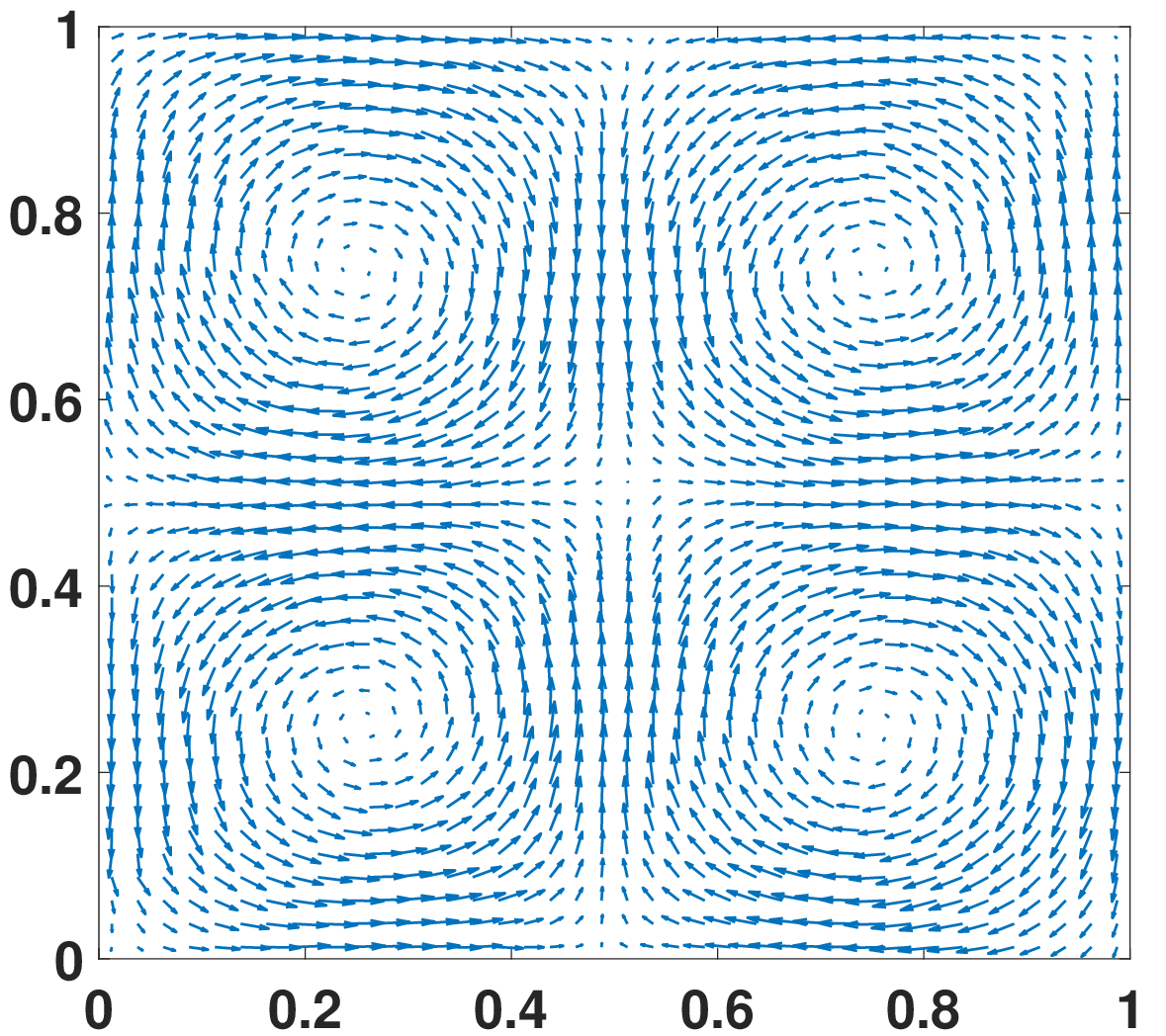}
\end{minipage}%
\begin{minipage}{0.333\textwidth}
\hspace{0.1cm}\includegraphics[scale=0.24]{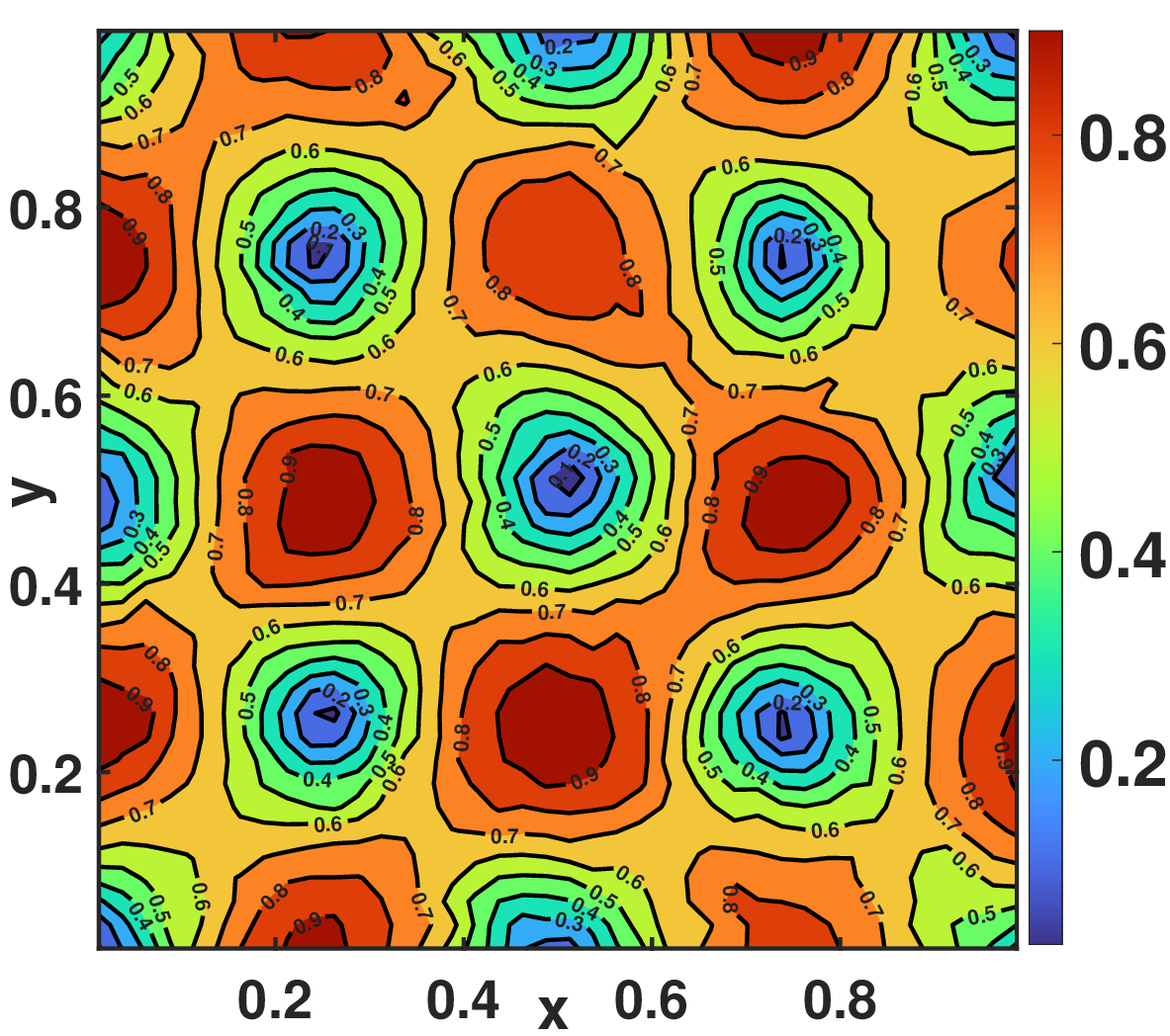}
\end{minipage}
\caption{\small Estimated state for the Taylor-Green vortex problem by the EnSF with $7\%$ observations and model uncertainty level $\tilde{W}_{t_n}^1$. (Left) Pressure field. (Center) Velocity field. (Right) Velocity magnitude.}
\label{TL_EstSol_omega1_BiH_7Obs}
\vspace{-0.4cm}
\end{figure}

\begin{figure}[h!]
\centering
\begin{minipage}{0.333\textwidth}
\includegraphics[scale=0.24]{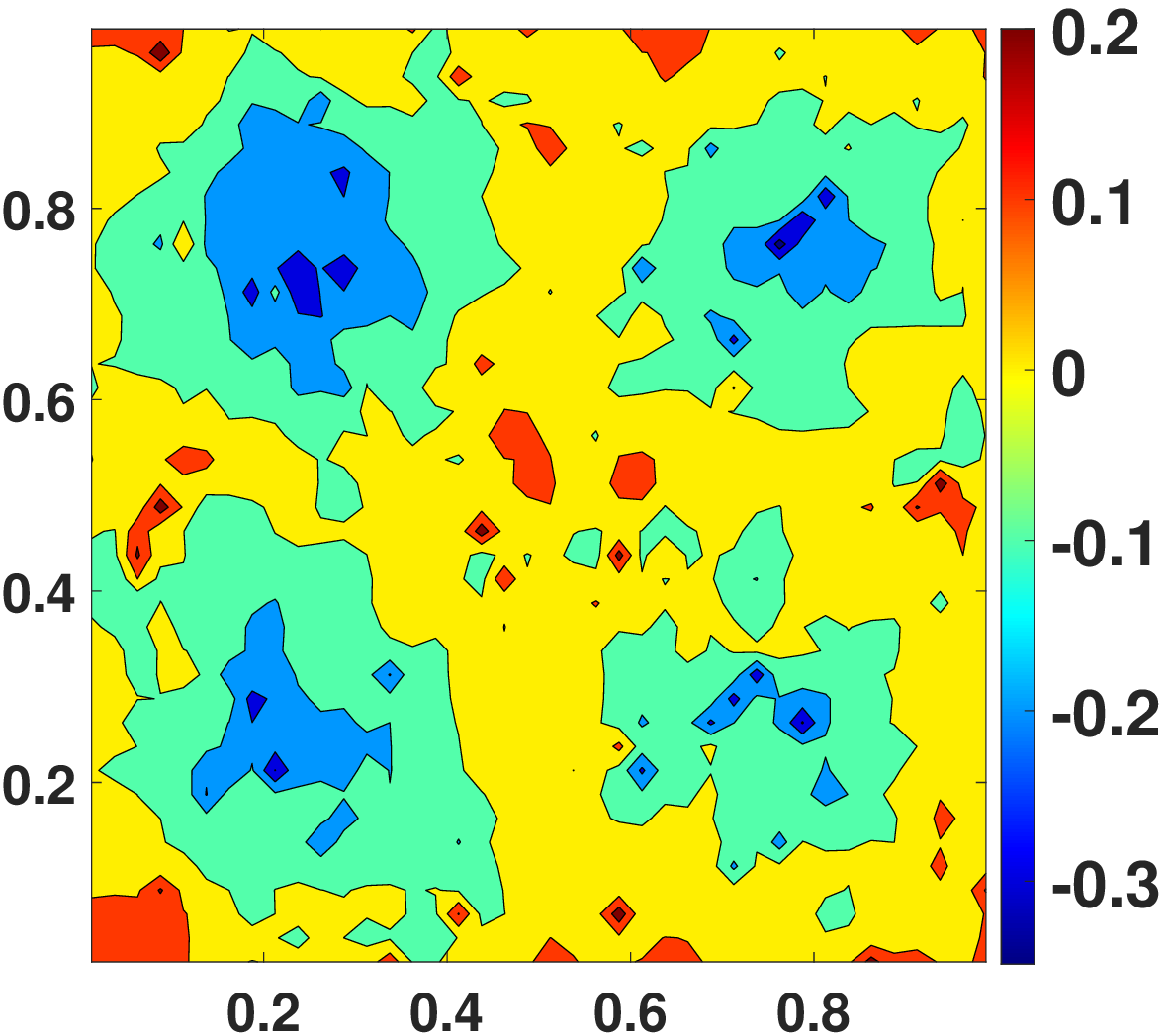}
\end{minipage}%
\begin{minipage}{0.333\textwidth}
\hspace{0.1cm}\includegraphics[scale=0.24]{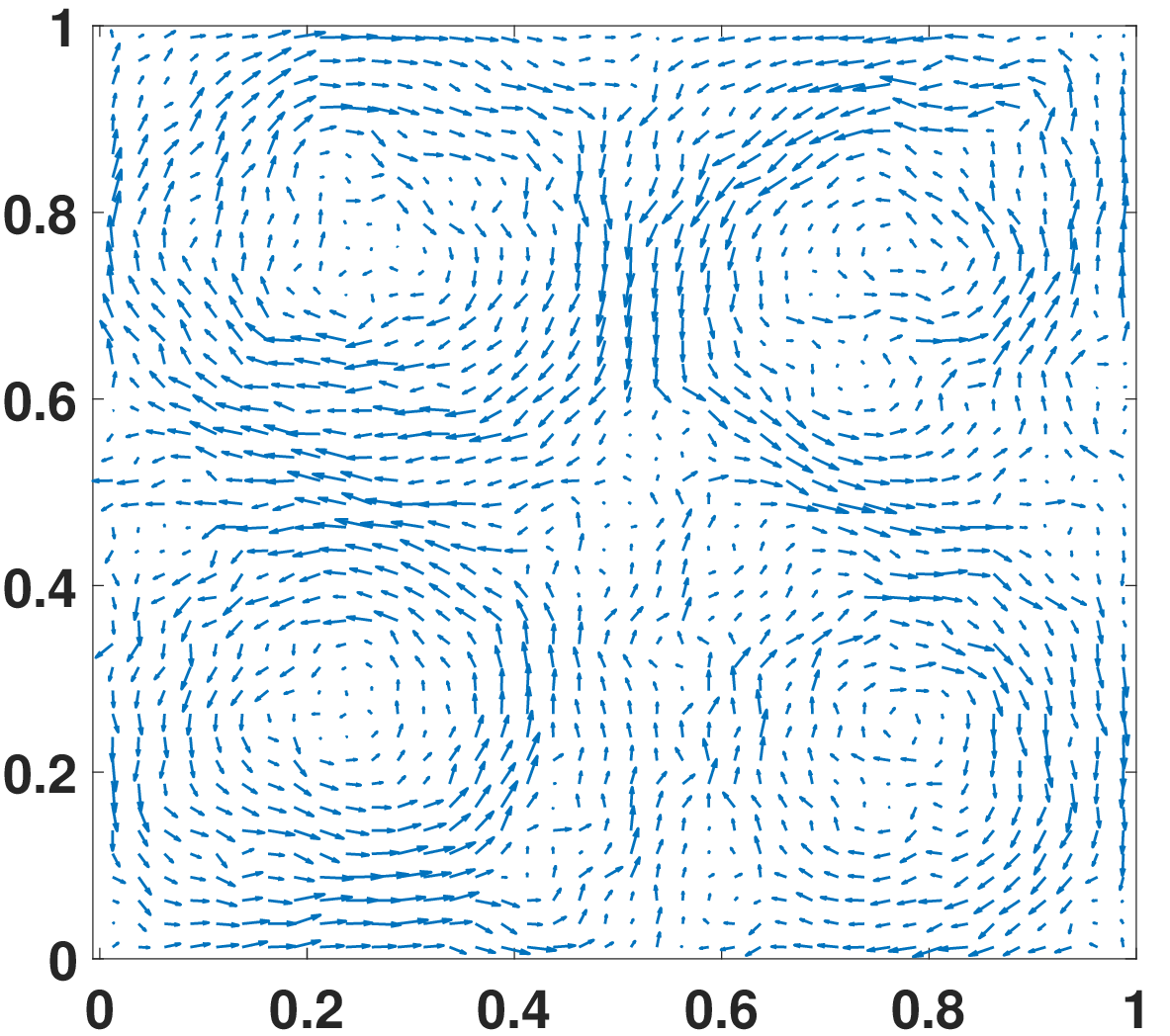}
\end{minipage}%
\begin{minipage}{0.333\textwidth}
\hspace{0.1cm}\includegraphics[scale=0.24]{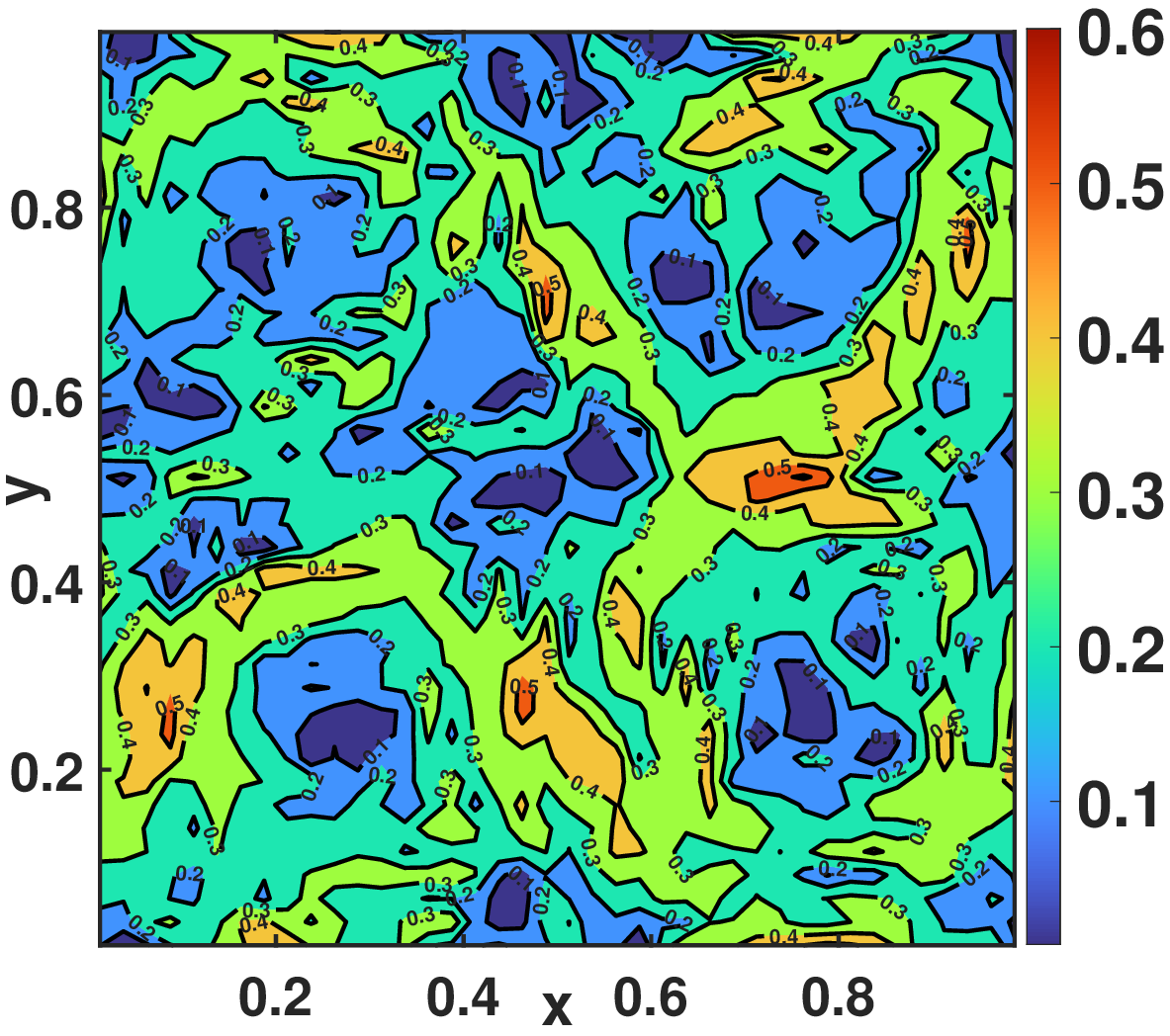}
\end{minipage}
\caption{\small Estimated state for the Taylor-Green vortex problem with $7\%$ observations by LETKF and model uncertainty level $\tilde{W}^1_{t_n}$. (Left) Pressure field. (Center) Velocity field. (Right) Velocity magnitude.}
\label{TL_EstSol_omega1_LETKF_7Obs}
\vspace{-0.4cm}
\end{figure}

\begin{figure}[h!]
\centering
\begin{minipage}{0.48\textwidth}
\hspace{1.5cm}\includegraphics[scale=0.25]{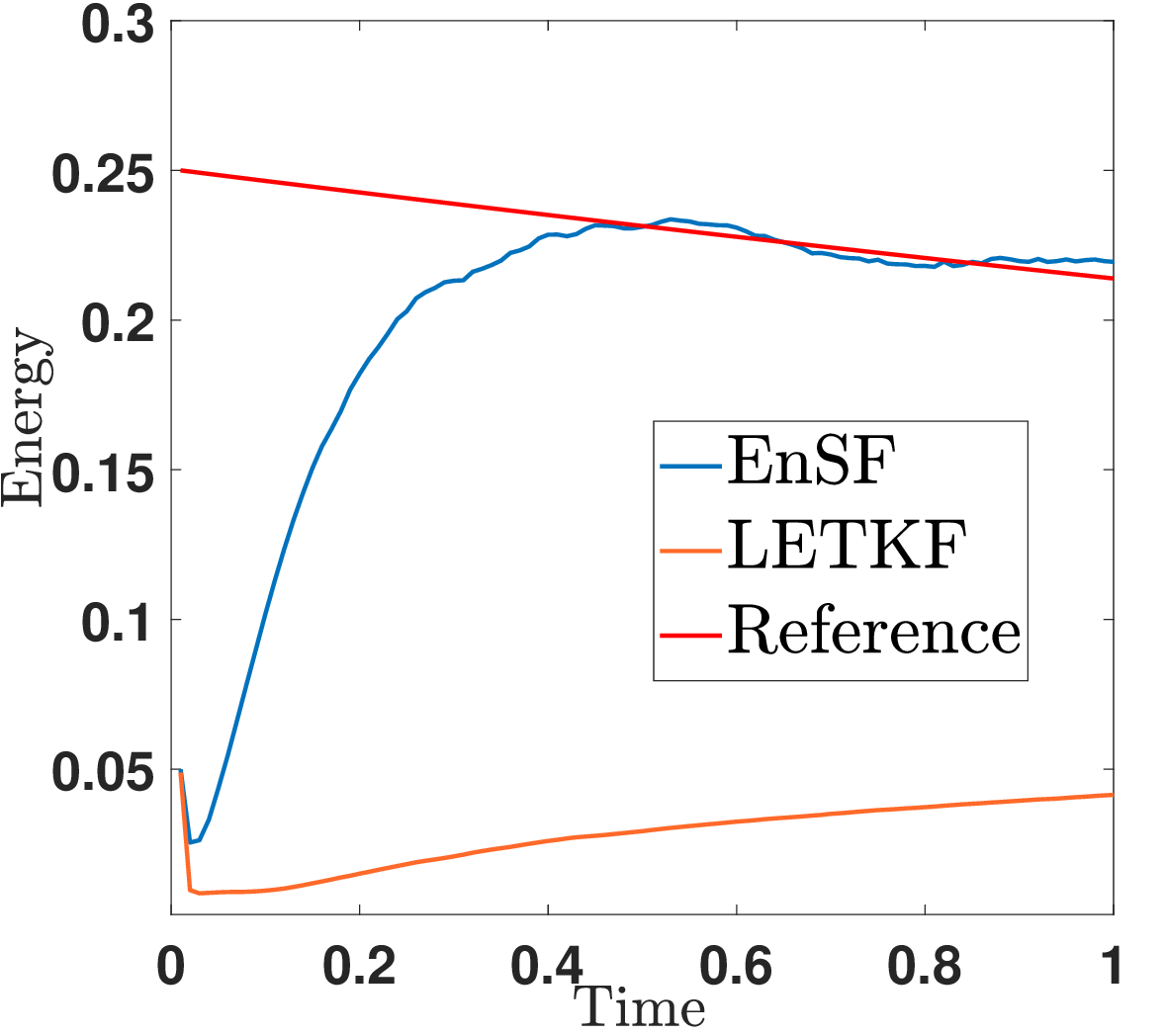}
\end{minipage}%
\begin{minipage}{0.48\textwidth}
\includegraphics[scale=0.26]{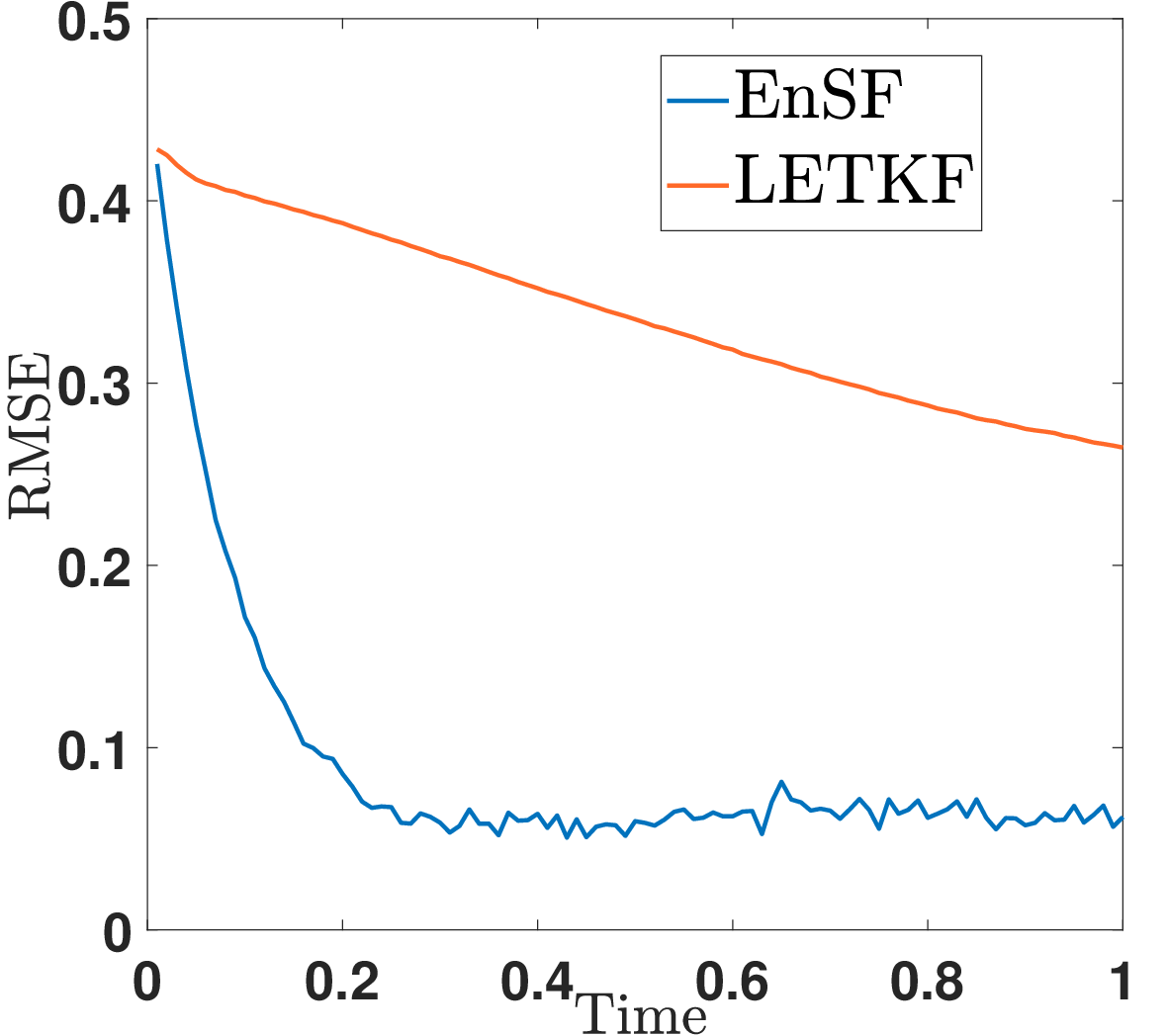}
\end{minipage}
\caption{\small  Estimated energies (Left) and RMSEs for state estimation (Right) with $7\%$ observations and noise $\tilde{W}^1_{t_n}$.}
\label{TL_Engery_omega1_7Obs}
\vspace{-0.4cm}
\end{figure}

\begin{figure}[h!]
\centering
\begin{minipage}{0.333\textwidth}
\includegraphics[scale=0.24]{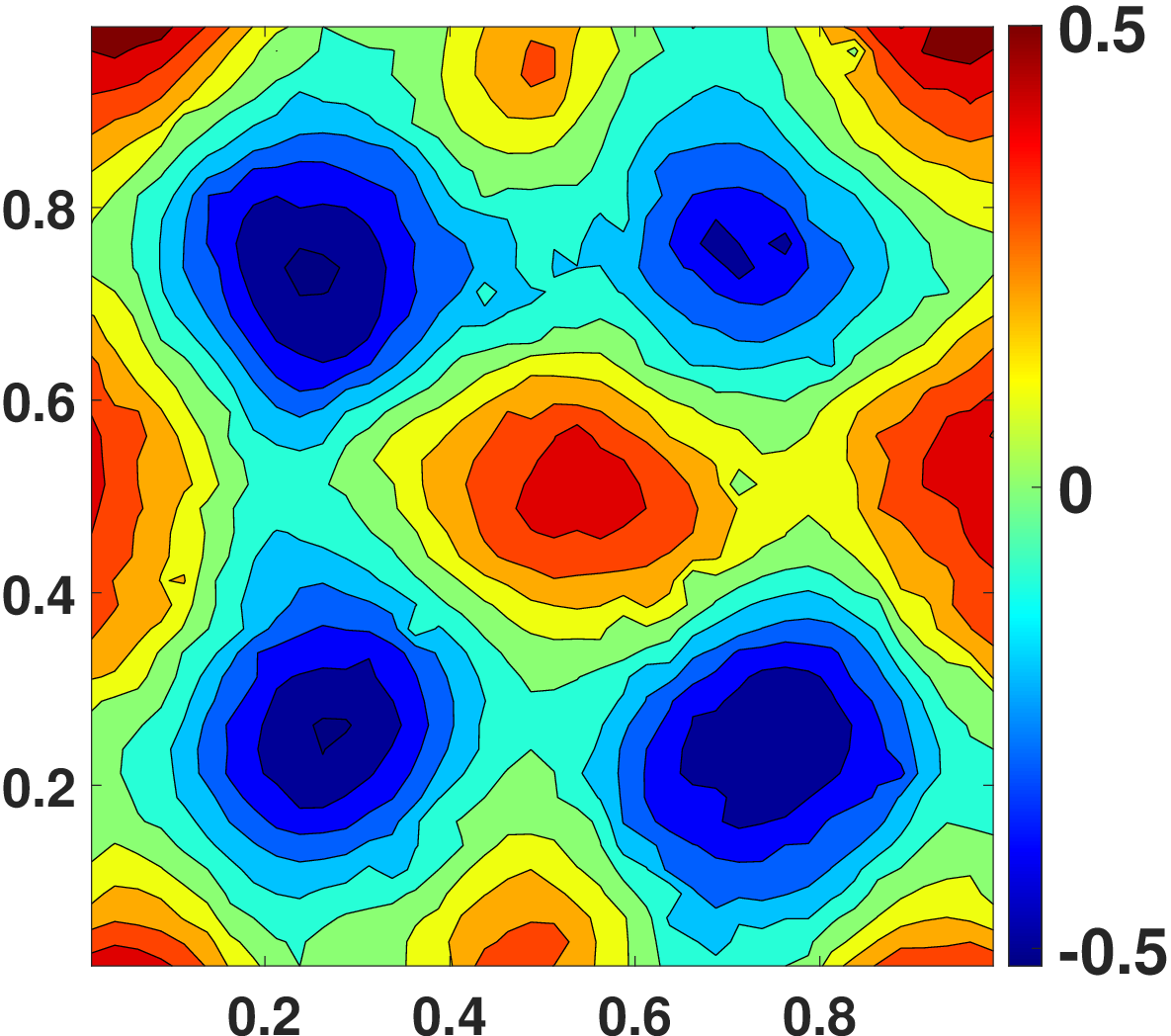}
\end{minipage}%
\begin{minipage}{0.333\textwidth}
\hspace{0.1cm}\includegraphics[scale=0.24]{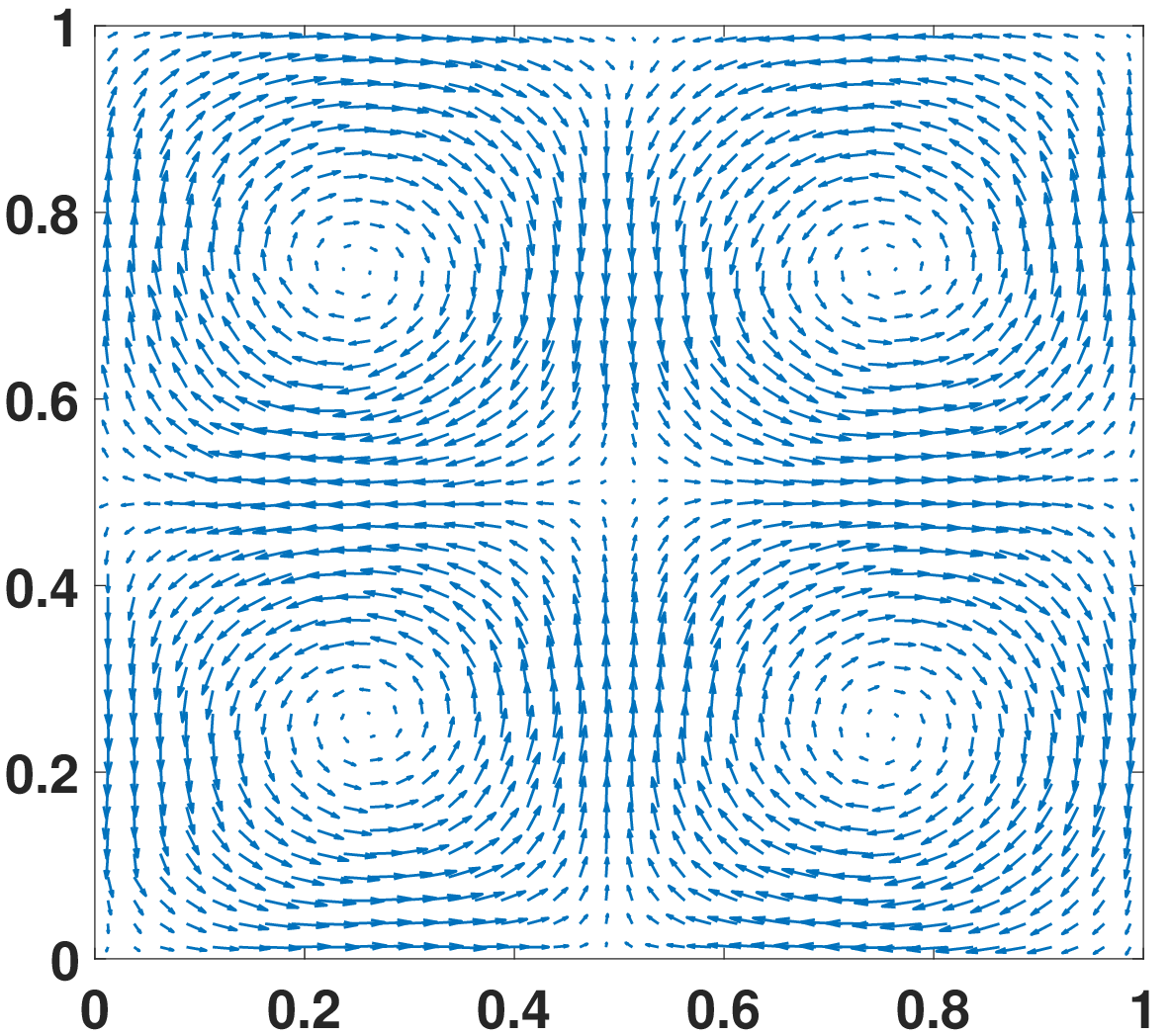}
\end{minipage}%
\begin{minipage}{0.333\textwidth}
\hspace{0.1cm}\includegraphics[scale=0.24]{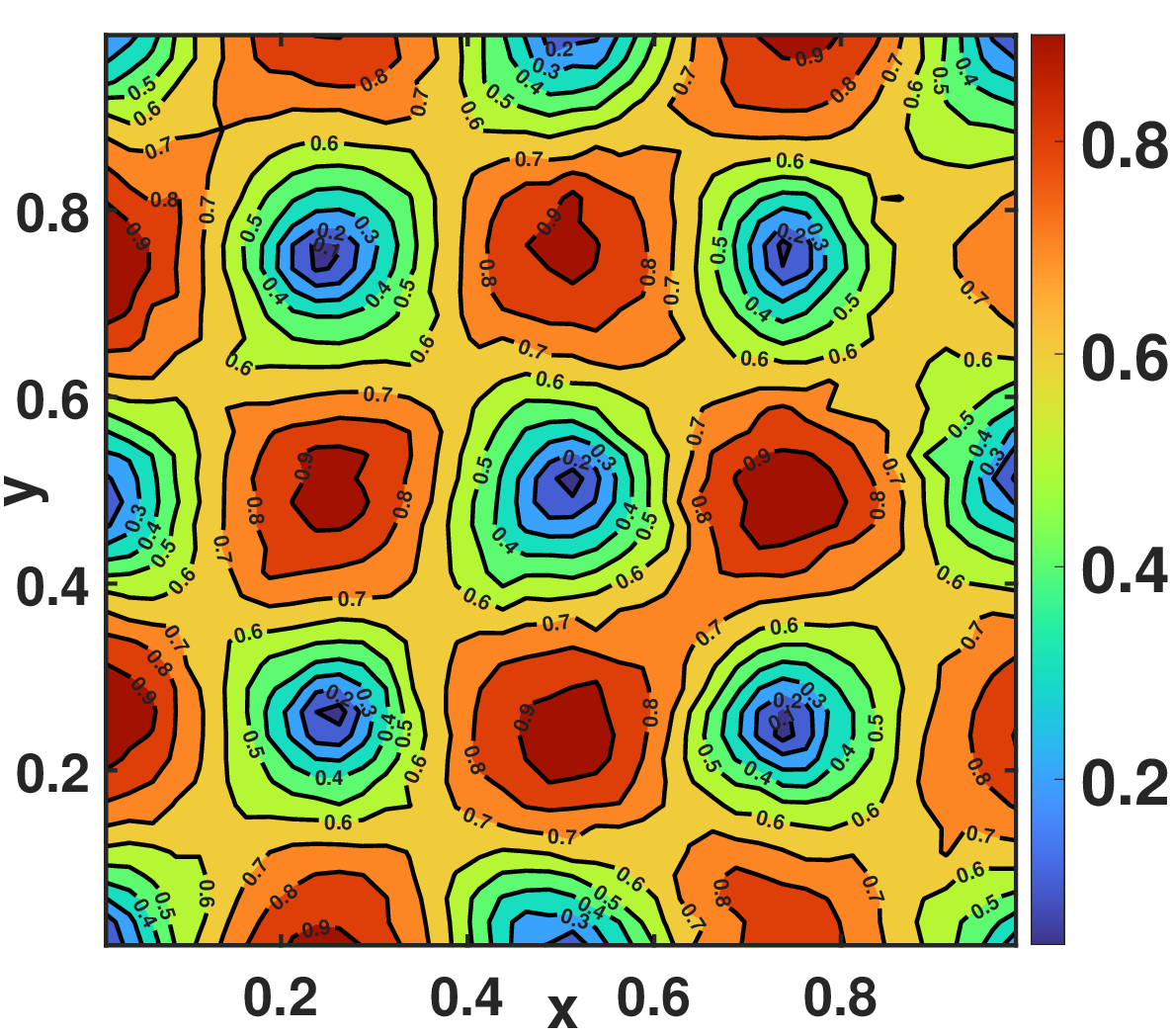}
\end{minipage}
\caption{\small Estimated state for the Taylor-Green vortex problem by the EnSF with $7\%$ observations and model uncertainty level $\tilde{W}_{t_n}^2$. (Left) Pressure field. (Center) Velocity field. (Right) Velocity magnitude. }
\label{TL_EstSol_omega2_BiH_7Obs}
\vspace{-0.4cm}
\end{figure}

\begin{figure}[h!]
\centering
\begin{minipage}{0.333\textwidth}
\includegraphics[scale=0.24]{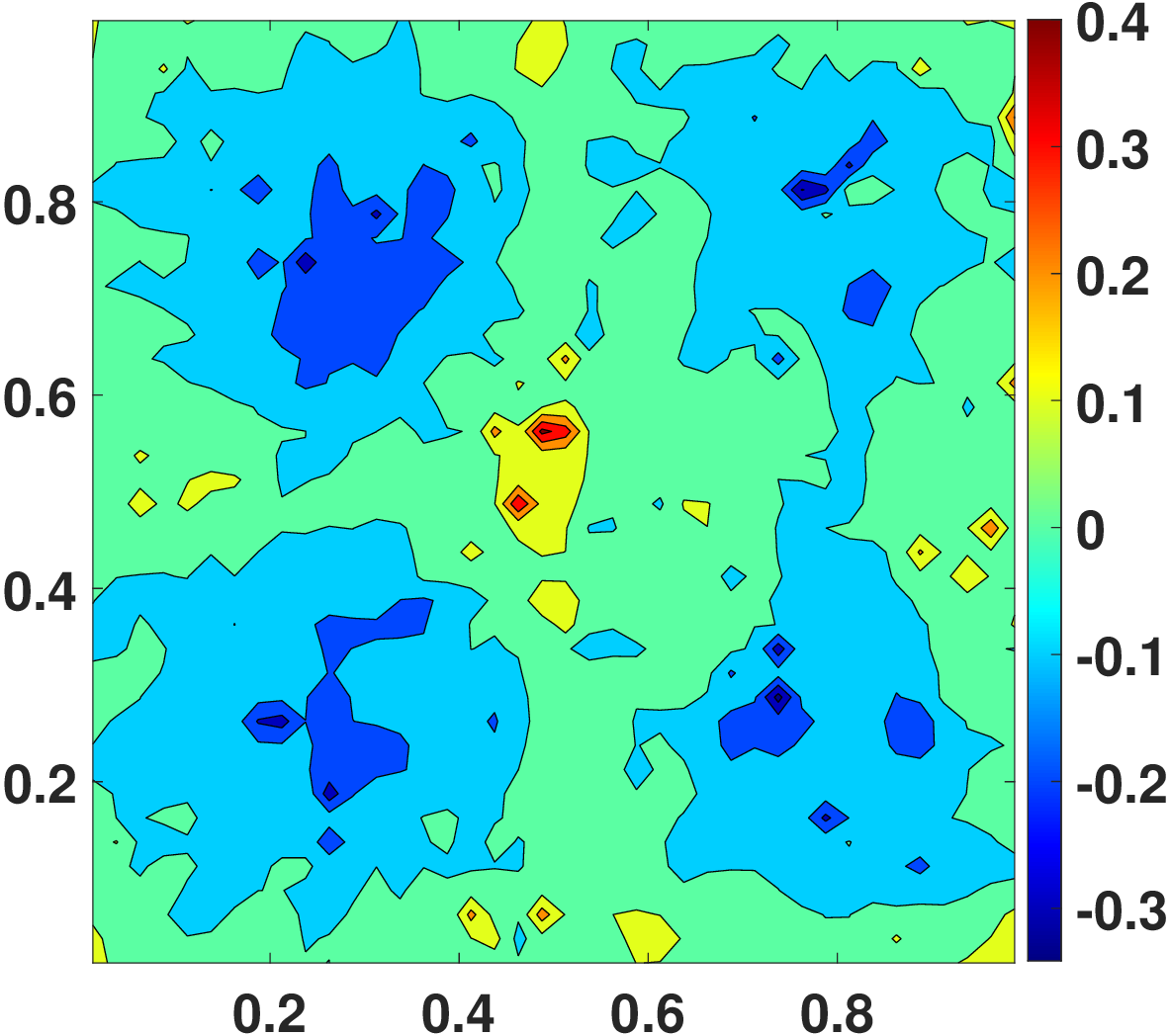}
\end{minipage}%
\begin{minipage}{0.333\textwidth}
\hspace{0.1cm}\includegraphics[scale=0.24]{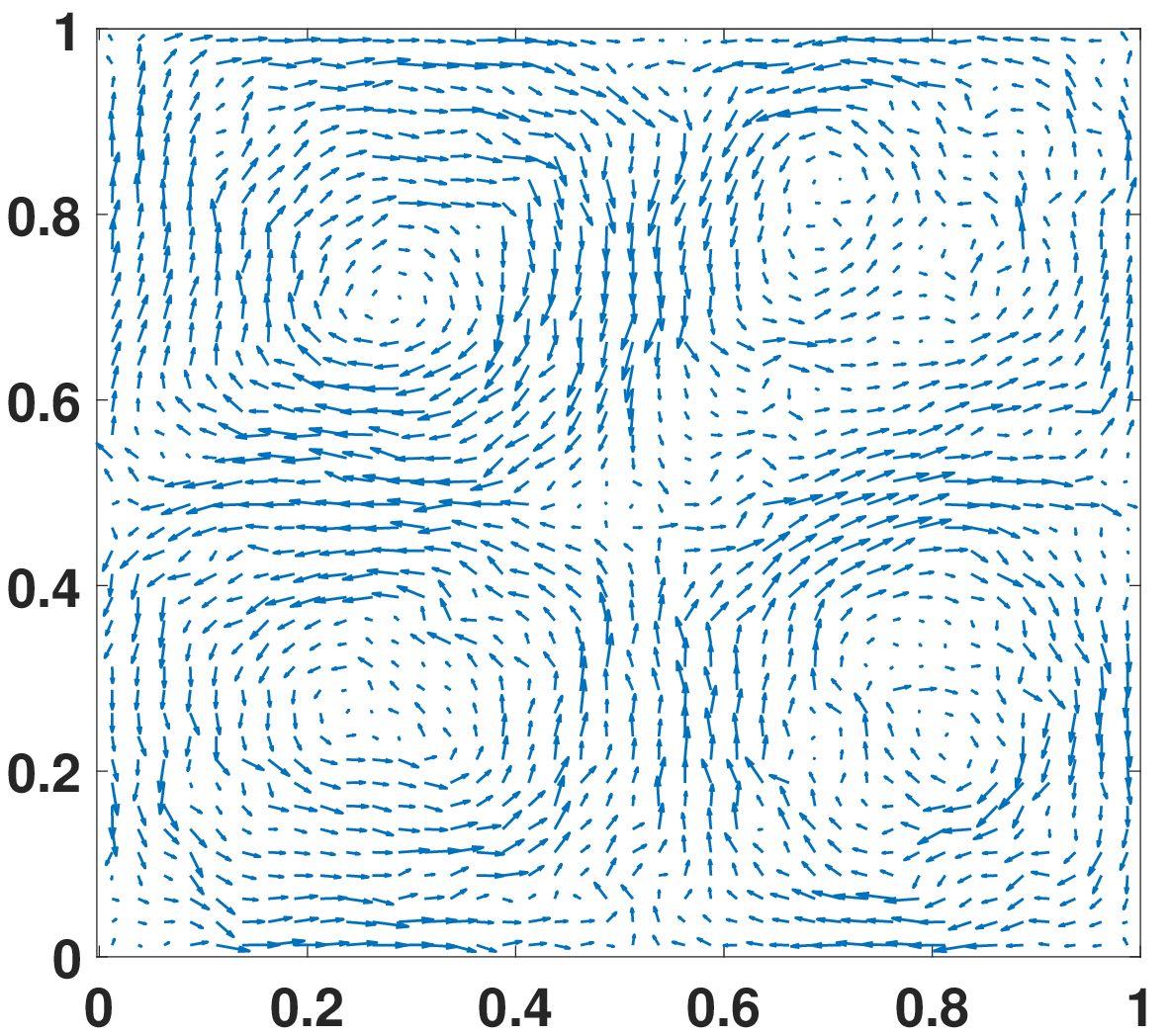}
\end{minipage}%
\begin{minipage}{0.333\textwidth}
\hspace{0.1cm}\includegraphics[scale=0.24]{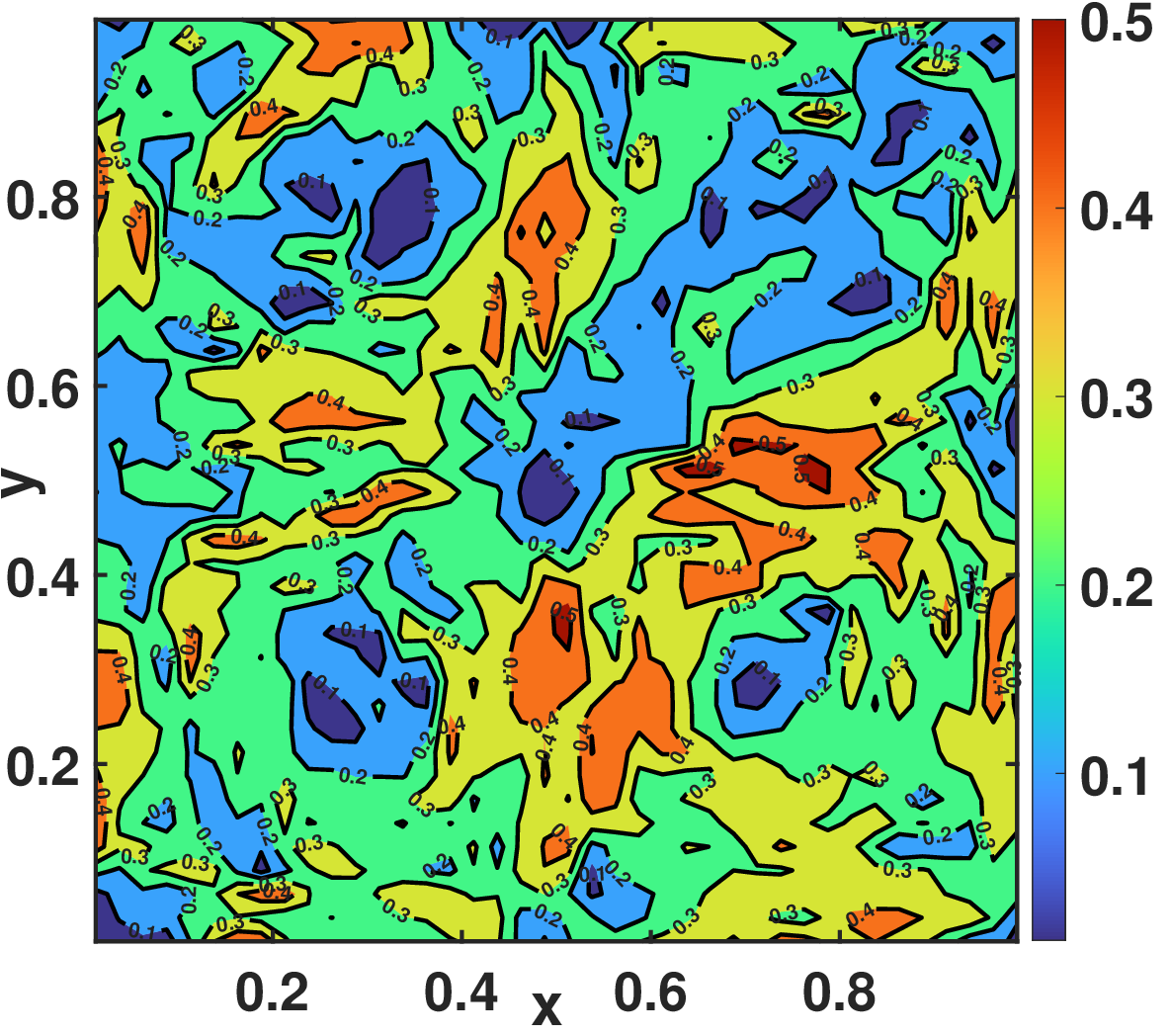}
\end{minipage}
\caption{\small Estimated state with $7\%$ observations by LETKF and noise $\tilde{W}_{t_n}^2$ (First) Pressure field. (Right) Velocity field. (Third) Velocity magnitude.}
\label{TL_EstSol_omega2_LETKF_7Obs}
\vspace{-0.4cm}
\end{figure}

\begin{figure}[h!]
\centering
\begin{minipage}{0.48\textwidth}
\hspace{1.5cm}\includegraphics[scale=0.25]{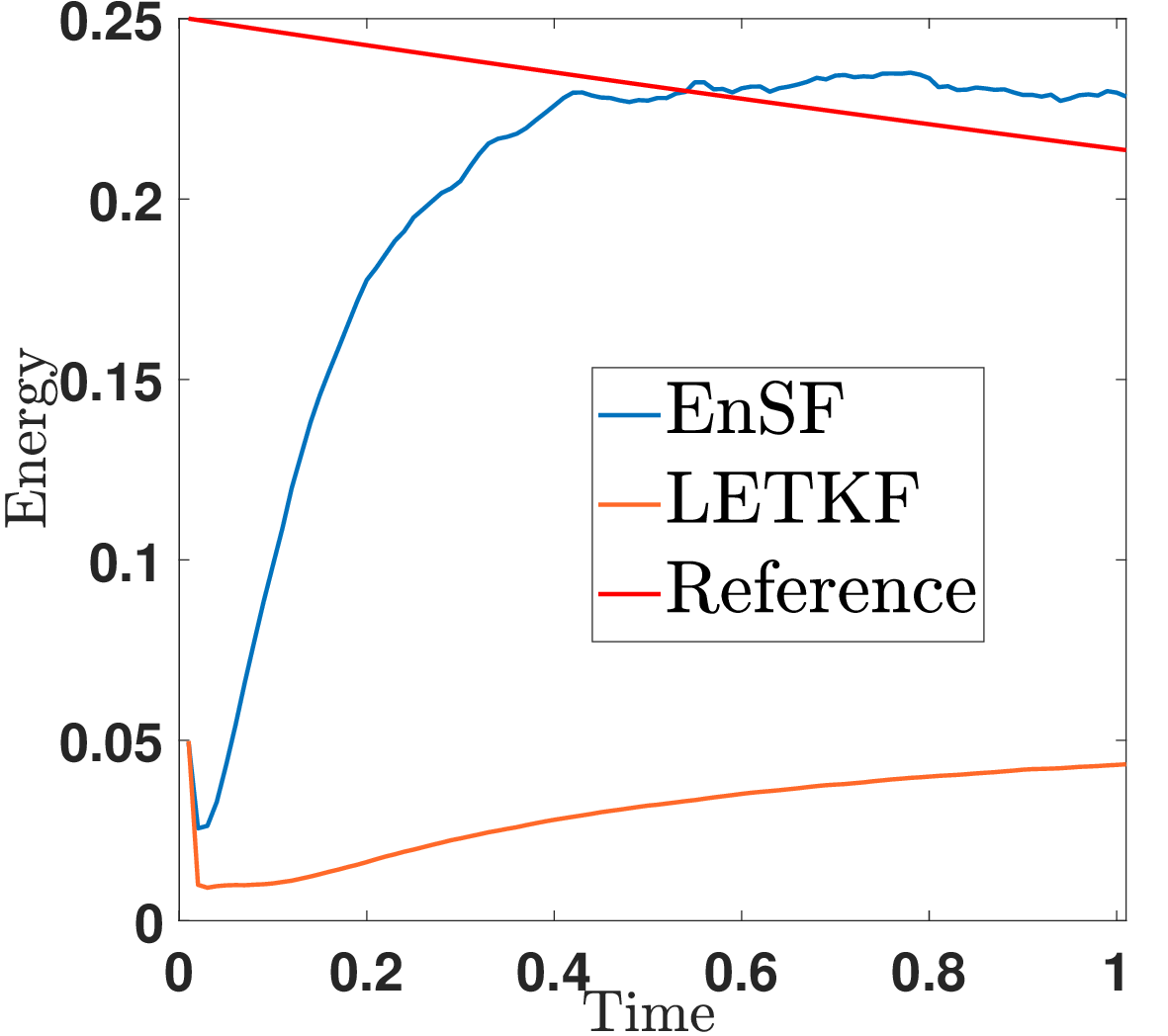}
\end{minipage}%
\begin{minipage}{0.48\textwidth}
\includegraphics[scale=0.26]{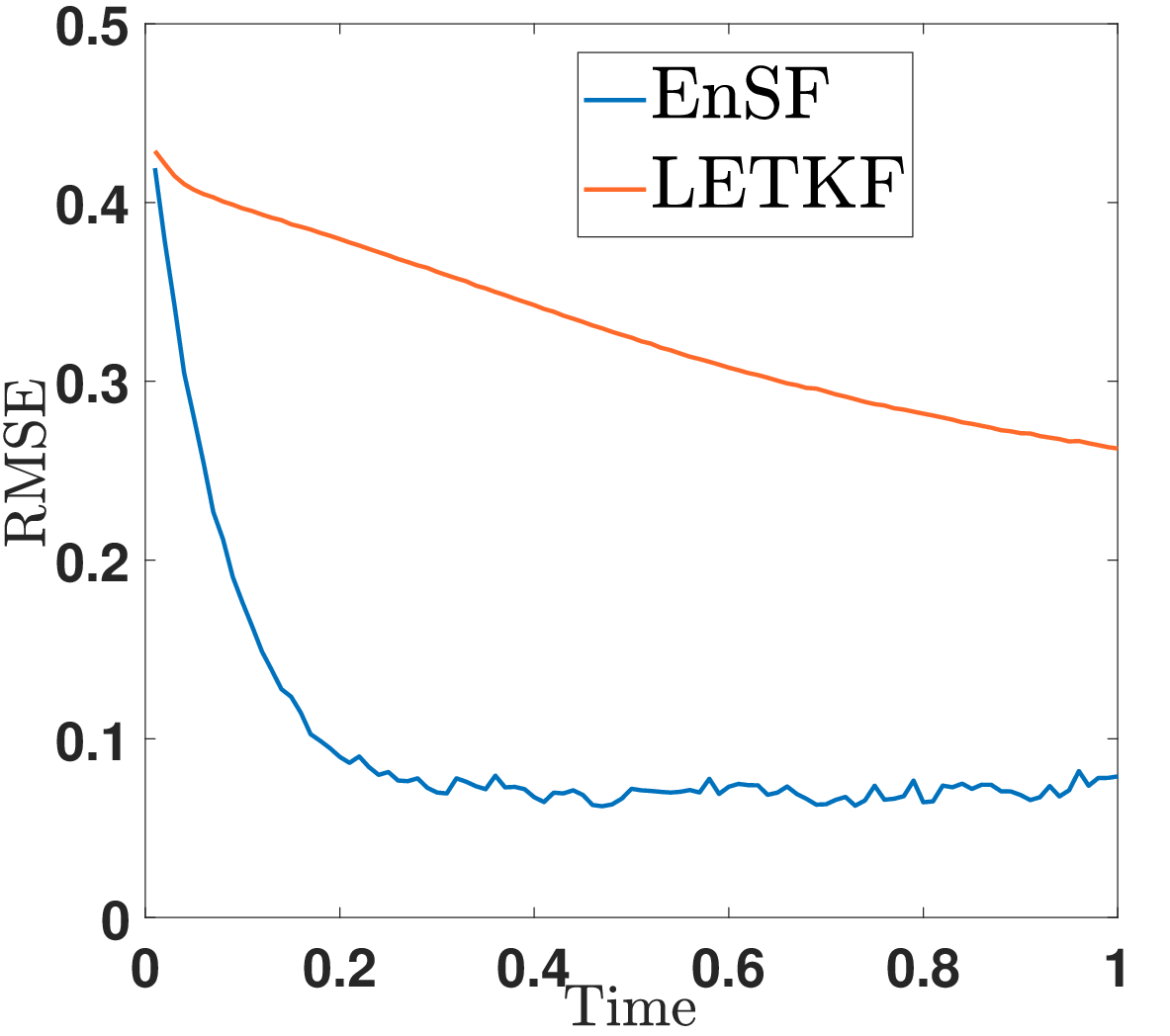}
\end{minipage}
\caption{\small Estimated energies (Left) and RMSEs for state estimations (Right) with $7\%$ observations and noise $\tilde{W}_{t_n}^2$.}
\label{TL_Engery_omega2_7Obs}
\vspace{-0.4cm}
\end{figure}

We conclude the Taylor-Green vortex test case by presenting the results with only $7\%$ observations. Due to the sparsity of observations, the Bi-harmonic inpainting technique was also applied to the EnSF in this experiment. The results corresponding to the noise $\tilde{W}^1_{t_n}$ are displayed in Figures~\ref{TL_EstSol_omega1_BiH_7Obs},~\ref{TL_EstSol_omega1_LETKF_7Obs}, and~\ref{TL_Engery_omega1_7Obs}. Specifically, Figure \ref{TL_EstSol_omega1_BiH_7Obs} illustrates the performance of the EnSF in state estimation, Figure \ref{TL_EstSol_omega1_LETKF_7Obs} shows the performance of the LETKF in state estimation, and Figure \ref{TL_Engery_omega1_7Obs} presents comparisons of the estimated energies and the RMSEs for state estimation. Similarly, results for the noise level $\tilde{W}^2_{t_n}$ are shown in Figures~\ref{TL_EstSol_omega2_BiH_7Obs},~\ref{TL_EstSol_omega2_LETKF_7Obs}, and~\ref{TL_Engery_omega2_7Obs}, following the same structure: EnSF estimates in Figure~\ref{TL_EstSol_omega2_BiH_7Obs}, LETKF estimates in Figure~\ref{TL_EstSol_omega2_LETKF_7Obs}, and energy/RMSE comparisons in Figure~\ref{TL_Engery_omega2_7Obs}.

\vspace{0.5em}

From the estimated solution figures (e.g., Figures~\ref{TL_EstSol_omega1_BiH_7Obs}, \ref{TL_EstSol_omega1_LETKF_7Obs}, \ref{TL_EstSol_omega2_BiH_7Obs}, and \ref{TL_EstSol_omega2_LETKF_7Obs}), we observe that the EnSF consistently produces accurate estimates, regardless of noise level. This accuracy is further confirmed by the RMSE plots in the second panels of Figures~\ref{TL_Engery_omega1_7Obs} and~\ref{TL_Engery_omega2_7Obs}, which show significantly lower errors for EnSF compared to LETKF. Specifically, the pressure and velocity fields estimated by the EnSF closely match the reference solution across most of the domain. In contrast, the LETKF provides only a coarse approximation of the velocity field and fails to recover the pressure field accurately. 

Finally, from the energy plots in the first panels of Figures~\ref{TL_Engery_omega1_7Obs} and~\ref{TL_Engery_omega2_7Obs}, it is clear that the LETKF fails to preserve the expected energy dissipation behavior. In contrast, the EnSF estimates closely follow the reference energy curve and exhibit correct dissipative trends. 


\subsubsection{External Force-Driven Flow}

\begin{figure}[h!]
\centering
\begin{minipage}{0.3\textwidth}
\includegraphics[scale=0.24]{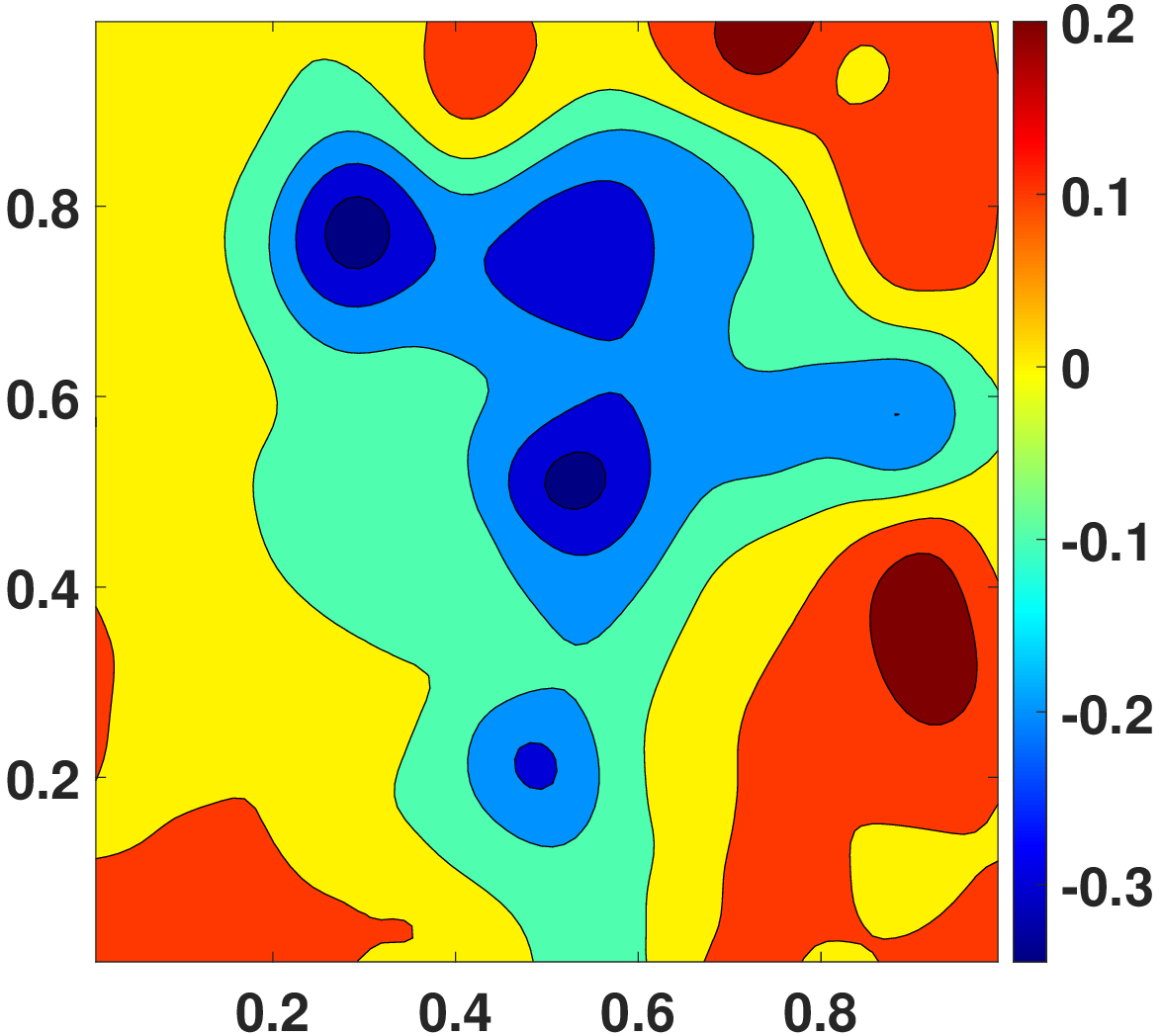}
\end{minipage}%
\begin{minipage}{0.3\textwidth}
\hspace{0.1cm}\includegraphics[scale=0.24]{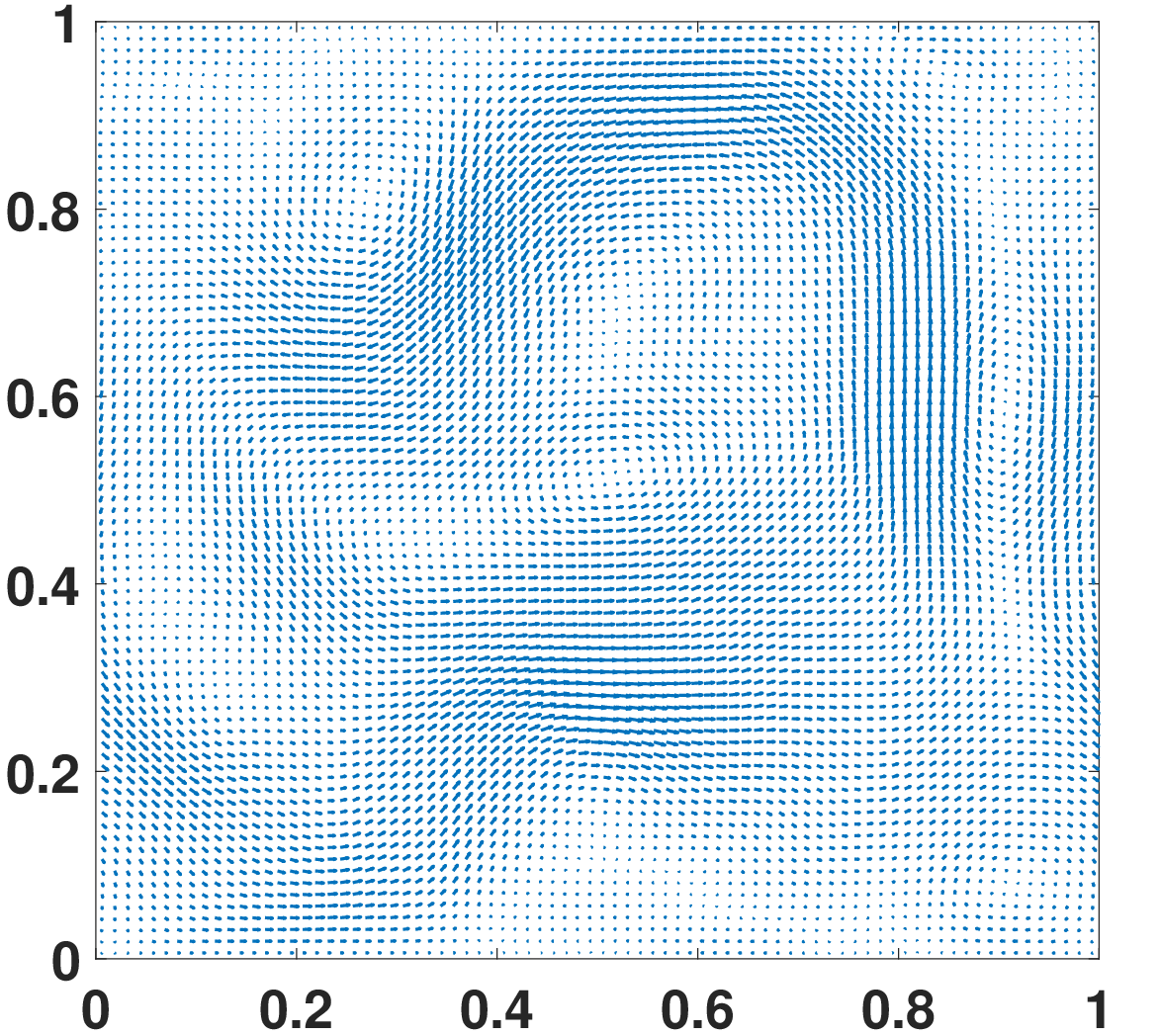}
\end{minipage}
\begin{minipage}{0.3\textwidth}
\hspace{0.1cm}\includegraphics[scale=0.24]{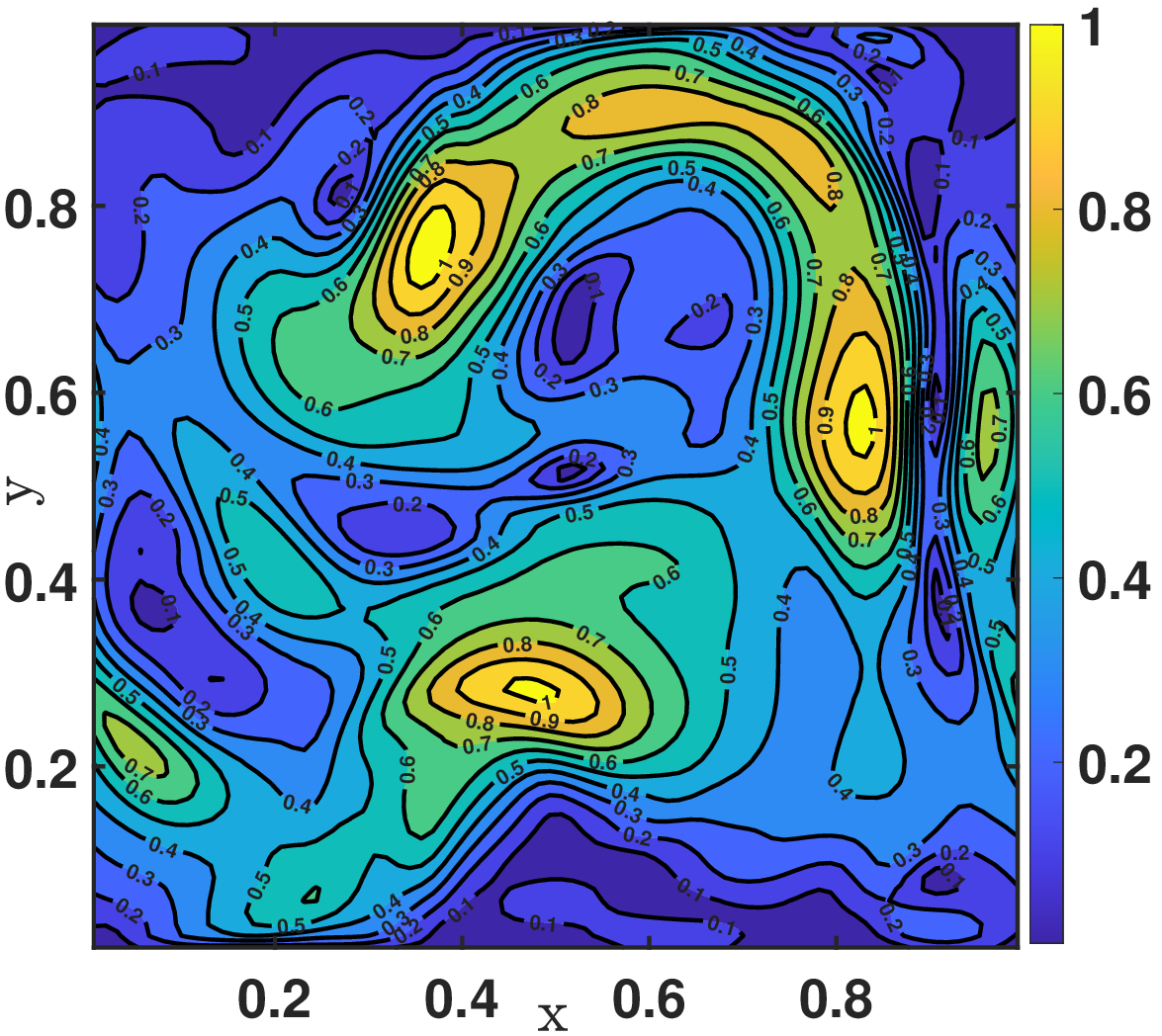}
\end{minipage}
\caption{\small Reference solution for the NS equation with external force-driven flow. (Left) Pressure field. (Center) Velocity field. (Right) Velocity magnitude.}
\label{RHSDriven_opt1_Ref}
\vspace{-0.2cm}
\end{figure}

In the second Navier-Stokes test case, we consider a more challenging setting by increasing the Reynolds number to  $\text{Re} = 2000$, which results in a more chaotic flow behavior. We consider the same domain $\Omega =(0, 1)^2$ with no-slip boundary conditions on the top and bottom edges and periodic boundary conditions on the two lateral sides. The external body force $\pmb{f}$ is given by $\pmb{f} = \left[f_1, f_2\right]^T$ where $f_1 = -0.5\cos(8\pi{x})$, $f_2 = 0.5\sin(8\pi{x})$. Throughout the experiment, we fix the mesh size at $h=1/80$ and $\theta = 100$. The final time $T$ is $T=15$, and the time step size for the reference solution is $\Delta{t} = T/8000$. The reference solutions are shown in Figure~\ref{RHSDriven_opt1_Ref}.

To carry out the prediction step in data assimilation, we solve the PDE using a coarser time step of $\Delta{t}_{\text{Filter}} = T / 2000$. The observational data coverage is set to $20\%$, and we assume an initial guess for the PDE model given by $\pmb{u}_0\sim 2\cdot N(0, \pmb{I}_d)$. At each data assimilation step, the external forcing term $\pmb{f}$ is perturbed by a small noise $\tilde{\omega}$, sampled from $0.0001 \cdot N(0, \pmb{I}_d)$. This test case differs slightly from the previous two, as it accounts not only for the additive term $\sigma_n\tilde{W}_{t_n}$, but also for noise in the external forcing. This introduces a minor modification to the model solver: at each filtering step, the solver applies the stiffness matrix to the perturbed right-hand side rather than the original forcing. The scaling factor $\sigma_n$ for the model uncertainty $\tilde{W}_{t_n}$ is set to $\sigma_n = 0.005$. 
\begin{figure}[h!]
\centering
\begin{minipage}{0.3\textwidth}
\includegraphics[scale=0.24]{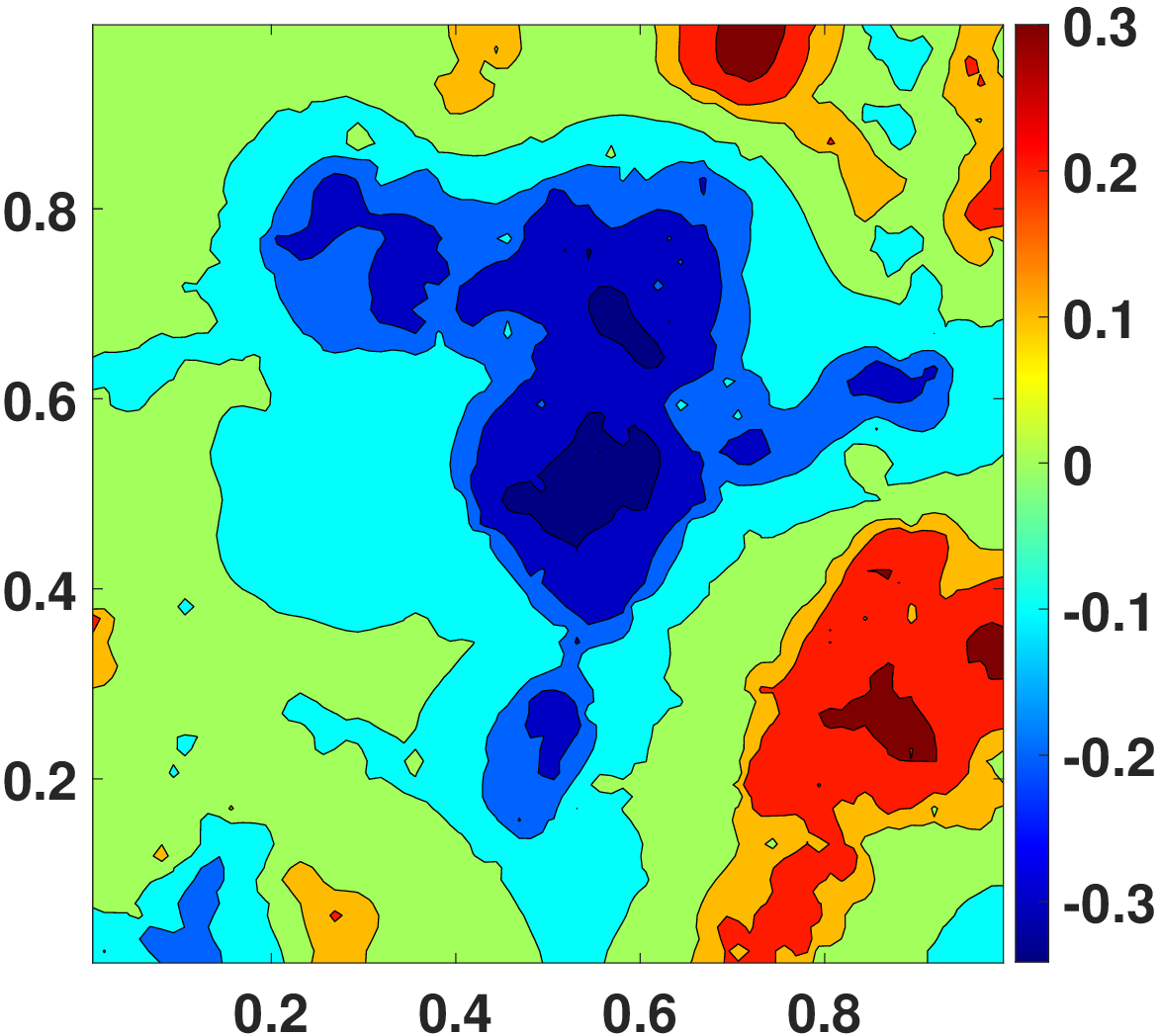}
\end{minipage}%
\begin{minipage}{0.3\textwidth}
\hspace{0.1cm}\includegraphics[scale=0.24]{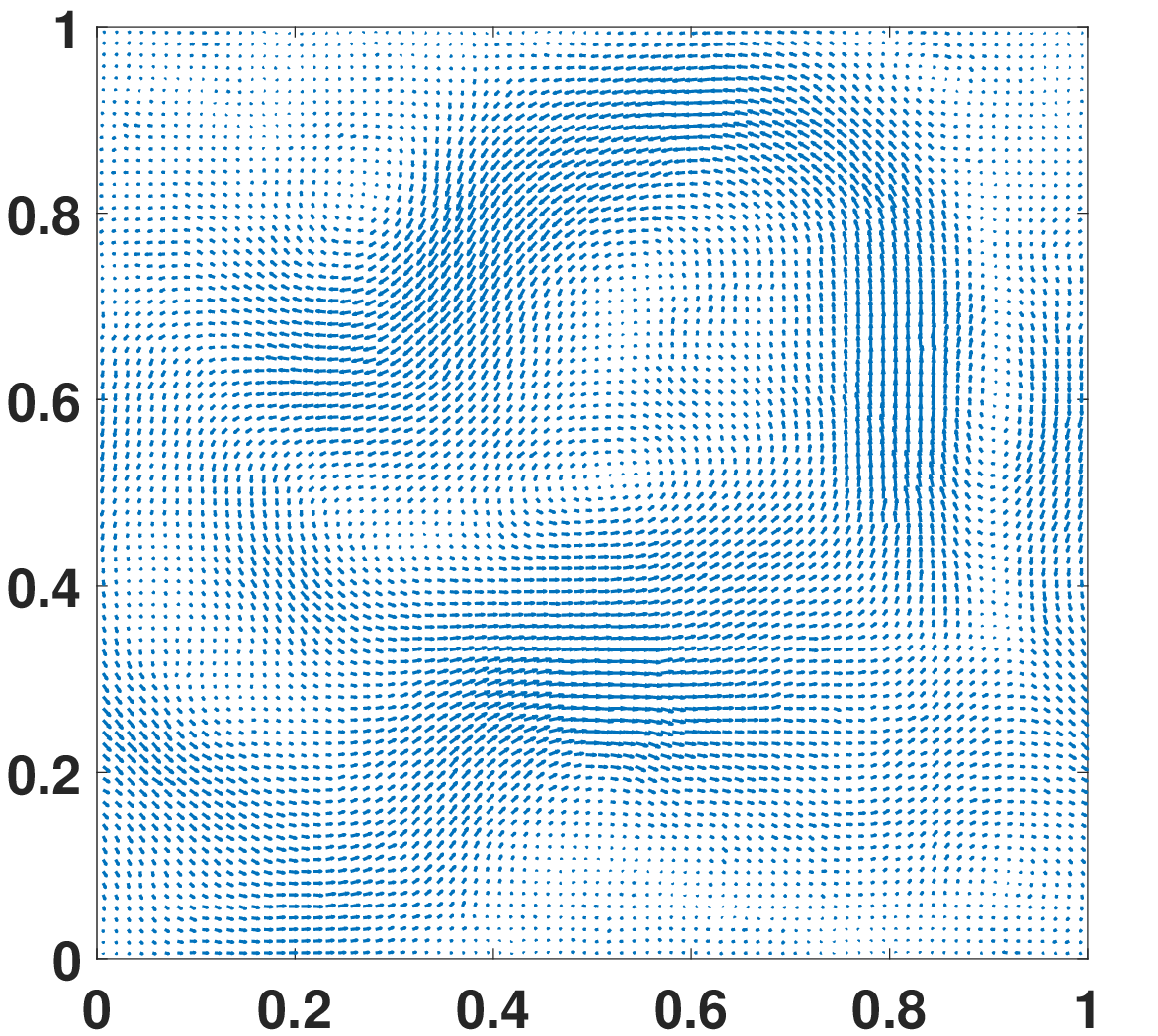}
\end{minipage}
\begin{minipage}{0.3\textwidth}
\hspace{0.1cm}\includegraphics[scale=0.24]{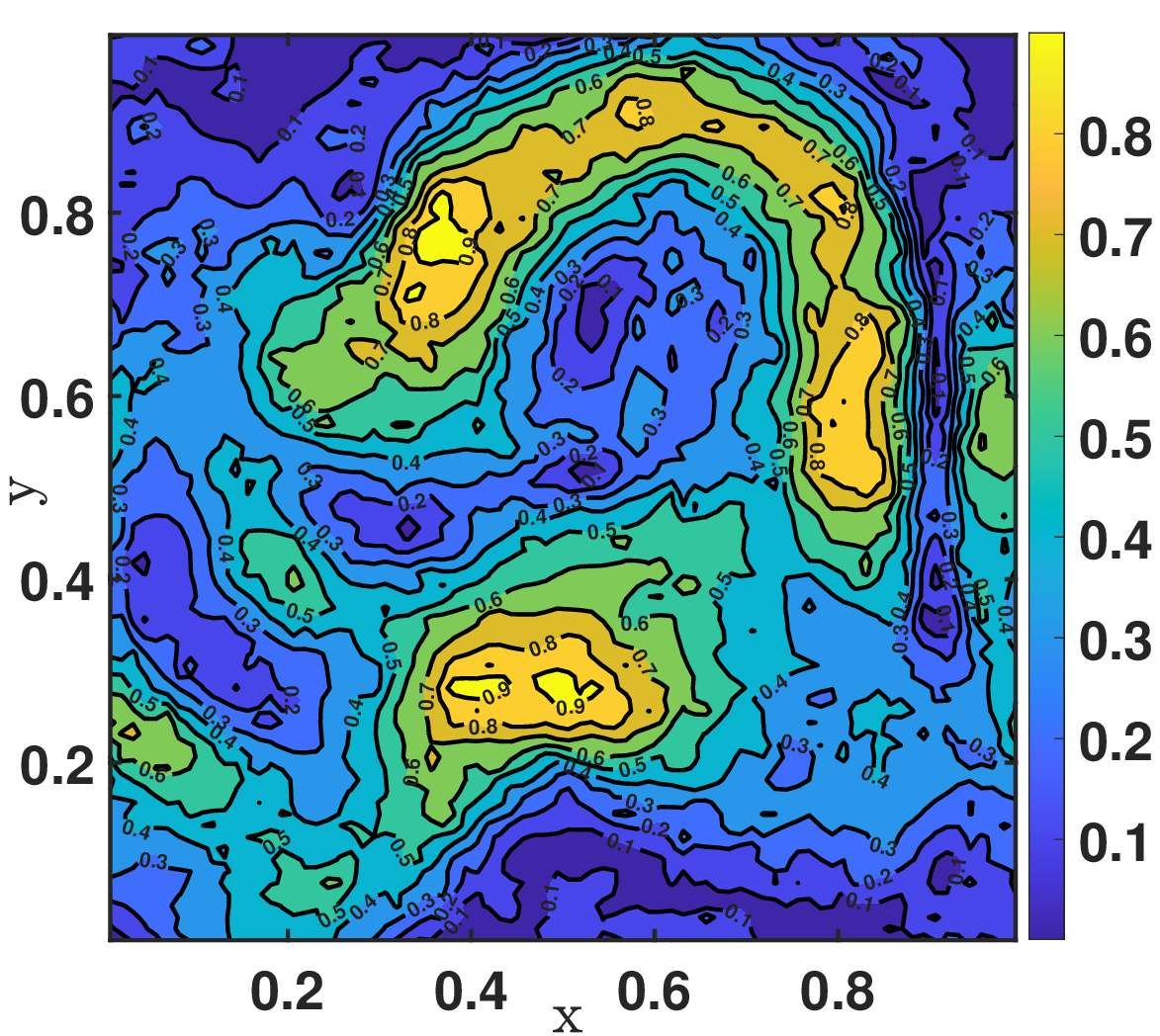}
\end{minipage}
\caption{\small Estimated state for the force-driven flow problem with $20\%$ observations by EnSF with inpainting. (First) Pressure field. (Right) Velocity field. (Third) Velocity magnitude.}
\label{RHSDriven_opt1_EnSF}
\vspace{-0.4cm}
\end{figure}
\begin{figure}[h!]
\centering
\begin{minipage}{0.3\textwidth}
\includegraphics[scale=0.24]{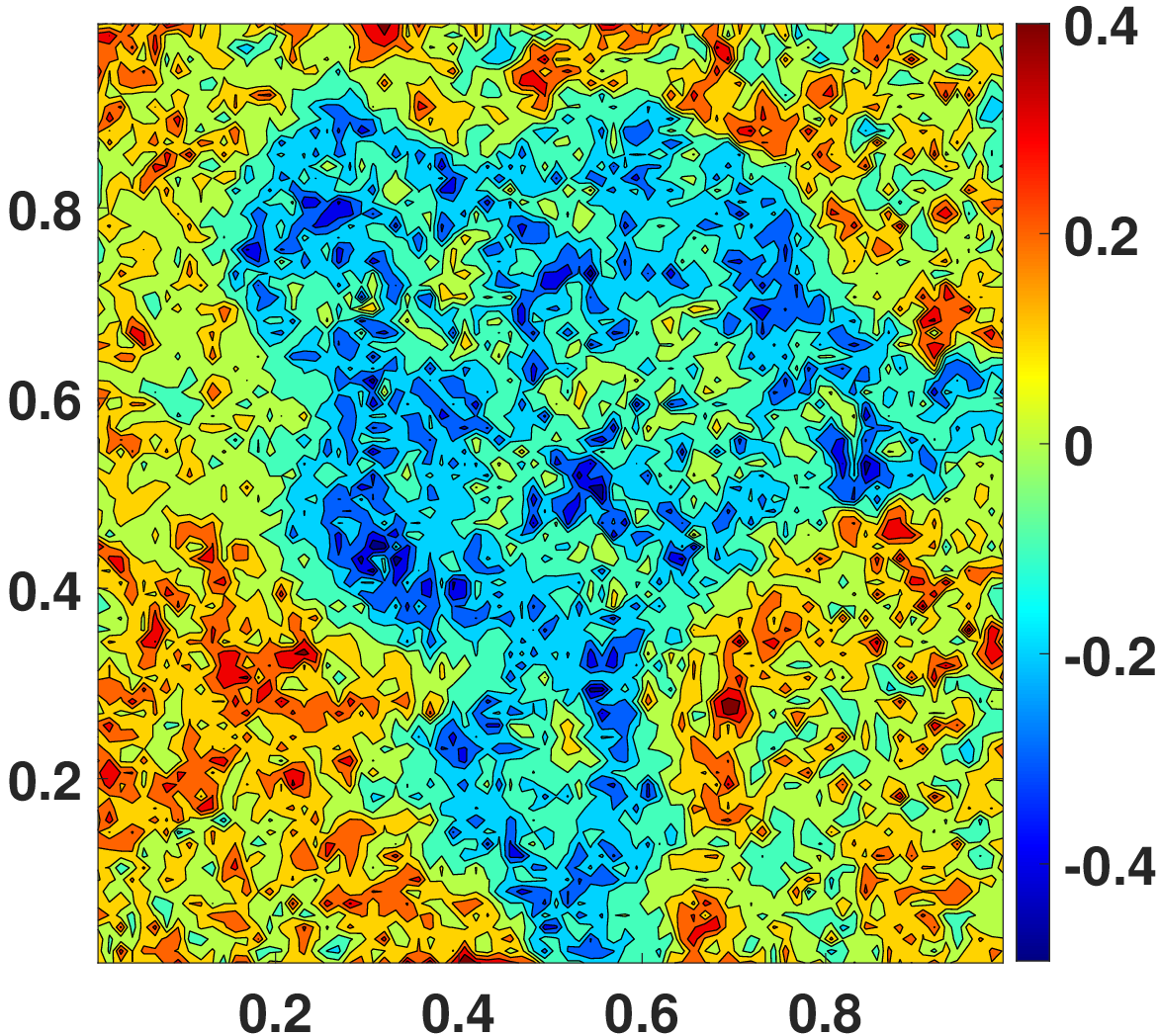}
\end{minipage}%
\begin{minipage}{0.3\textwidth}
\hspace{0.1cm}\includegraphics[scale=0.24]{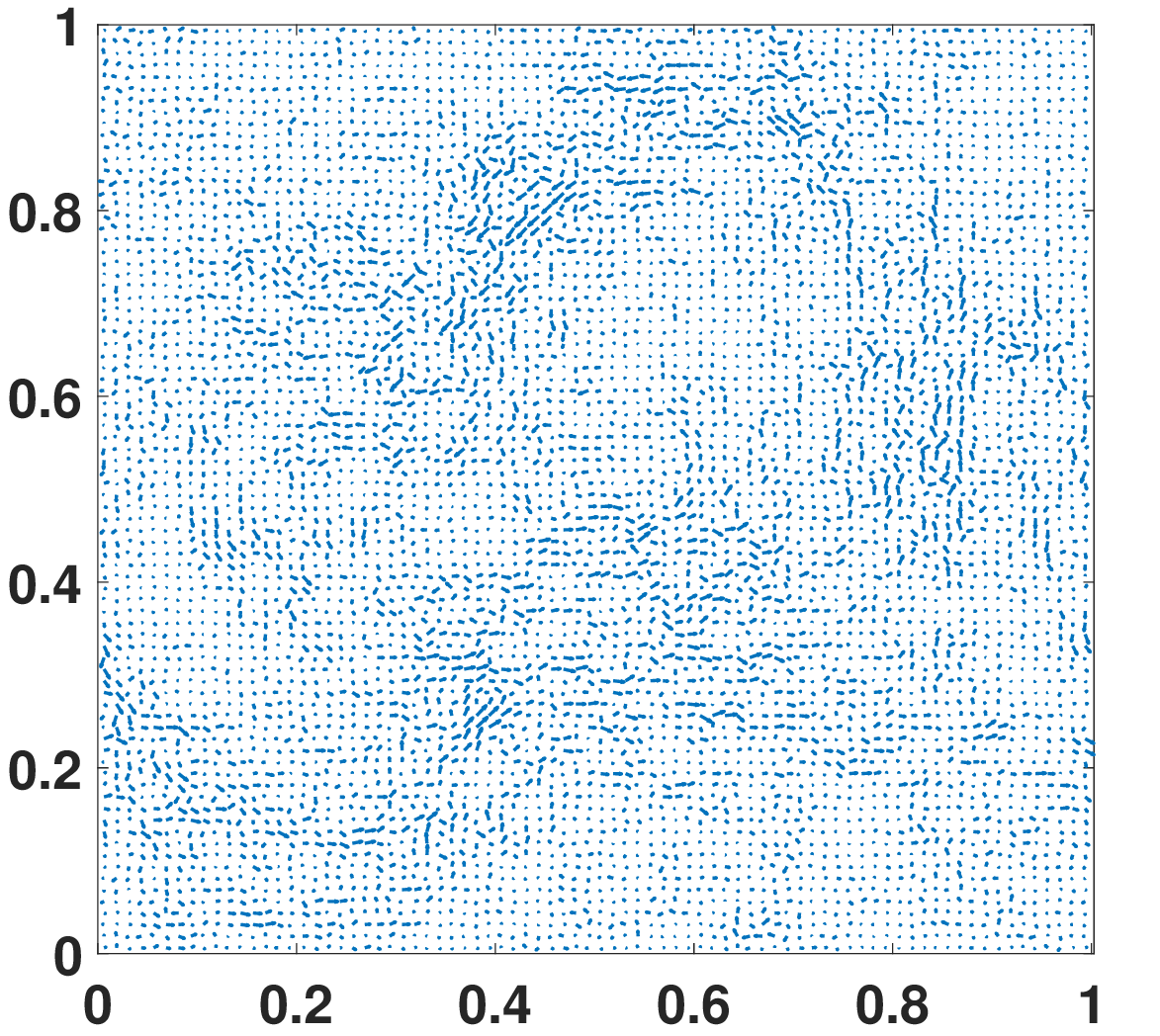}
\end{minipage}
\begin{minipage}{0.3\textwidth}
\hspace{0.1cm}\includegraphics[scale=0.24]{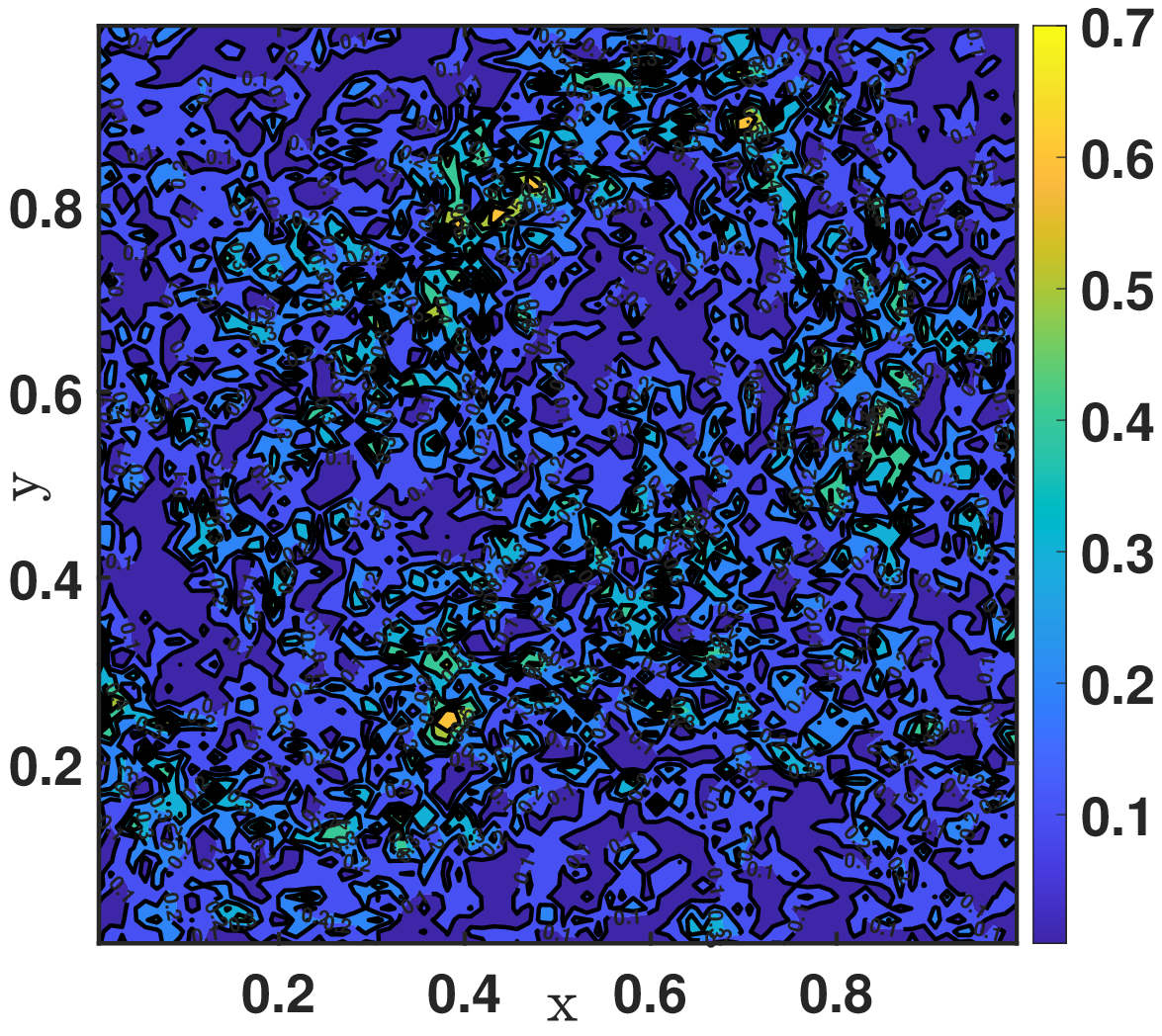}
\end{minipage}
\caption{\small Estimated state for the force-driven flow problem with $20\%$ observations by LETKF. (First) Pressure field. (Right) Velocity field. (Third) Velocity magnitude.}
\label{RHSDriven_opt1_LETKF}
\vspace{-0.3cm}
\end{figure}
\begin{figure}[h!]
    \centering
\includegraphics[scale=0.27]{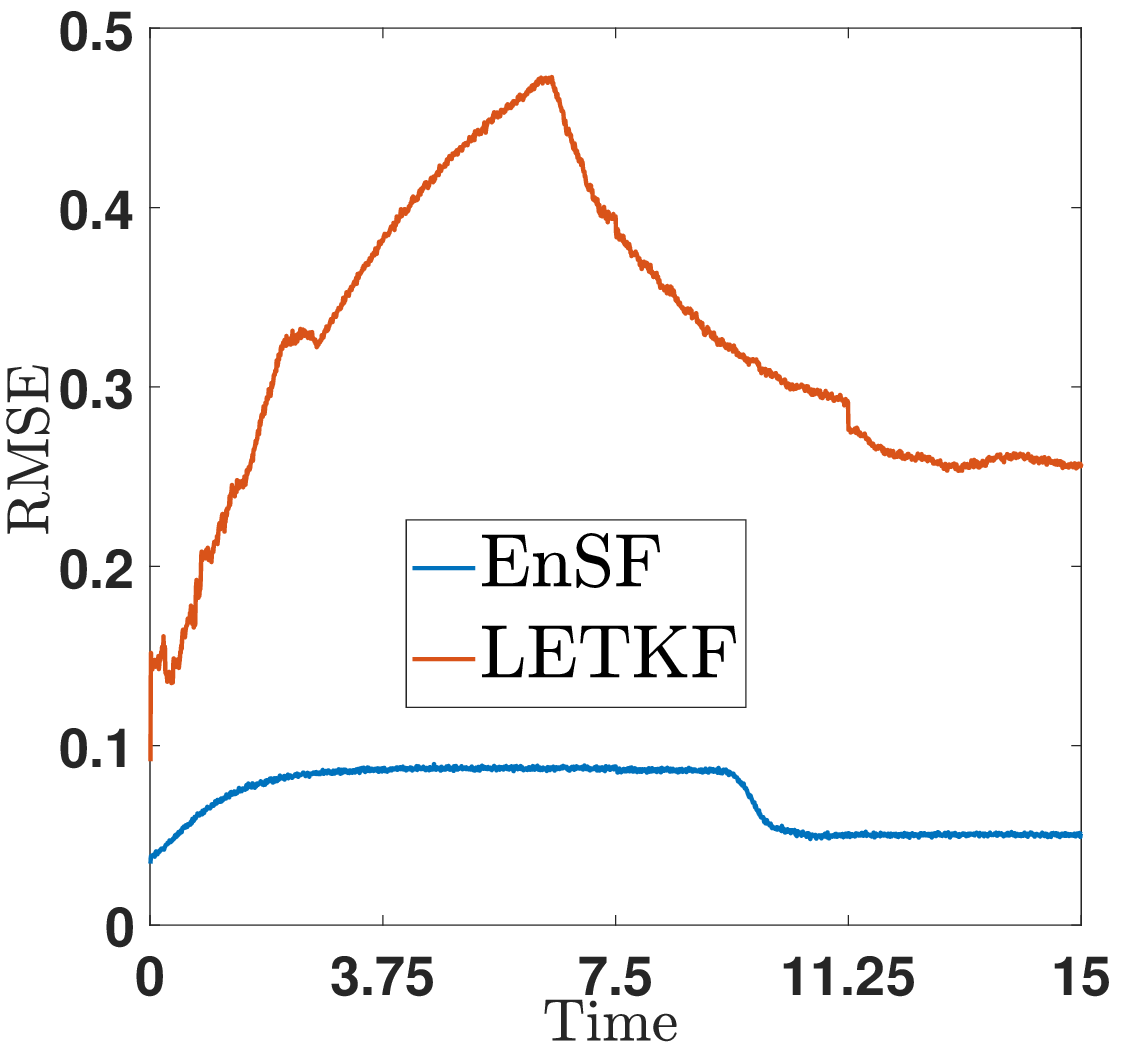}
    \caption{\small RMSE for State estimations.}
    \label{RMSE_opt1_RHSdriven}
\end{figure}

The pressure fields, the velocity fields, and the velocity magnitudes obtained from the EnSF and from the LETKF are shown in Figures~\ref{RHSDriven_opt1_EnSF} and~\ref{RHSDriven_opt1_LETKF}, respectively. These figures demonstrate that our approach recovers the reference solution over nearly the entire region, whereas the LETKF results remain very noisy despite roughly capturing the overall shape. Finally, the RMSEs for both methods are reported in Figure~\ref{RMSE_opt1_RHSdriven}, which clearly shows that the EnSF yields stable errors over time that remain consistently and substantially lower than those of the LETKF.

\subsection{Allen-Cahn equation}

The final example that we use to examine the performance of our adaptive learning procedure is the Allen-Cahn equation with general mobility:
\begin{equation}
\label{AllenCahn}
    \left\{\begin{array}{ll}
    \dfrac{\partial{\phi}}{\partial{t}} = -M\left(\phi\right)\mu, \; & (\pmb{x}, t) \in \Omega \times (0, T], \vspace{0.1cm} \\
    \mu = -\varepsilon^2\nabla{\phi} + F'(\phi), \; & (\pmb{x}, t) \in \Omega \times (0, T], \vspace{0.1cm} \\
    \phi(\pmb{x}, 0) = \phi_0(\pmb{x}), \; & \pmb{x} \in \Omega
    \end{array}\right.
\end{equation}
and subject to the homogeneous Neumann boundary condition, where $\Omega = (-0.5, 0.5)^2$, $T>0$ is the terminal time, $\varepsilon >0$ represents the interfacial width parameter, $M(\phi) \geq 0$ is a general mobility function, and $F(\phi) = \frac{1}{4}\left(1-\phi^2\right)^2$ is the double-well potential function. For all numerical simulations, we choose $\varepsilon = 0.01$.

\vspace{0.25em}

We adopt the grain coarsening dynamics test case from~\cite{Hou2023} with three different scenarios for the mobility: 
\begin{enumerate}
    \item (Case 1) Constant mobility, given by $M(\phi) = M_1(\phi) \equiv 1$.
    \item (Case 2) Non-constant but solution-independent mobility, defined as $M(\phi) = M_2(\phi) = \max\{1+\xi_t, 0\}$, where $\xi_t$ is a perturbed noise.
    \item (Case 3) Solution-dependent mobility with noise $\xi_t$, defined as $M(\phi) = M_3(\phi) = \max\{1-\phi^2+\xi_t, 0\}$.
\end{enumerate}
In Case 2 and Case 3, the noise term $\xi_t$ represents the uncertainty in mobility. To generate the synthetic ``true solution'', we use $\xi_t \sim 0.5\cdot N(0, \pmb{I}_d)$ in Case 2 and $\xi_t \sim 2 \cdot N(0, \pmb{I}_d)$ in Case 3. However, in the prediction of estimated solutions without observational data — reflecting limited knowledge of the true physical behavior — we intentionally use mis-specified noises: $\xi_t \sim 0.1\cdot N(0, \pmb{I}_d)$ in Case 2 and $\xi_t \sim 0.4\cdot N(0, \pmb{I}_d)$ in Case 3.

To generate high-fidelity ground truth, we solve the system \eqref{AllenCahn} numerically on the interval $[0, T]$ with $T=10$ using the linear, second-order BDF scheme scheme developed in \cite{Hou2023}. In particular, we discretize the system~\eqref{AllenCahn} in space via central finite differences and advance in time with BDF2. This exact solver also serves as the model solver in the prediction step of our data assimilation framework. To reduce computational cost during prediction, one may alternatively use the first-order BDF method proposed in~\cite{Shen2016}. We note that solutions of the Allen–Cahn equation \eqref{AllenCahn} typically require a long evolution time to reach their steady state and exhibit two distinct phases: an initial rapid transition followed by a much slower convergence to equilibrium~\cite{Hou2023}. Therefore, the algorithm from \cite{Hou2023} is constructed to provide a robust time-stepping framework that allows non-uniform time-stepping to effectively handle such behavior. However, this is not the primary focus of our work. Instead, we adopt a fixed and sufficiently small time step $\Delta{t} = T/1000 = 0.01$ to obtain reliable reference solutions. It has been shown numerically in~\eqref{AllenCahn} choosing $\Delta{t} = 0.01$ uniformly yields comparable results to those obtained with a nonuniform time-stepping strategy. The time step size for the data assimilation process in all cases is fixed at $\Delta{t}_{\text{Filter}} = T/250$. The spatial mesh size is fixed at $h=1/128$ for both ground-truth generation and filtering process. Finally, the initial conditions for all cases are generated by randomly sampling the values uniformly from $-0.9$ to $0.9$

For all three test scenarios, we consider varying amounts of observational data, with the specific configurations tailored to each case. The observational data are received through the \textbf{\textit{arctangent operator}}. We account for two primary sources of uncertainty: incomplete knowledge of the mobility and model error, the latter represented by the term $\sigma_n\tilde{W}_{t_n}$ in~\eqref{SPDE:scheme_simplifed}. Consequently, because noise is injected into the prediction at each filtering step and different noise intensities are used in computing the stiffness matrix within the data assimilation solver, the resulting solution corresponds to an SPDE whose sample paths generally diverge from those of the reference solution.  In all experiments, we fix $\sigma_n = 0.01$, while the level of uncertainty in the mobility varies between scenarios. In addition to the state estimation task, we examine both the discrete energy functional and the supremum norm of the estimates, since it is well-known that the solution to the Allen-Cahn equation satisfies an energy‐dissipation law and the maximum‐bound principle~\cite{Hou2023}. Detail of the implementation for the Allen-Cahn equations can be found in~\url{https://github.com/Toanhuynh997/StateEst_PDEs/tree/main/AllenCahn}.

\subsubsection{Case 1: Constant Mobility}

We first present the results for Case 1. To account for mobility uncertainty, we introduce a small perturbation at each filtering time step, defined as $\eta \sim 0.05\cdot N(0, \pmb{I}_d)$, which is added to $M_1(\phi)$. In this test case, we examine three levels of observational coverage: $100\%$, $70\%$, and $10\%$. For the $10\%$ observation scenario, we additionally compare our approach with the LETKF.

\begin{figure}[h!]
\begin{minipage}{0.22\textwidth}
    \includegraphics[scale = 0.2]{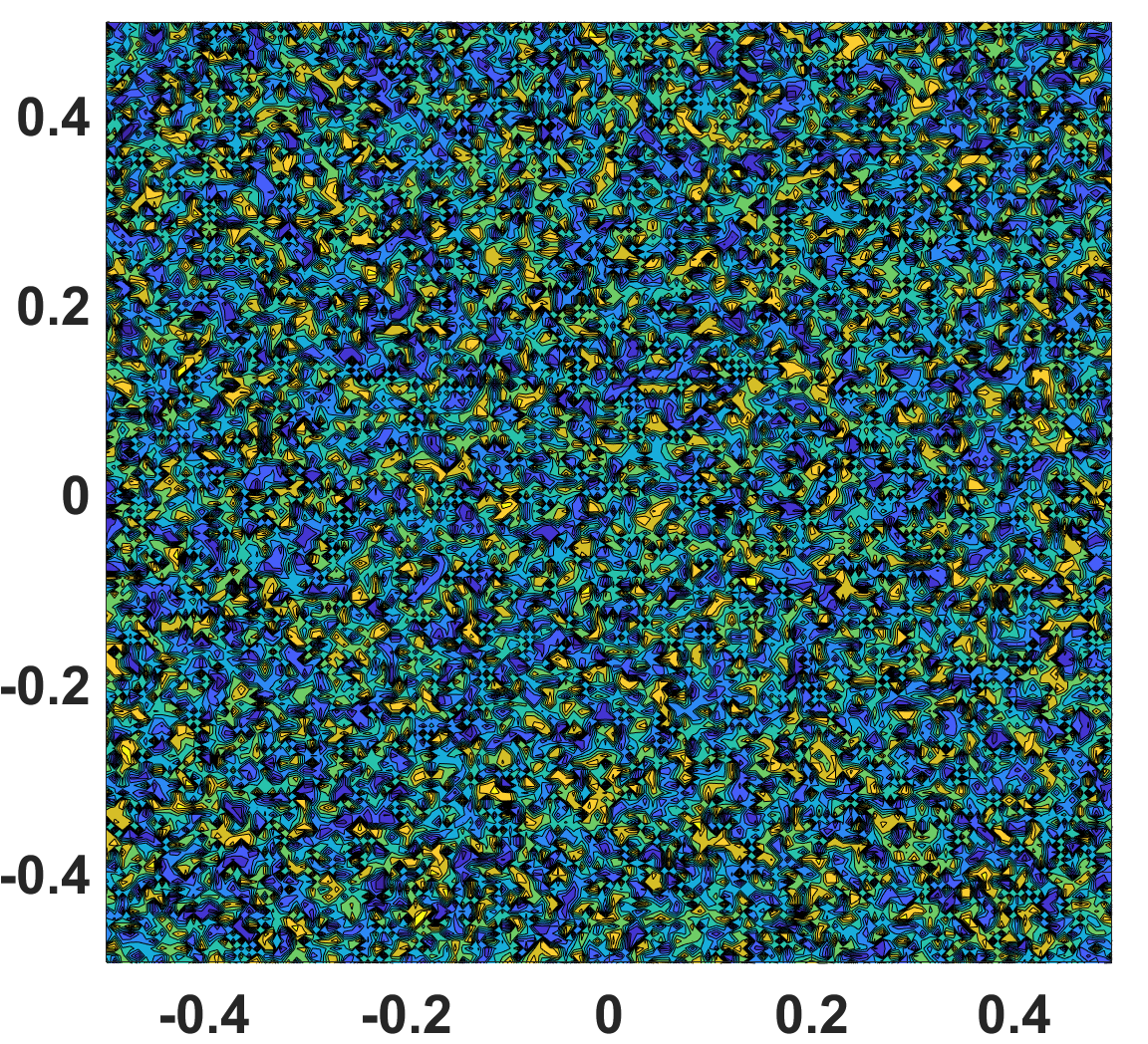}
\end{minipage}%
\begin{minipage}{0.22\textwidth}
    \includegraphics[scale = 0.2]{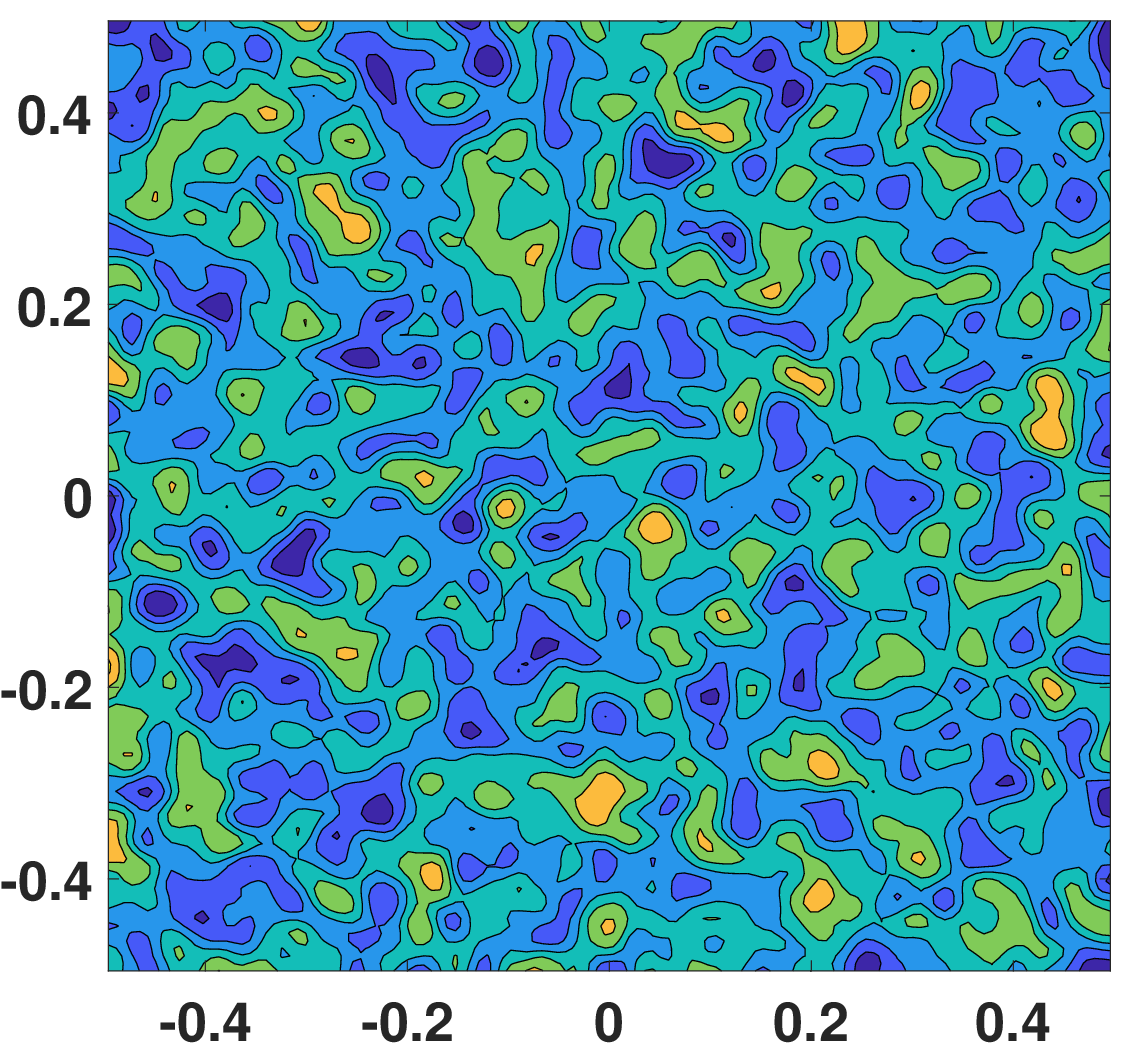}
\end{minipage}%
\begin{minipage}{0.22\textwidth}
    \includegraphics[scale = 0.2]{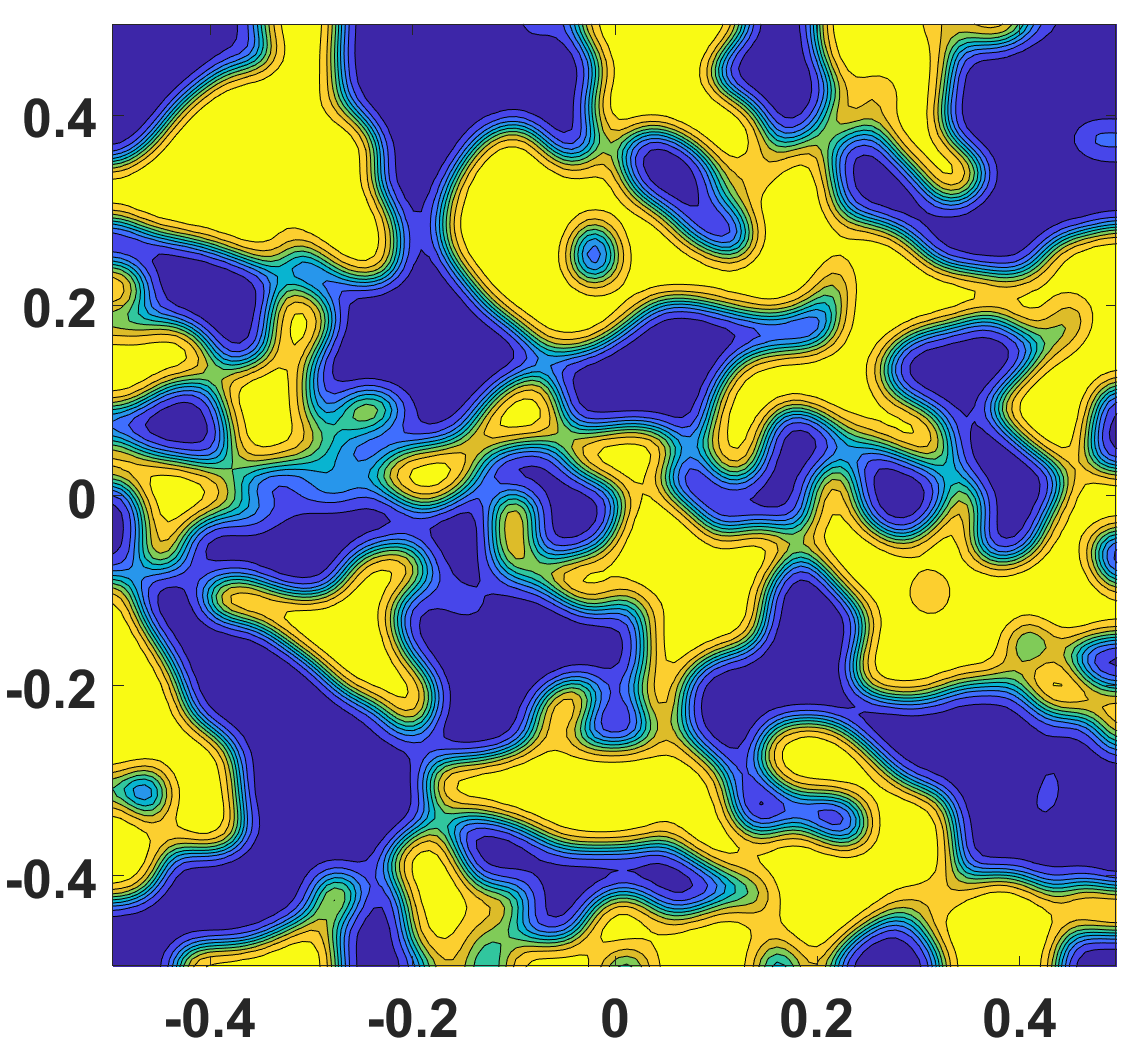}
\end{minipage}%
\begin{minipage}{0.22\textwidth}
    \includegraphics[scale = 0.2]{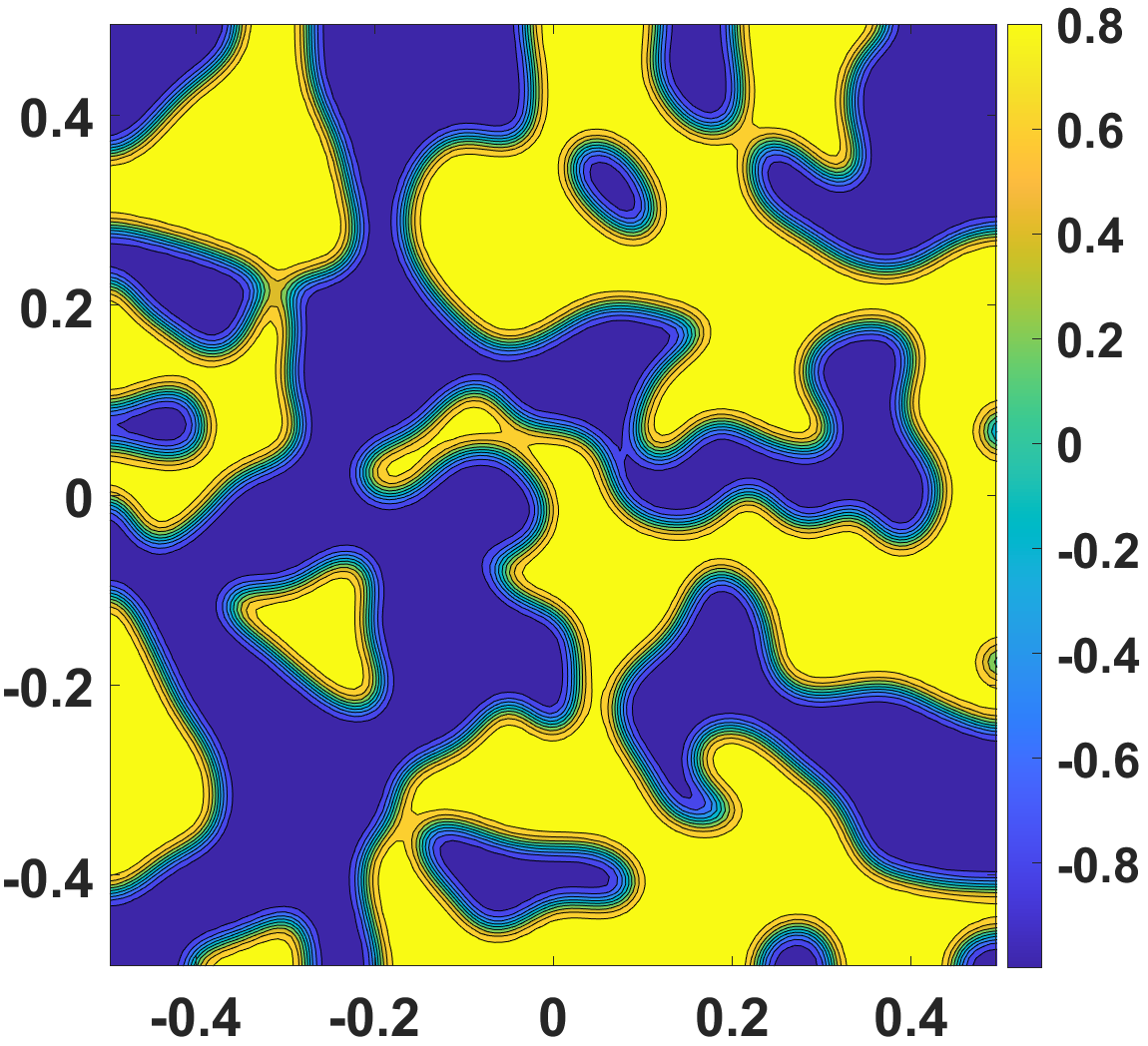}
\end{minipage}

\begin{minipage}{0.22\textwidth}
    \includegraphics[scale = 0.2]{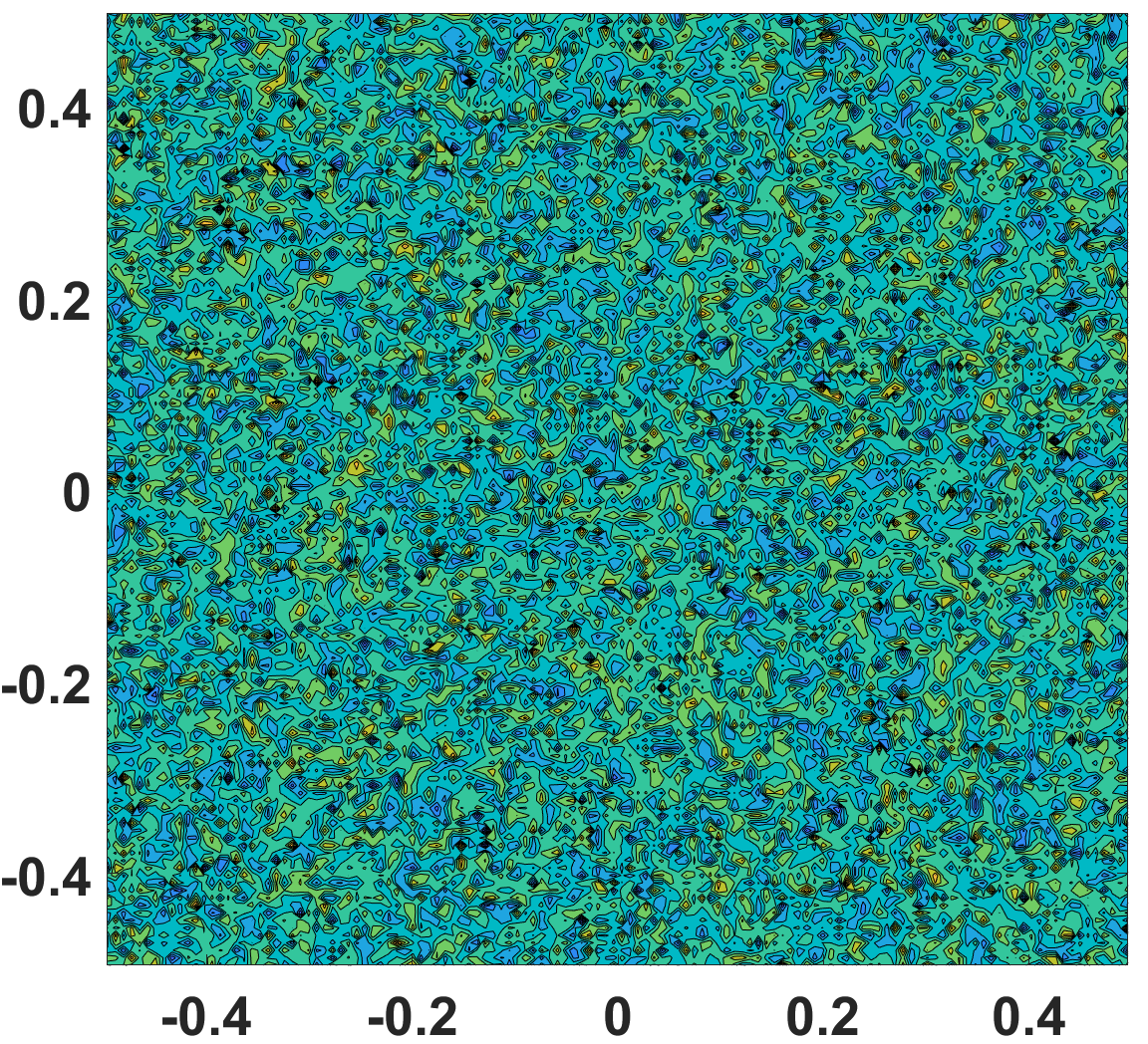}
\end{minipage}%
\begin{minipage}{0.22\textwidth}
    \includegraphics[scale = 0.2]{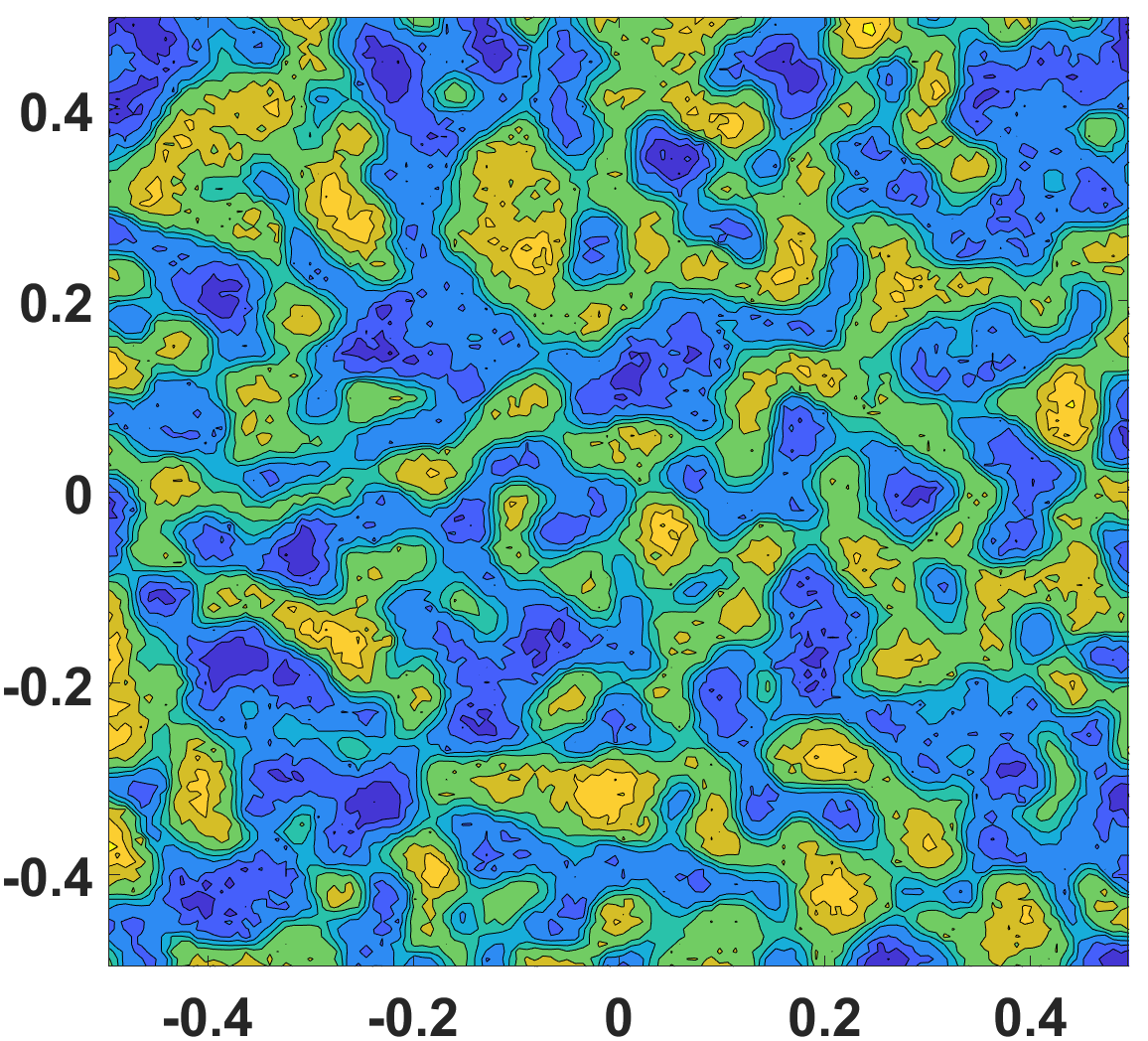}
\end{minipage}%
\begin{minipage}{0.22\textwidth}
    \includegraphics[scale = 0.2]{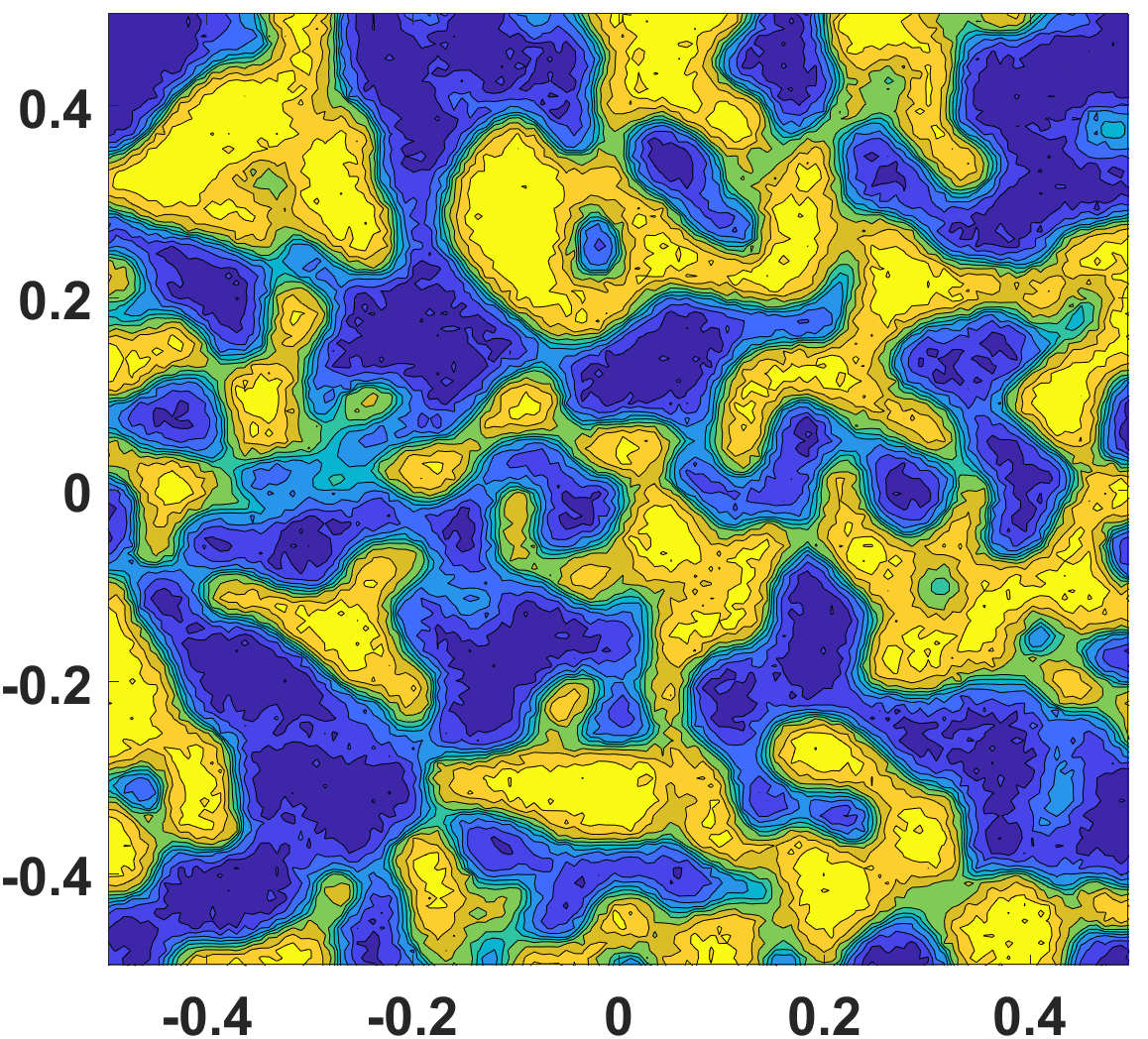}
\end{minipage}%
\begin{minipage}{0.22\textwidth}
    \includegraphics[scale = 0.2]{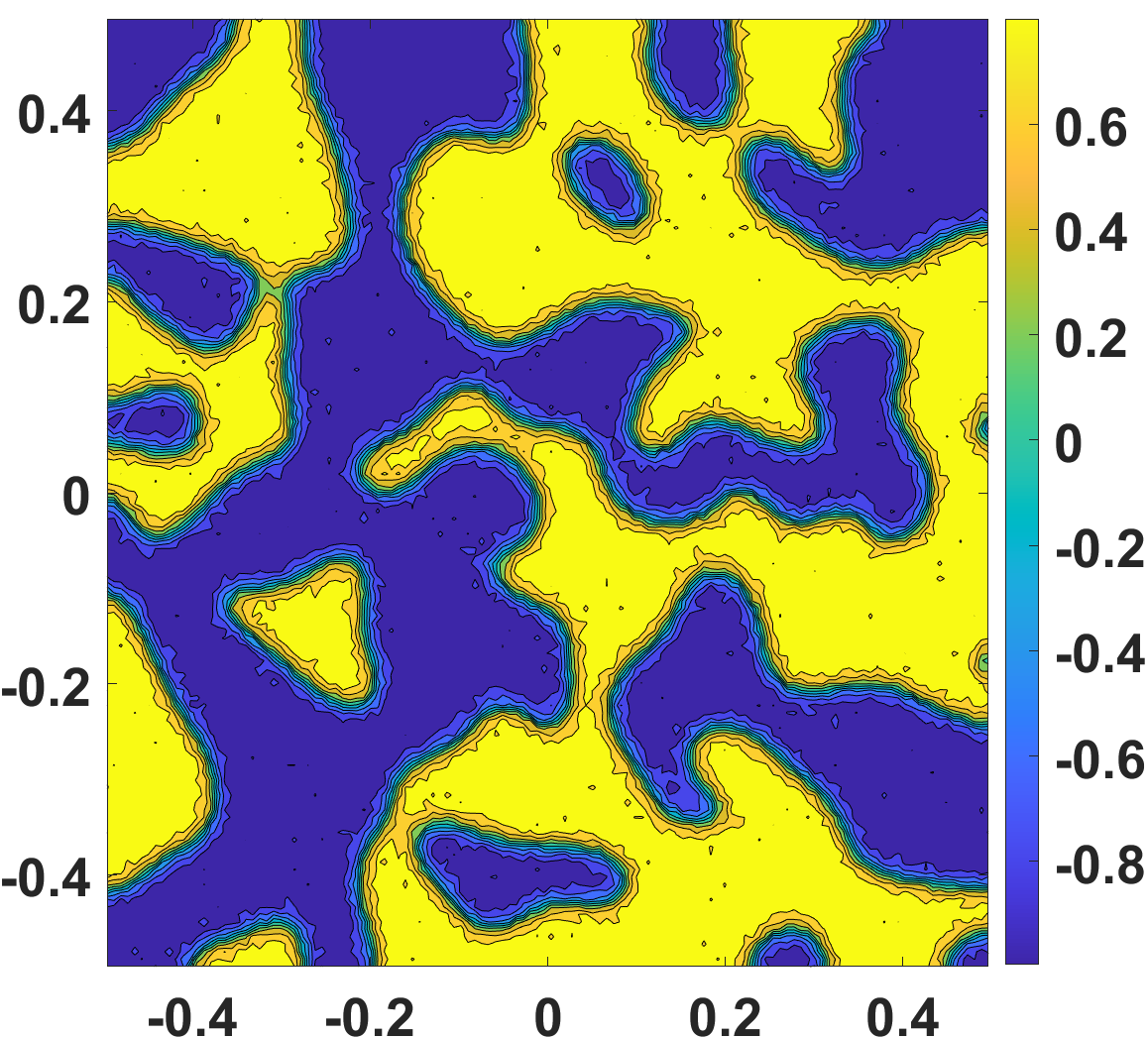}
\end{minipage}

\begin{minipage}{0.22\textwidth}
    \includegraphics[scale = 0.2]{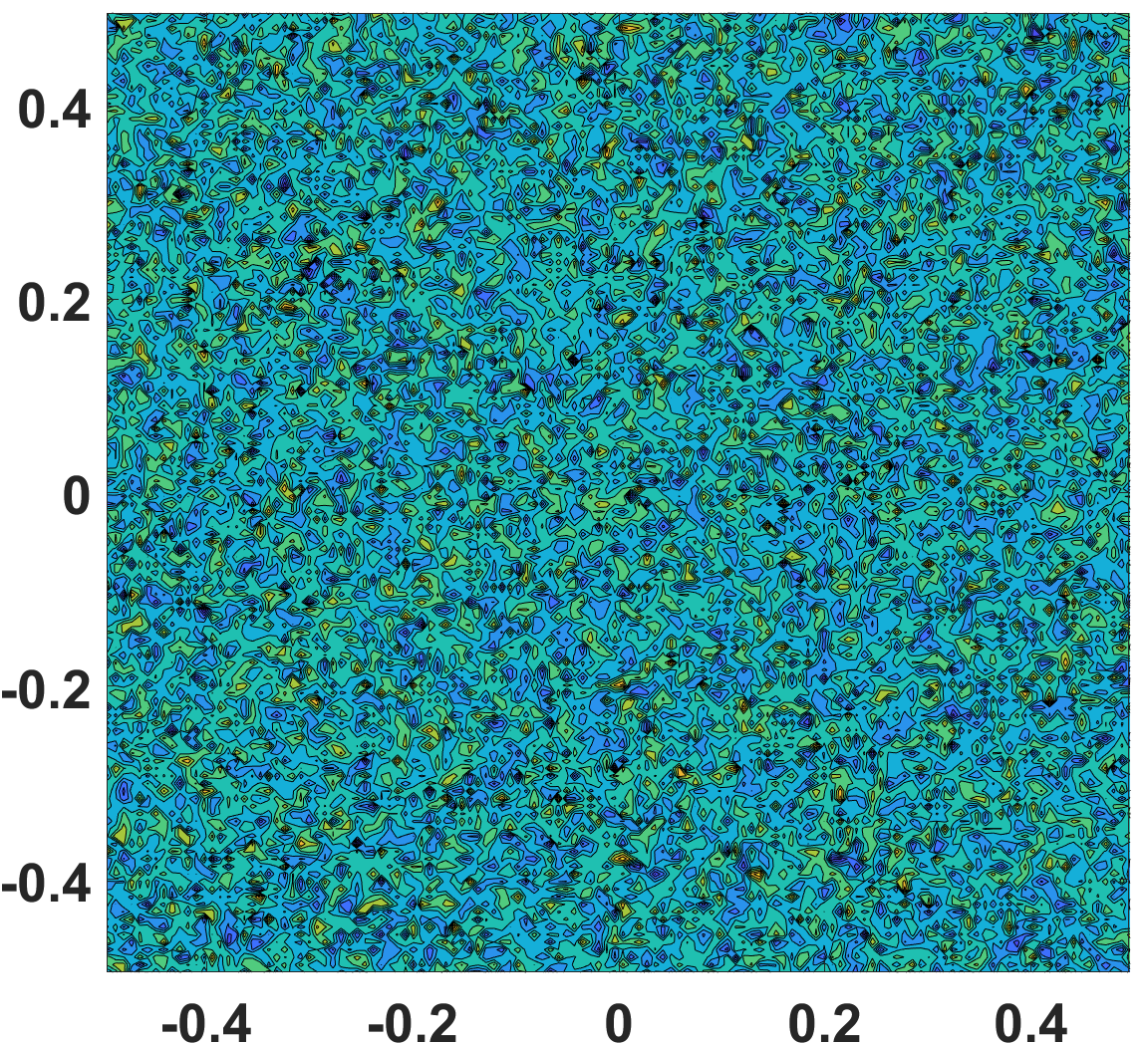}
\end{minipage}%
\begin{minipage}{0.22\textwidth}
    \includegraphics[scale = 0.2]{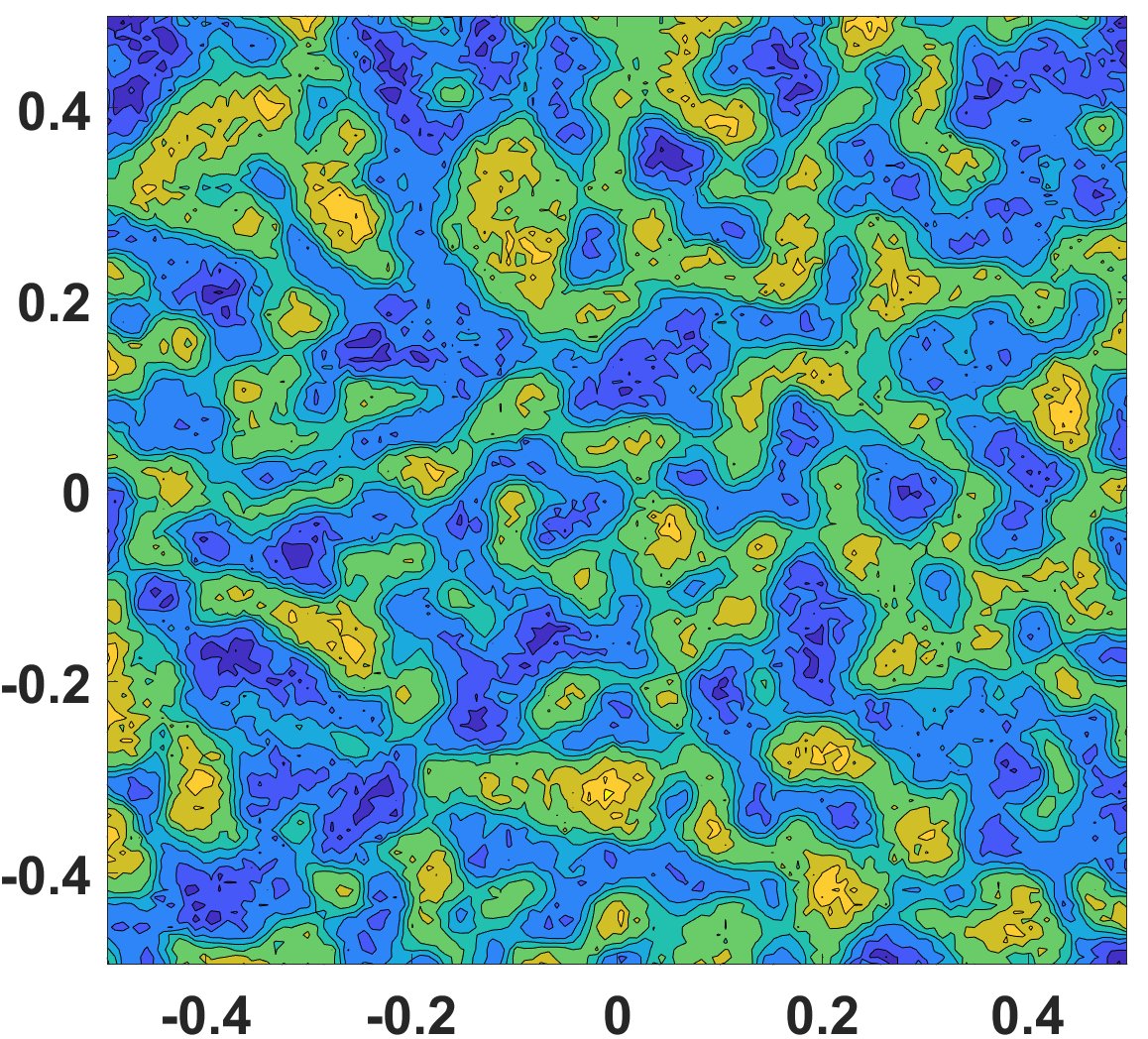}
\end{minipage}%
\begin{minipage}{0.22\textwidth}
    \includegraphics[scale = 0.2]{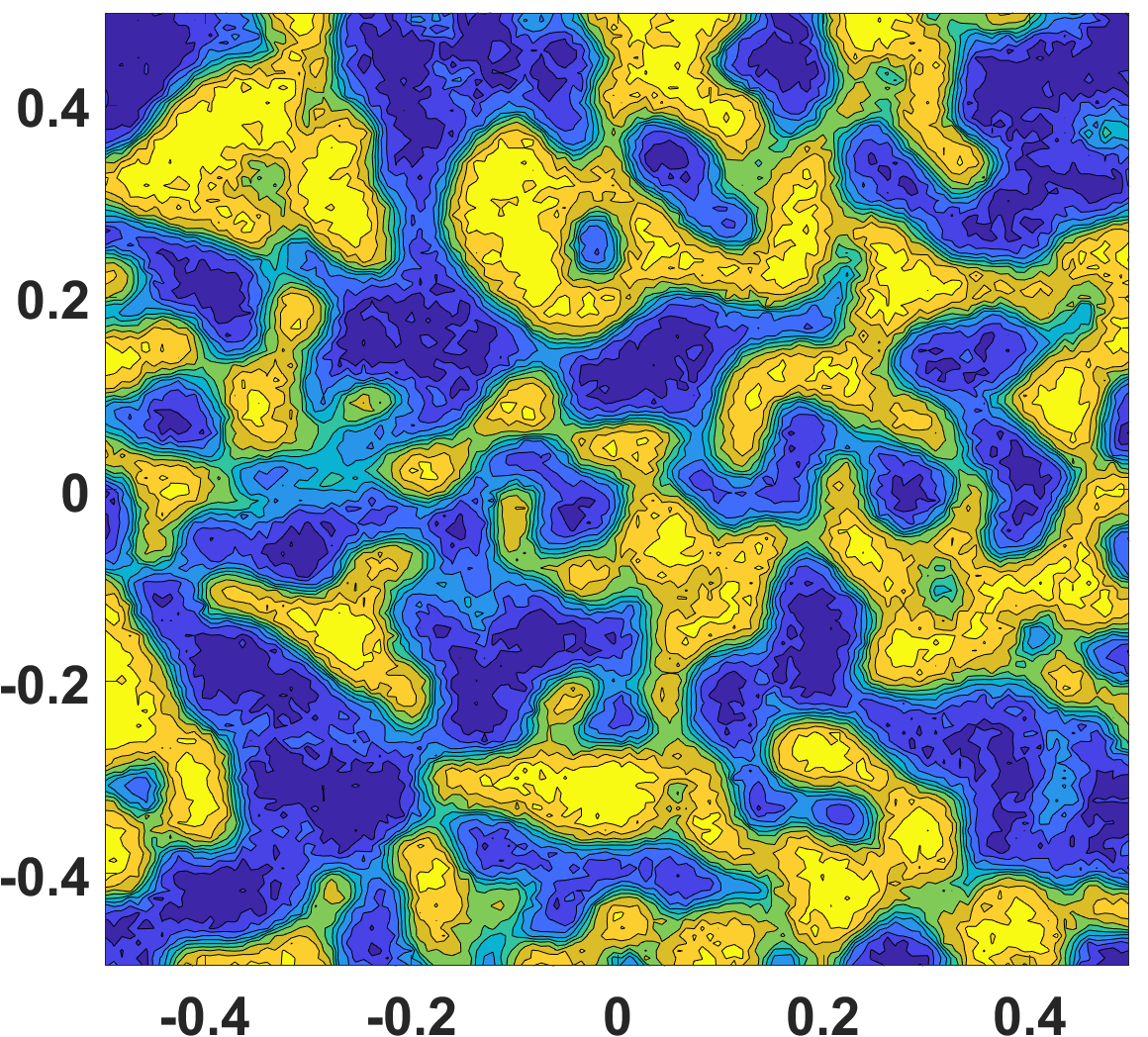}
\end{minipage}%
\begin{minipage}{0.22\textwidth}
    \includegraphics[scale = 0.2]{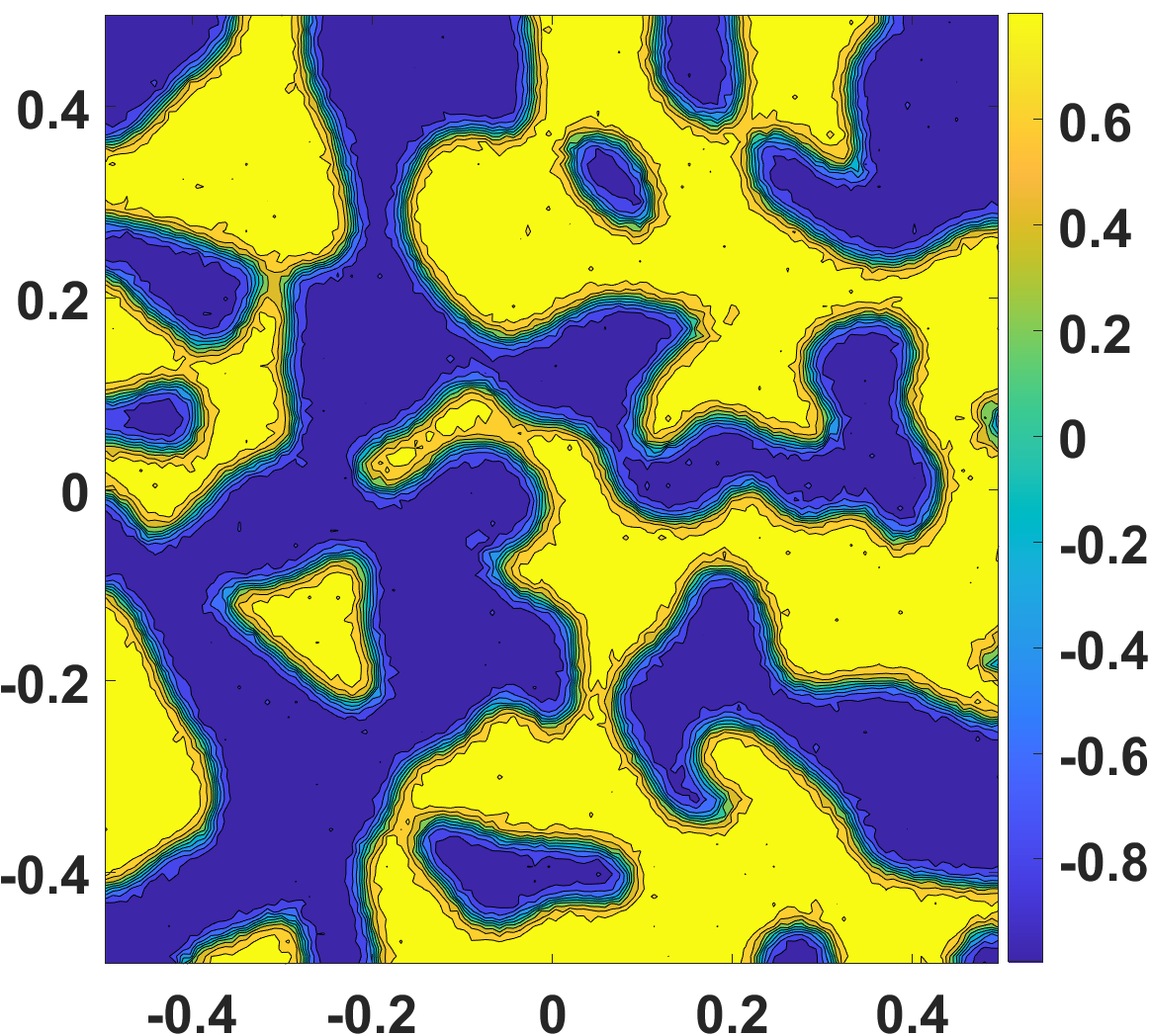}
\end{minipage}
\caption{\small Solutions of the Allen–Cahn equation in Case 1 at time $t=0$, $t= \f{T}{2}$, $t= \f{2T}{3}$, and $t=T$. (First row) Reference solution. (Second row) Estimated solution with $100\%$ observations. (Third row) Estimated solution with $70\%$ observations.}
\label{ConsMob_100_70Obs}
\vspace{-0.2cm}
\end{figure}

\begin{figure}[h!]
\vspace{-0.1cm}
   \begin{minipage}{0.3\textwidth}
    \includegraphics[scale = 0.22]{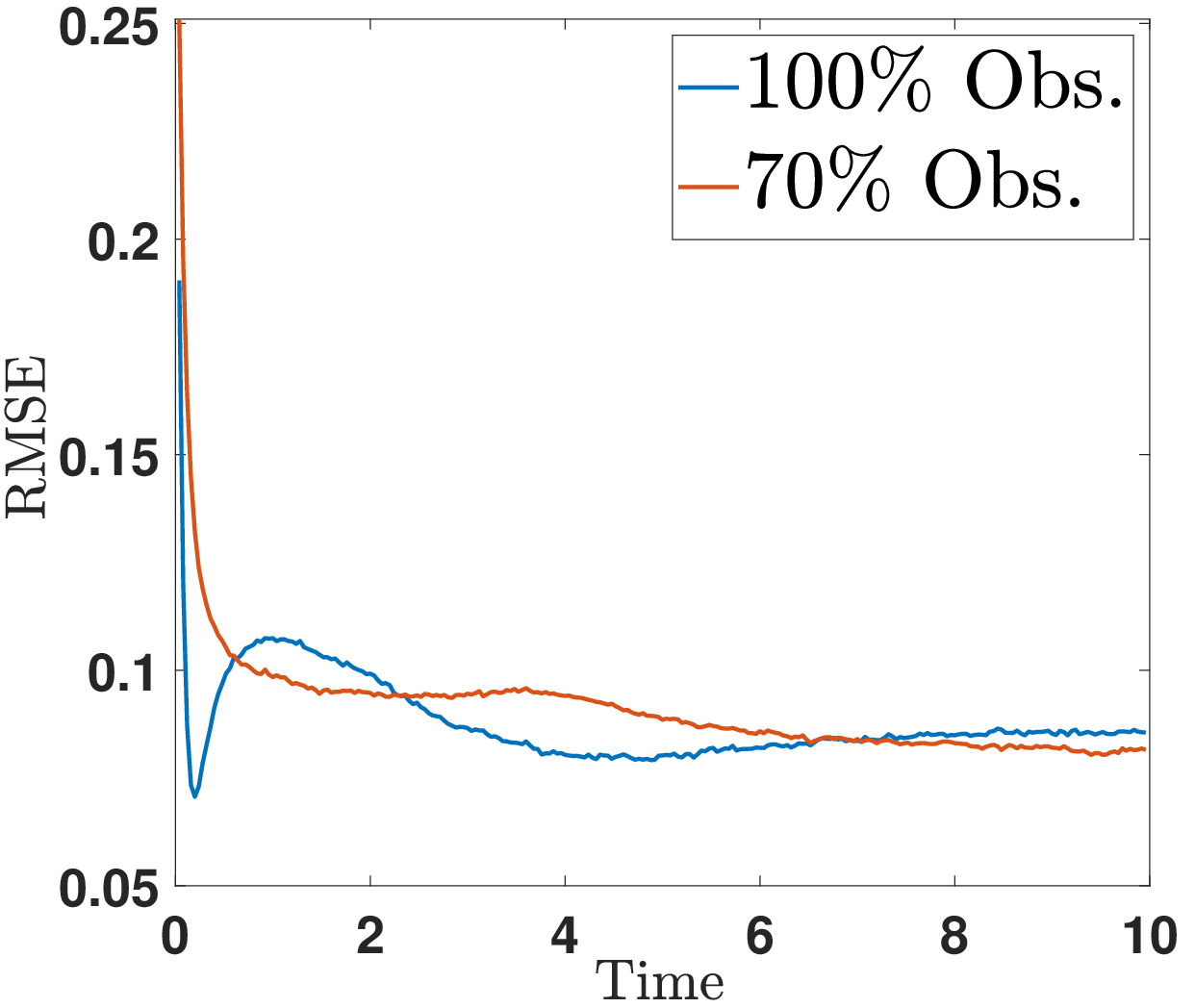}
\end{minipage}%
\begin{minipage}{0.3\textwidth}
    \includegraphics[scale = 0.22]{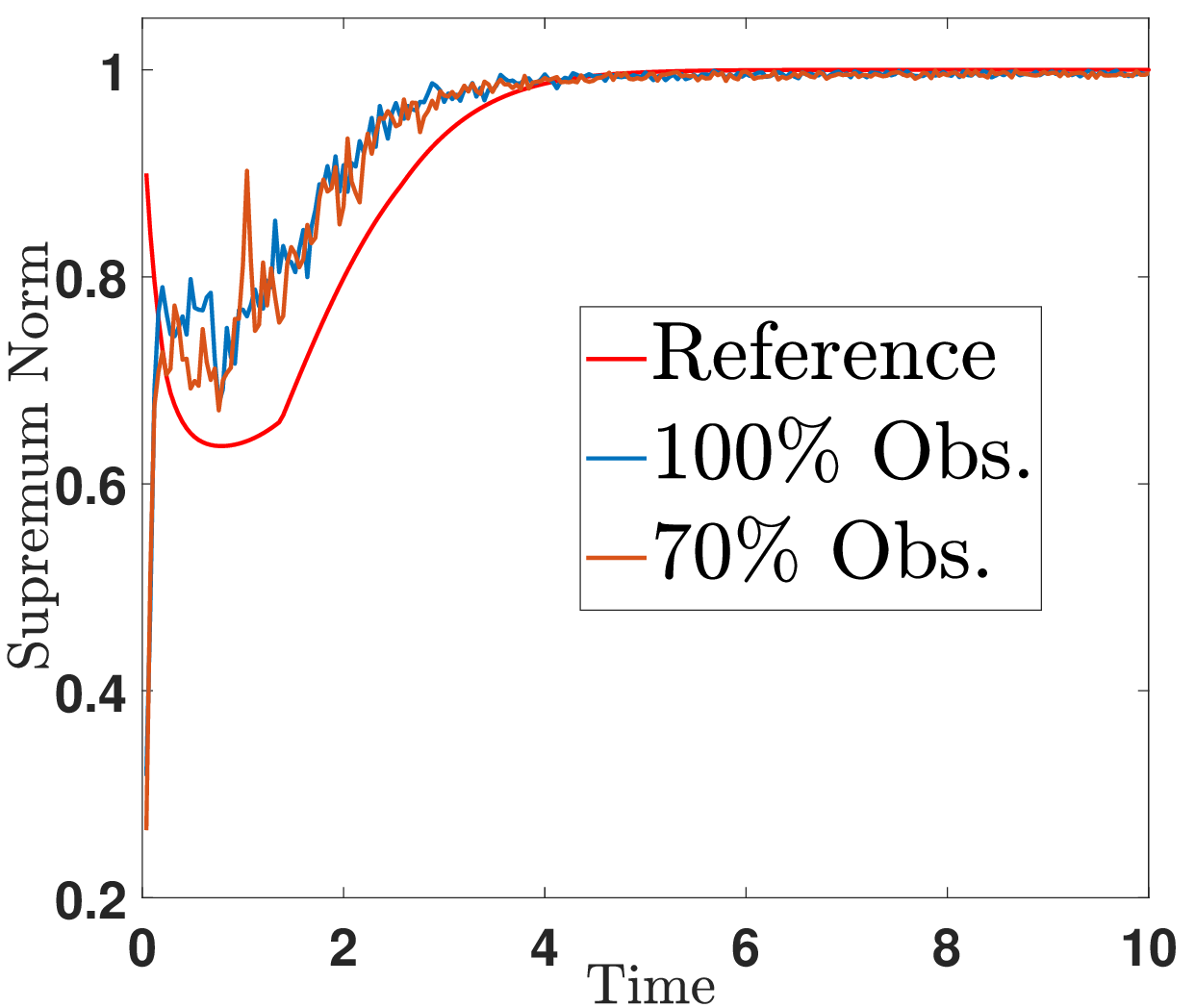}
\end{minipage}%
\begin{minipage}{0.3\textwidth}
    \includegraphics[scale = 0.22]{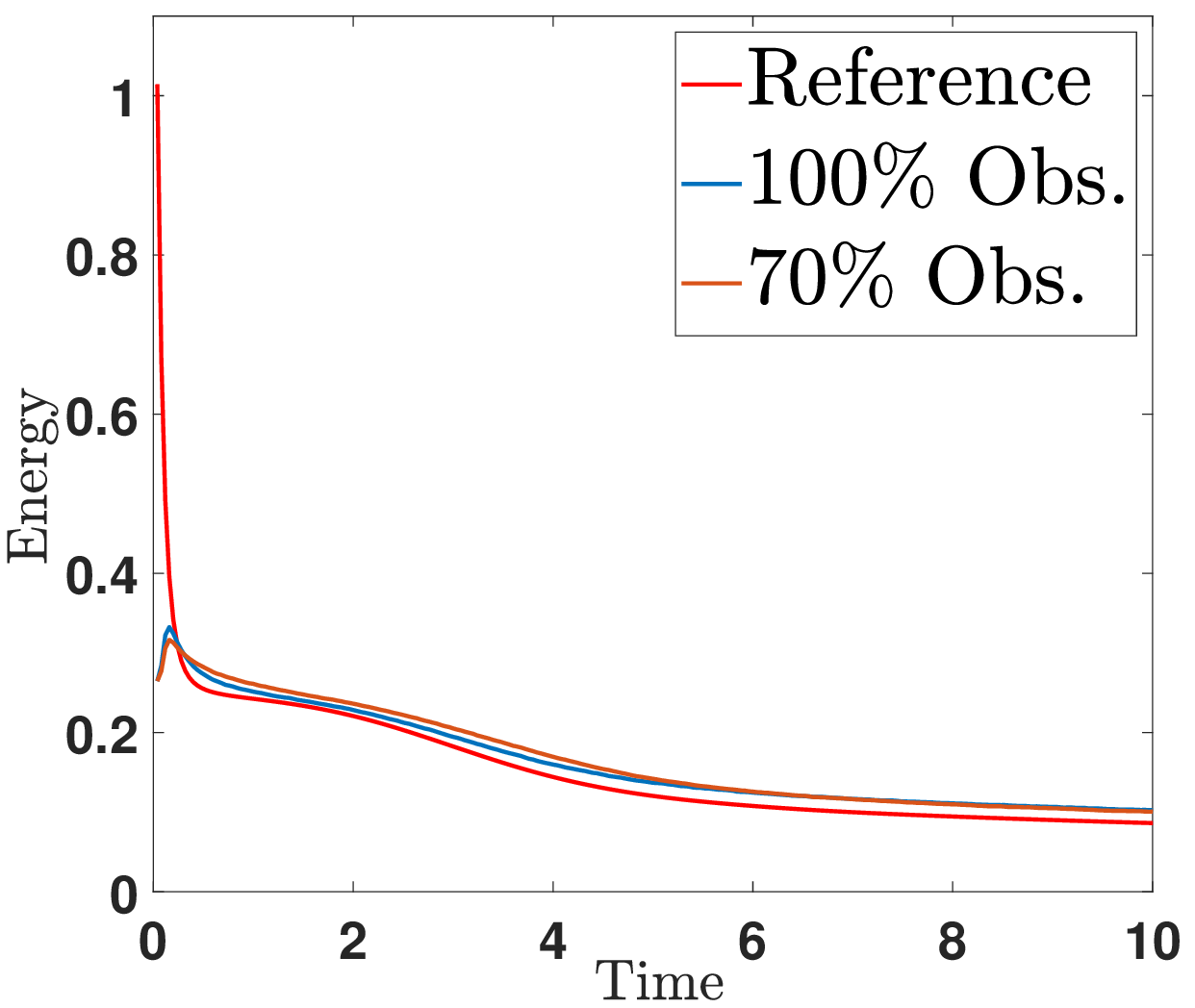}
\end{minipage}
\caption{\small Illustration of RMSEs, supremum norms, and discrete energies with $100\%$ and $70\%$ observations for Case 1. (Left) RMSE. (Center) Supremum Norm. (Right) Energy.}
\label{RMSE_Mass_Energy_100_70Obs}
\vspace{-0.3cm}
\end{figure}

Figure~\ref{ConsMob_100_70Obs} shows the evolution of the estimated solutions for the $100\%$ and $70\%$ observation scenarios at times $t = 0$, $t = T/2$, $t = 2T/3$, and $t = T$. The corresponding RMSEs, supremum norms, and energy values are presented in Figure~\ref{RMSE_Mass_Energy_100_70Obs}.
As shown in Figure~\ref{ConsMob_100_70Obs}, the estimated solutions gradually improve over time and accurately capture the sharp interface of the reference solution, despite starting from different initial conditions. This level of precision is further supported by the RMSE result in the first panel of Figure~\ref{RMSE_Mass_Energy_100_70Obs}, which shows a steady decline over time. This demonstrates the effectiveness of our method, particularly given the sensitivity of the Allen–Cahn equation to initial conditions. Finally, the last two panels of Figure~\ref{RMSE_Mass_Energy_100_70Obs} confirm that the estimated solutions adhere to both the maximum bound principle and the energy dissipation property. 

\begin{figure}[h!]
\begin{minipage}{0.22\textwidth}
    \includegraphics[scale = 0.2]{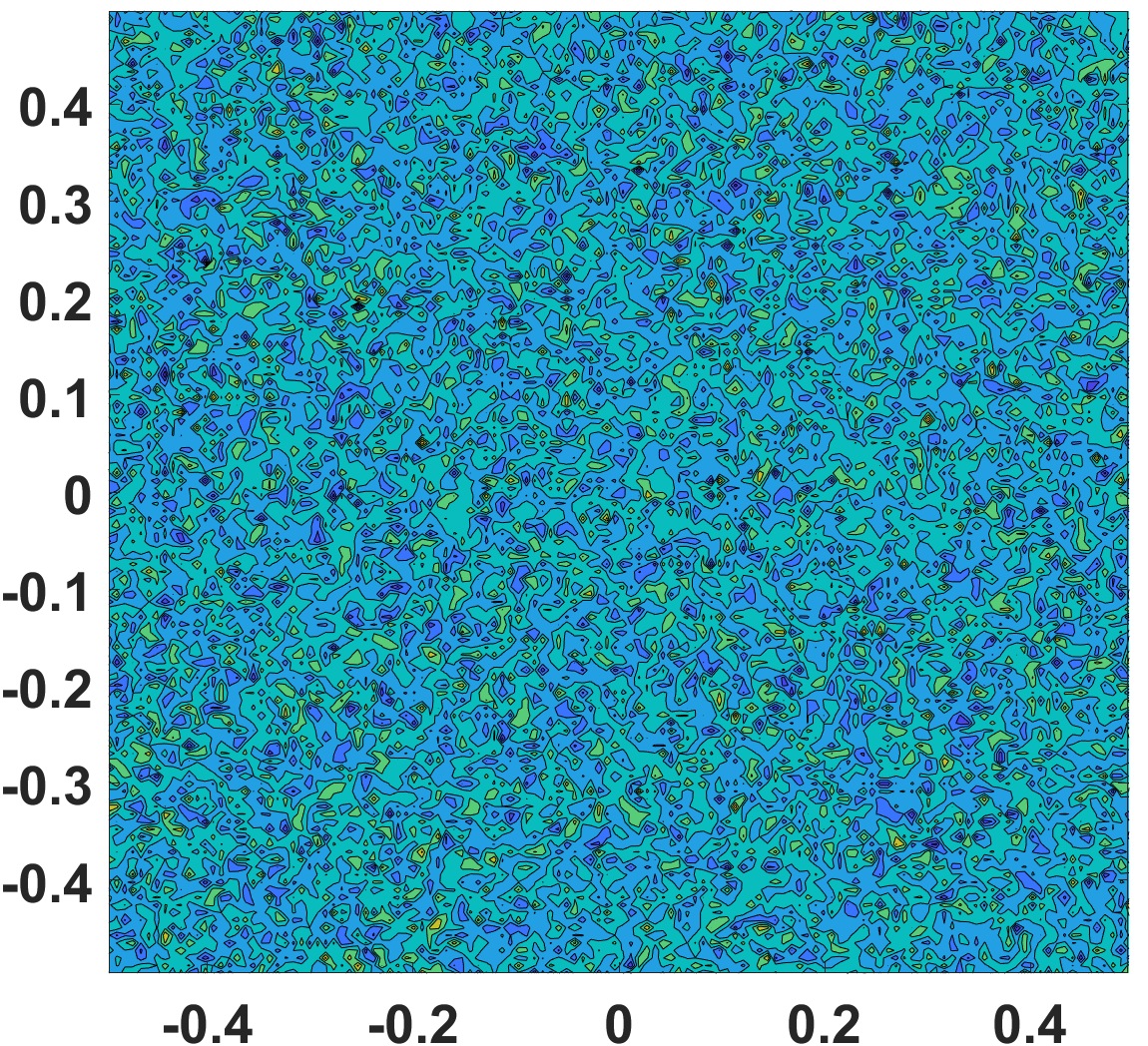}
\end{minipage}%
\begin{minipage}{0.22\textwidth}
    \includegraphics[scale = 0.2]{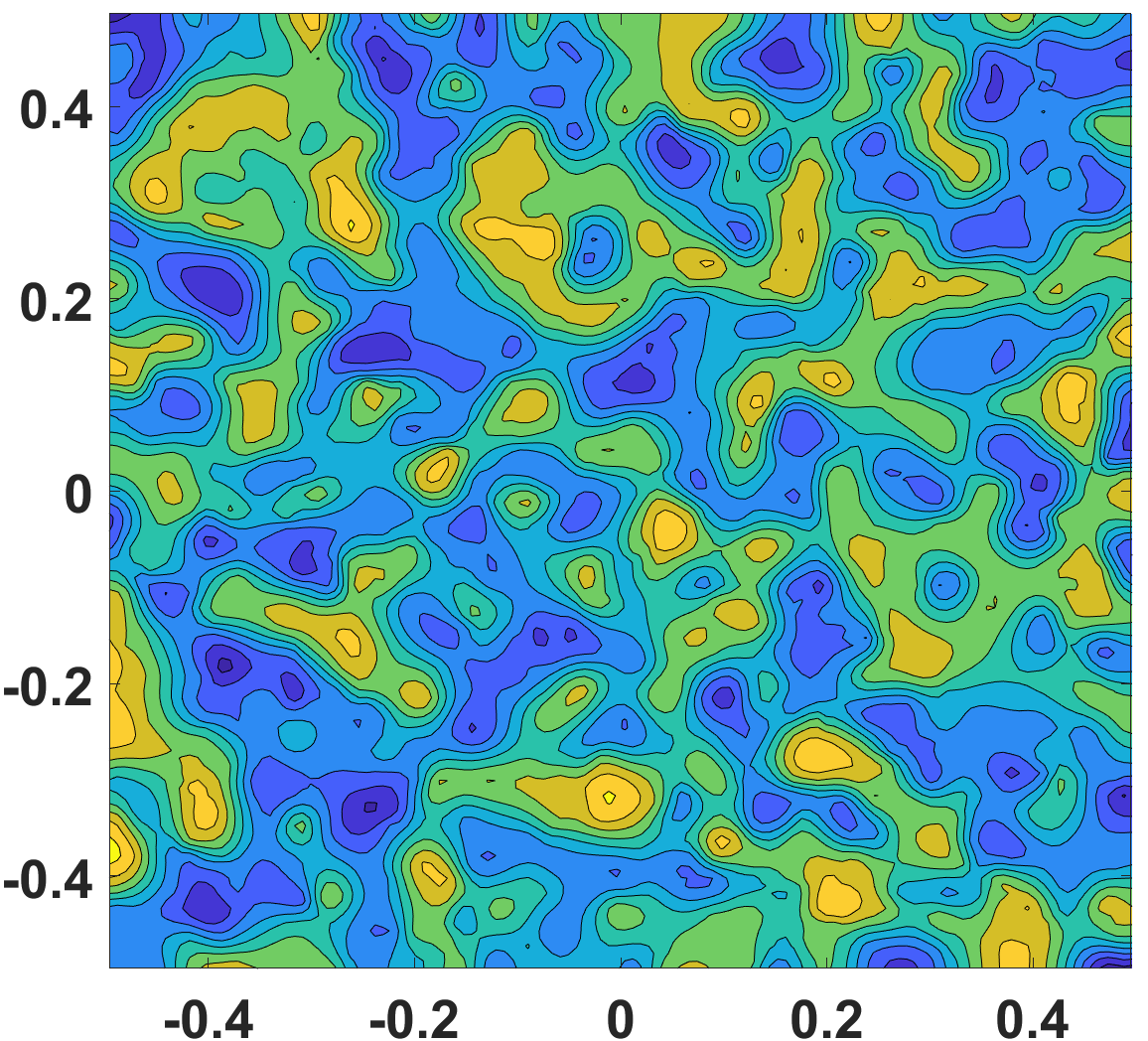}
\end{minipage}%
\begin{minipage}{0.22\textwidth}
    \includegraphics[scale = 0.2]{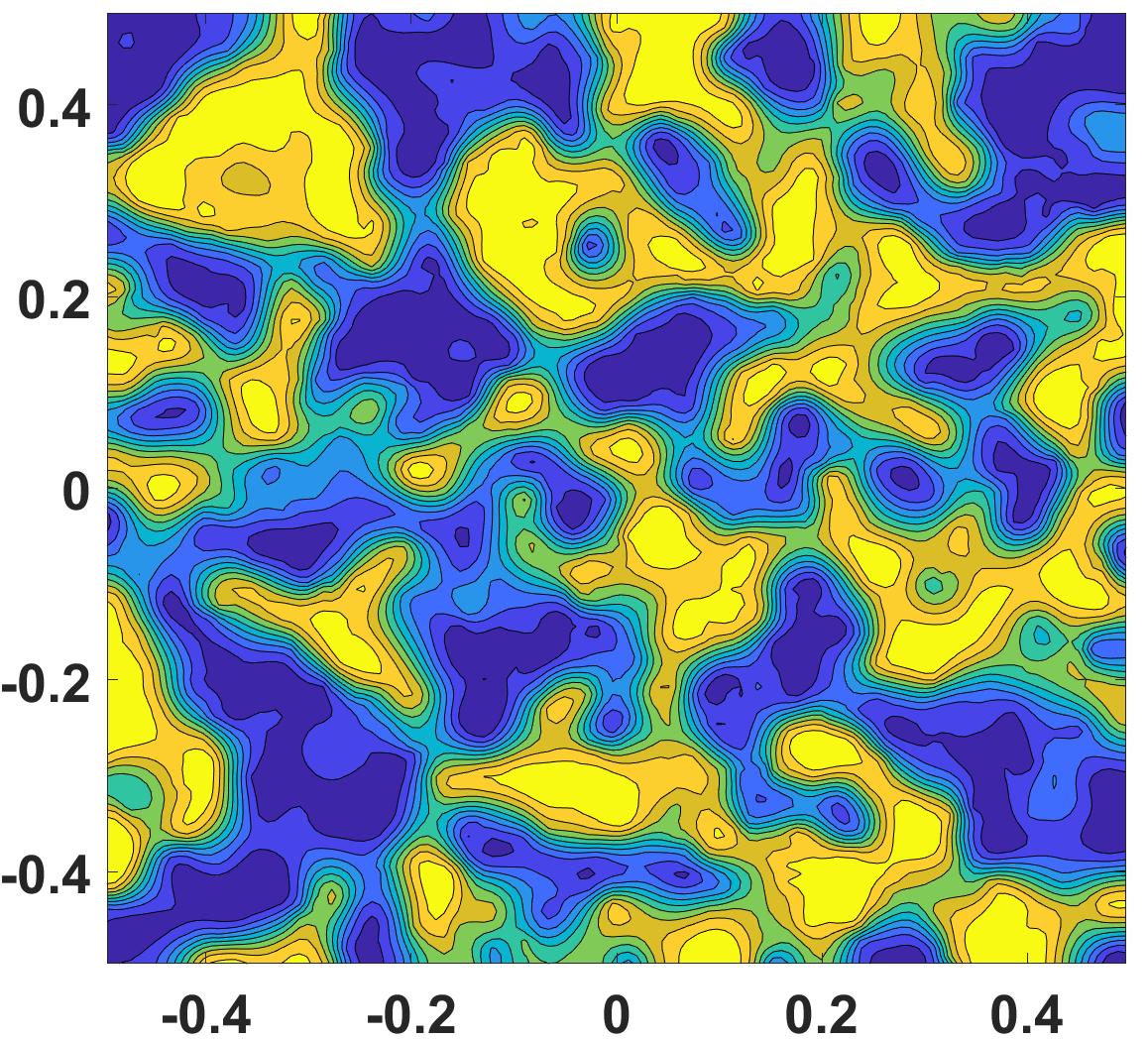}
\end{minipage}%
\begin{minipage}{0.22\textwidth}
    \includegraphics[scale = 0.2]{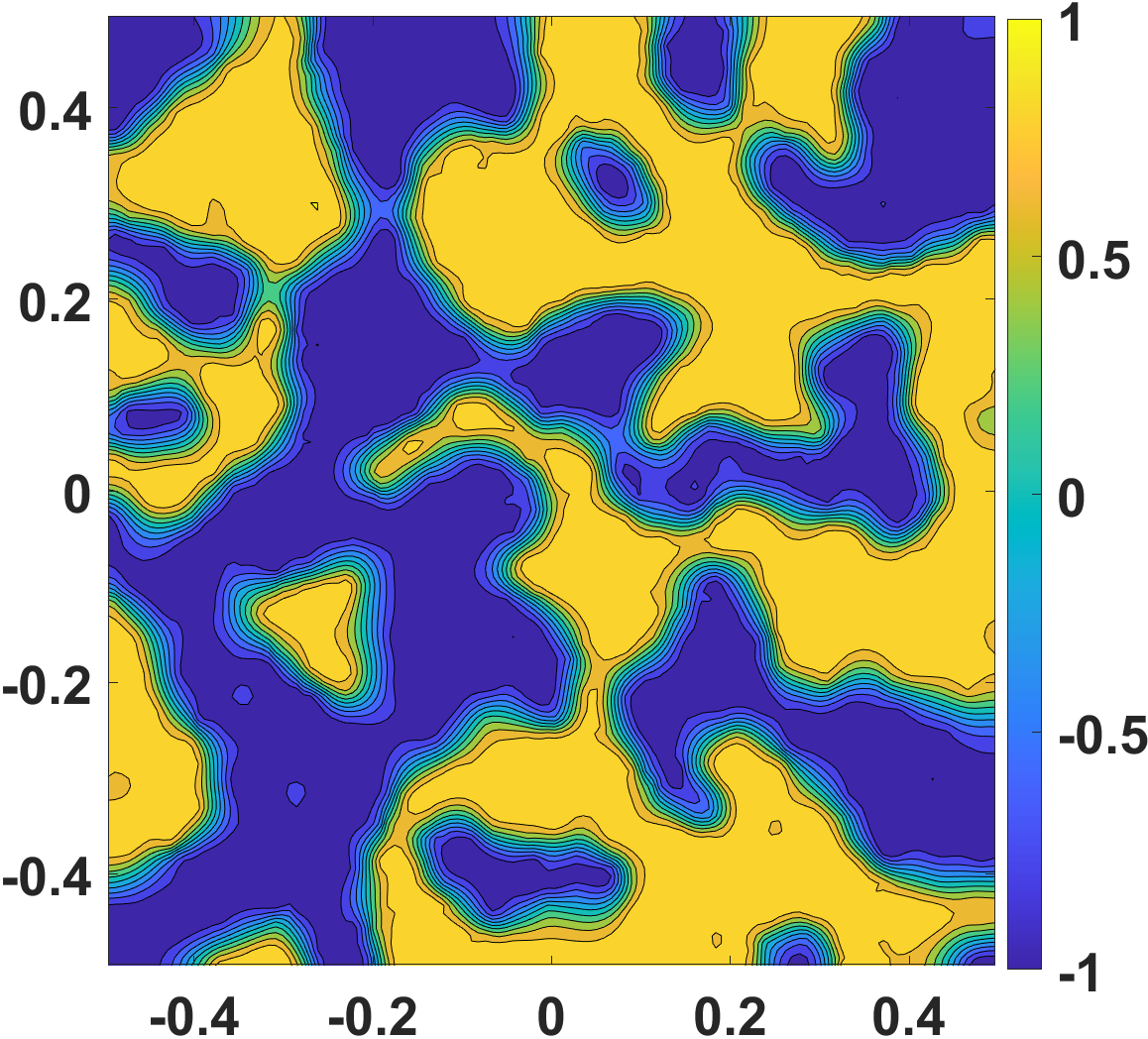}
\end{minipage}

\begin{minipage}{0.22\textwidth}
    \includegraphics[scale = 0.2]{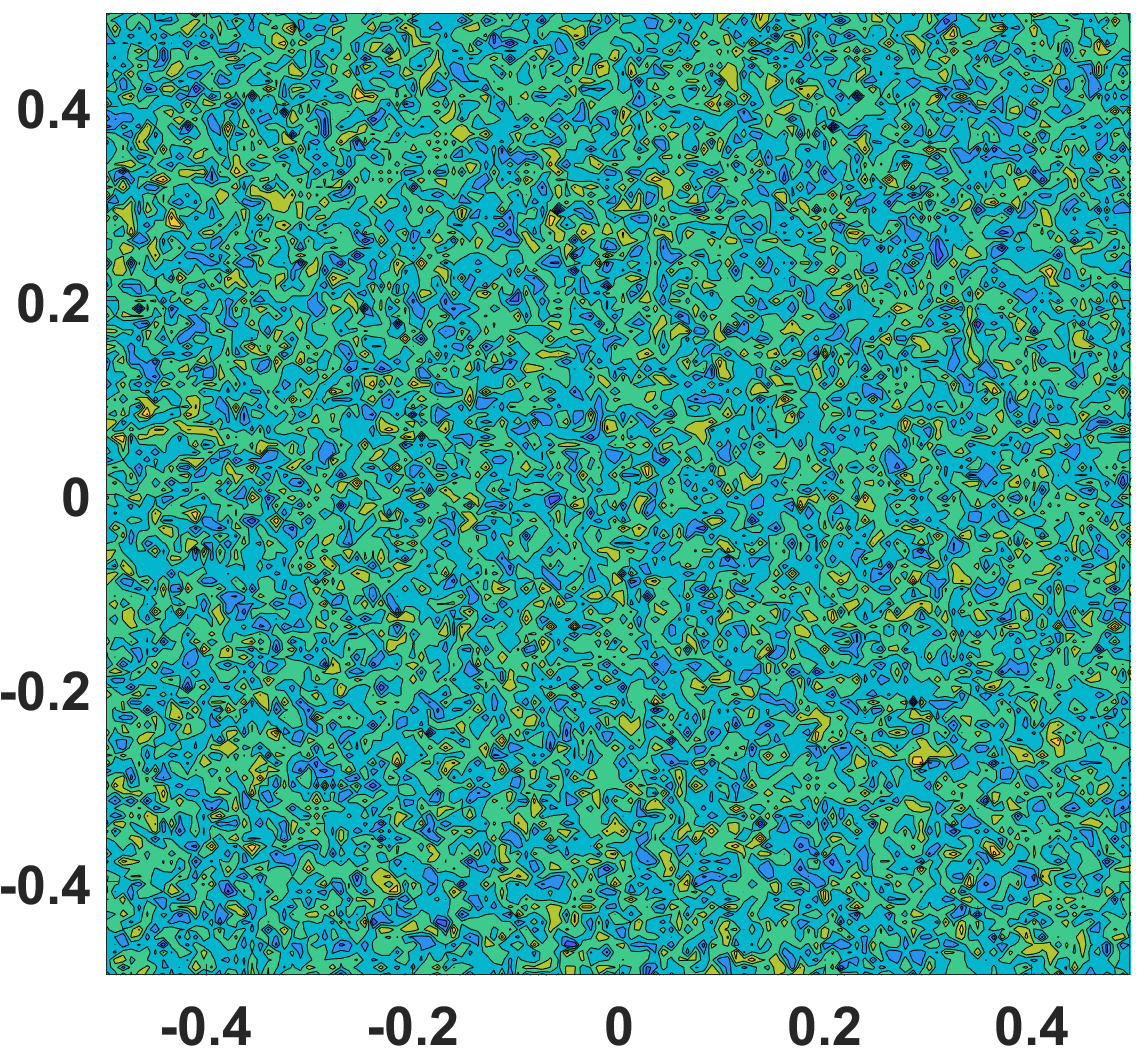}
\end{minipage}%
\begin{minipage}{0.22\textwidth}
    \includegraphics[scale = 0.2]{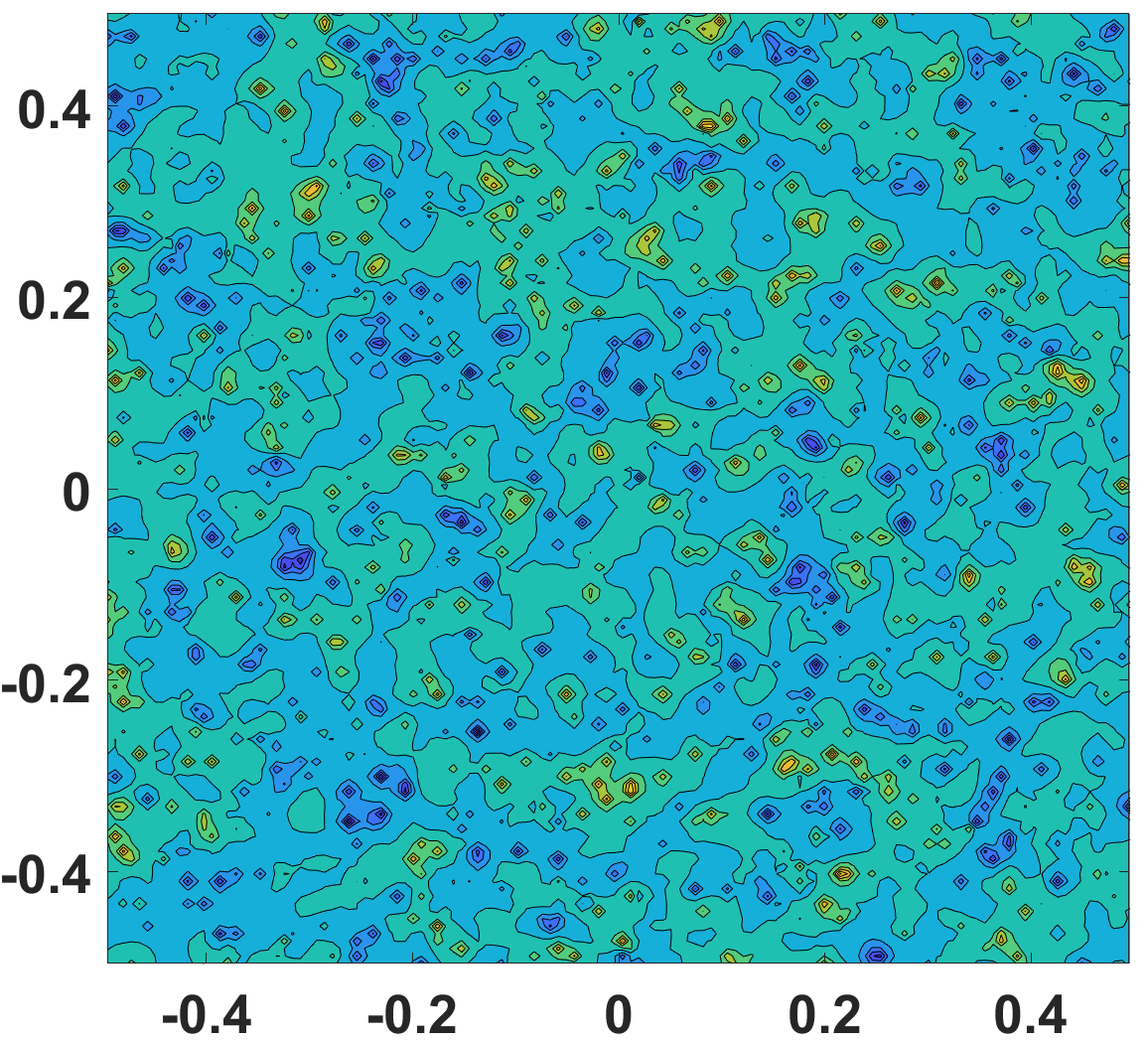}
\end{minipage}%
\begin{minipage}{0.22\textwidth}
    \includegraphics[scale = 0.2]{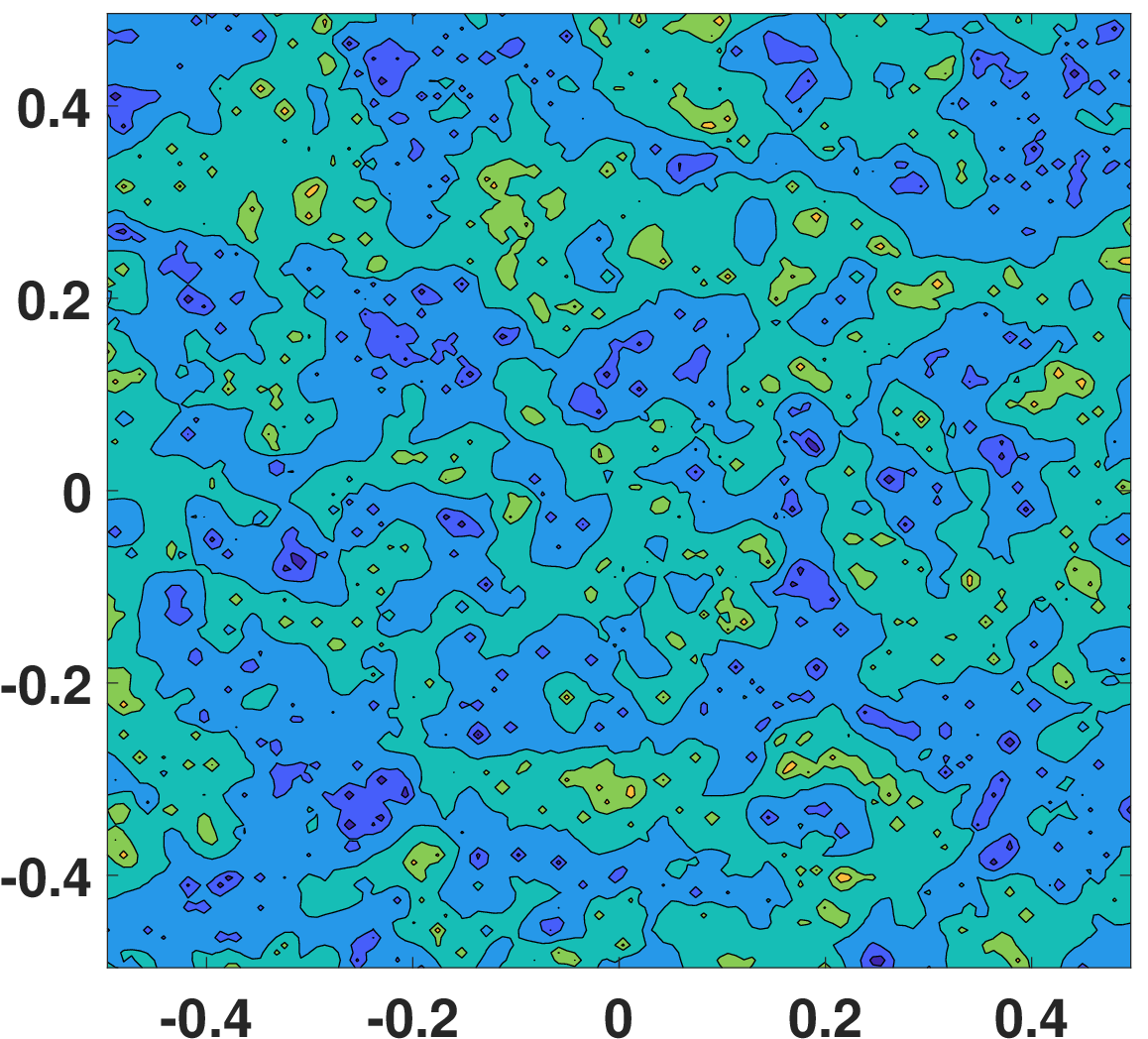}
\end{minipage}%
\begin{minipage}{0.22\textwidth}
    \includegraphics[scale = 0.2]{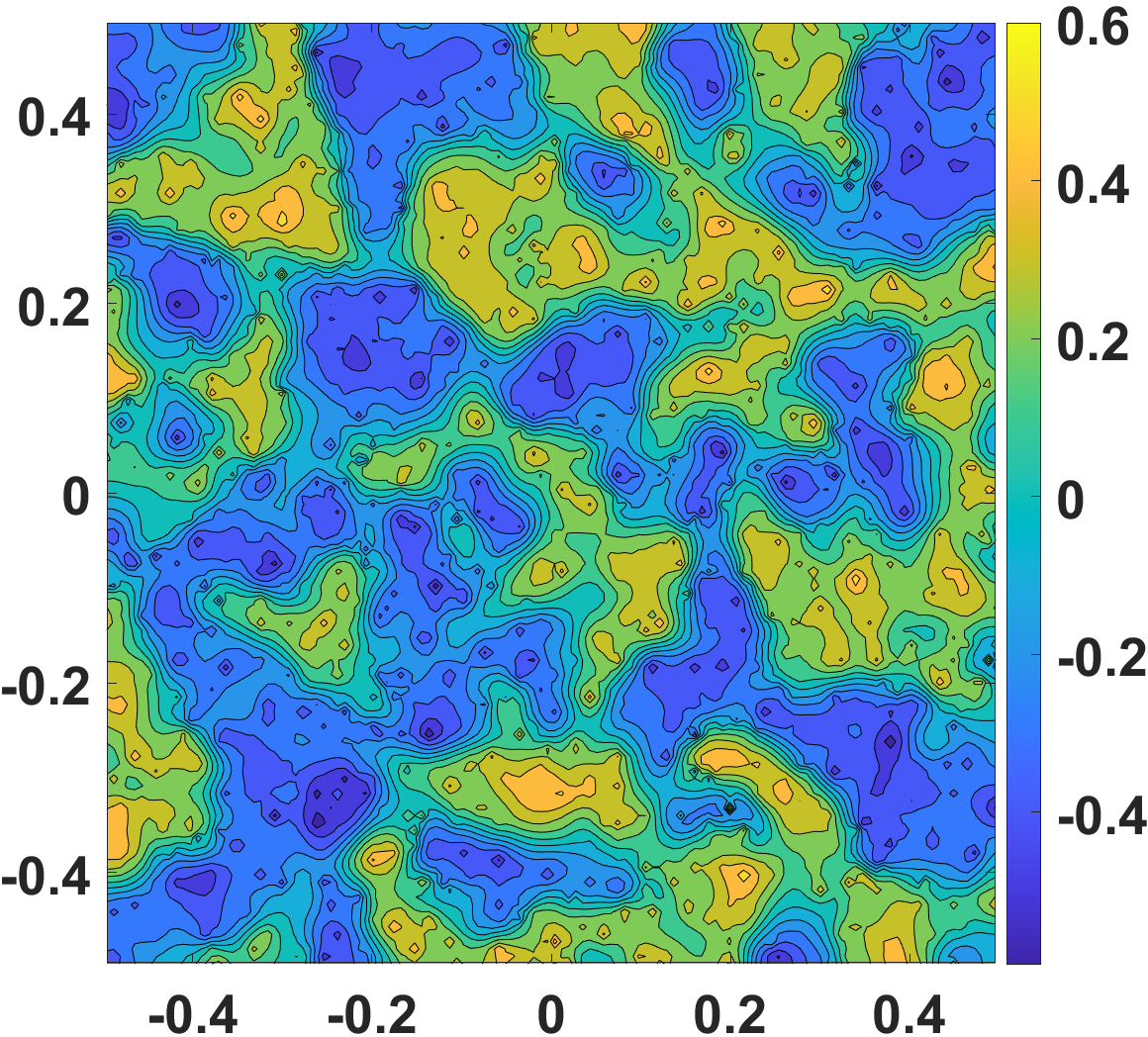}
\end{minipage}
\caption{\small Solutions of the Allen–Cahn equation in Case 1 at time $t=0$, $t= \f{T}{2}$, $t= \f{2T}{3}$, and $t=T$ with $10\%$ observations. (First row) Estimated solution by EnSF. (Second row) Estimated solution by LETKF.}
\label{ConsMob_10Obs}
\vspace{-0.2cm}
\end{figure}

\begin{figure}[h!]
\vspace{-0.2cm}
\centering
\includegraphics[scale = 0.42]{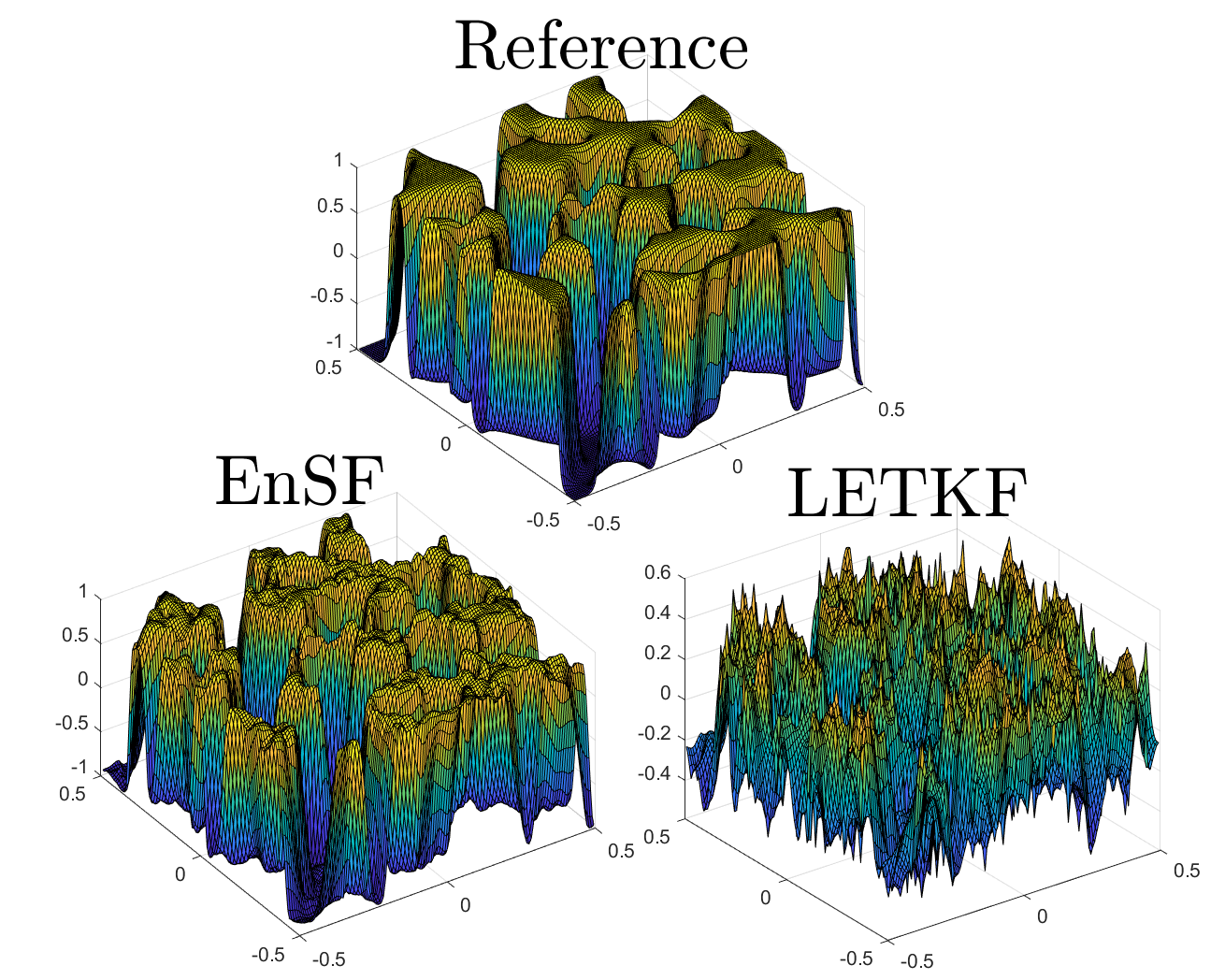}
\caption{\small 3D view of the reference and the estimated solutions for the Allen-Cahn equation in Case 1 at final time T with $10\%$ observations.}
\label{Cons_10Obs_3D}
\vspace{-0.15cm}
\end{figure}

\begin{figure}[h!]
\vspace{-0.15cm}
   \begin{minipage}{0.33\textwidth}
    \includegraphics[scale = 0.22]{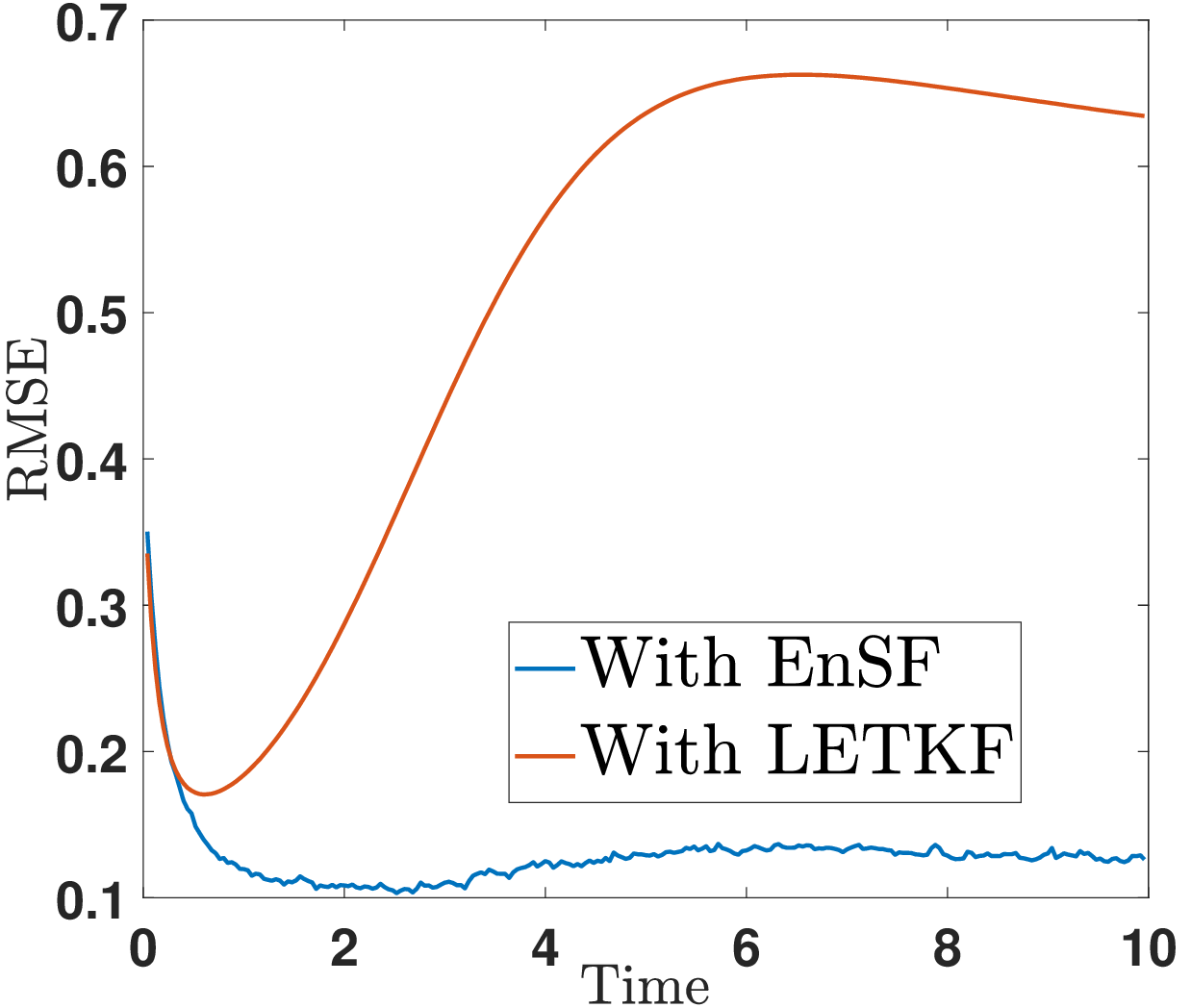}
\end{minipage}%
\begin{minipage}{0.33\textwidth}
    \includegraphics[scale = 0.22]{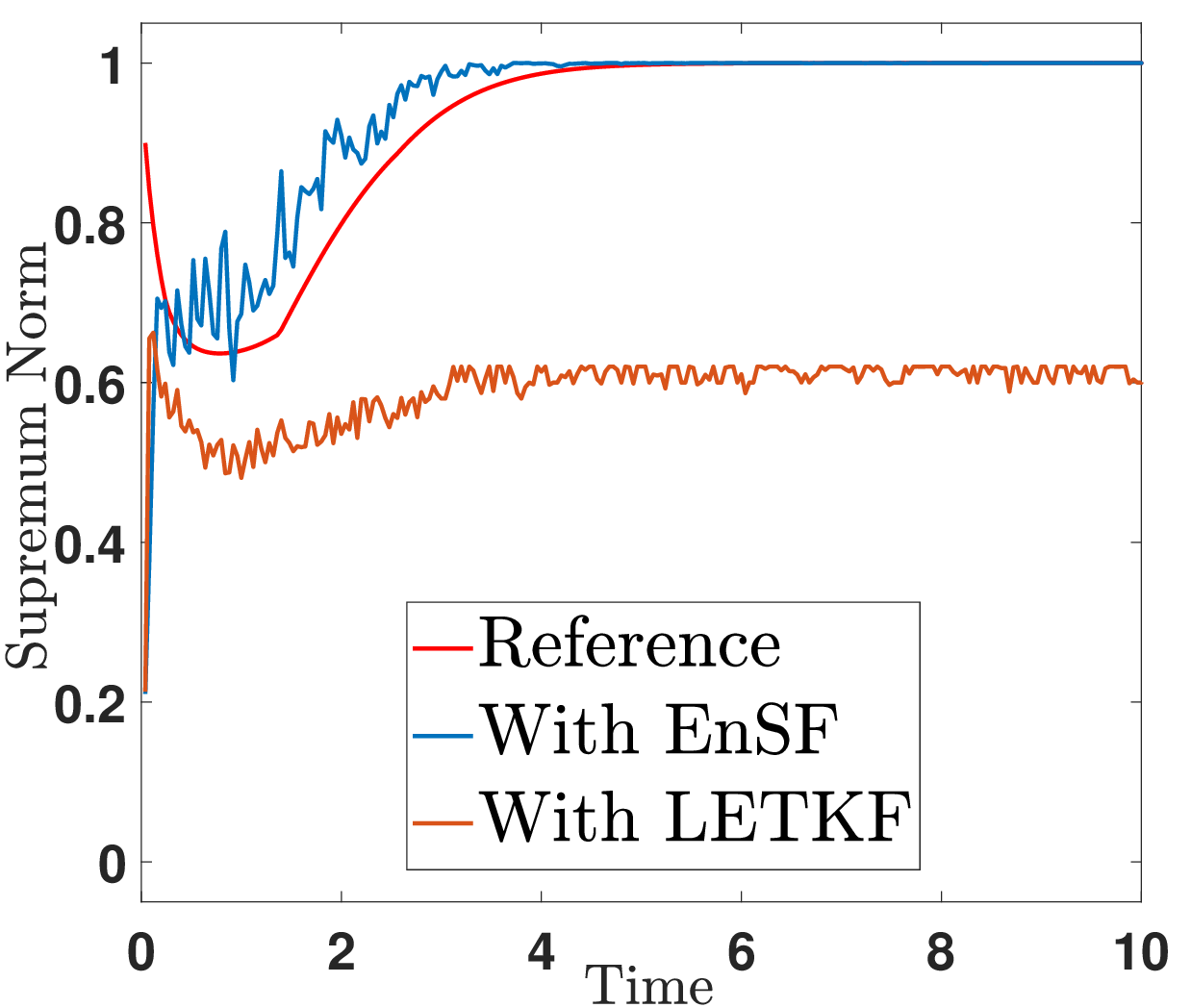}
\end{minipage}%
\begin{minipage}{0.33\textwidth}
    \includegraphics[scale = 0.22]{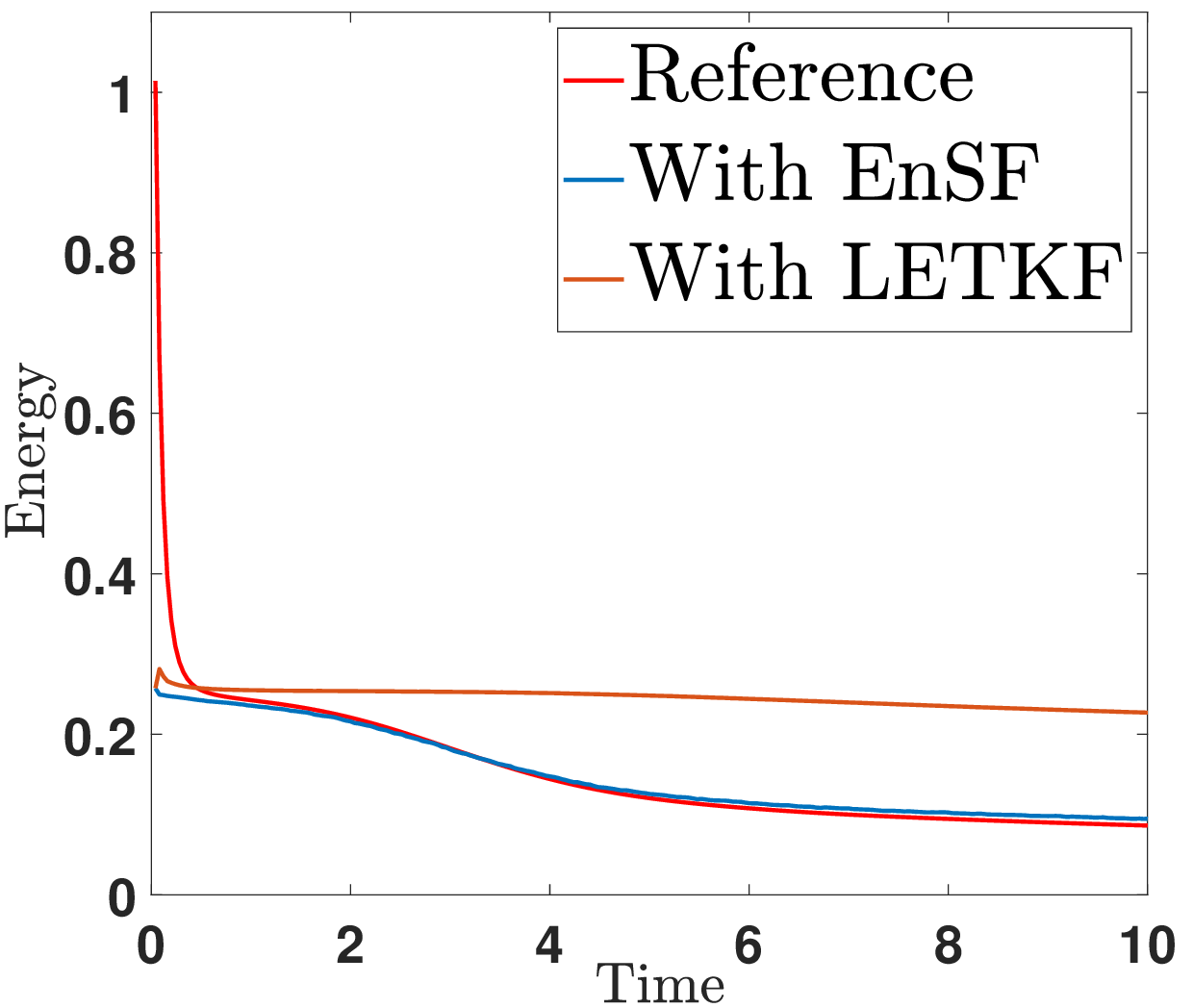}
\end{minipage}
\caption{\small Illustration of RMSEs, supremum norms, and discrete energies with $10\%$ observations for Case 1. (First) RMSE. (Second) Supremum Norm. (Third) Energy. }
\label{Cons_10Obs_RMSE_Mass_Energy}
\vspace{-0.2cm}
\end{figure}

We conclude Case 1 by presenting the results for the $10\%$ observation scenario, comparing estimates obtained using EnSF with inpainting and the LETKF. The evolution of the estimated solutions is shown in Figure~\ref{ConsMob_10Obs}.  It can be observed that our method produces more accurate estimates (top row): the final panel displays a much sharper solution that effectively filters out uncertainty, whereas the LETKF result at time $T$ remains visibly noisy. This observation is further supported by the 3D solution estimation figure depicted in Figure~\ref{Cons_10Obs_3D}. Finally, we plot Figure~\ref{Cons_10Obs_RMSE_Mass_Energy} and compare the estimated solutions in terms of RMSE, supremum norm, and energy. While both methods satisfy the desired theoretical properties, our method outperforms LETKF: it achieves smaller RMSE and its supremum‐norm and energy trajectories lie closer to those of the reference solution.

\subsubsection{Case 2: Non-constant mobility with uncertainty}

To account for the mobility uncertainty in Case 2, we define the mobility as $M_2(\phi) = \max\{1+ \xi_t, 0\}$, where $ \xi_t \sim 0.1 \cdot N(0, \pmb{I}_d)$ is the noise used during prediction. The experimental setup for this case follows the same structure as in Case 1. Specifically, we evaluate the method under three levels of observational coverage: $100\%$, $70\%$, and $10\%$. For the $10\%$ scenario, we also compare the performance of our method with that of the LETKF.

\begin{figure}[h!]
\begin{minipage}{0.22\textwidth}
    \includegraphics[scale = 0.2]{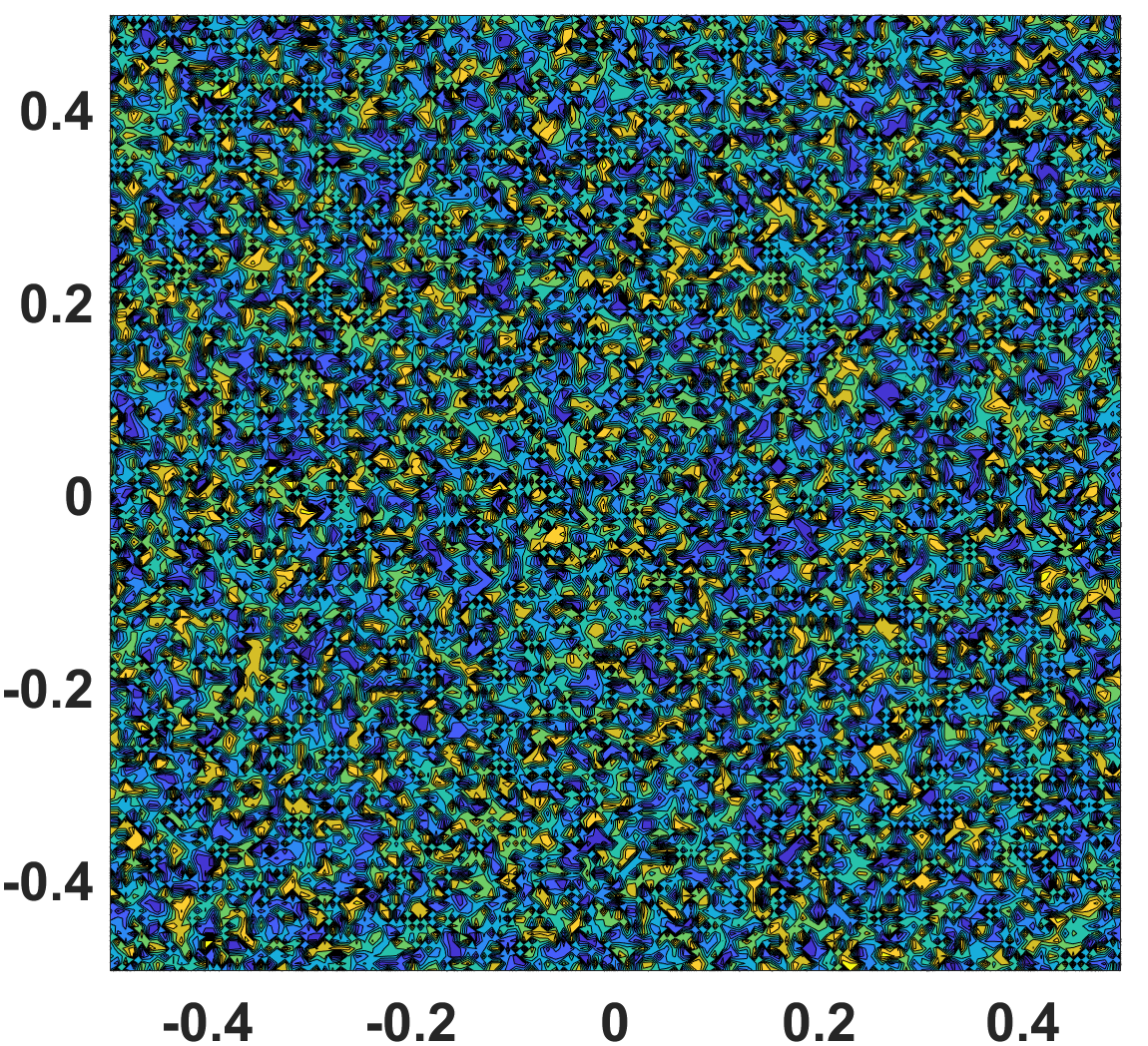}
\end{minipage}%
\begin{minipage}{0.22\textwidth}
    \includegraphics[scale = 0.2]{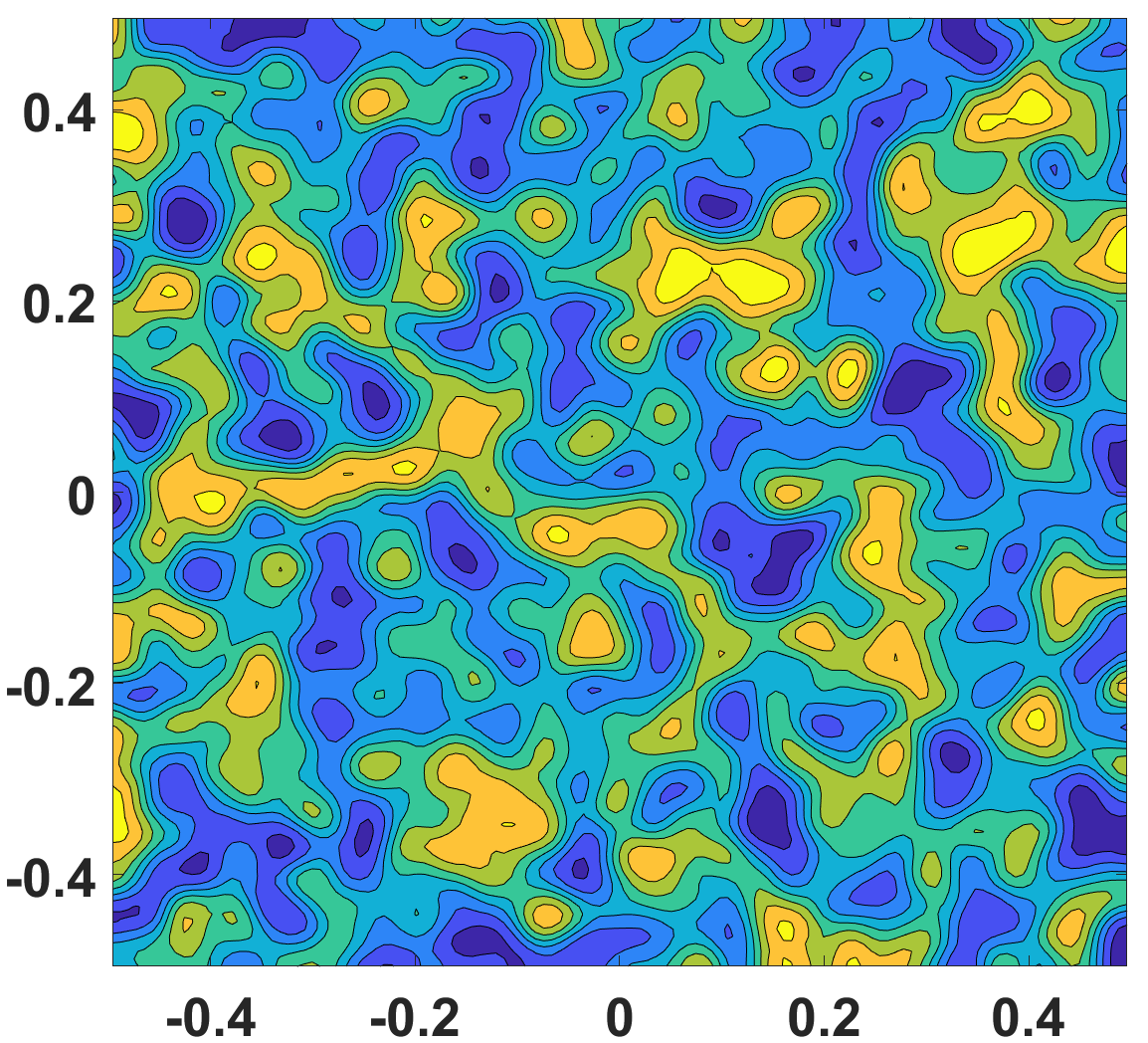}
\end{minipage}%
\begin{minipage}{0.22\textwidth}
    \includegraphics[scale = 0.2]{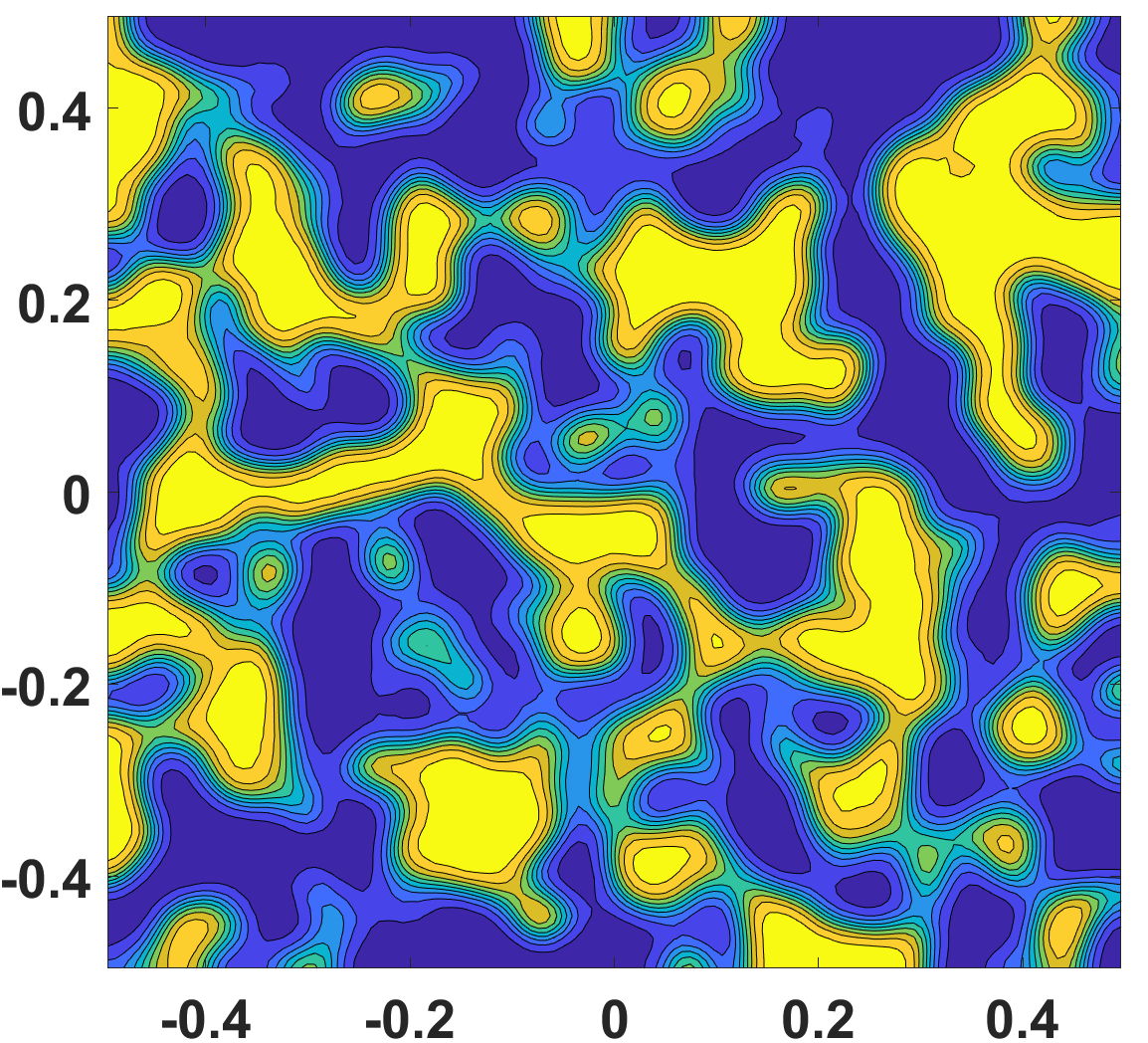}
\end{minipage}%
\begin{minipage}{0.22\textwidth}
    \includegraphics[scale = 0.2]{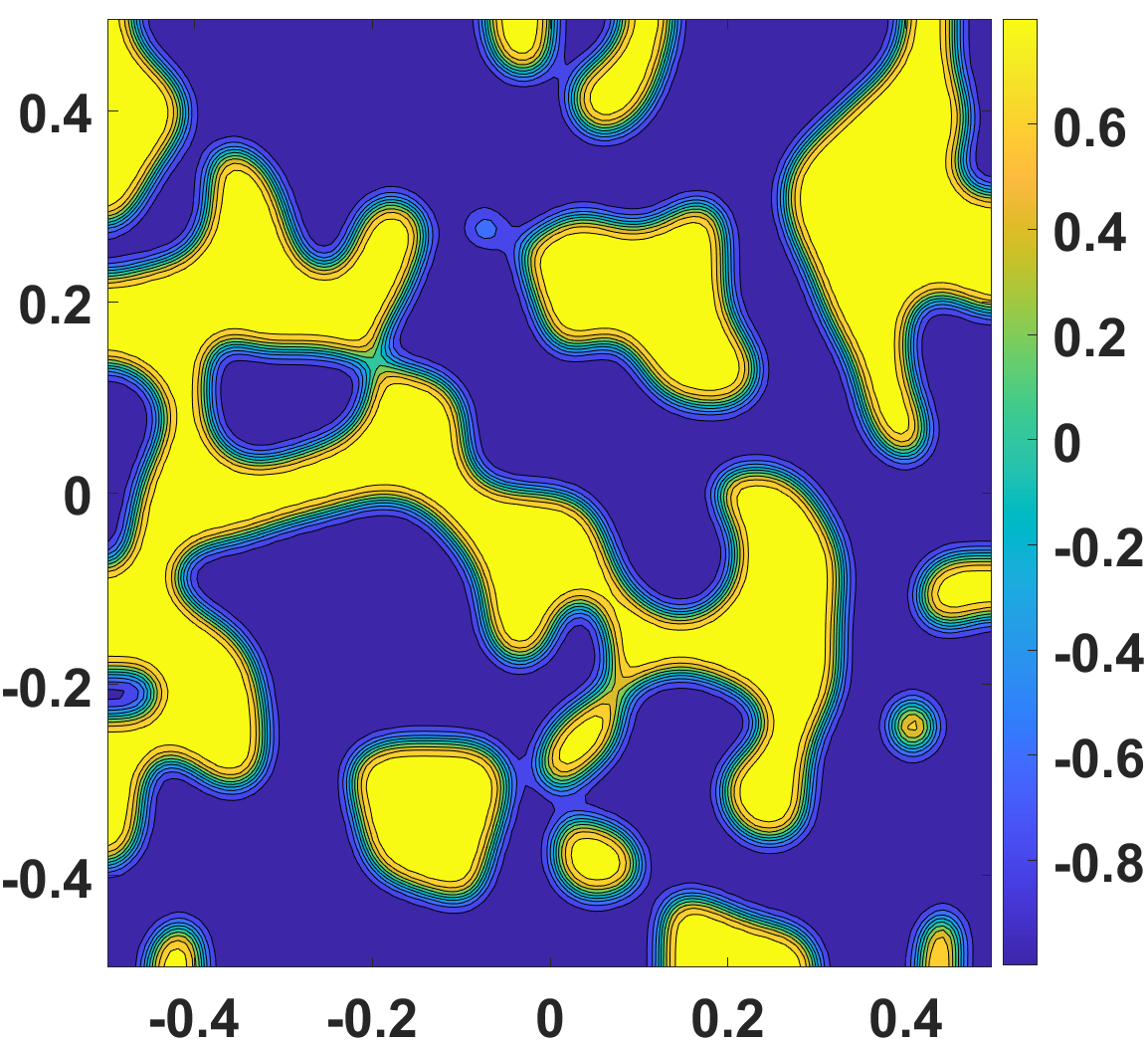}
\end{minipage}

\begin{minipage}{0.22\textwidth}
    \includegraphics[scale = 0.2]{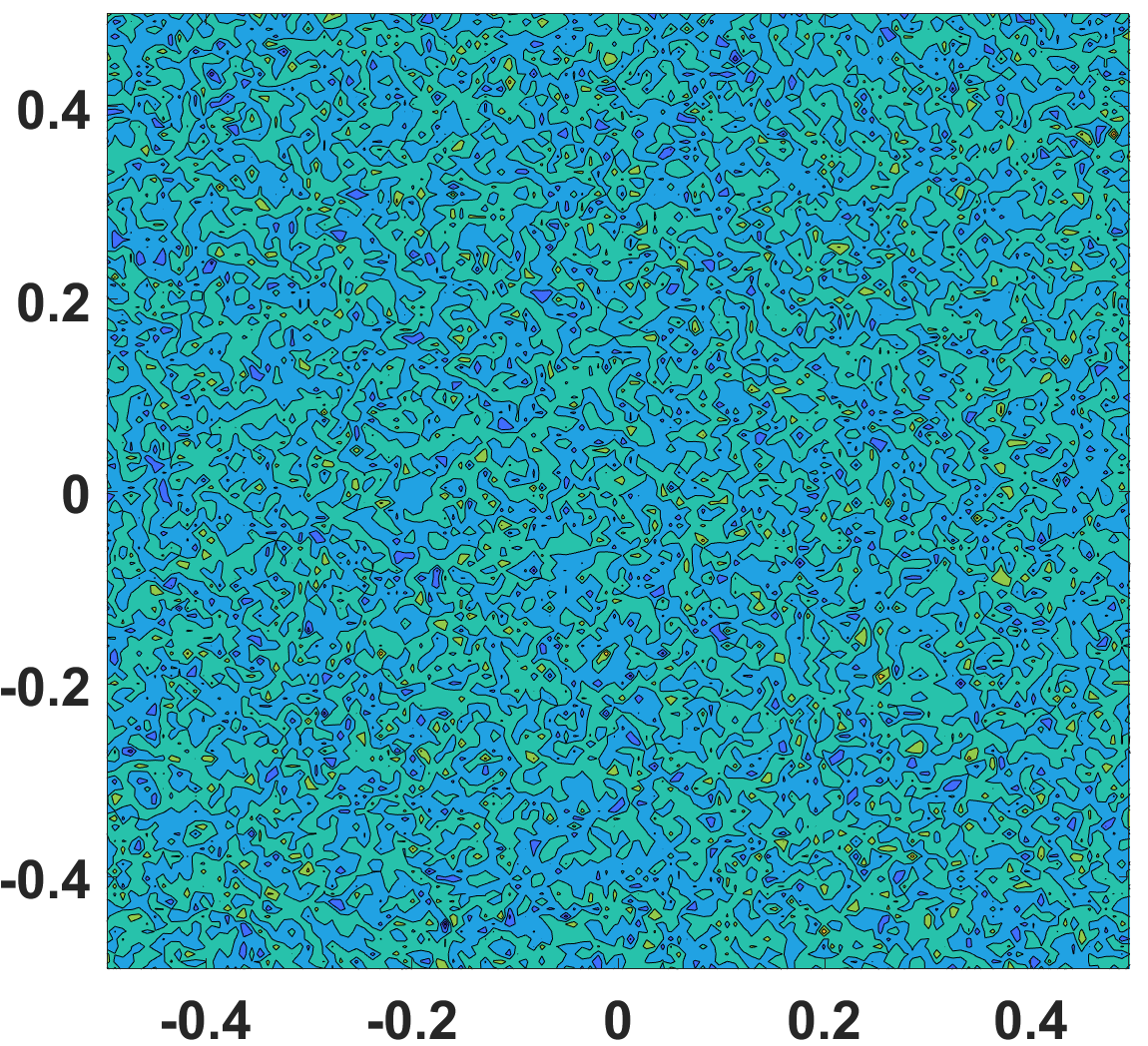}
\end{minipage}%
\begin{minipage}{0.22\textwidth}
    \includegraphics[scale = 0.2]{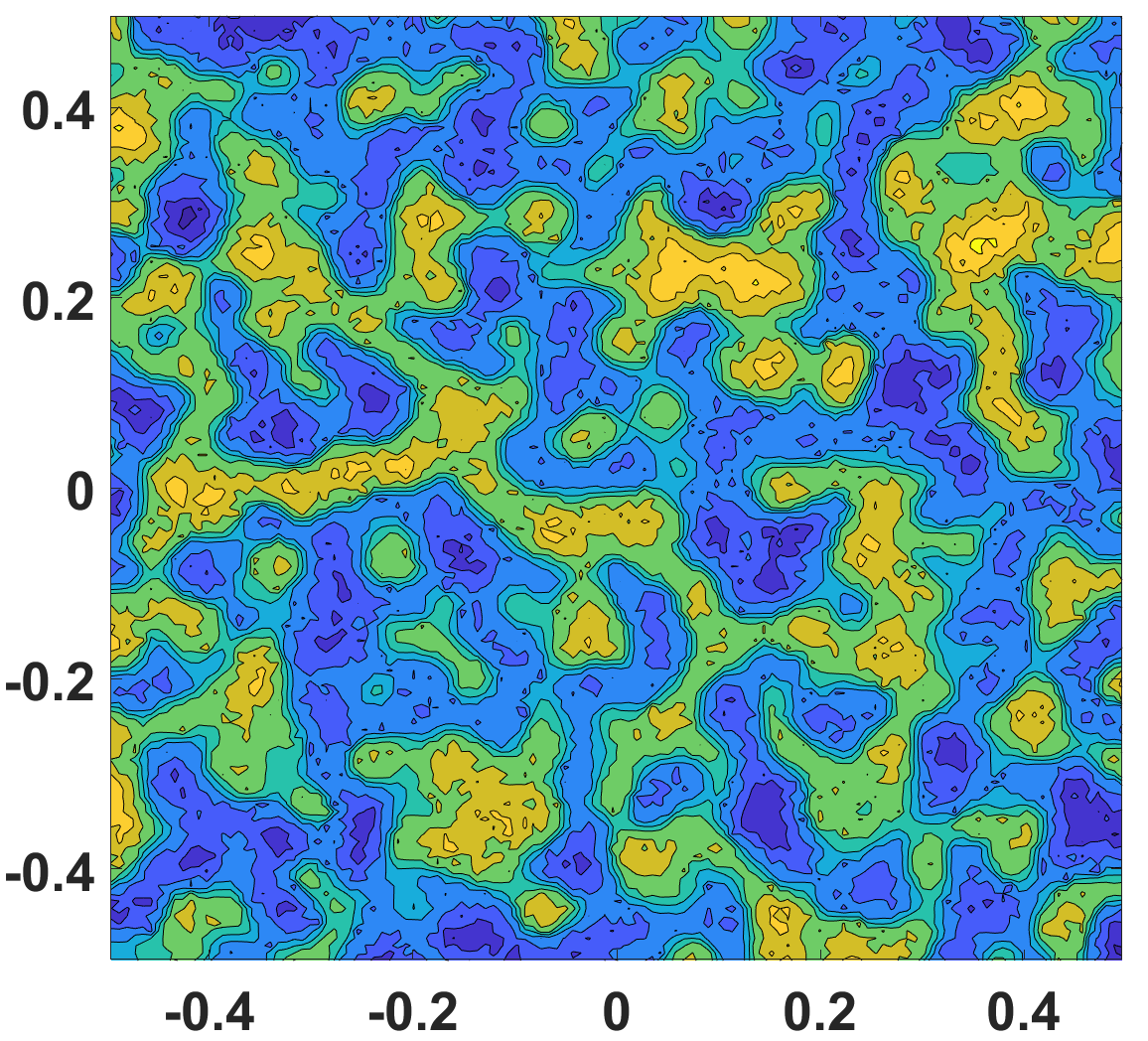}
\end{minipage}%
\begin{minipage}{0.22\textwidth}
    \includegraphics[scale = 0.2]{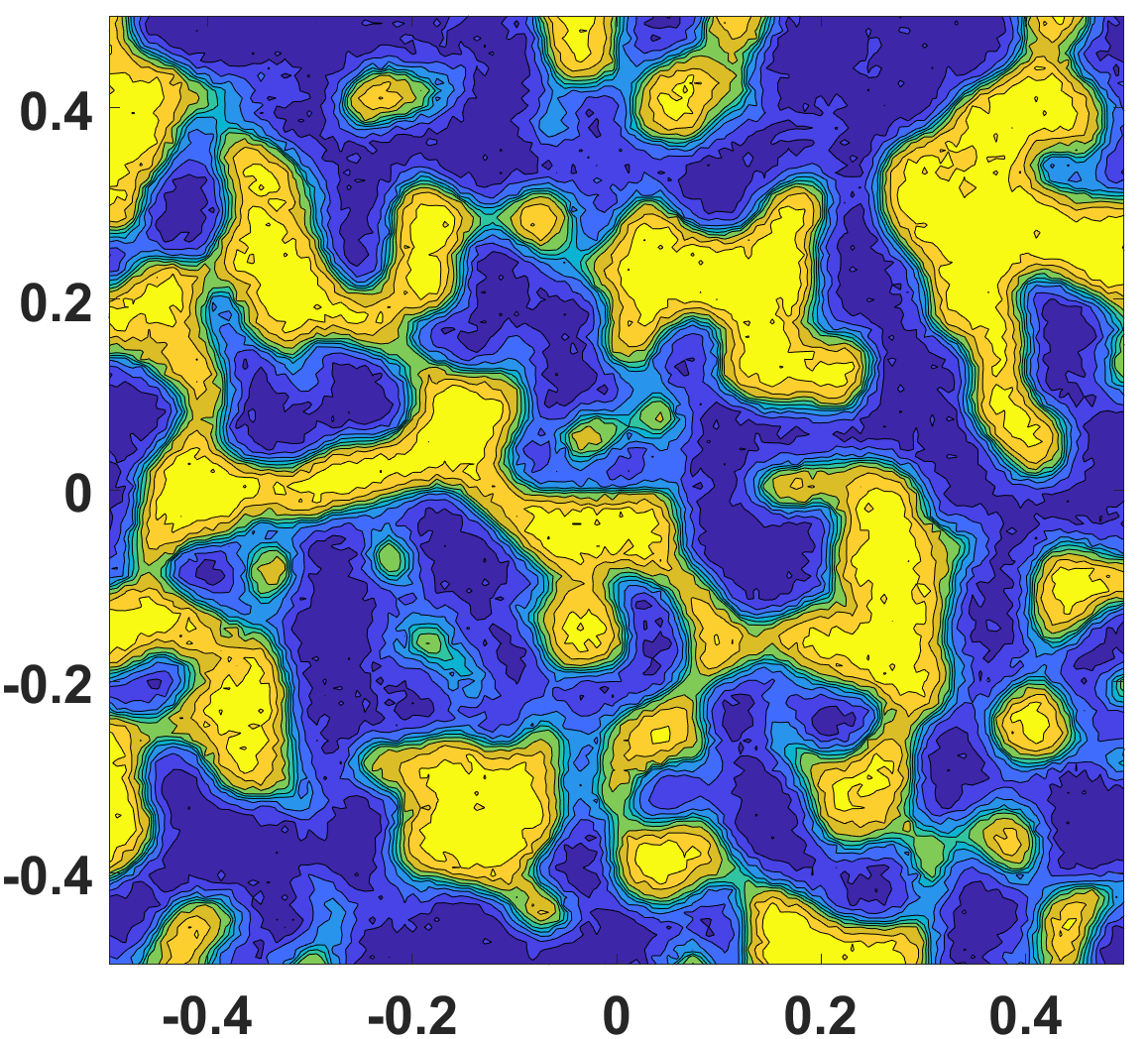}
\end{minipage}%
\begin{minipage}{0.22\textwidth}
    \includegraphics[scale = 0.2]{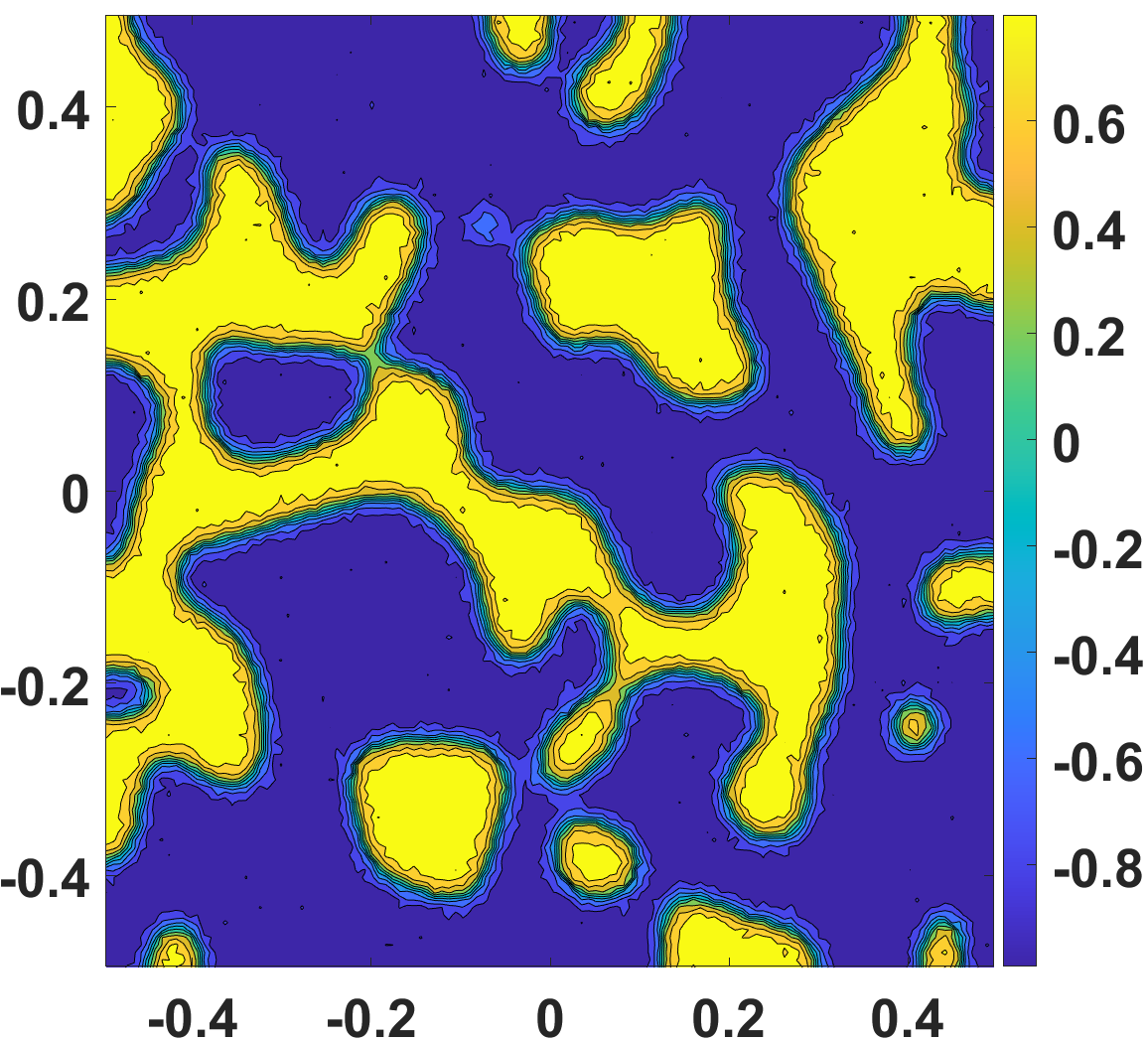}
\end{minipage}

\begin{minipage}{0.22\textwidth}
    \includegraphics[scale = 0.2]{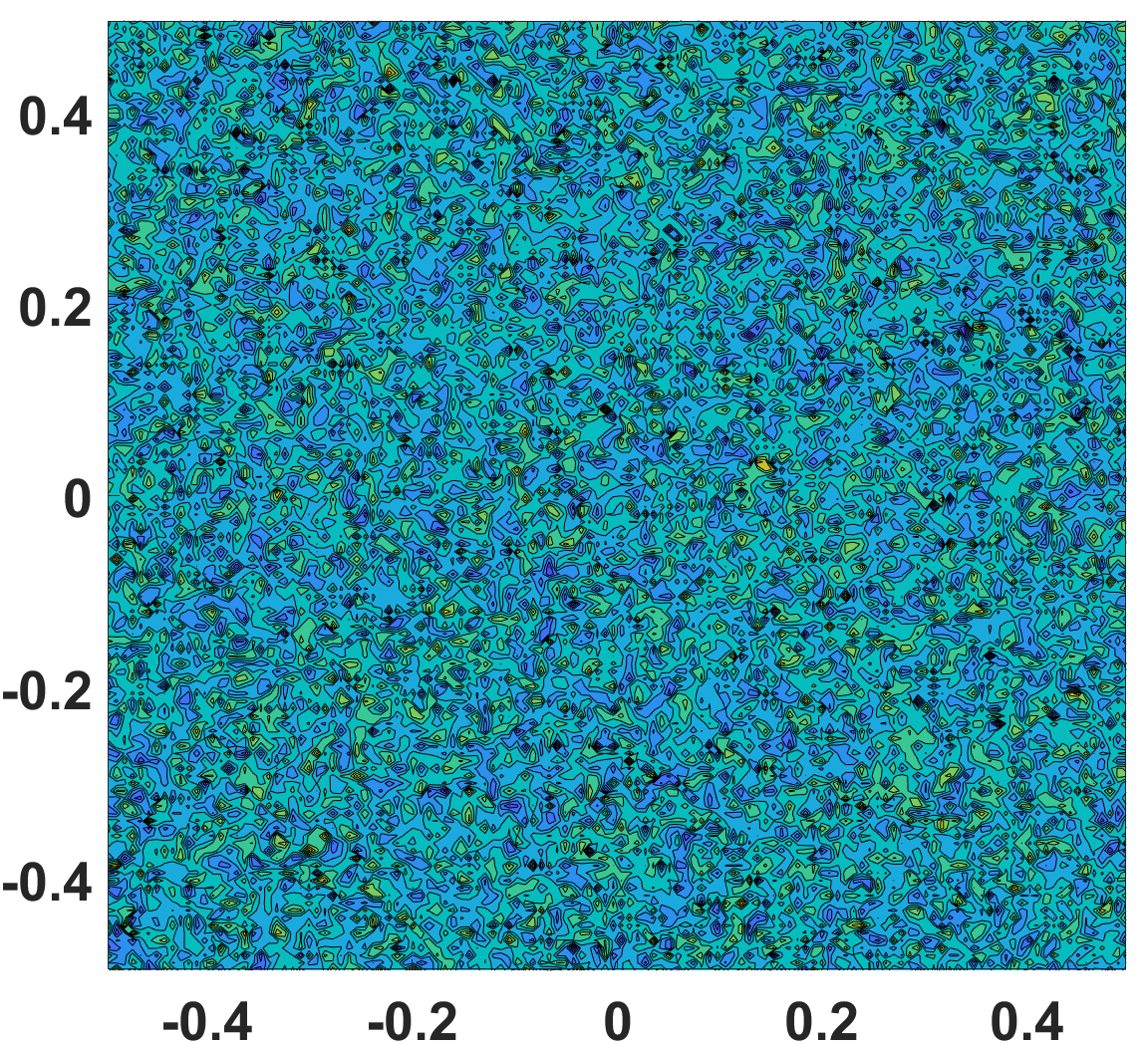}
\end{minipage}%
\begin{minipage}{0.22\textwidth}
    \includegraphics[scale = 0.2]{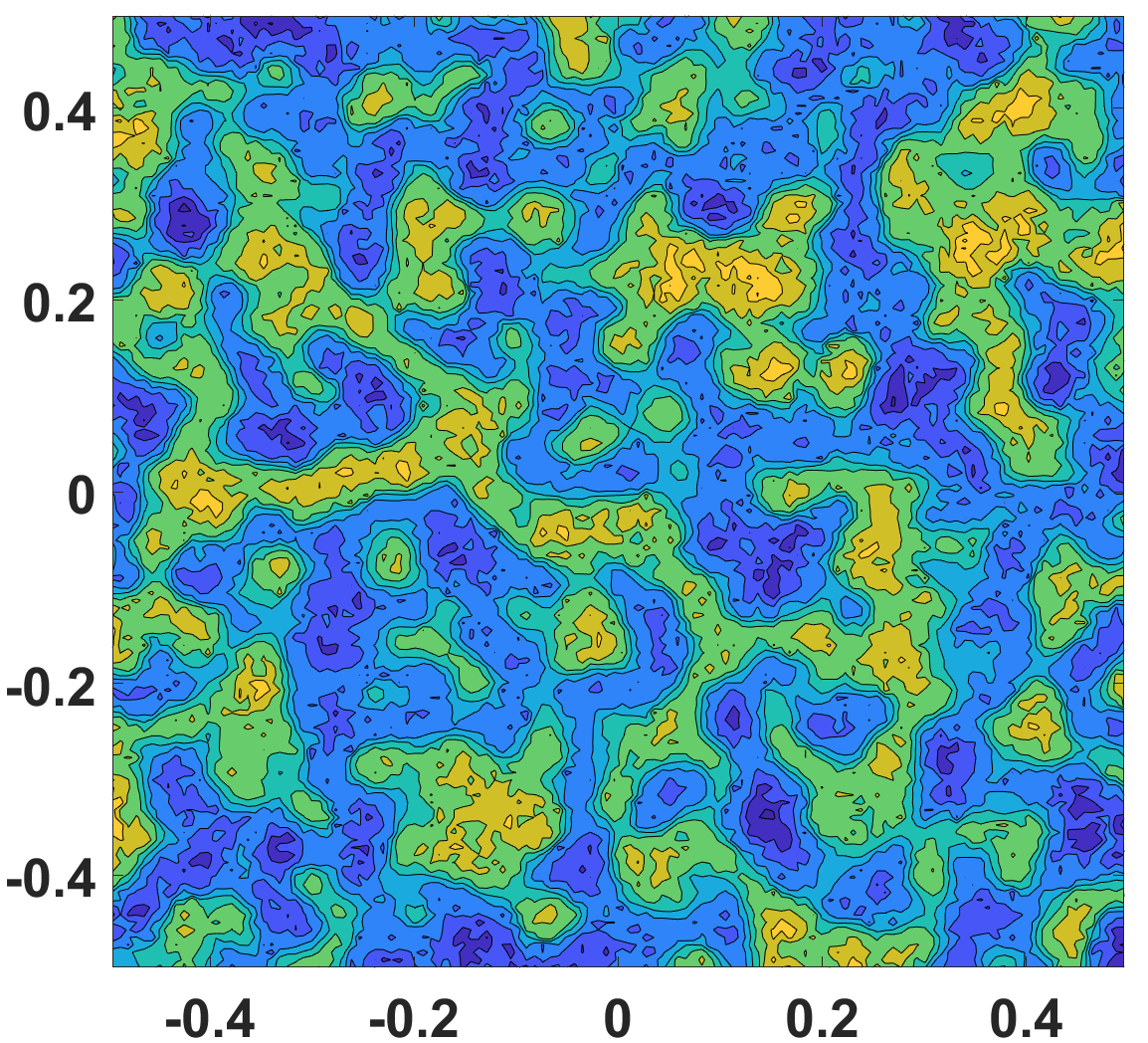}
\end{minipage}%
\begin{minipage}{0.22\textwidth}
    \includegraphics[scale = 0.2]{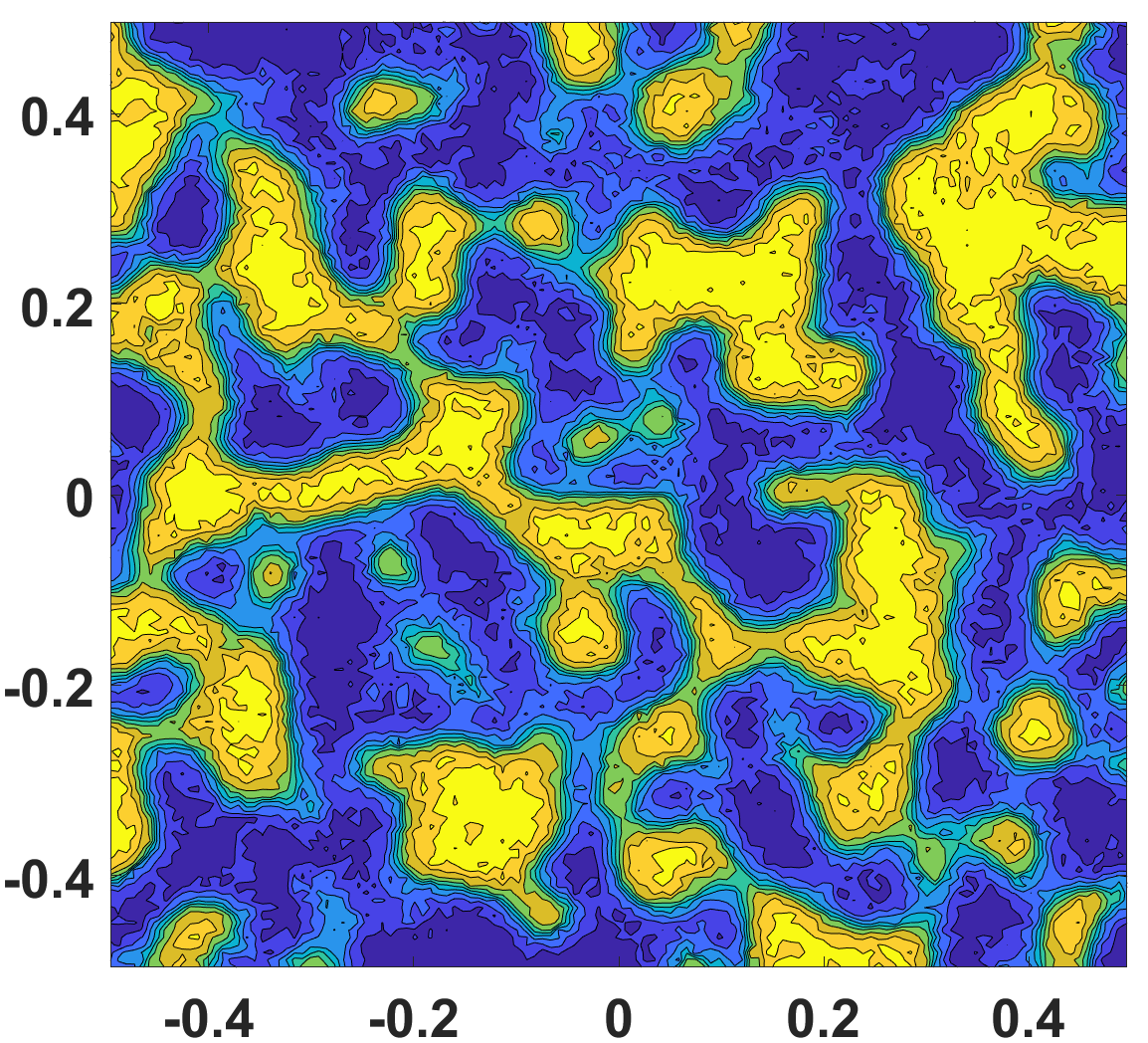}
\end{minipage}%
\begin{minipage}{0.22\textwidth}
    \includegraphics[scale = 0.2]{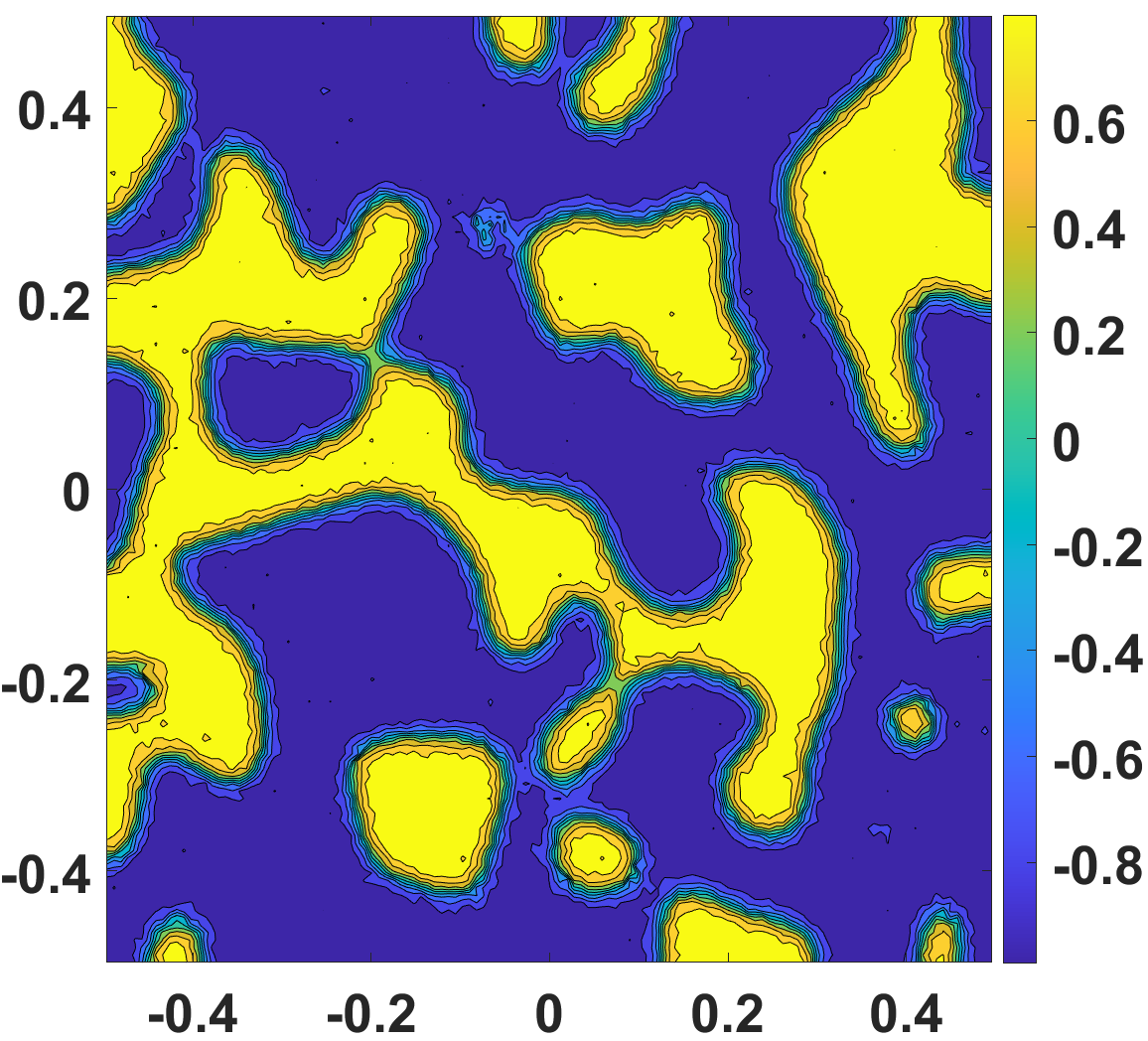}
\end{minipage}
\caption{\small Solutions of the Allen–Cahn equation in Case 2 at time $t=0, \f{T}{2}, \f{2T}{3}, \; T$. (First row) Reference. (Second row) Estimated solution with $100\%$ observations. (Third row) Estimated solution with $70\%$ observations.}
\label{NonConsM_100_70Obs}
\end{figure}

\begin{figure}[h!]
   \begin{minipage}{0.33\textwidth}
    \includegraphics[scale = 0.22]{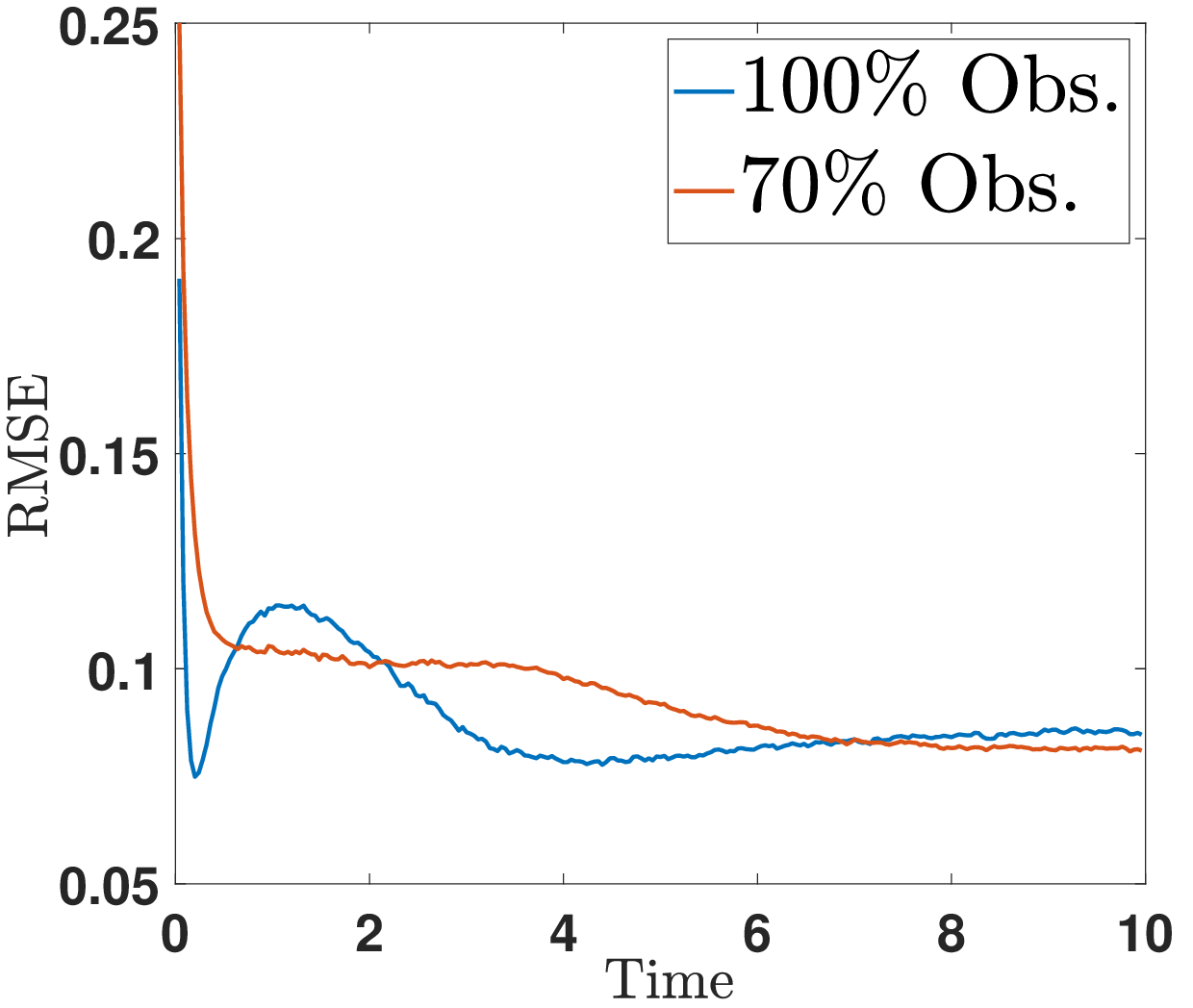}
\end{minipage}%
\begin{minipage}{0.33\textwidth}
    \includegraphics[scale = 0.22]{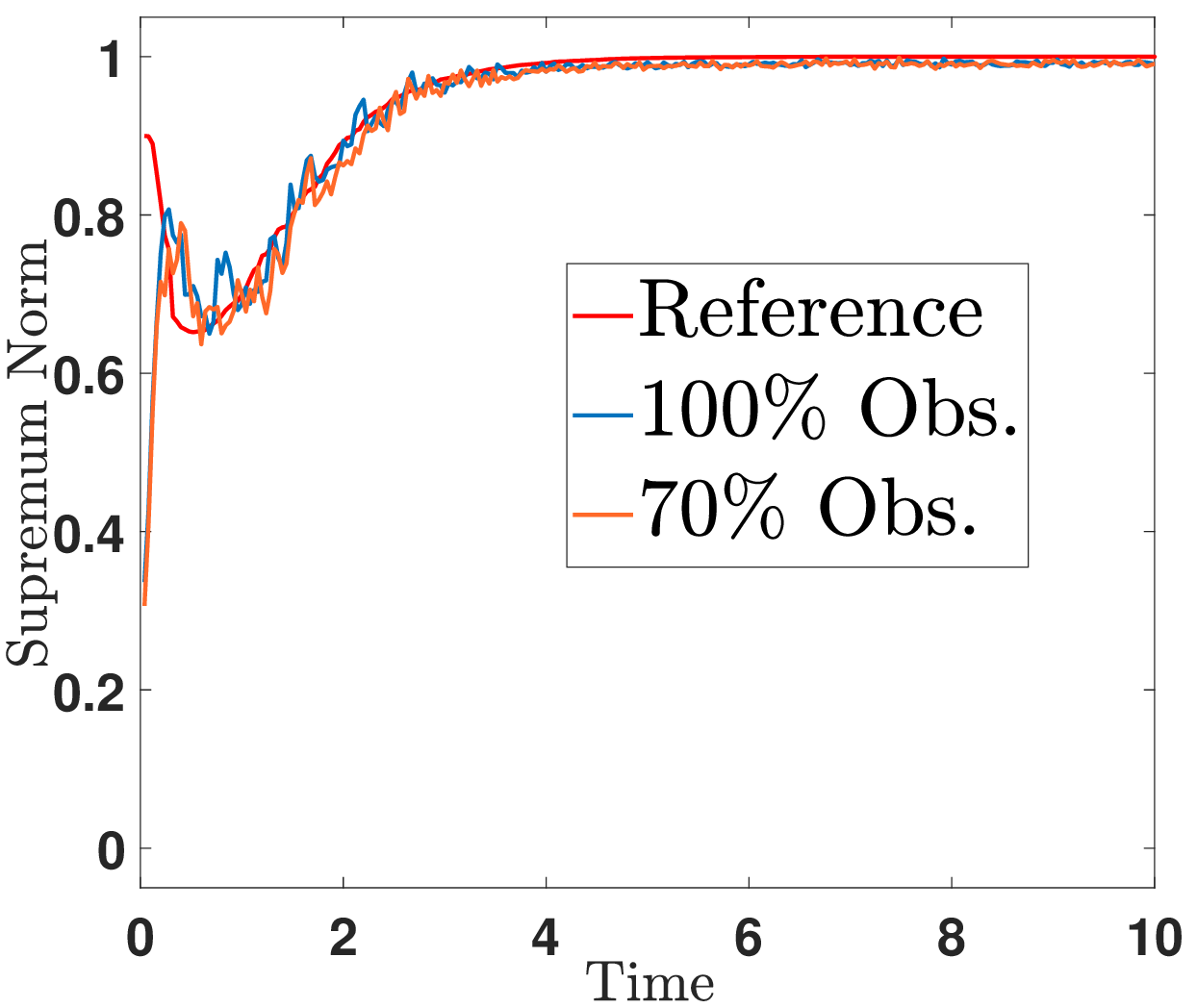}
\end{minipage}%
\begin{minipage}{0.33\textwidth}
    \includegraphics[scale = 0.22]{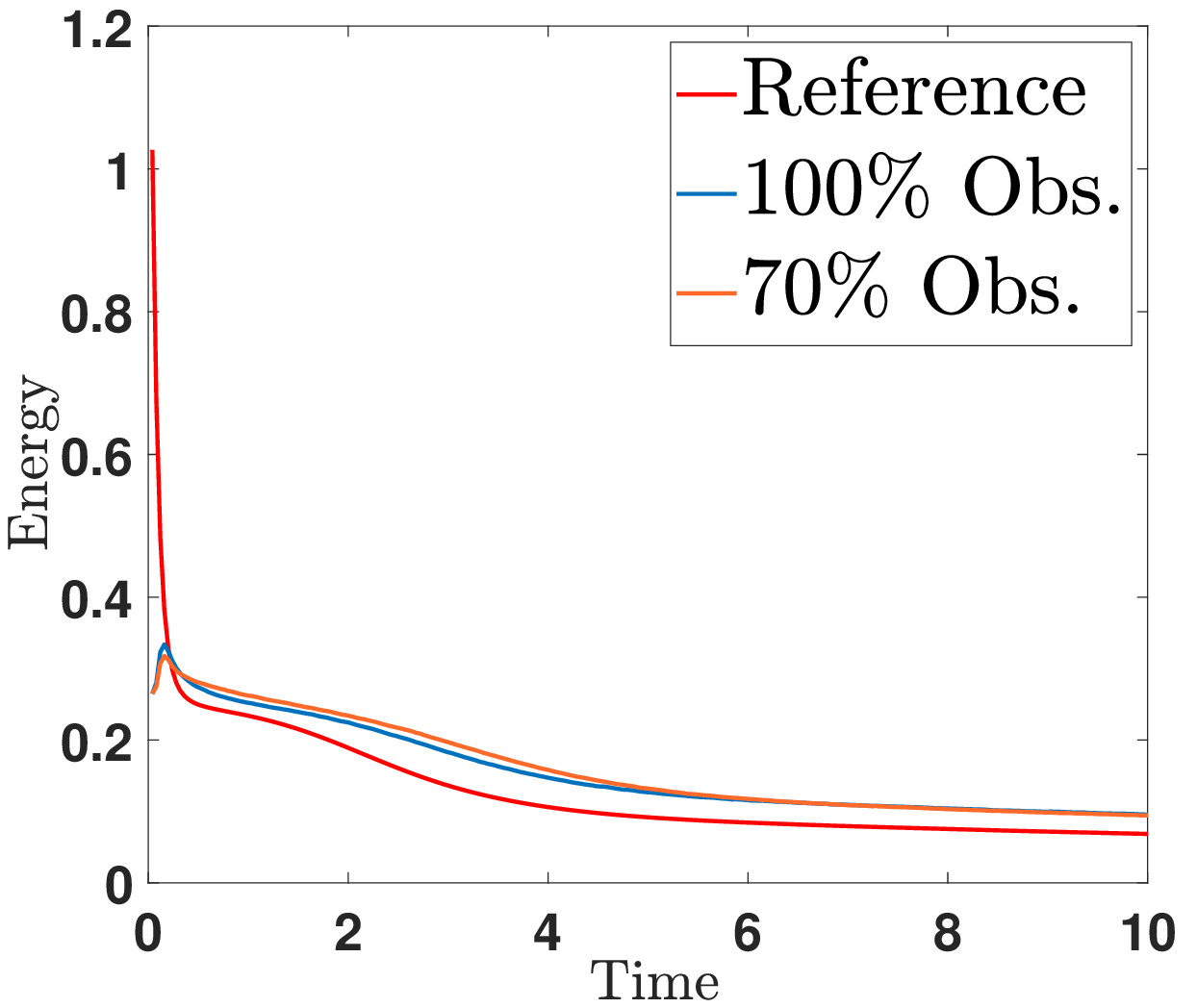}
\end{minipage}
\caption{\small Illustration of RMSEs, supremum norms, and discrete energies with $100\%$ and $70\%$ observations for Case 2. (Left) RMSE.  (Center) Supremum Norm. (Right) Energy. }
\label{NonCons_100_70Obs_RMSE_Mass_Energy}
\vspace{-0.2cm}
\end{figure}

The evolution of the estimated solutions with $100\%$ and $70\%$ observations are shown in Figure~\ref{NonConsM_100_70Obs}. Despite differences in the initial guess and the presence of various uncertainties in the PDE model, we obtain a consistent state estimation performance in Case 2. In particular, the estimated solution progressively improves over time and accurately captures the transition interfaces of the reference solution.

The RMSEs, the supremum norms, and the discrete energies of the estimates are displayed in Figure~\ref{NonCons_100_70Obs_RMSE_Mass_Energy}. The first panel of Figure~\ref{NonCons_100_70Obs_RMSE_Mass_Energy} confirms that the accuracies for both levels of observational data are nearly the same. From the last two panels, it is clear that we still achieve the maximum bound principle and the energy dissipation law for the estimated states.

\begin{figure}[h!]
\begin{minipage}{0.22\textwidth}
    \includegraphics[scale = 0.2]{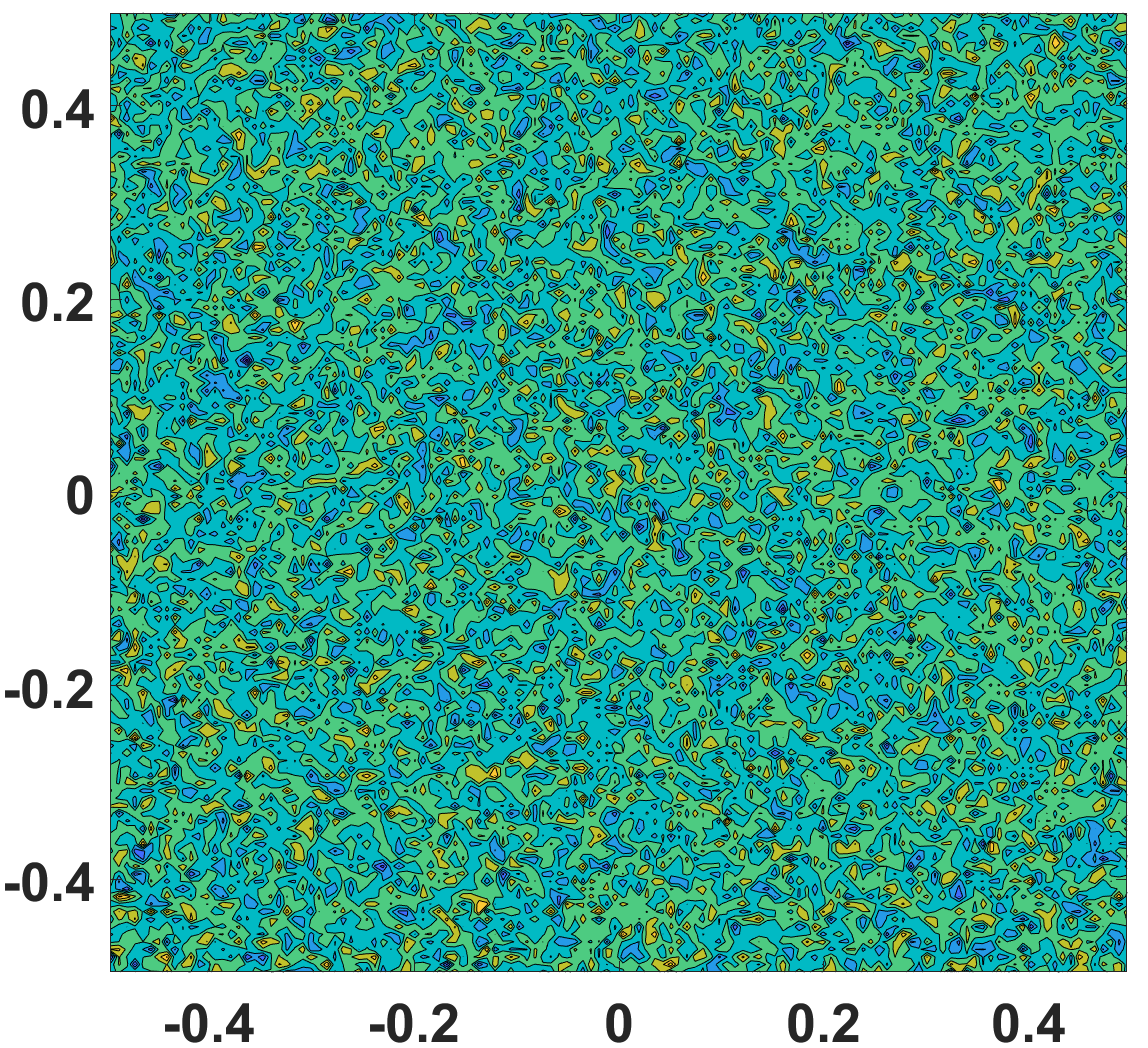}
\end{minipage}%
\begin{minipage}{0.22\textwidth}
    \includegraphics[scale = 0.2]{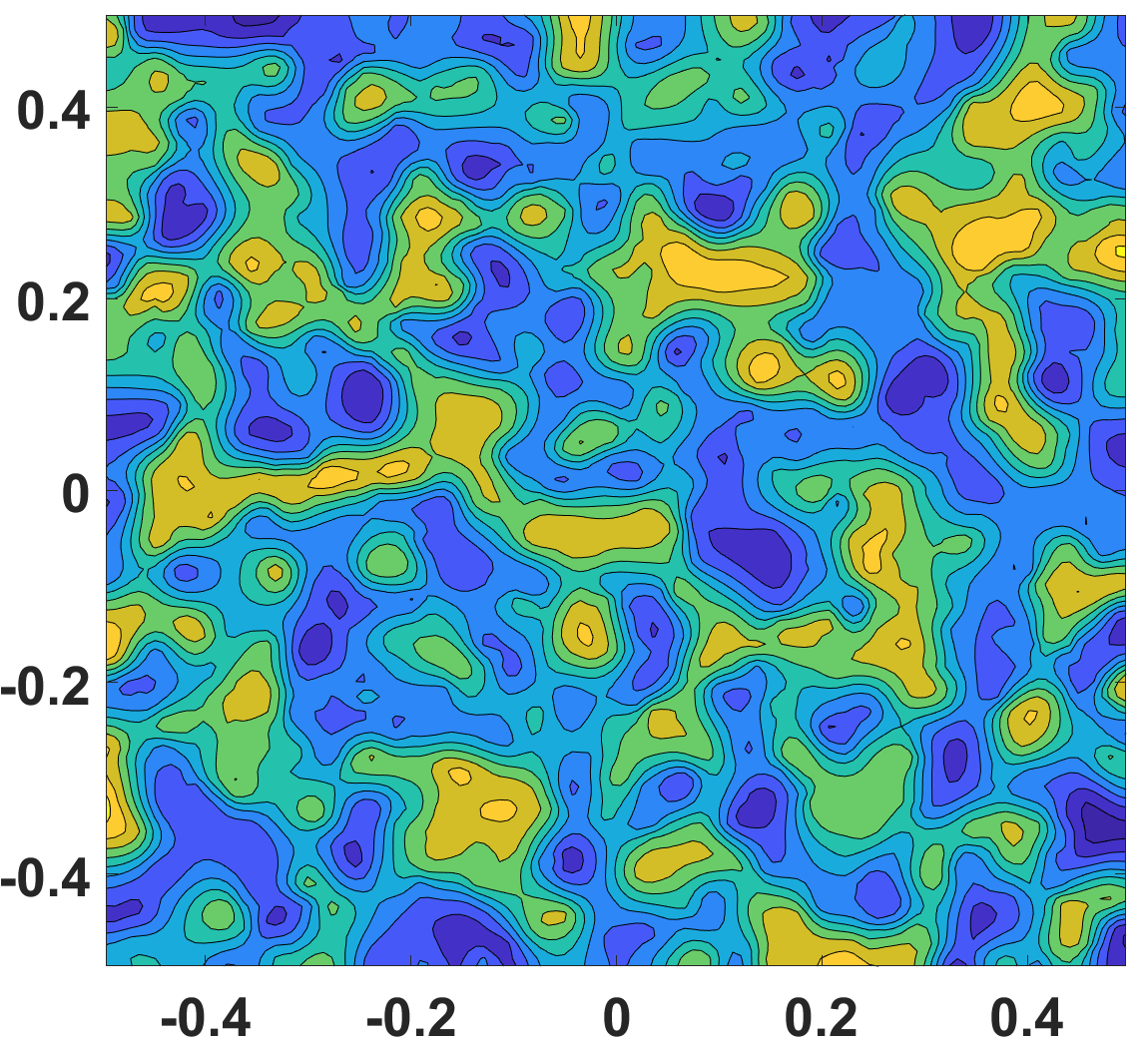}
\end{minipage}%
\begin{minipage}{0.22\textwidth}
    \includegraphics[scale = 0.2]{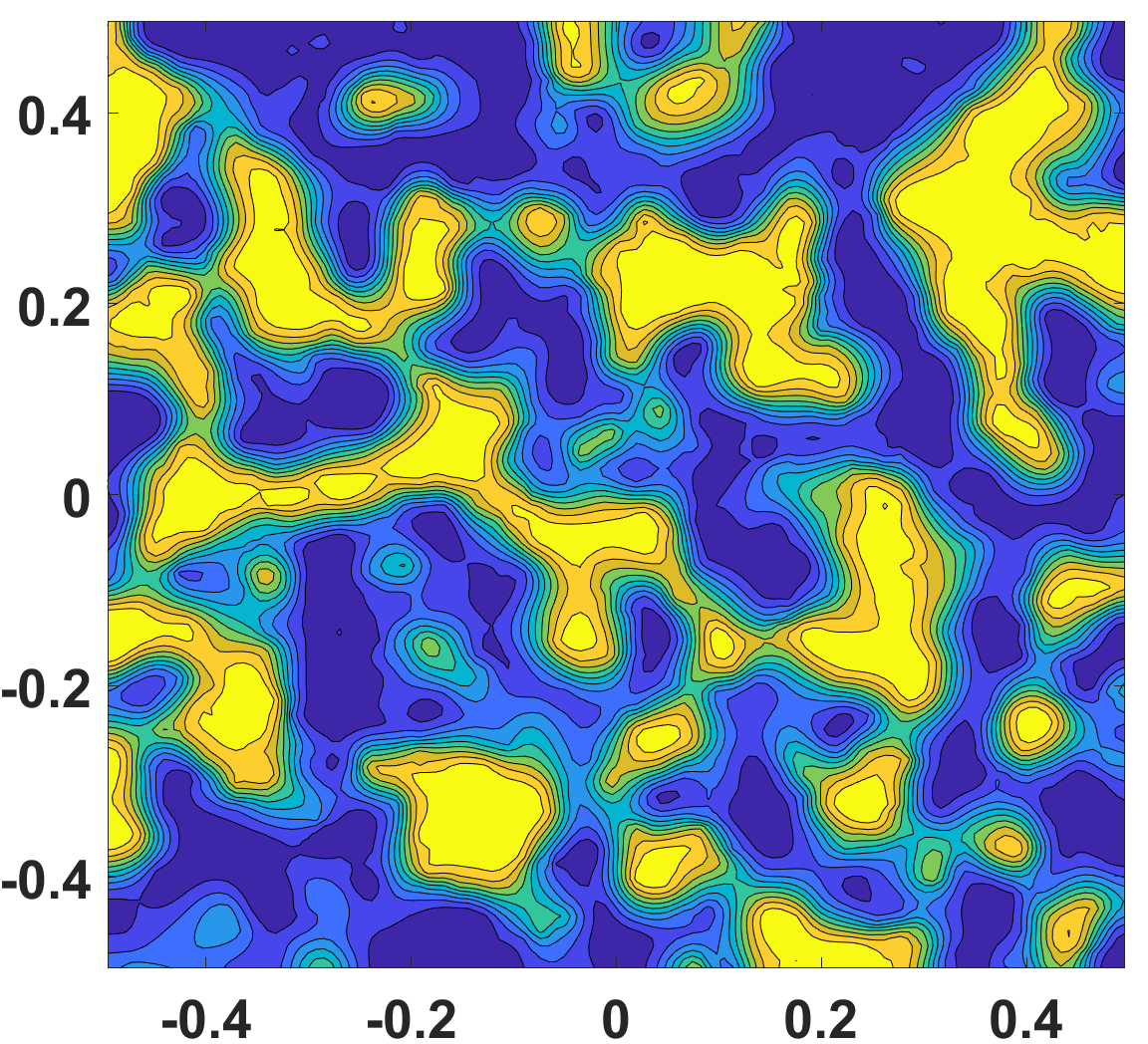}
\end{minipage}%
\begin{minipage}{0.22\textwidth}
    \includegraphics[scale = 0.2]{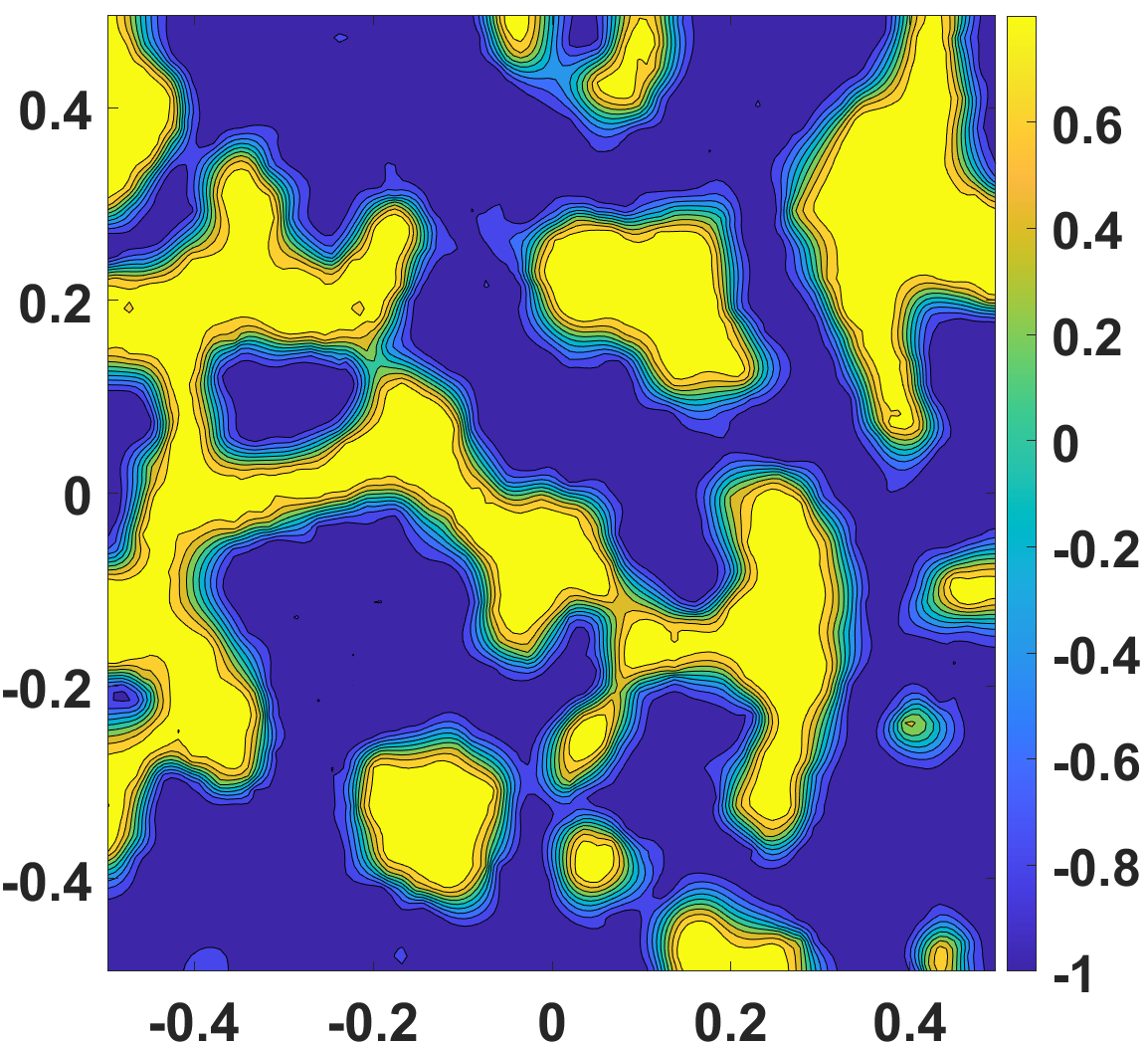}
\end{minipage}

\begin{minipage}{0.22\textwidth}
    \includegraphics[scale = 0.2]{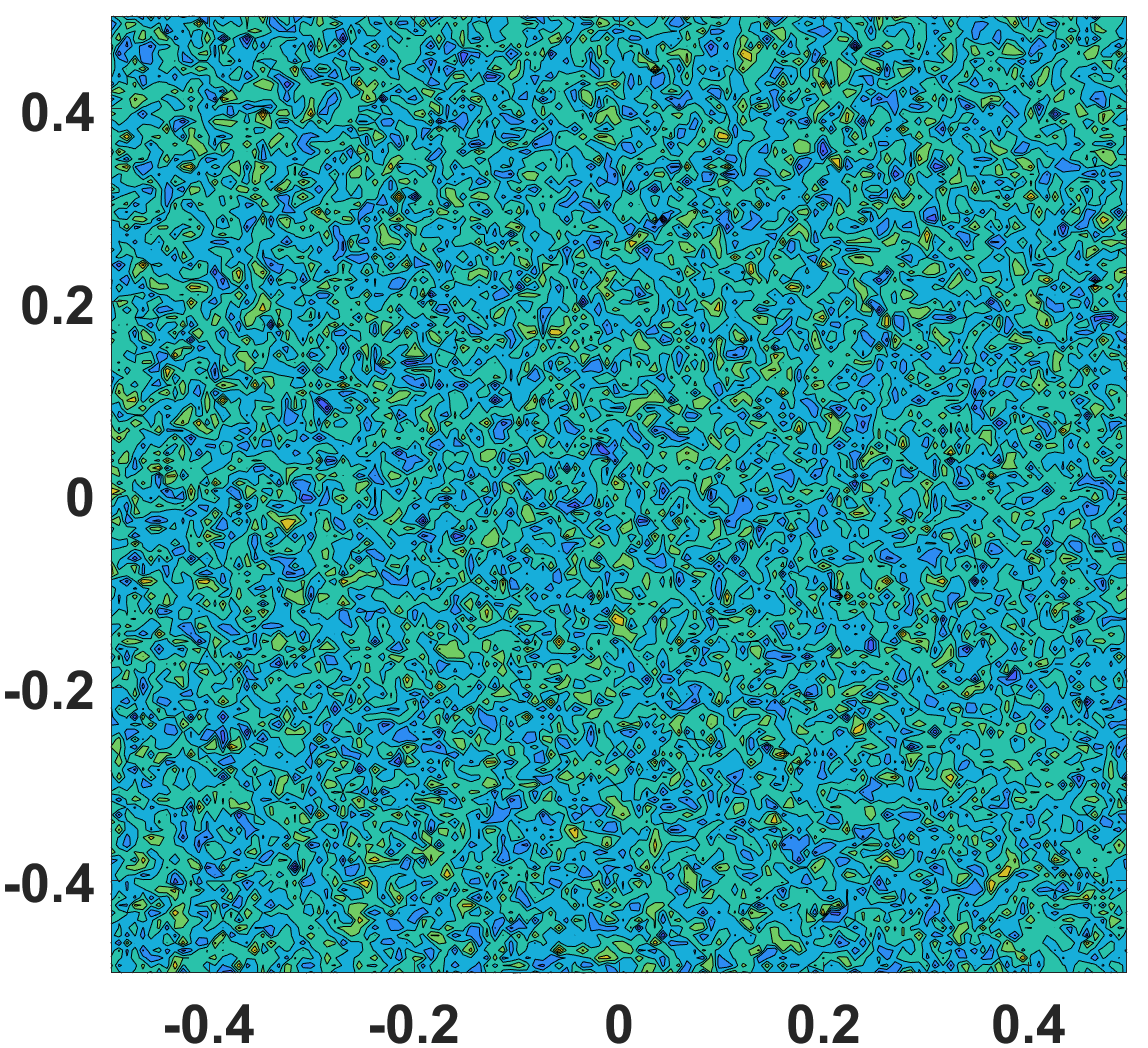}
\end{minipage}%
\begin{minipage}{0.22\textwidth}
    \includegraphics[scale = 0.2]{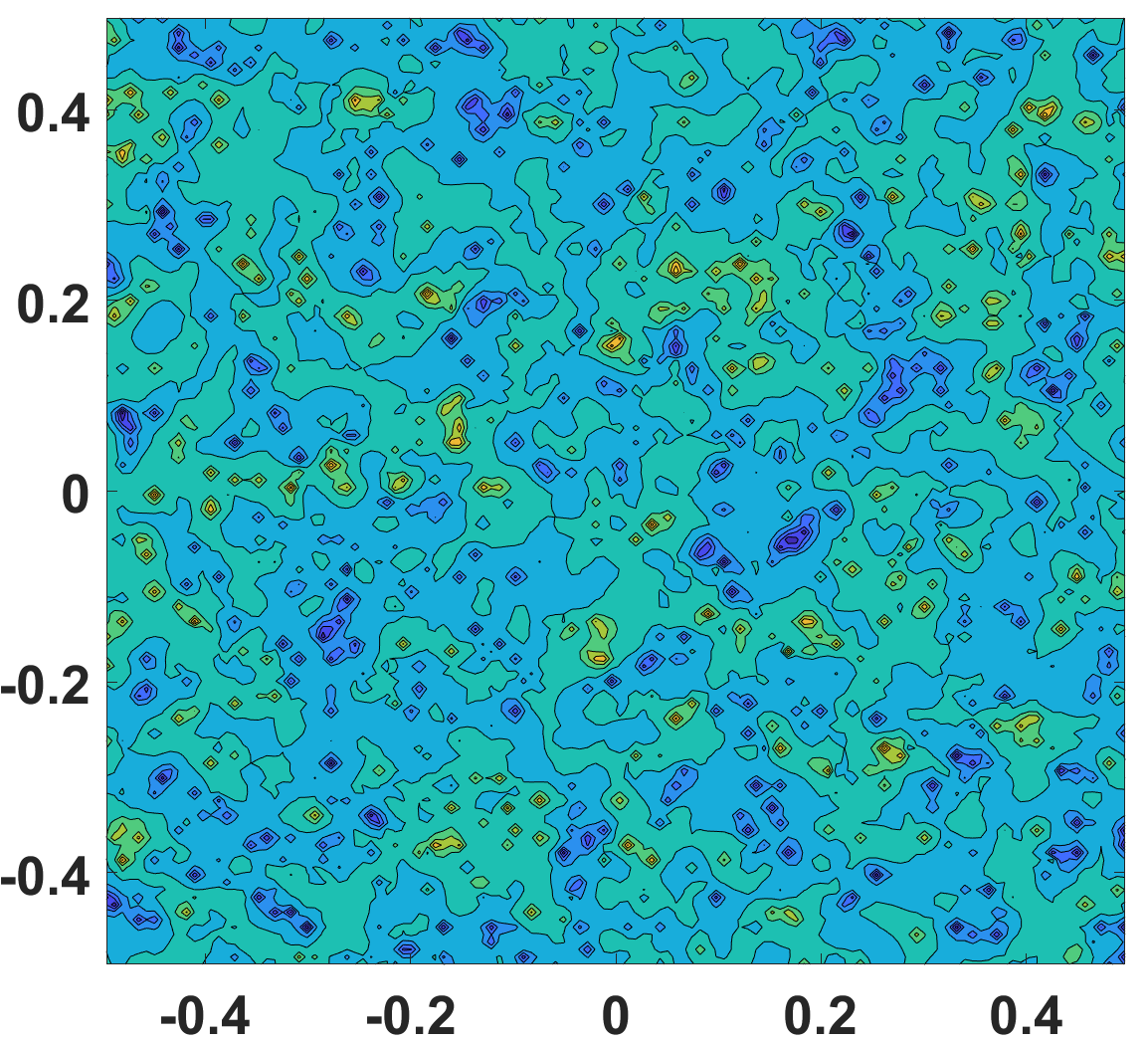}
\end{minipage}%
\begin{minipage}{0.22\textwidth}
    \includegraphics[scale = 0.2]{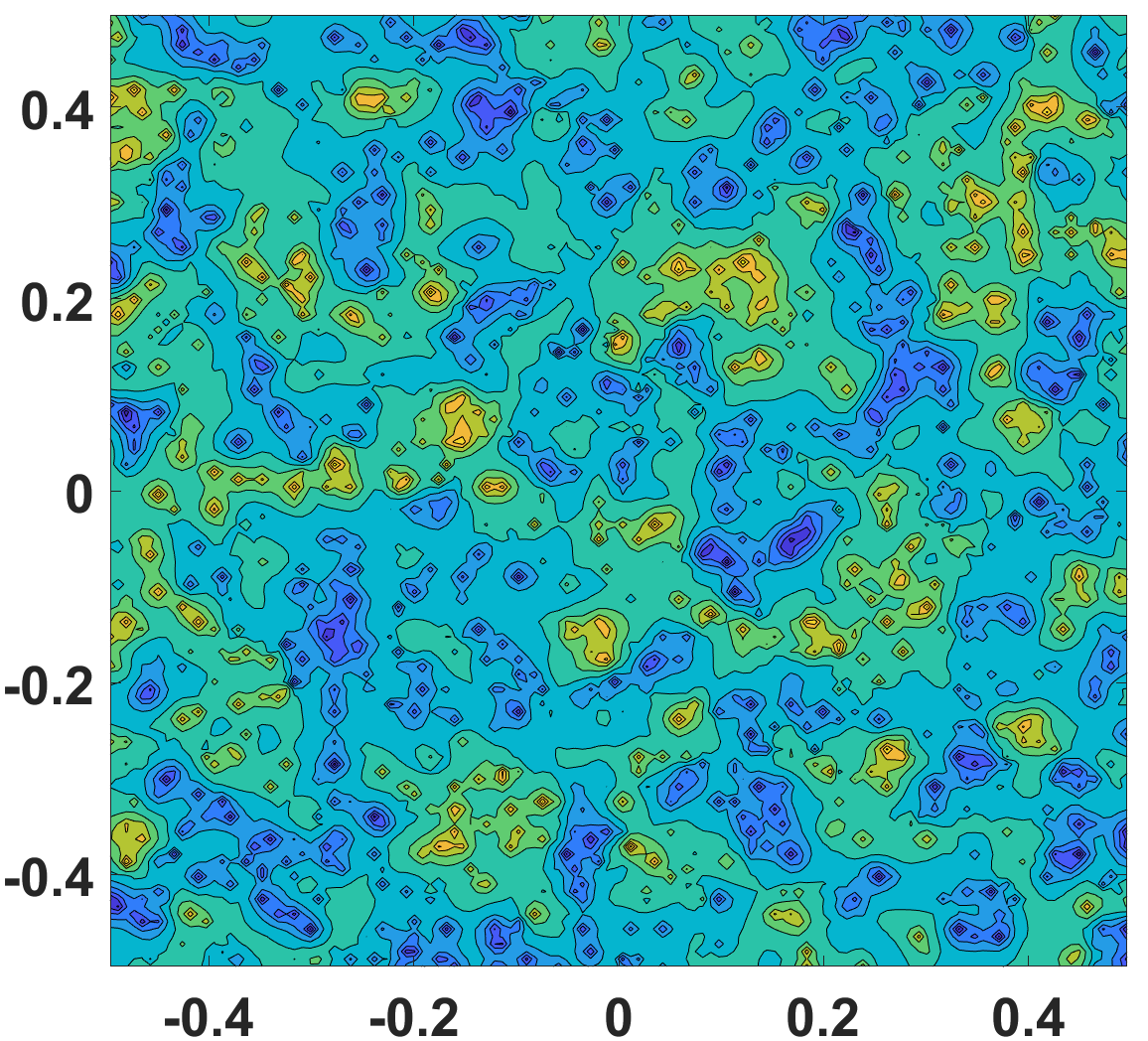}
\end{minipage}%
\begin{minipage}{0.22\textwidth}
    \includegraphics[scale = 0.2]{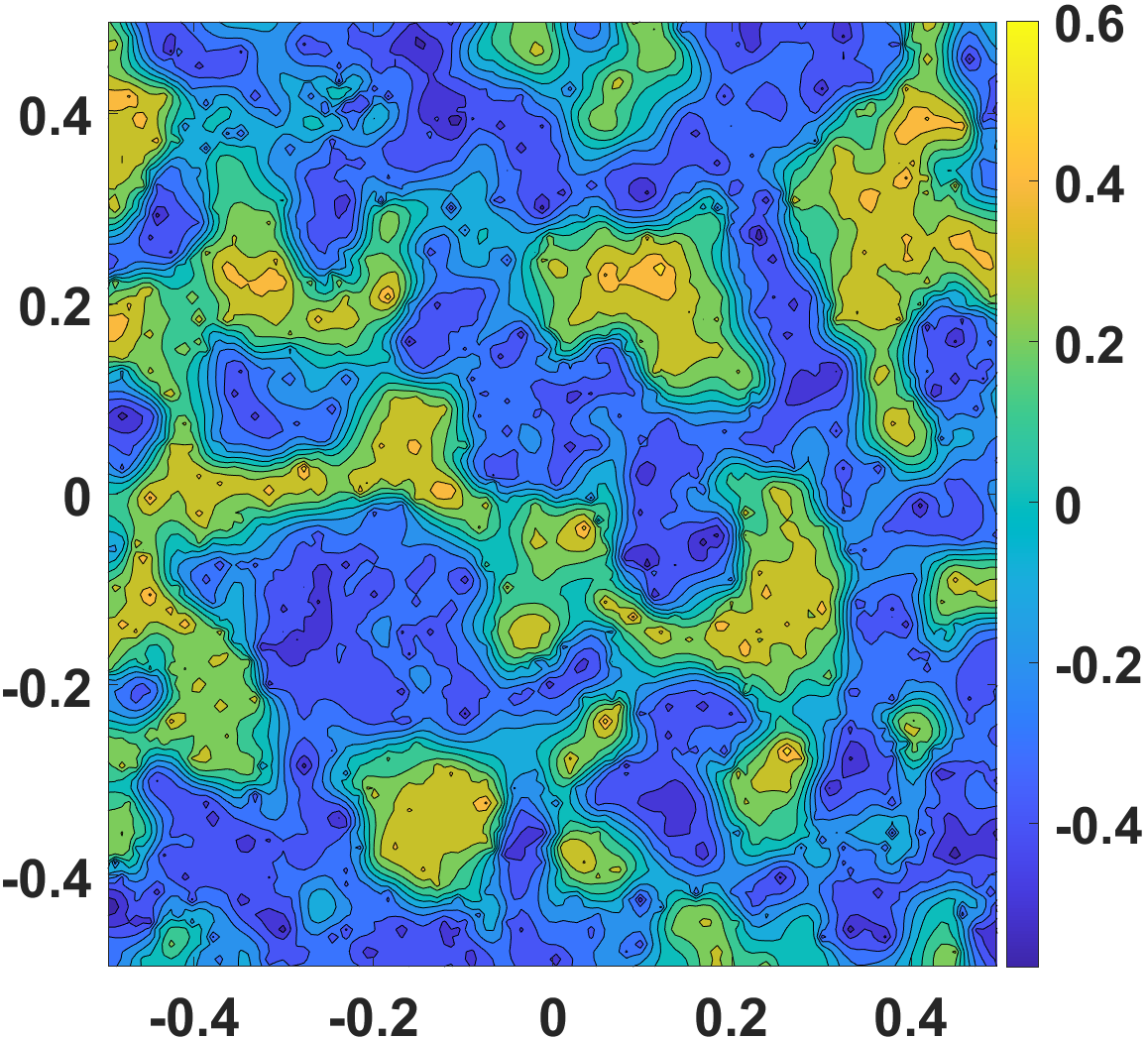}
\end{minipage}
\caption{\small Solutions of the Allen–Cahn equation in Case 2 at time $t=0, \f{T
}{2}, \f{2T}{3}, T$ with $10\%$ observations. (First row) Estimated solution by EnSF. (Second row) Estimated solution by LETKF.}
\label{NonConsMob_10Obs}
\vspace{-0.2cm}
\end{figure}
\begin{figure}[h!]
\vspace{-0.2cm}
\centering
\includegraphics[scale = 0.42]{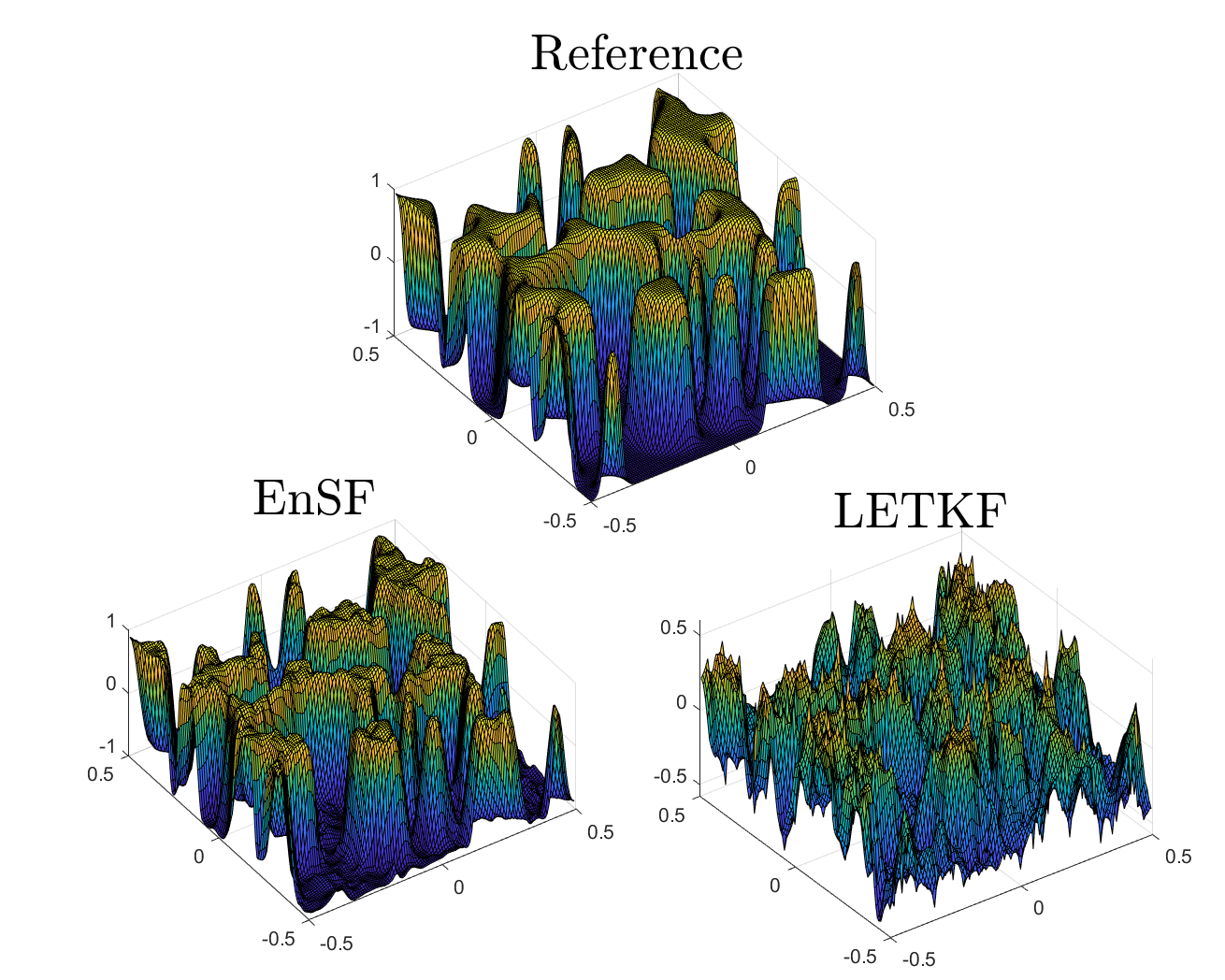}
\caption{\small 3D view of the reference and the estimated solutions for the Allen-Cahn equation in Case 2 at final time T with $10\%$ observations.}
\label{NonCons_10Obs_3D}
\vspace{-0.3cm}
\end{figure}

\begin{figure}[h!]
\vspace{-0.1cm}
   \begin{minipage}{0.33\textwidth}
    \includegraphics[scale = 0.22]{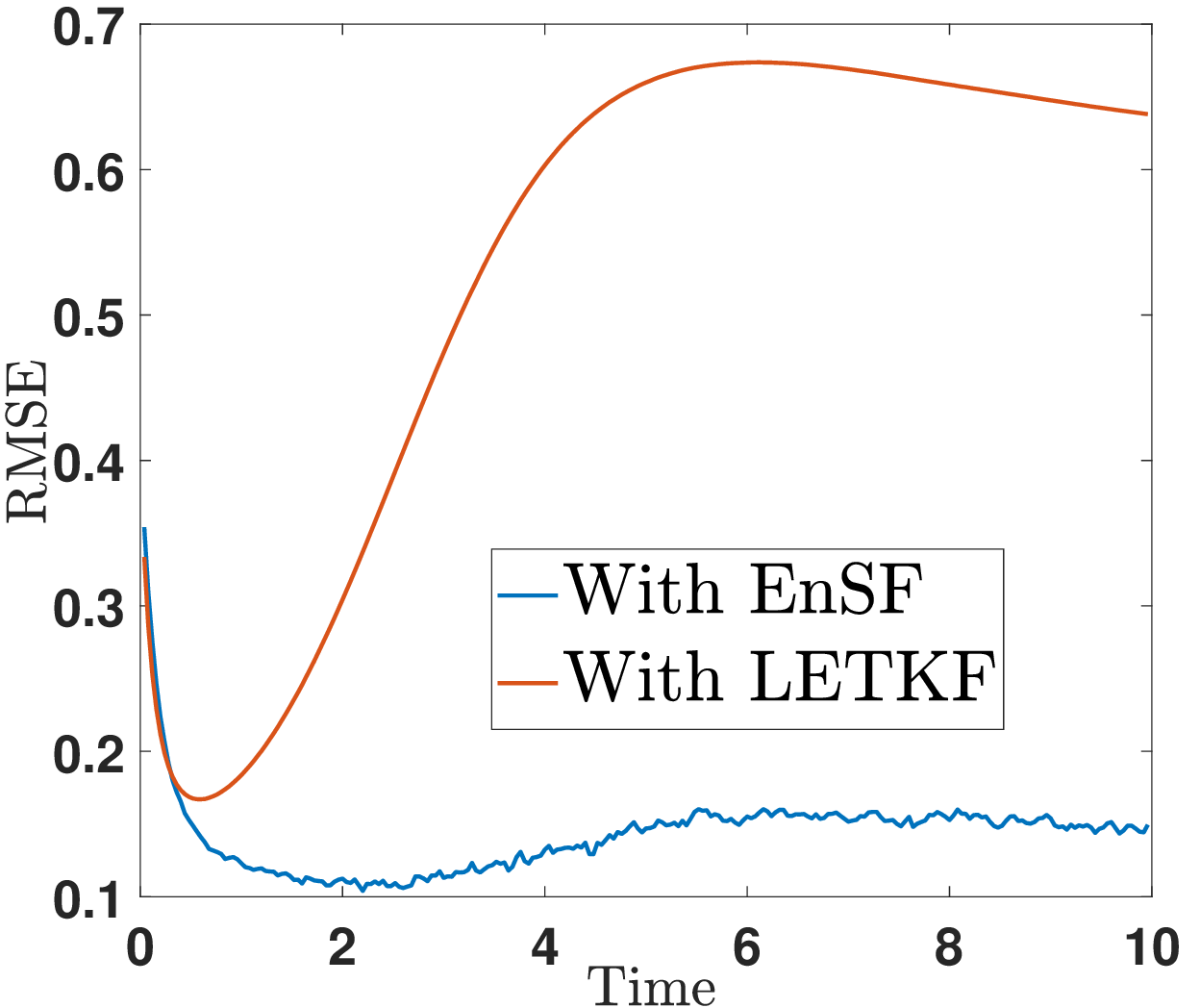}
\end{minipage}%
\begin{minipage}{0.33\textwidth}
    \includegraphics[scale = 0.22]{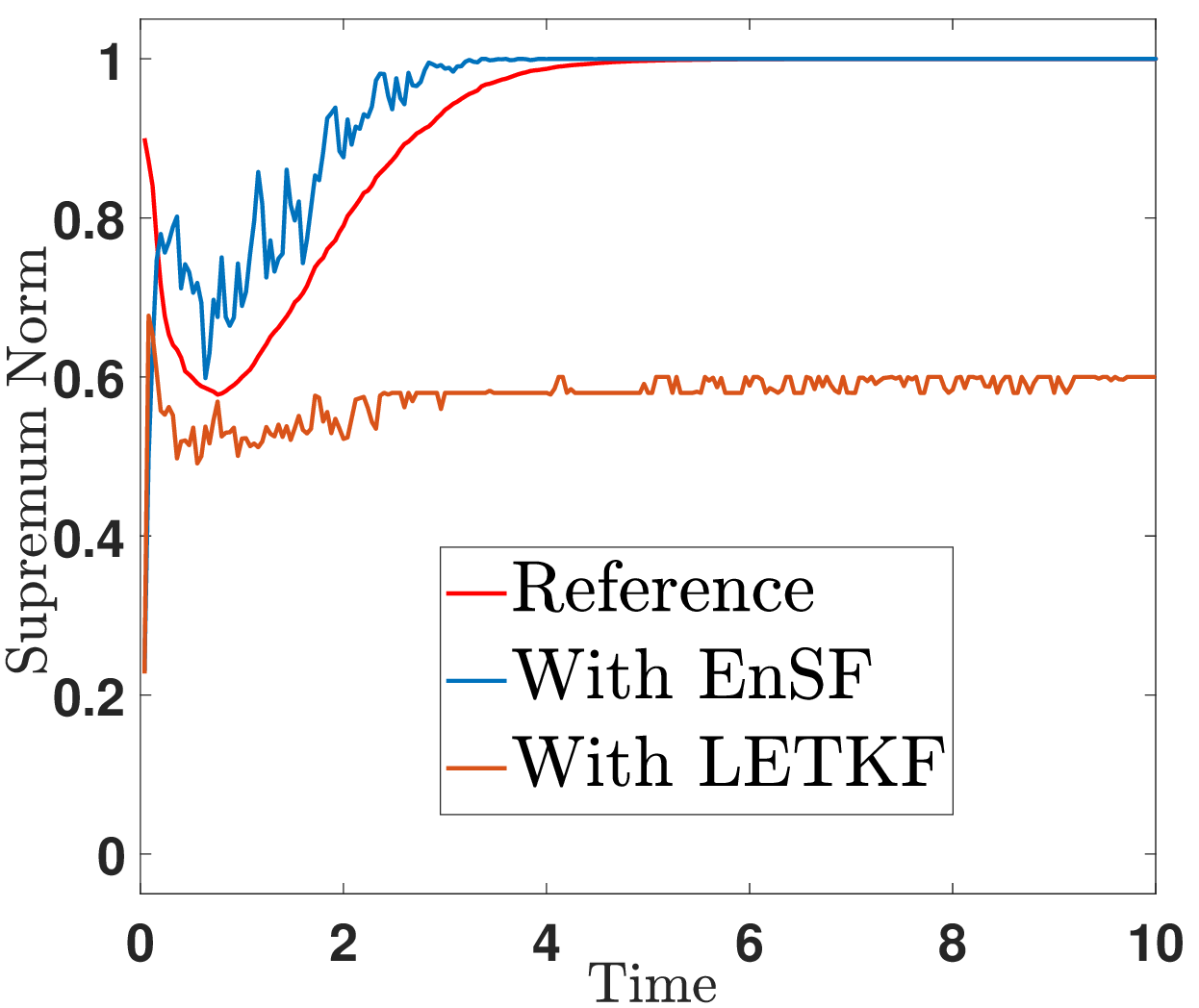}
\end{minipage}%
\begin{minipage}{0.33\textwidth}
    \includegraphics[scale = 0.22]{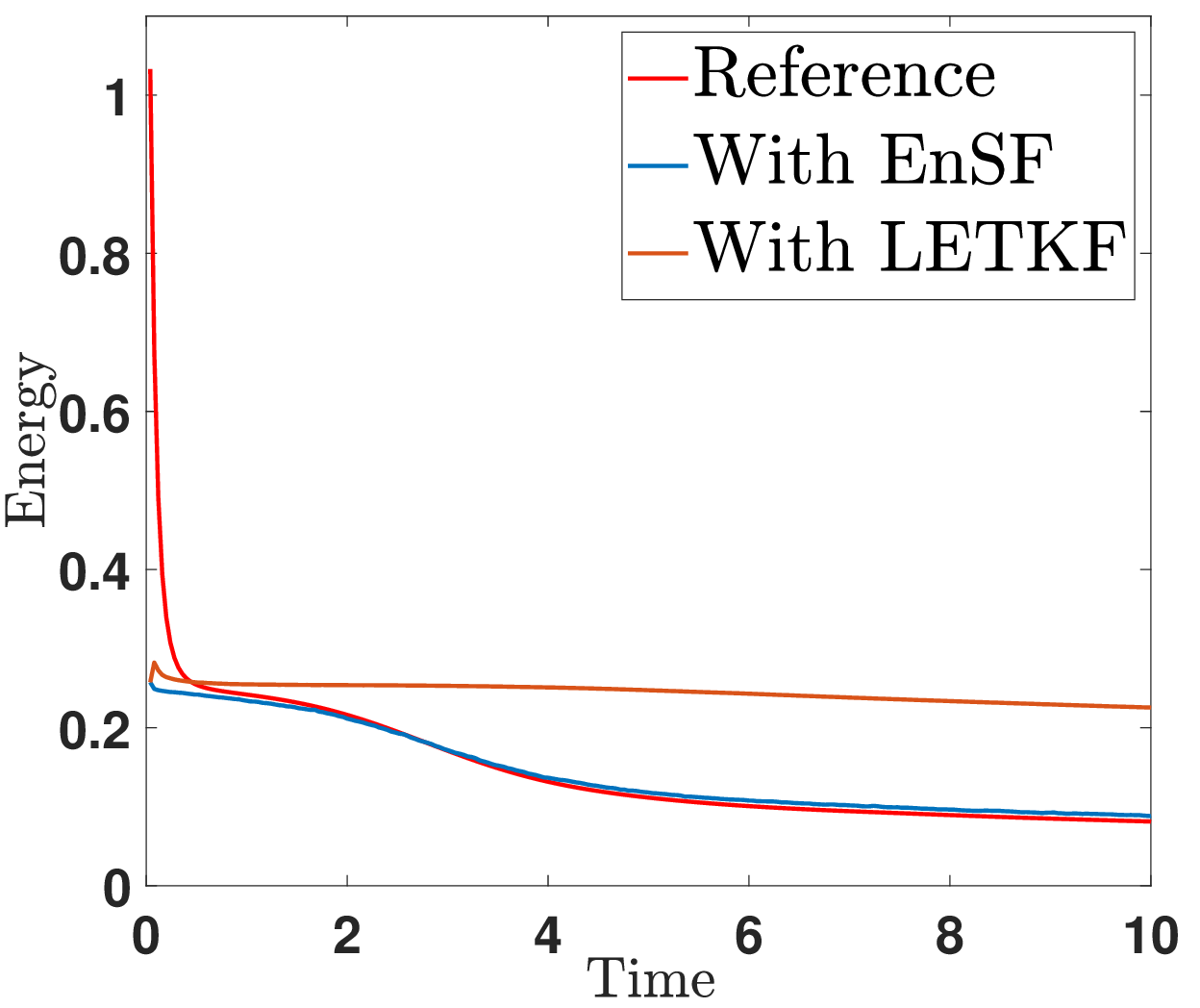}
\end{minipage}
\caption{\small Illustration of RMSEs, supremum norms, and discrete energies with $10\%$ observations for Case 2. (Left) RMSE. (Center) Supremum Norm. (Right) Energy.}
\label{NonCons_10Obs_RMSE_Mass_Energy}
\vspace{-0.1cm}
\end{figure}

Finally, we report the results with $10\%$ observations. Estimates produced by the EnSF with inpainting and by the LETKF are shown in Figure~\ref{NonConsMob_10Obs}, while the 3D views of the estimates at the final time $T$ are plotted in Figure~\ref{NonCons_10Obs_3D}. The corresponding RMSEs, supremum norms, and the discrete energies are depicted in Figure~\ref{NonCons_10Obs_RMSE_Mass_Energy}. These figures again confirm that our method outperforms the LETKF in terms of accuracy. Moreover, the estimated solution by our method yields more accurate supremum norm discrete energy. 

\subsubsection{Solution-dependent mobility with uncertainty}
We conclude the numerical experiments section with Case 3, which involves a solution-dependent mobility subject to noise perturbations. The prediction step in data assimilation is carried out by solving Eq.~\eqref{AllenCahn} with the perturbed mobility function $M_3(\phi) = \max\{1-\phi^2+ \xi_t, 0\}$, where $ \xi_t \sim 0.4 \cdot N(0, \pmb{I}_d)$. The experimental setup mirrors that of the previous cases, with observational coverage levels of $100\%$, $70\%$ and $10\%$. We evaluate four key quantities: the estimated solution, root-mean-square error (RMSE), supremum norm, and discrete energy.  For the $10\%$ observation scenario, results from our method are again compared with those obtained using the LETKF.

\begin{figure}[h!]
\vspace{-0.3cm}
\begin{minipage}{0.22\textwidth}
    \includegraphics[scale = 0.2]{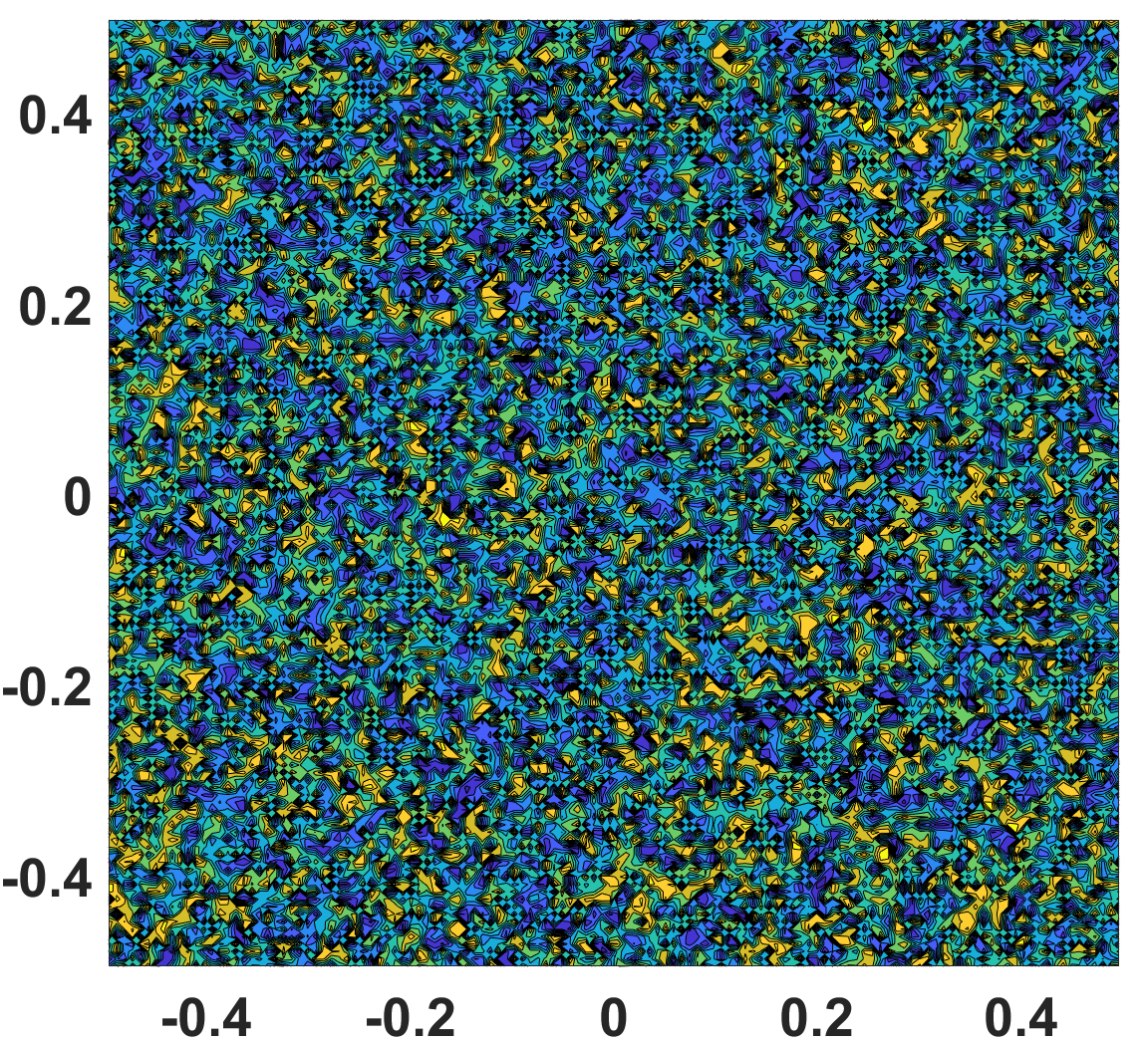}
\end{minipage}%
\begin{minipage}{0.22\textwidth}
    \includegraphics[scale = 0.2]{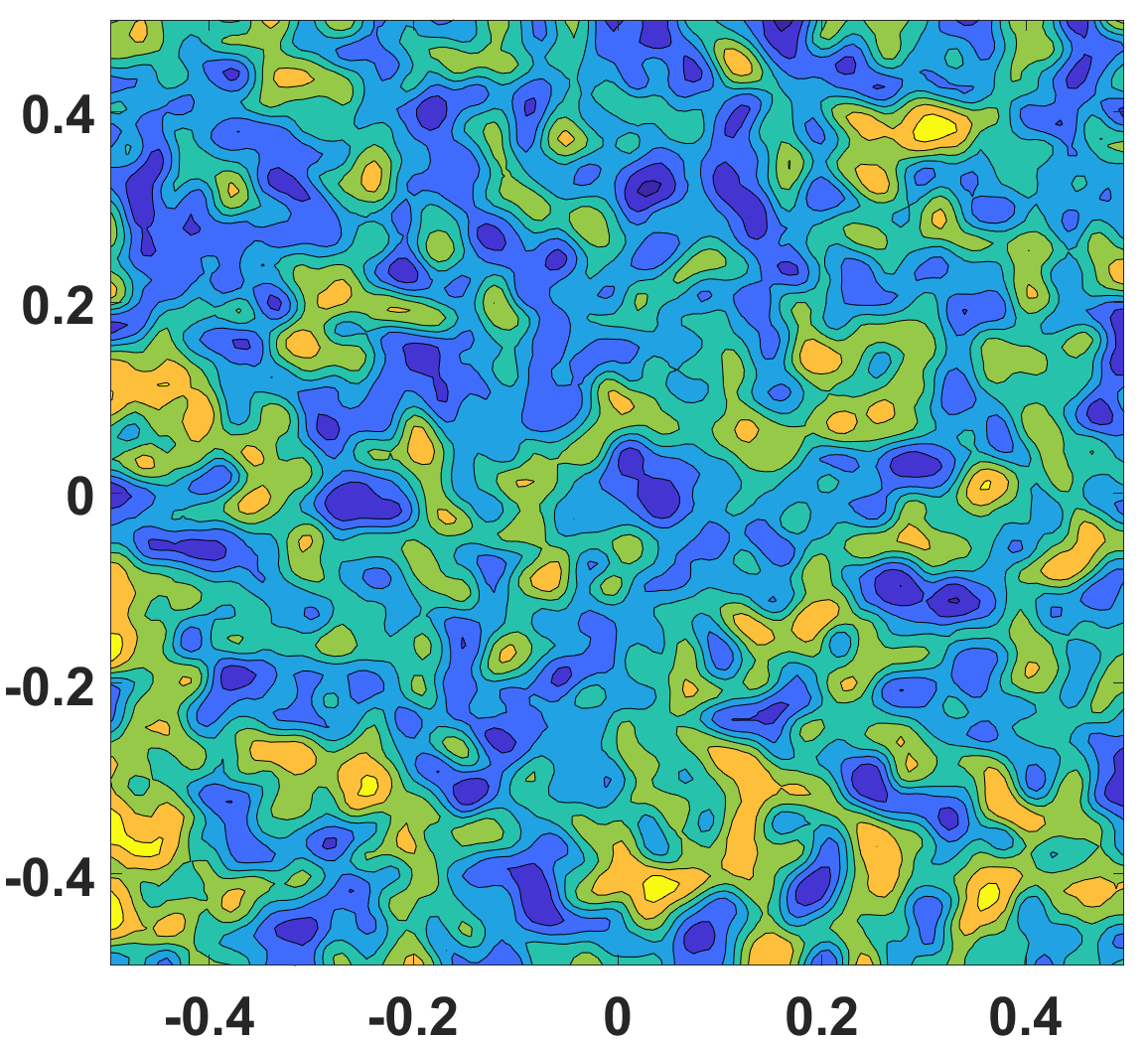}
\end{minipage}%
\begin{minipage}{0.22\textwidth}
    \includegraphics[scale = 0.2]{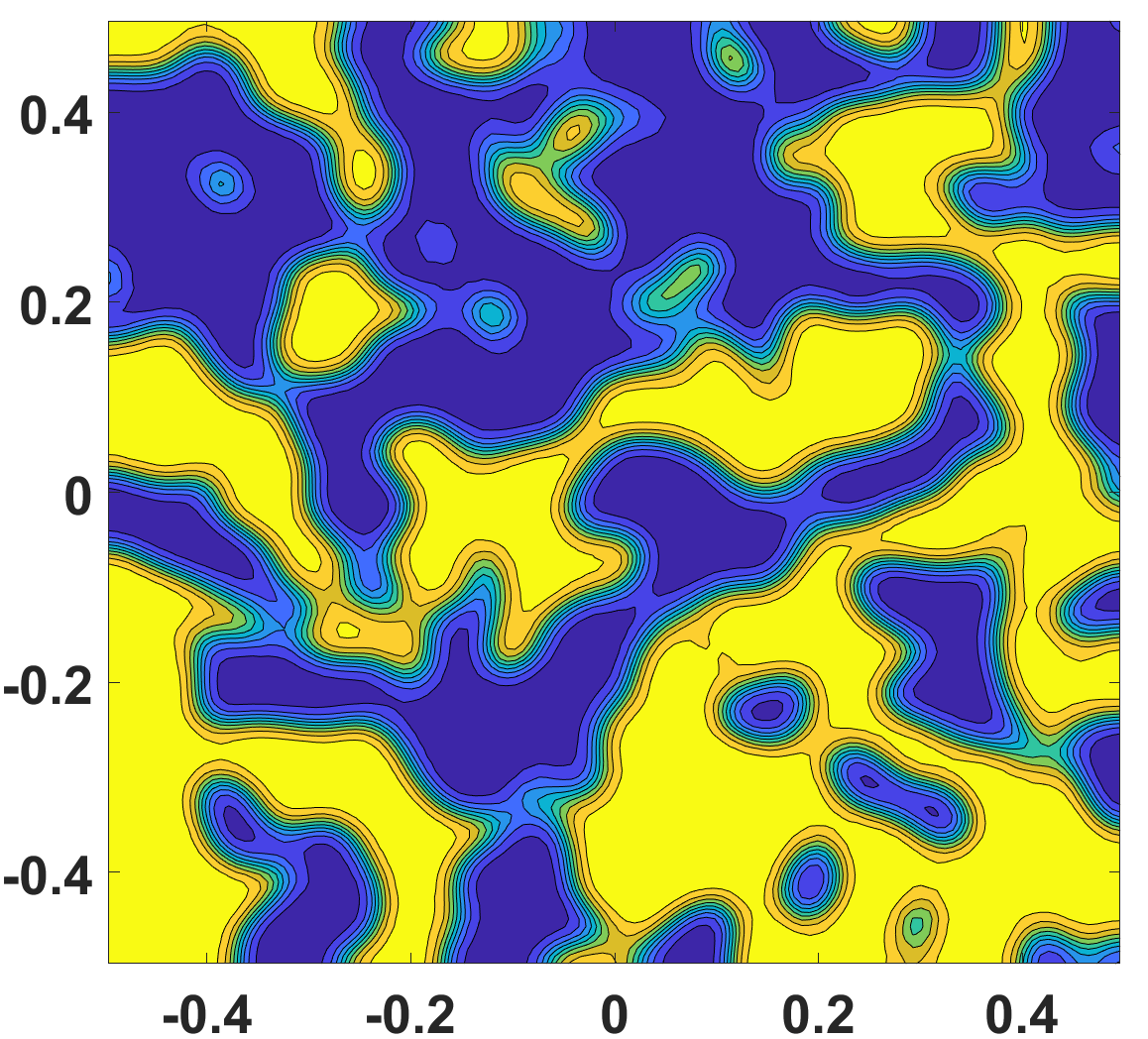}
\end{minipage}%
\begin{minipage}{0.22\textwidth}
    \includegraphics[scale = 0.2]{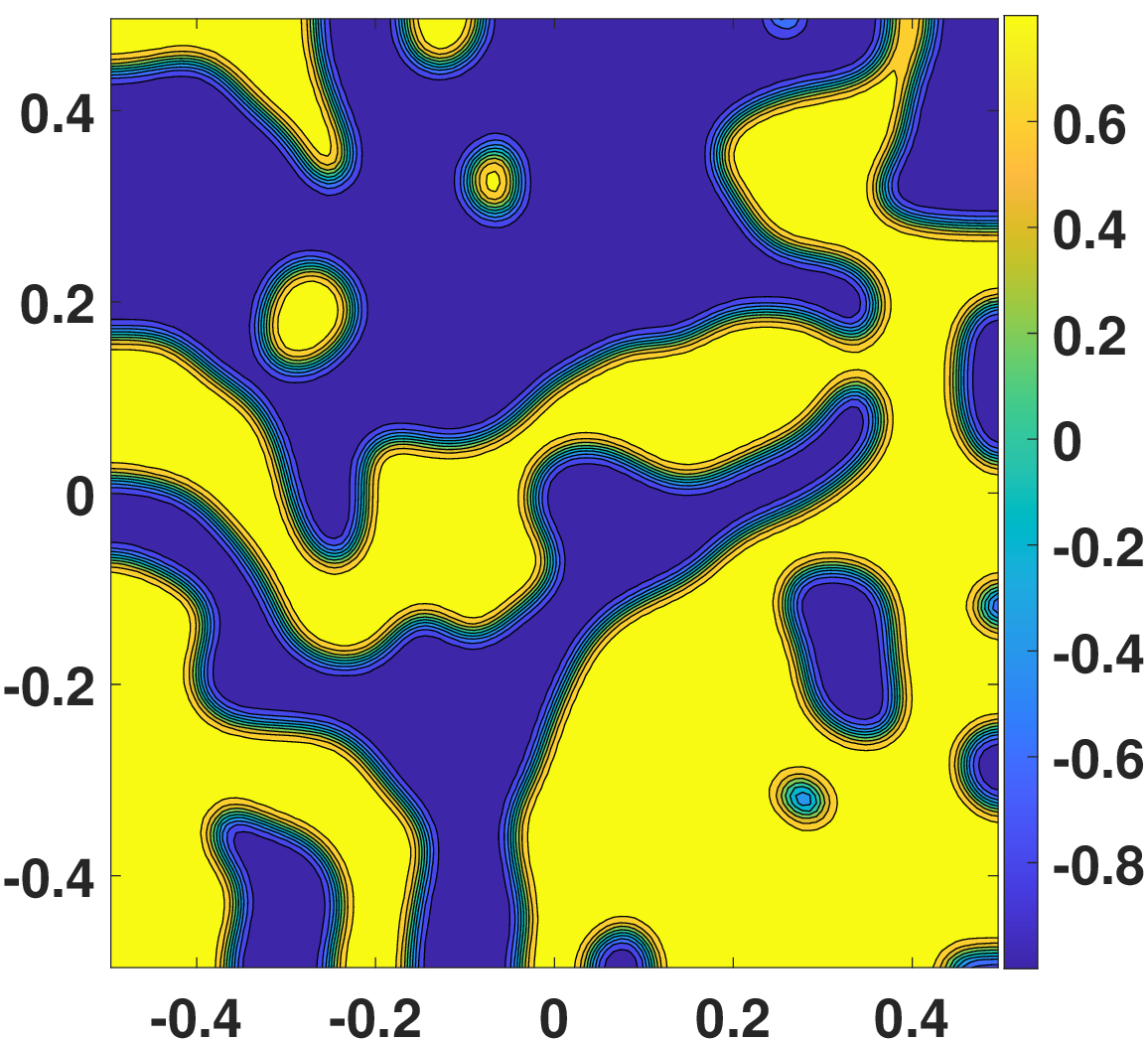}
\end{minipage}

\begin{minipage}{0.22\textwidth}
    \includegraphics[scale = 0.2]{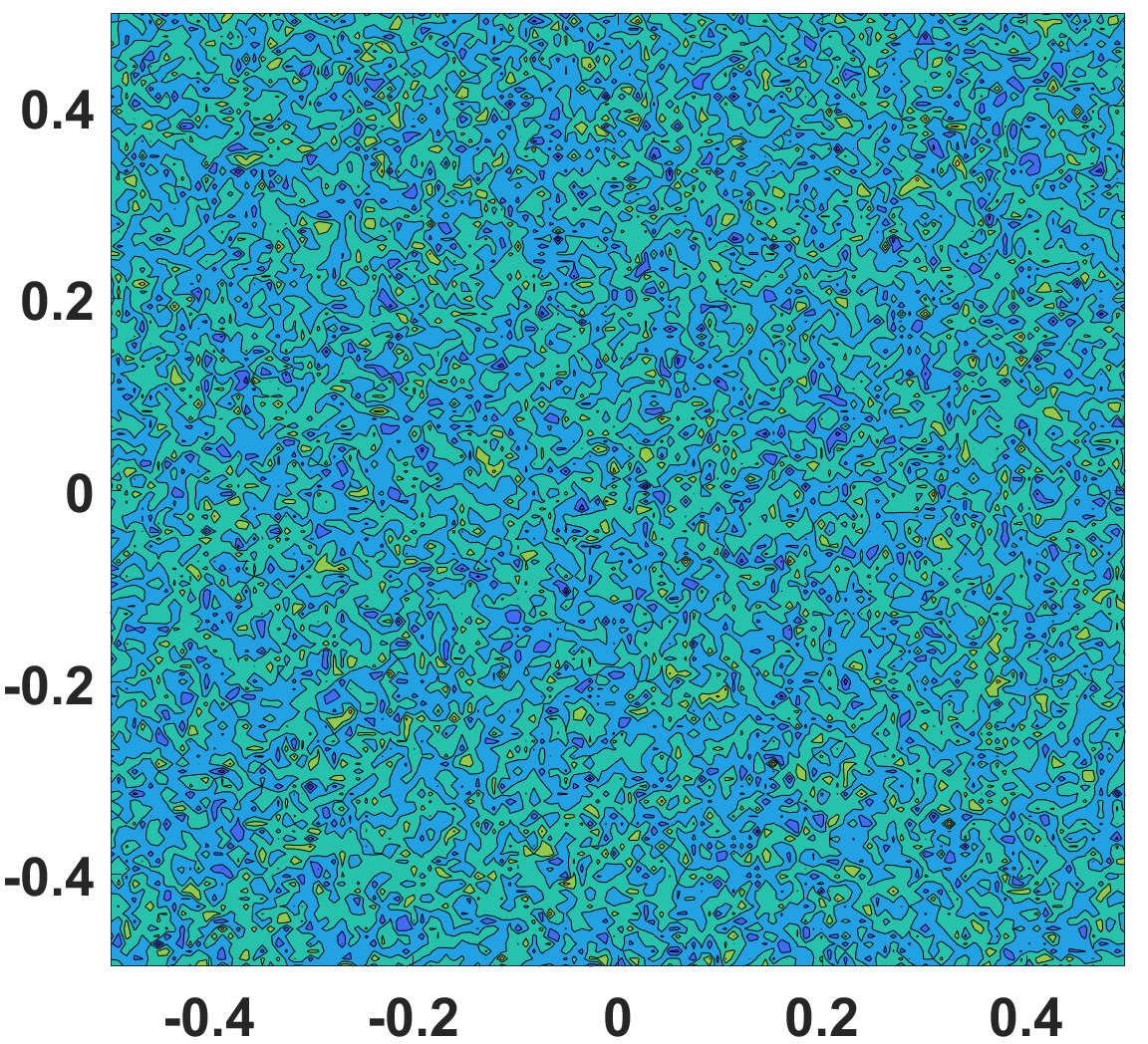}
\end{minipage}%
\begin{minipage}{0.22\textwidth}
    \includegraphics[scale = 0.2]{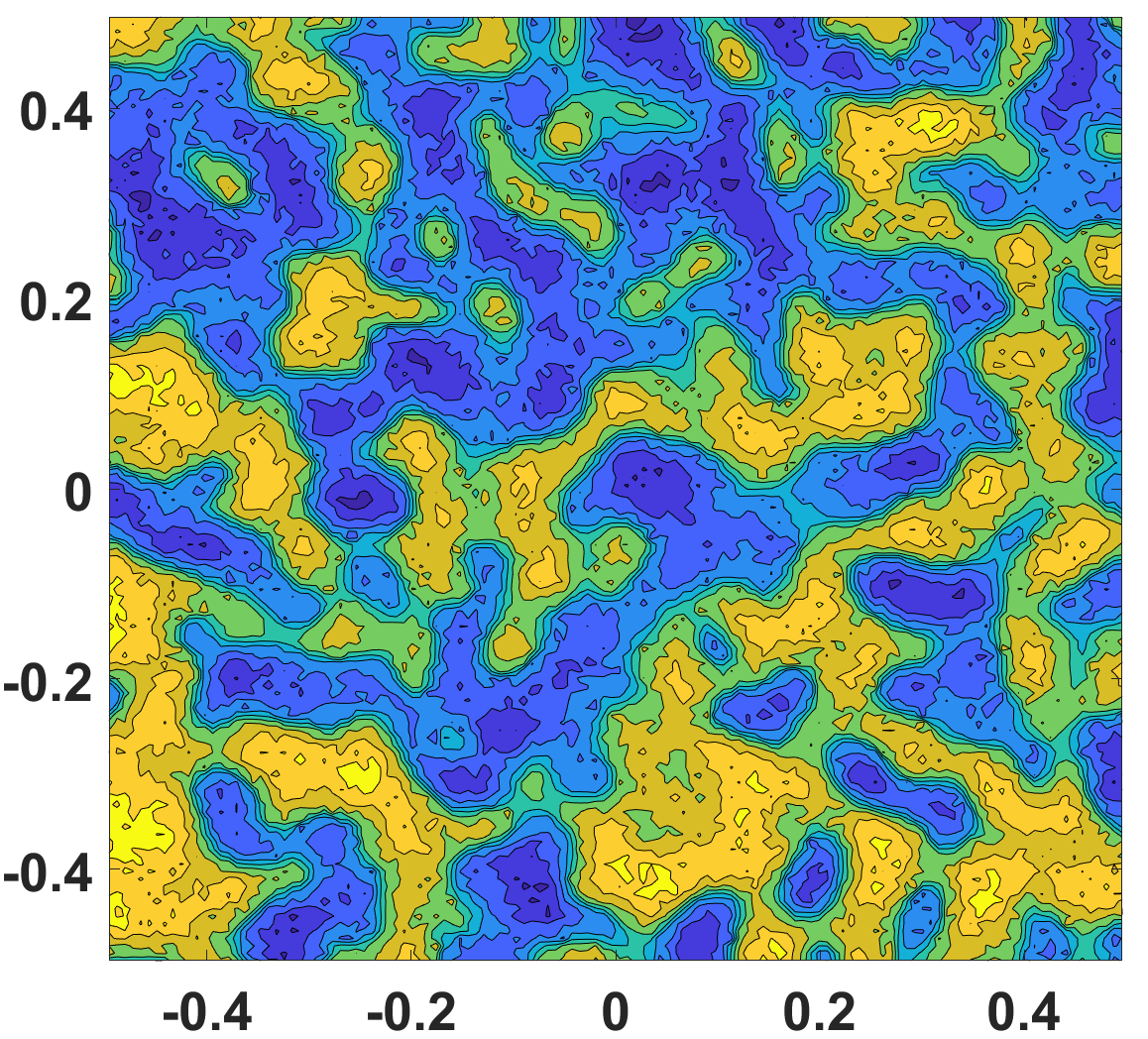}
\end{minipage}%
\begin{minipage}{0.22\textwidth}
    \includegraphics[scale = 0.2]{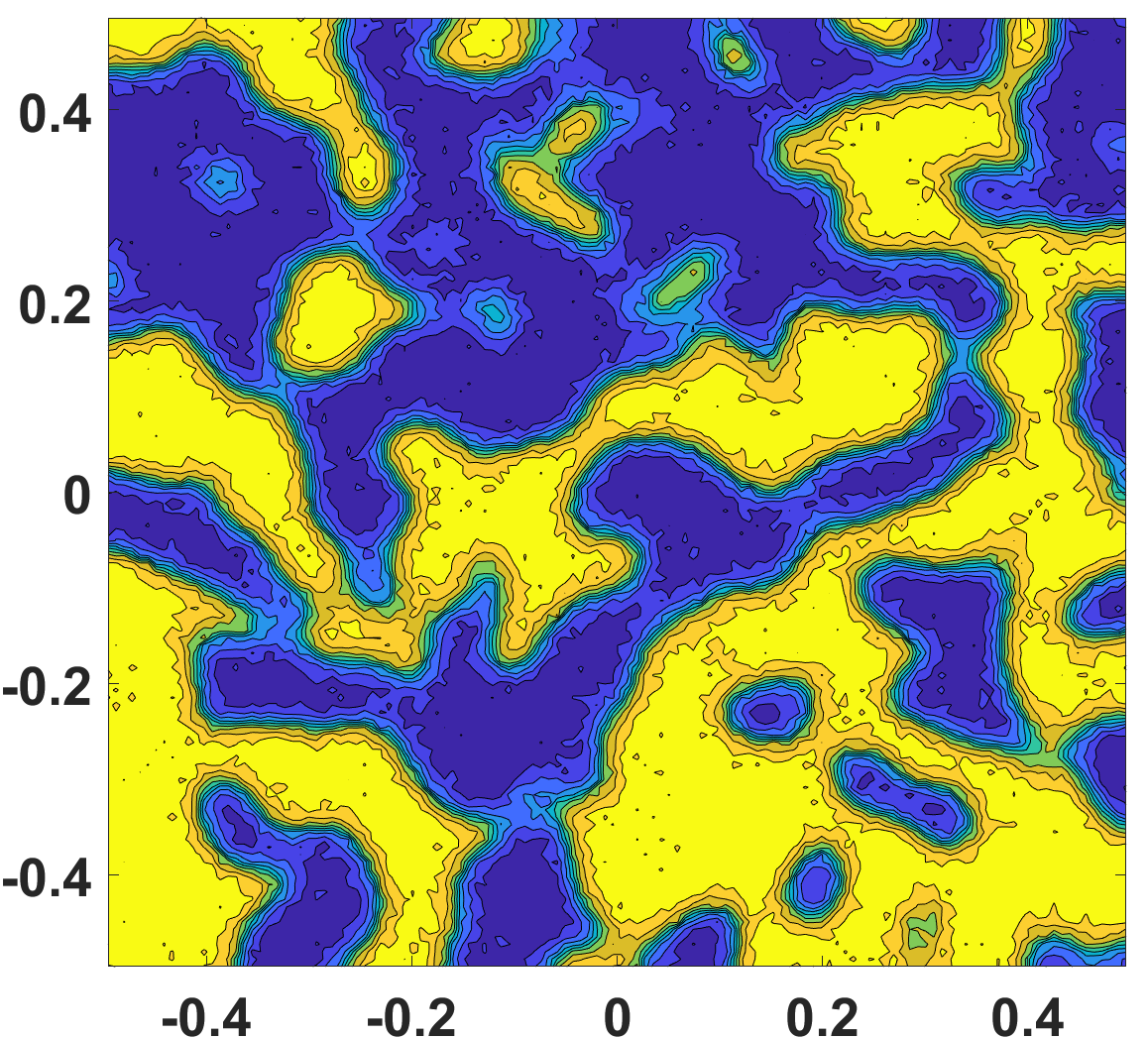}
\end{minipage}%
\begin{minipage}{0.22\textwidth}
    \includegraphics[scale = 0.2]{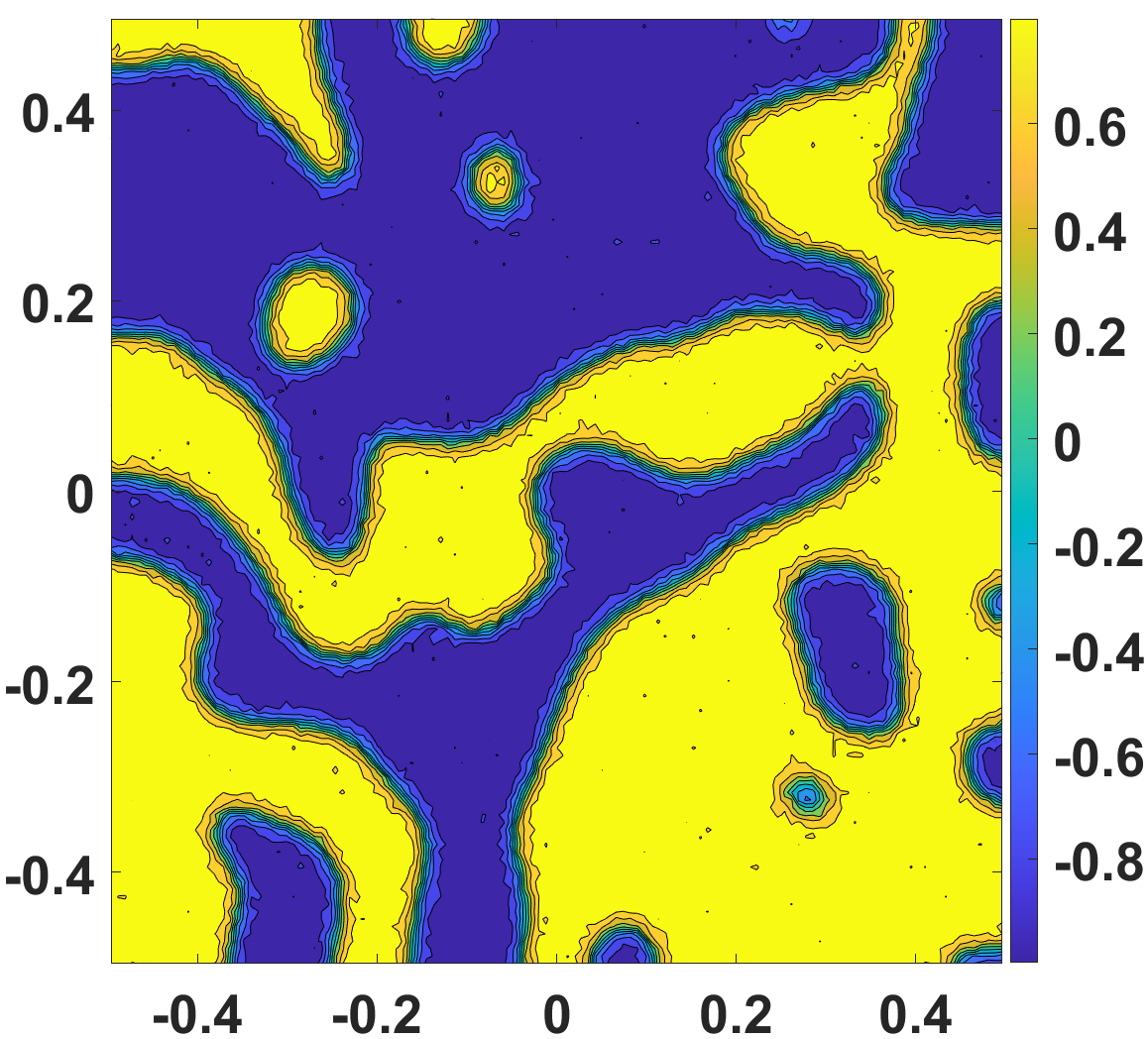}
\end{minipage}

\begin{minipage}{0.22\textwidth}
    \includegraphics[scale = 0.2]{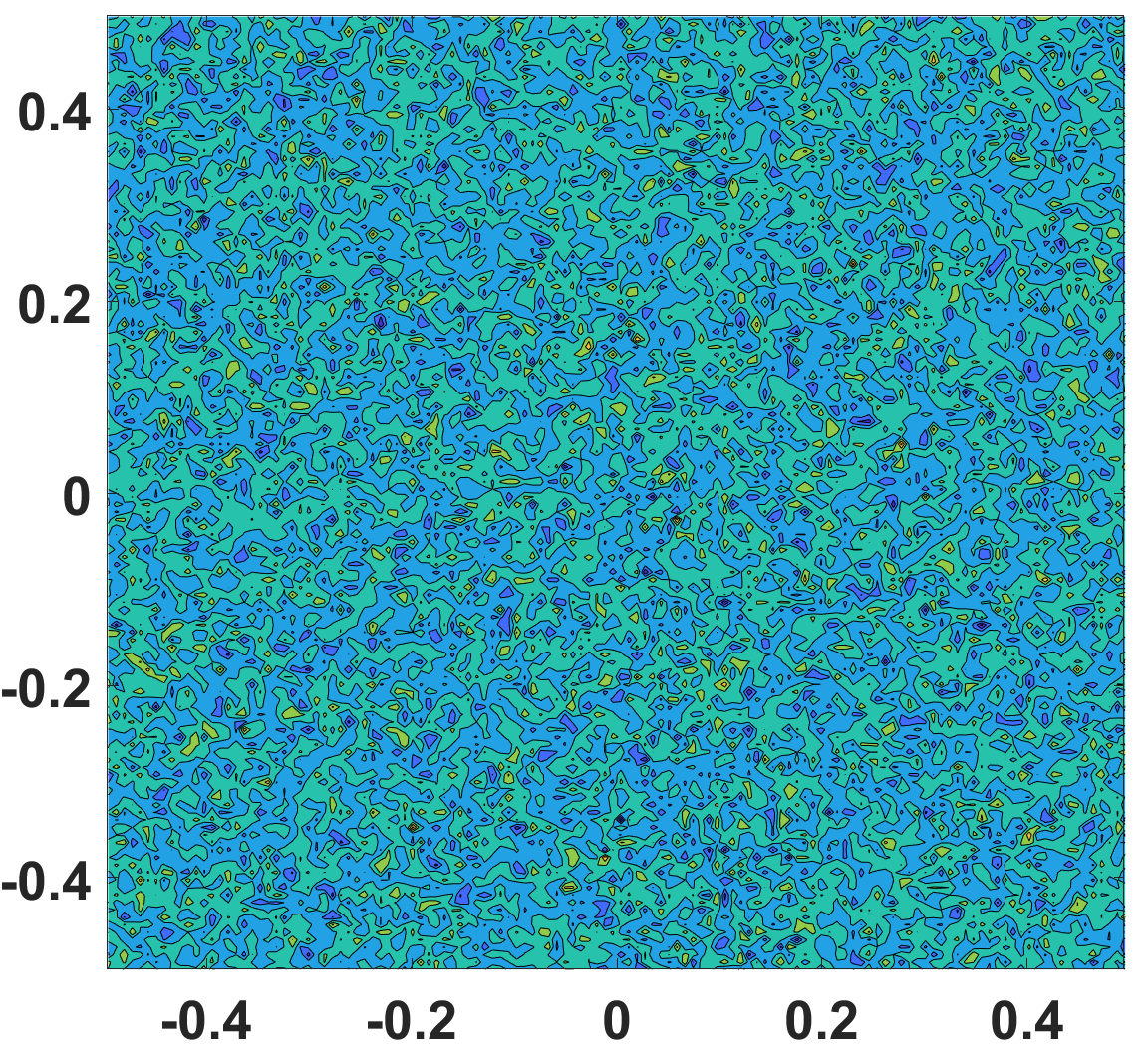}
\end{minipage}%
\begin{minipage}{0.22\textwidth}
    \includegraphics[scale = 0.2]{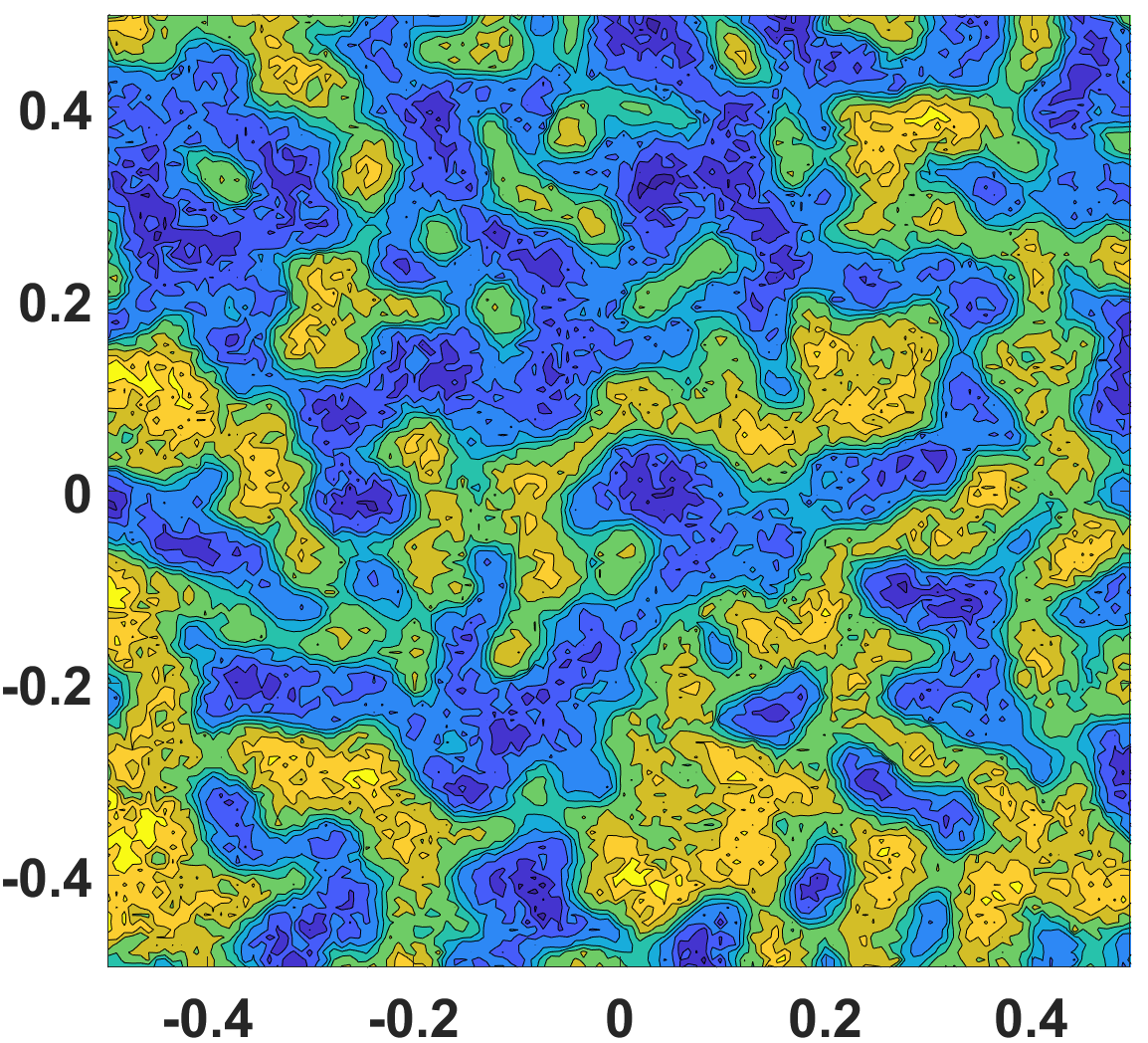}
\end{minipage}%
\begin{minipage}{0.22\textwidth}
    \includegraphics[scale = 0.2]{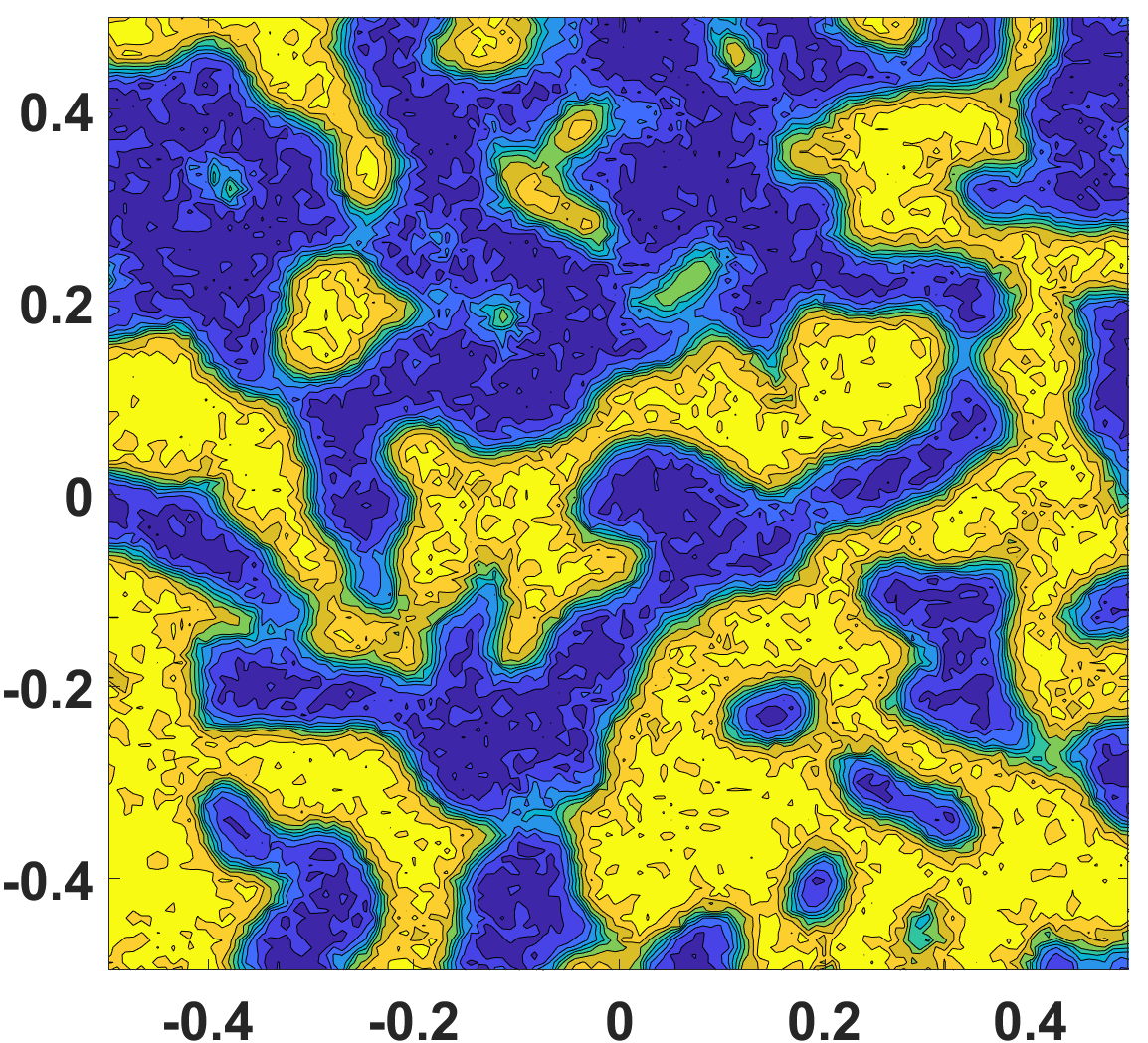}
\end{minipage}%
\begin{minipage}{0.22\textwidth}
    \includegraphics[scale = 0.2]{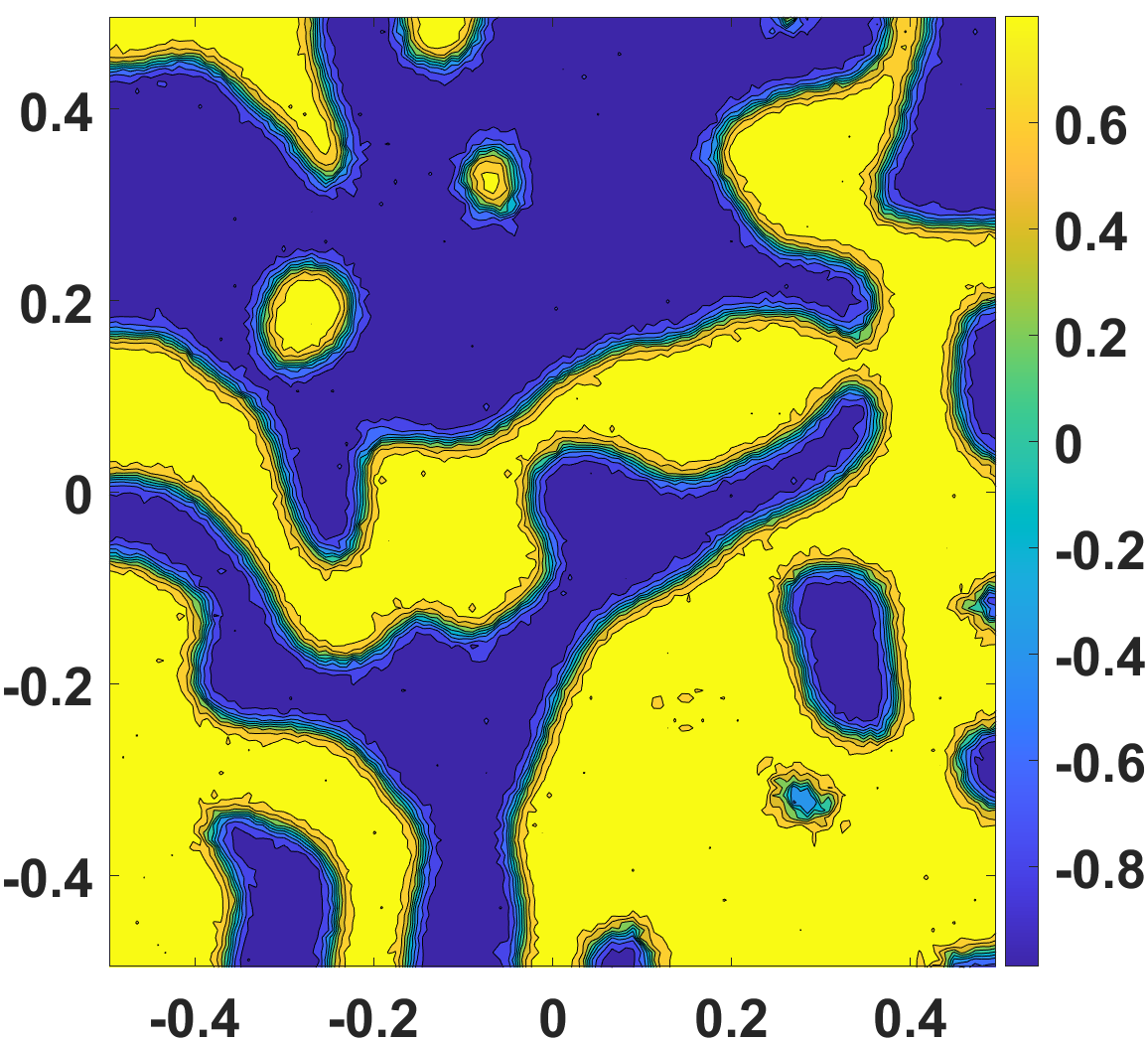}
\end{minipage}
\caption{\small Solutions of the Allen–Cahn equation in Case 3 at time $t=0, T/2, 2T/3, T$. (First row) Reference. (Second row) Estimated solution with $100\%$ observations.
(Third row) Estimated solution with $70\%$ observations.}
\label{DegMob_100_70Obs}
\end{figure}

\begin{figure}[h!]
   \begin{minipage}{0.33\textwidth}
    \includegraphics[scale = 0.22]{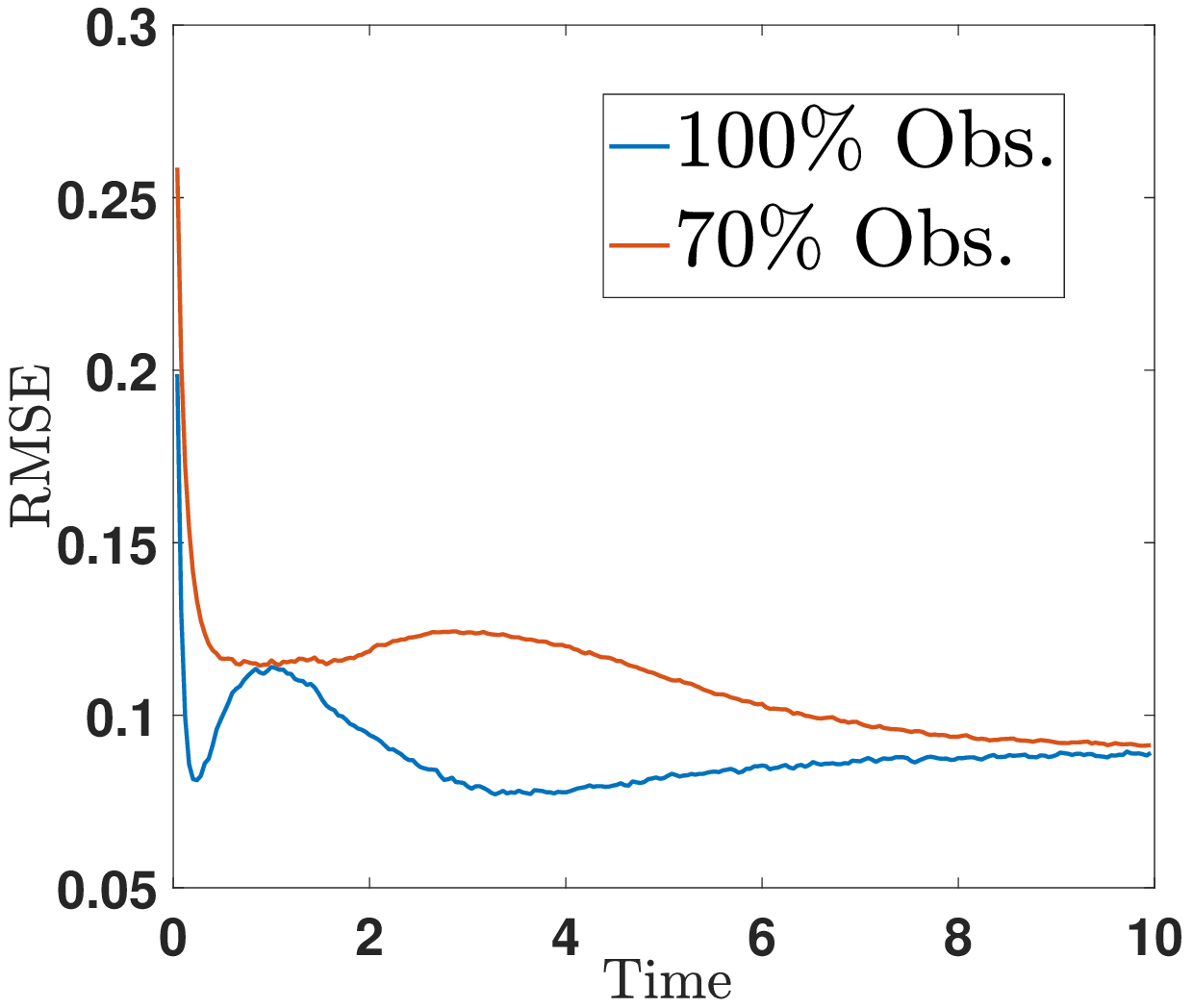}
\end{minipage}%
\begin{minipage}{0.33\textwidth}
    \includegraphics[scale = 0.22]{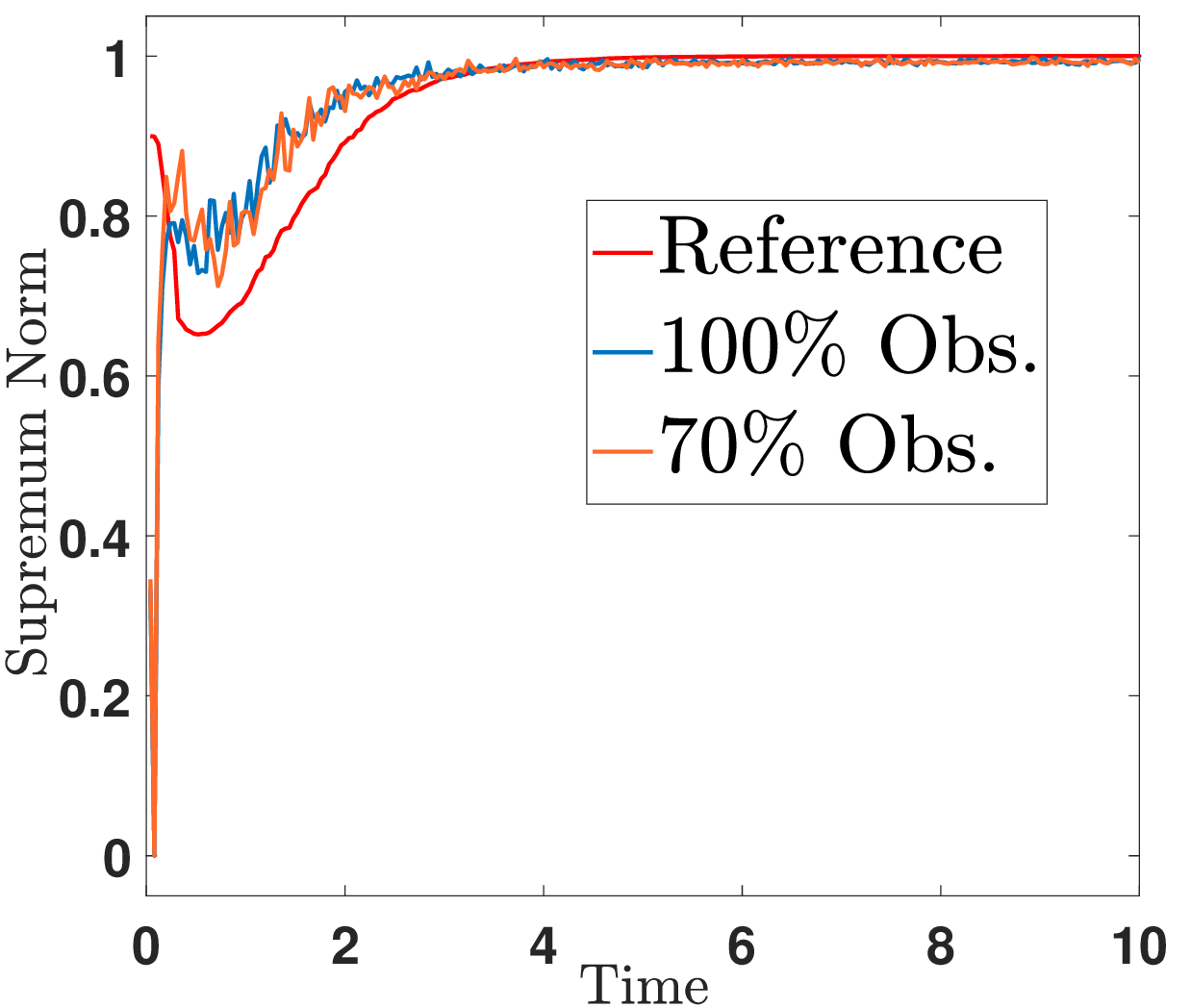}
\end{minipage}%
\begin{minipage}{0.33\textwidth}
    \includegraphics[scale = 0.22]{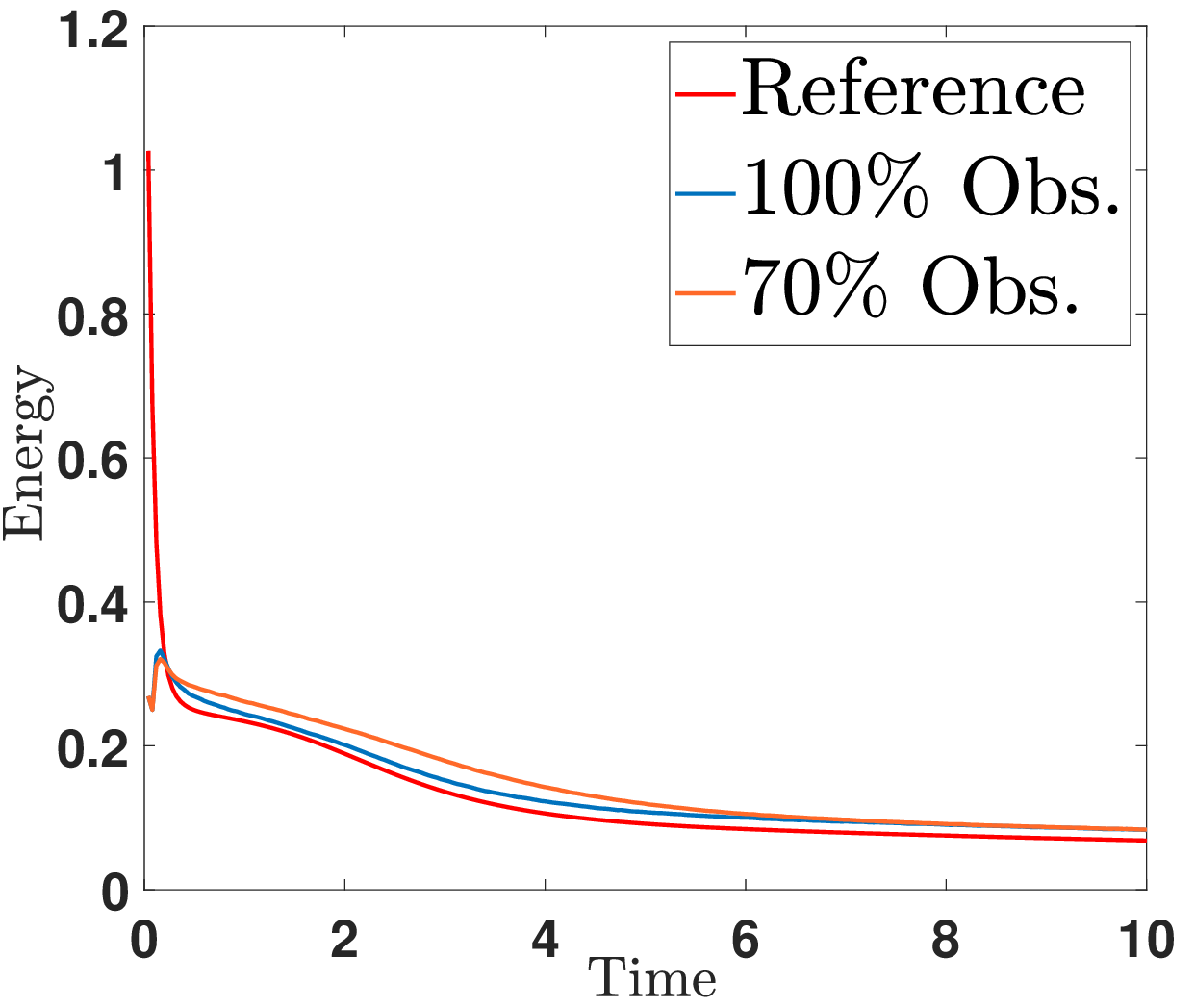}
\end{minipage}
\caption{\small Illustration of RMSEs, supremum norms, and discrete energies with $100\%$ and $70\%$ observations for Case 3. (Left) RMSE.  (Center) Supremum Norm. (Right) Energy.}
\label{DegCons_100_70Obs_RMSE_Mass_Energy}
\end{figure}

\begin{figure}[h!]
\begin{minipage}{0.22\textwidth}
    \includegraphics[scale = 0.2]{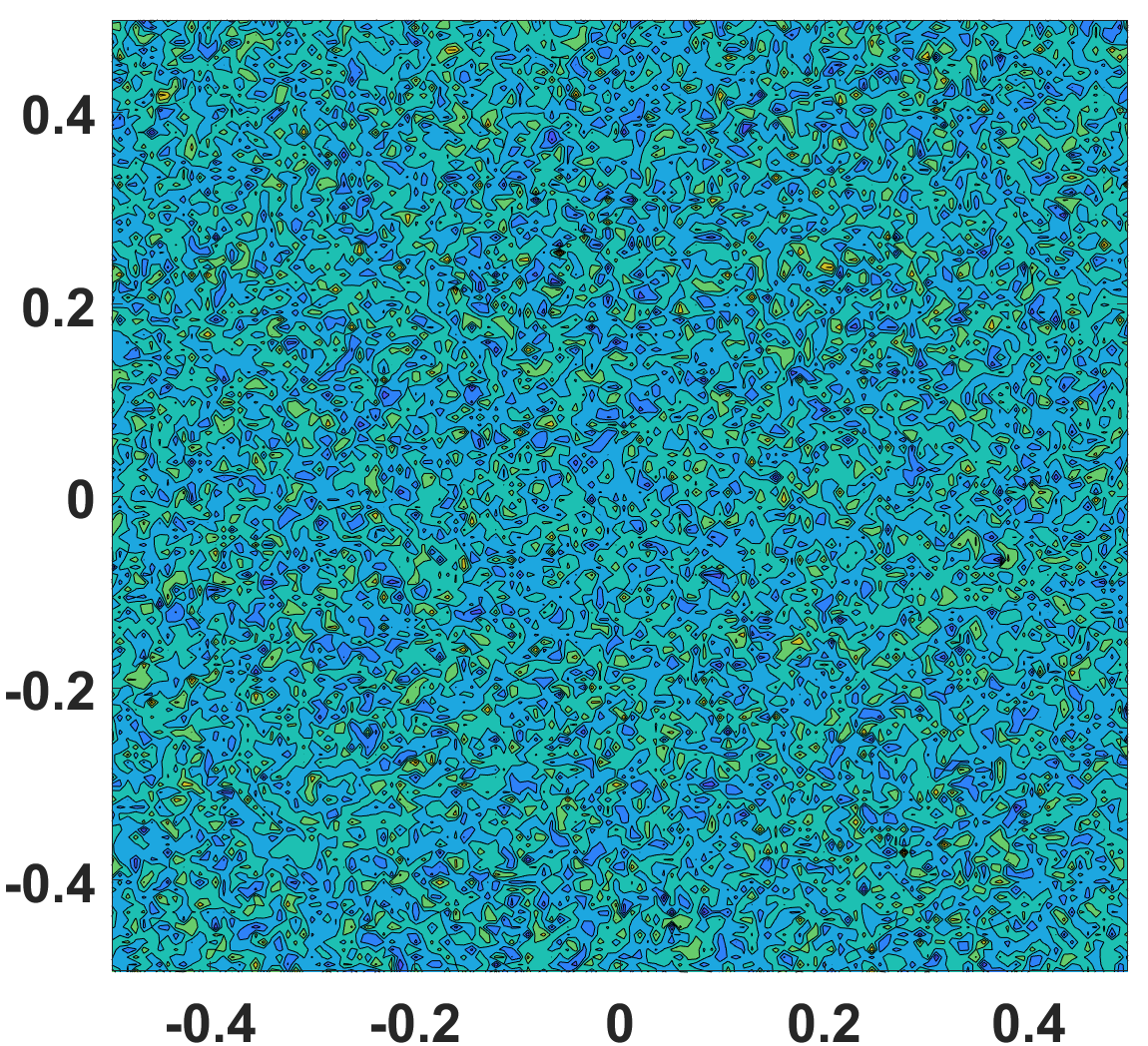}
\end{minipage}%
\begin{minipage}{0.22\textwidth}
    \includegraphics[scale = 0.2]{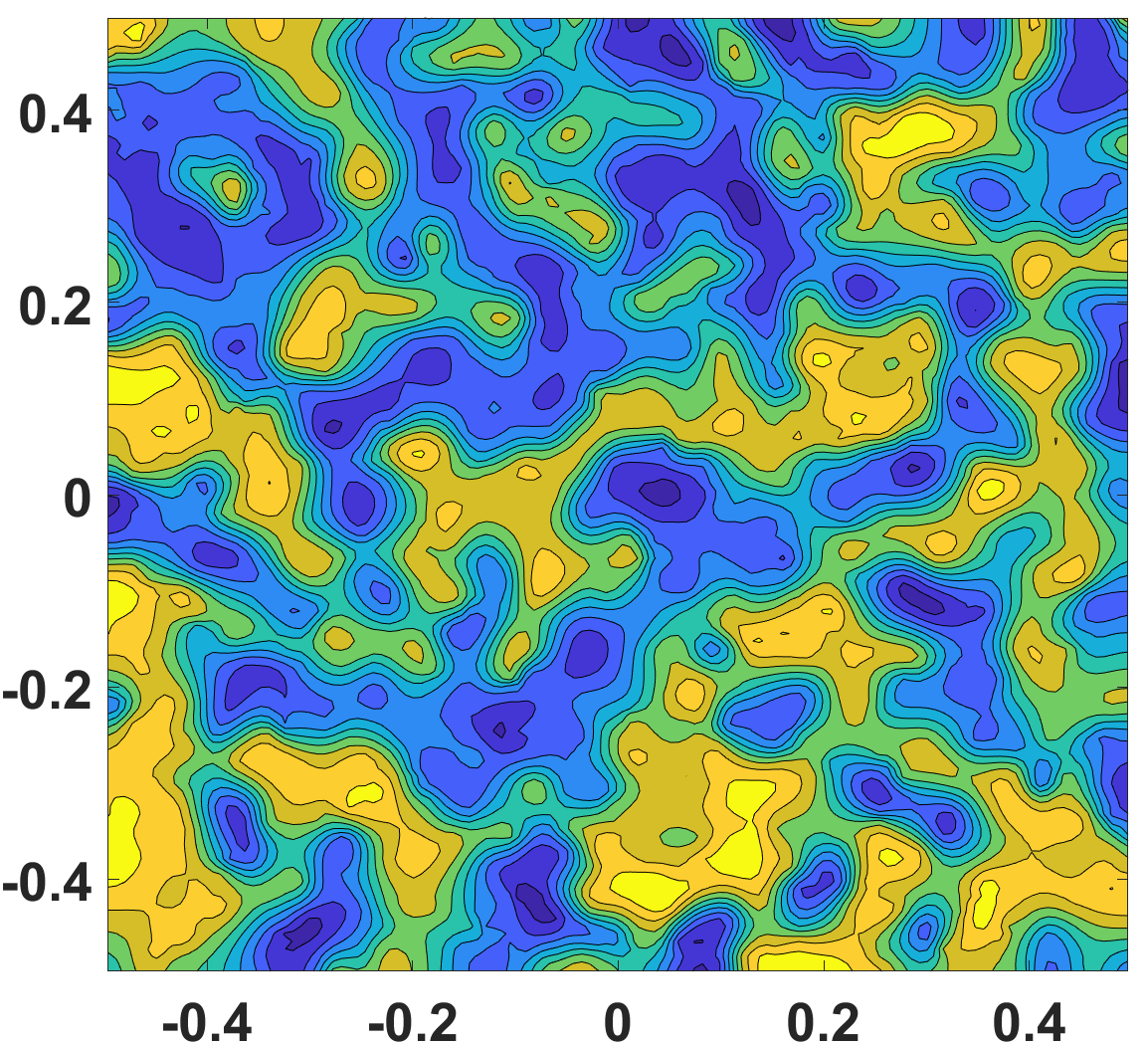}
\end{minipage}%
\begin{minipage}{0.22\textwidth}
    \includegraphics[scale = 0.2]{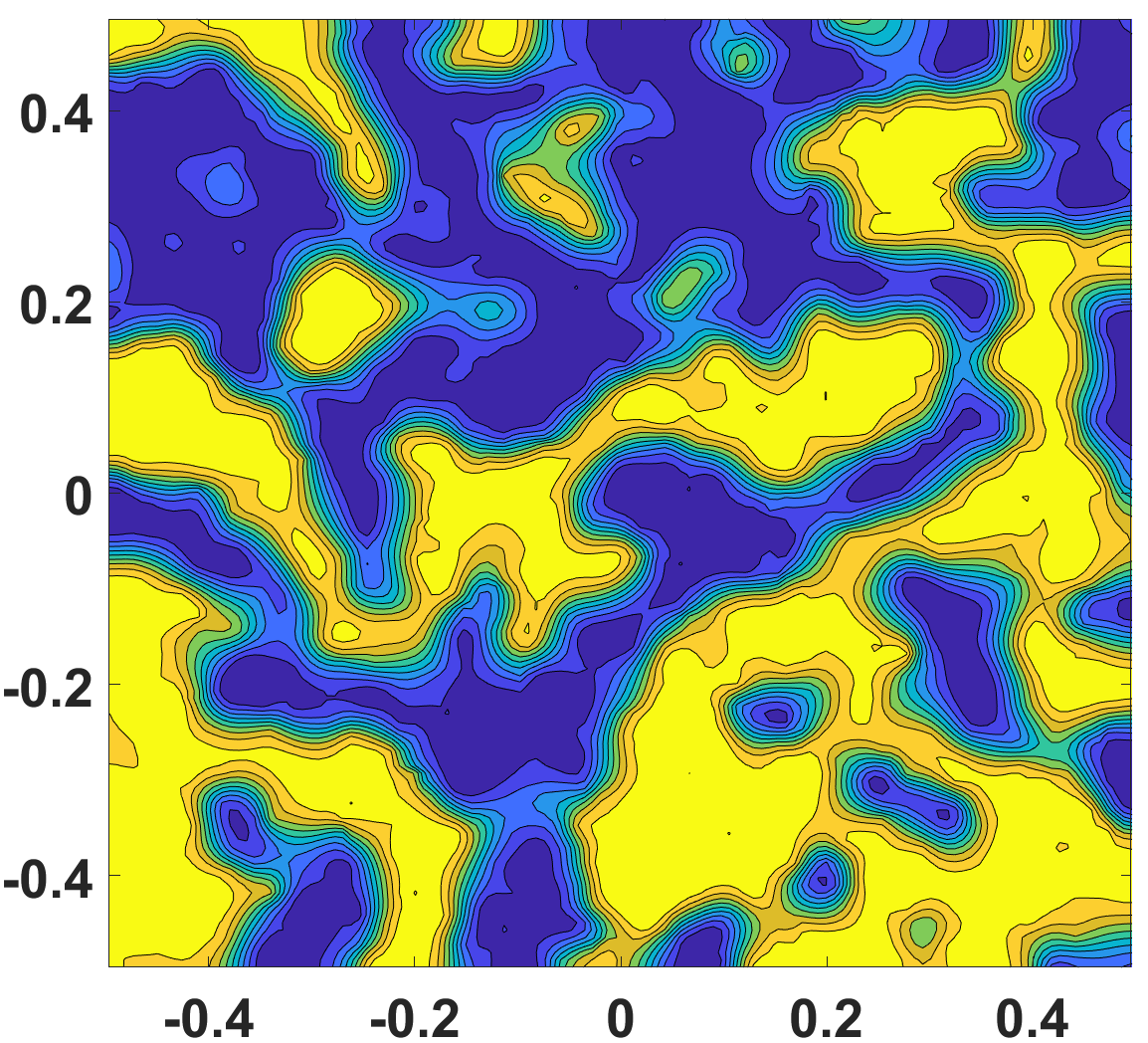}
\end{minipage}%
\begin{minipage}{0.22\textwidth}
    \includegraphics[scale = 0.2]{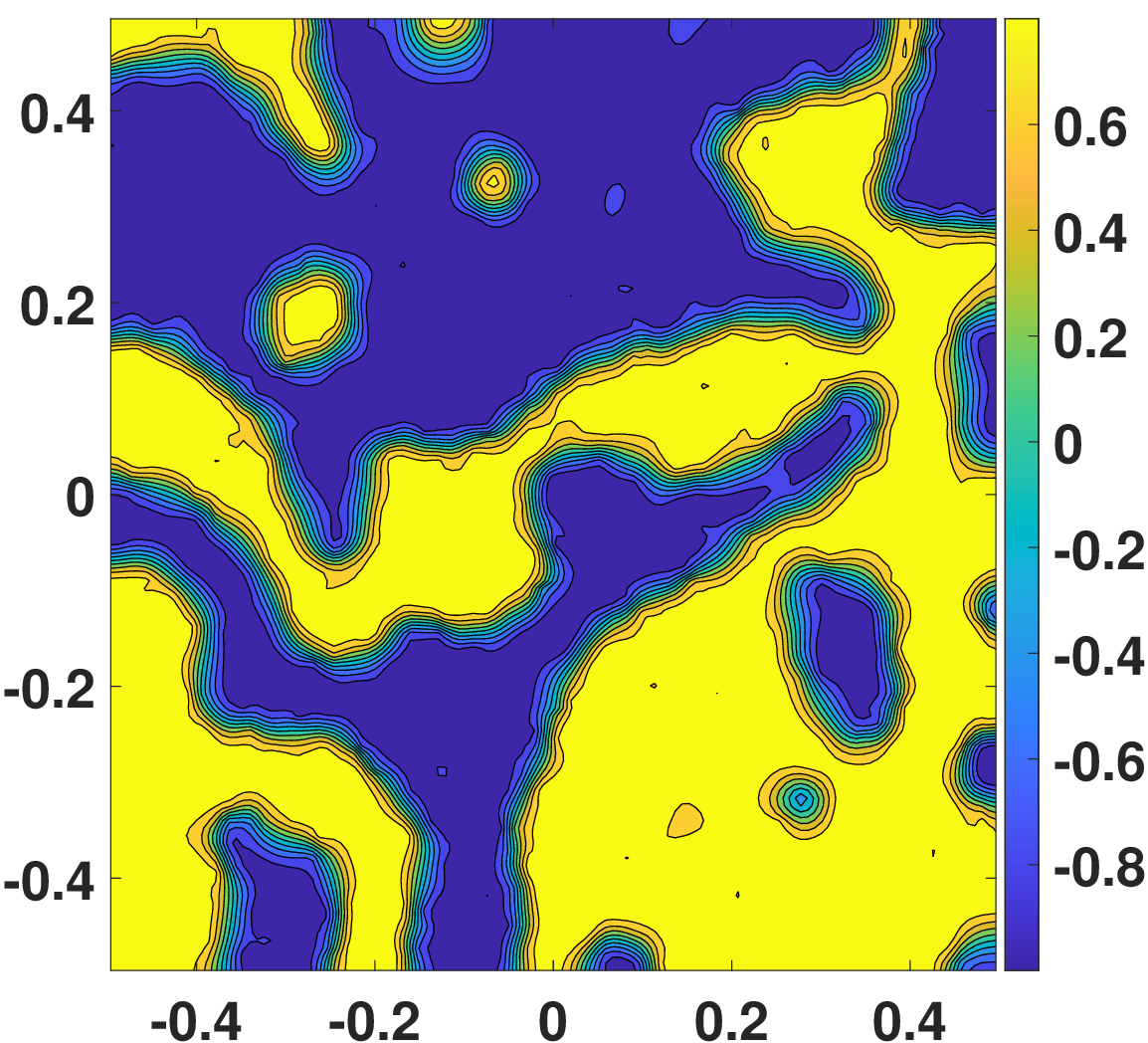}
\end{minipage}

\begin{minipage}{0.22\textwidth}
    \includegraphics[scale = 0.2]{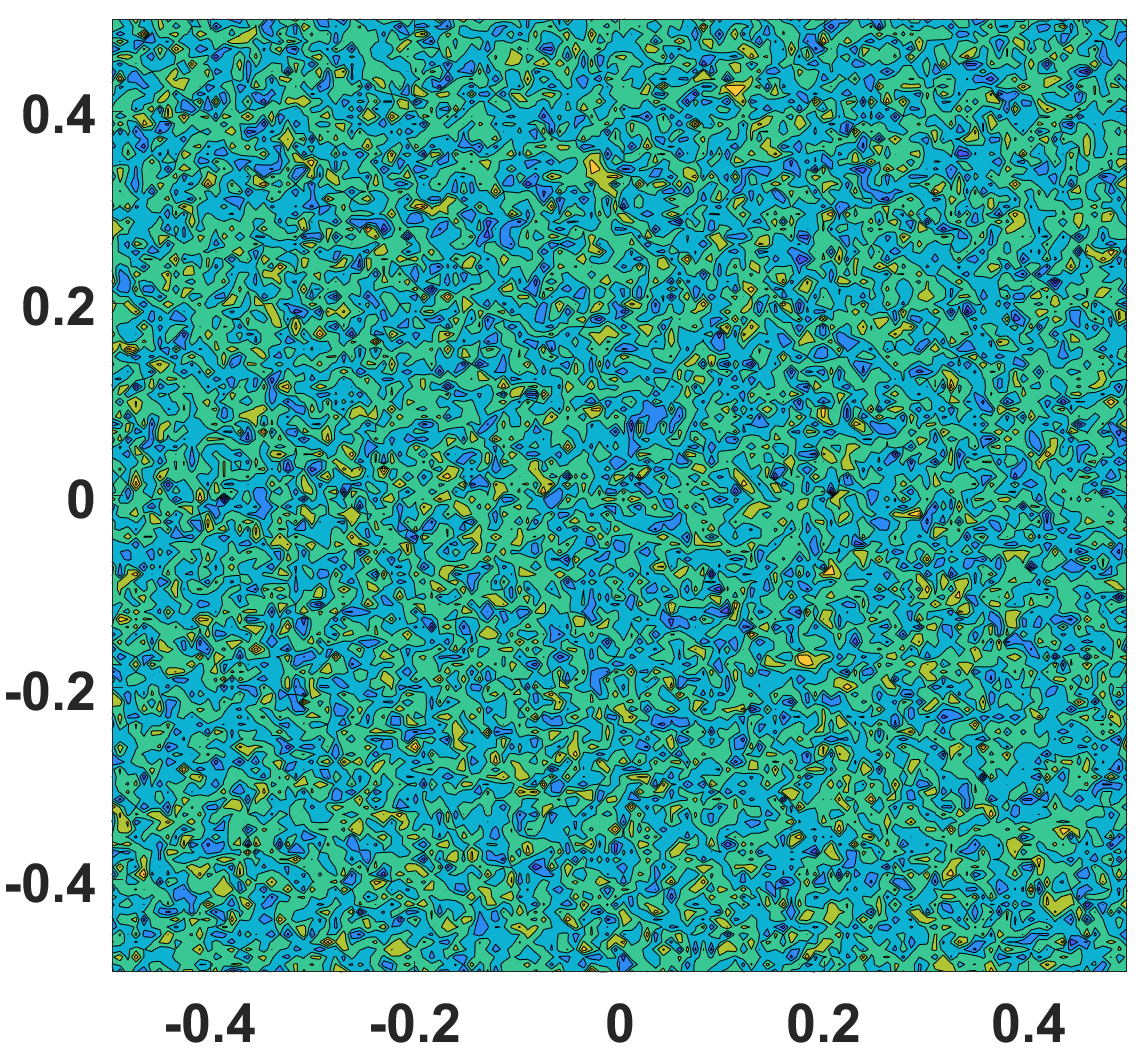}
\end{minipage}%
\begin{minipage}{0.22\textwidth}
    \includegraphics[scale = 0.2]{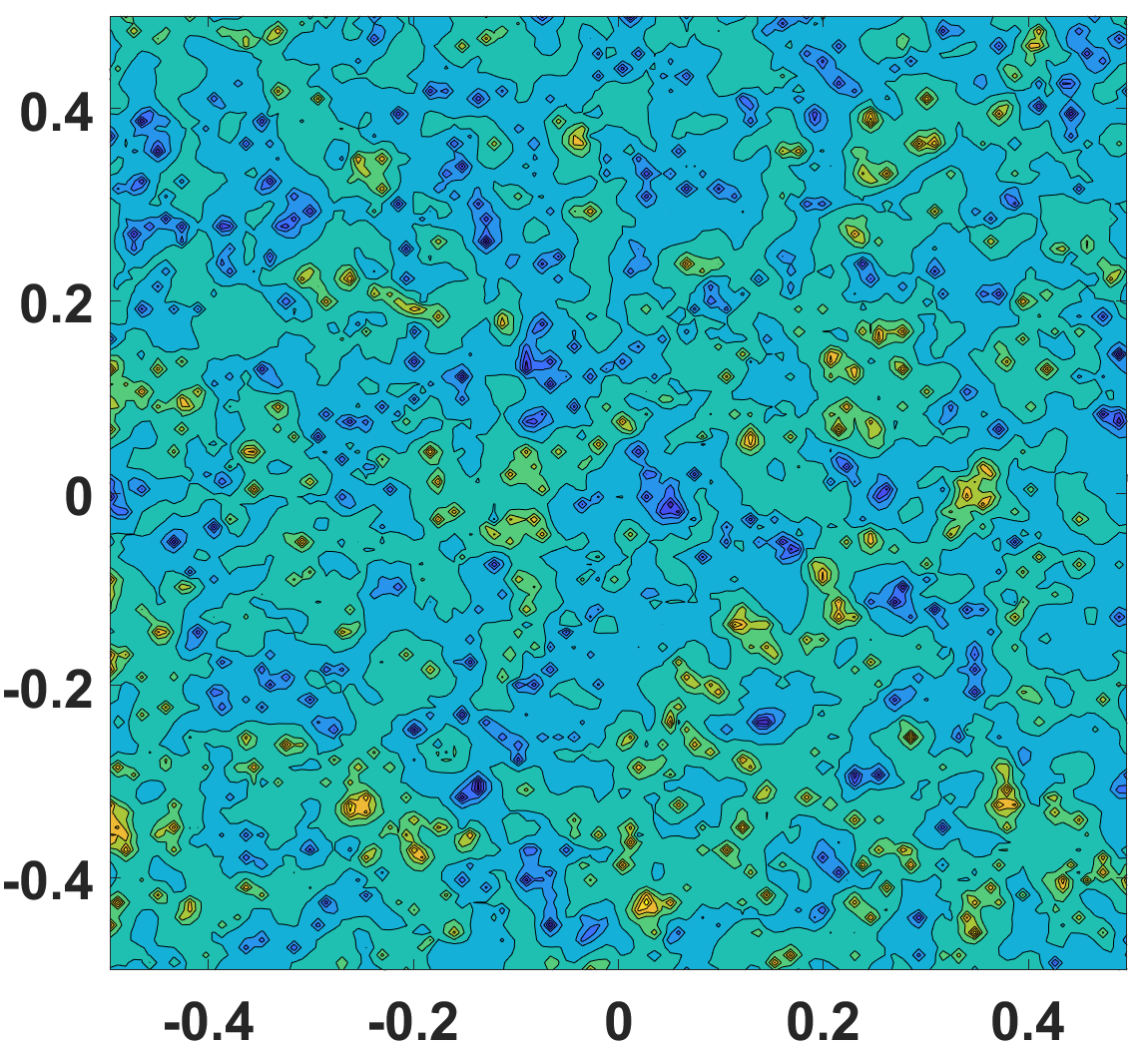}
\end{minipage}%
\begin{minipage}{0.22\textwidth}
    \includegraphics[scale = 0.2]{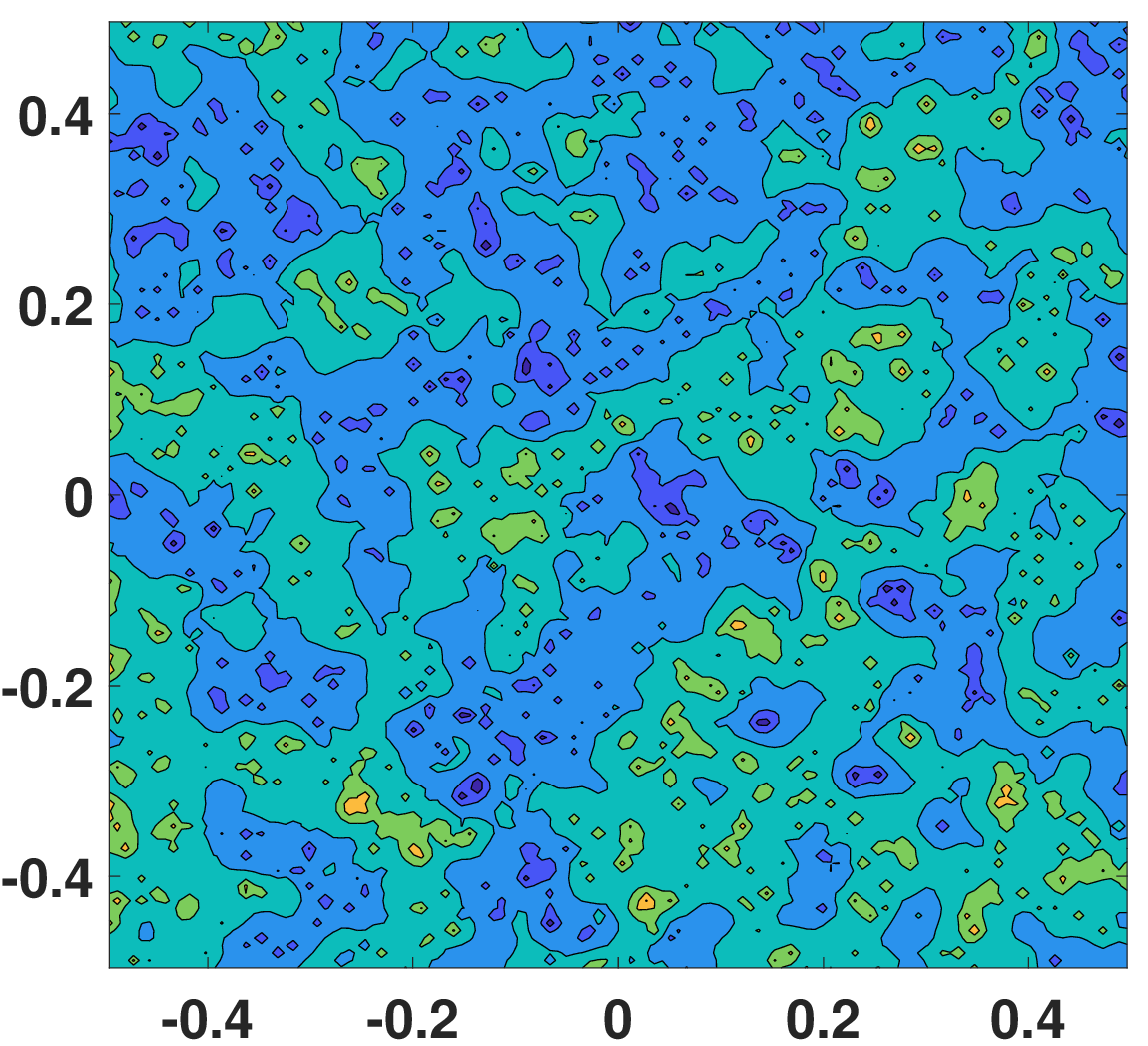}
\end{minipage}%
\begin{minipage}{0.22\textwidth}
    \includegraphics[scale = 0.2]{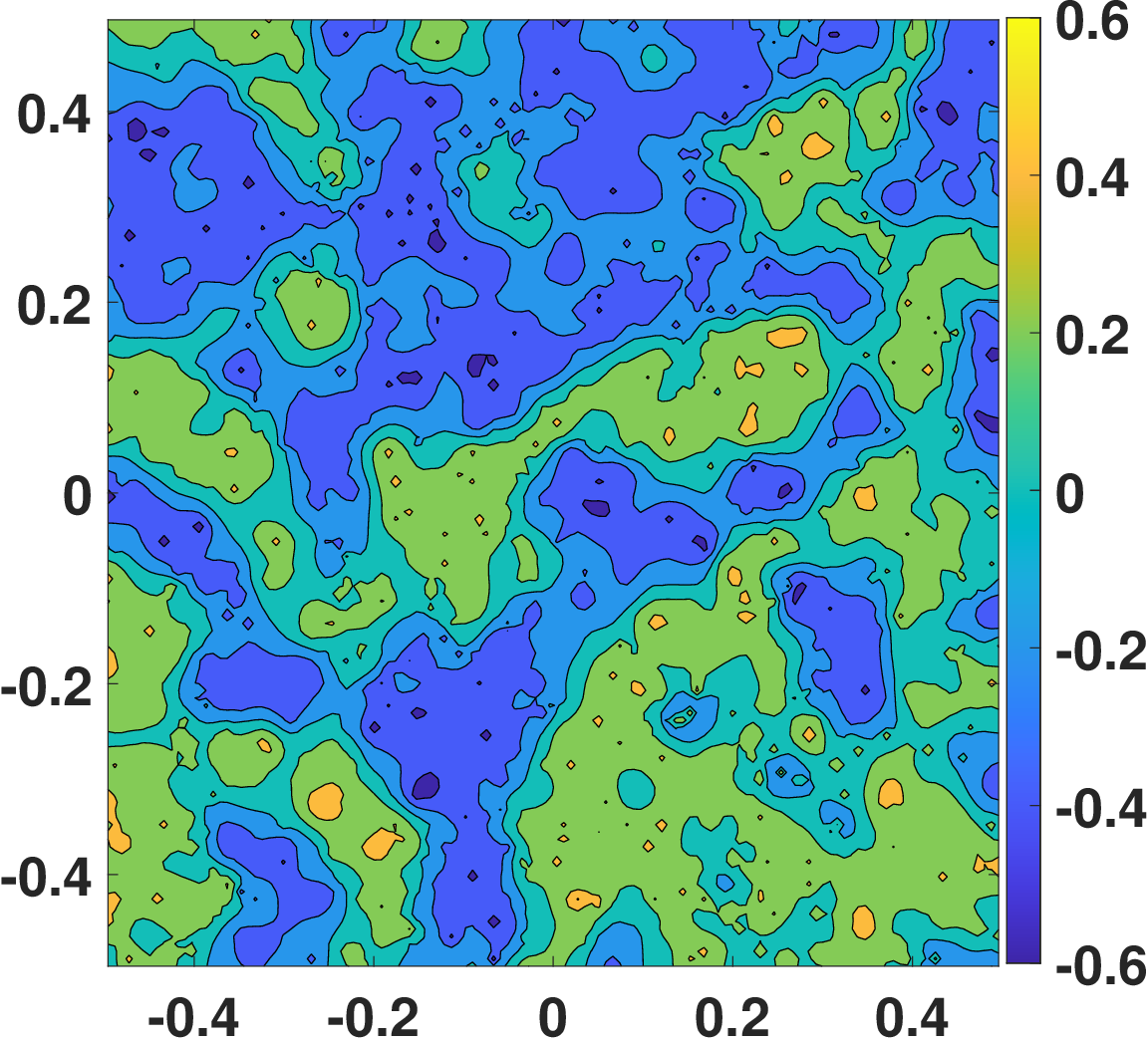}
\end{minipage}
\caption{\small Solutions of the Allen–Cahn equation in Case 3 at time $t=0, \f{T}{2}, \f{2T}{3}, T$ with $10\%$ observations. (First row) Estimated solution by EnSF. (Second row) Estimated solution by LETKF.}
\label{DegMob_10Obs}
\vspace{-0.2cm}
\end{figure}

\begin{figure}[h!]
\vspace{-0.2cm}
\centering
\includegraphics[scale = 0.42]{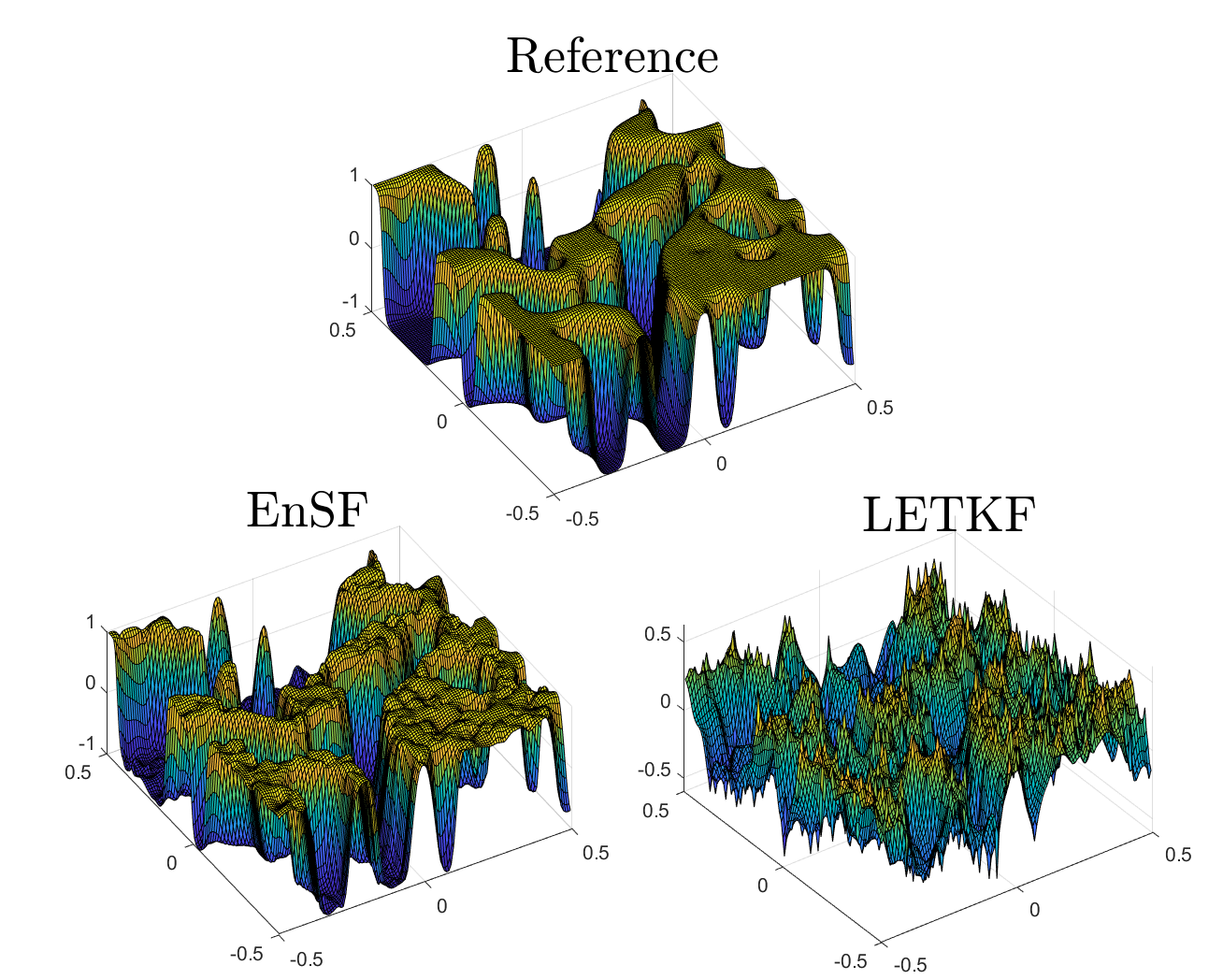}
\caption{\small 3D view of the reference and the estimated solutions for the Allen-Cahn equation in Case 3 at final time T with $10\%$ observations.}
\label{DegCons_10Obs_3D}
\end{figure}

\begin{figure}[h!]
   \begin{minipage}{0.33\textwidth}
    \includegraphics[scale = 0.22]{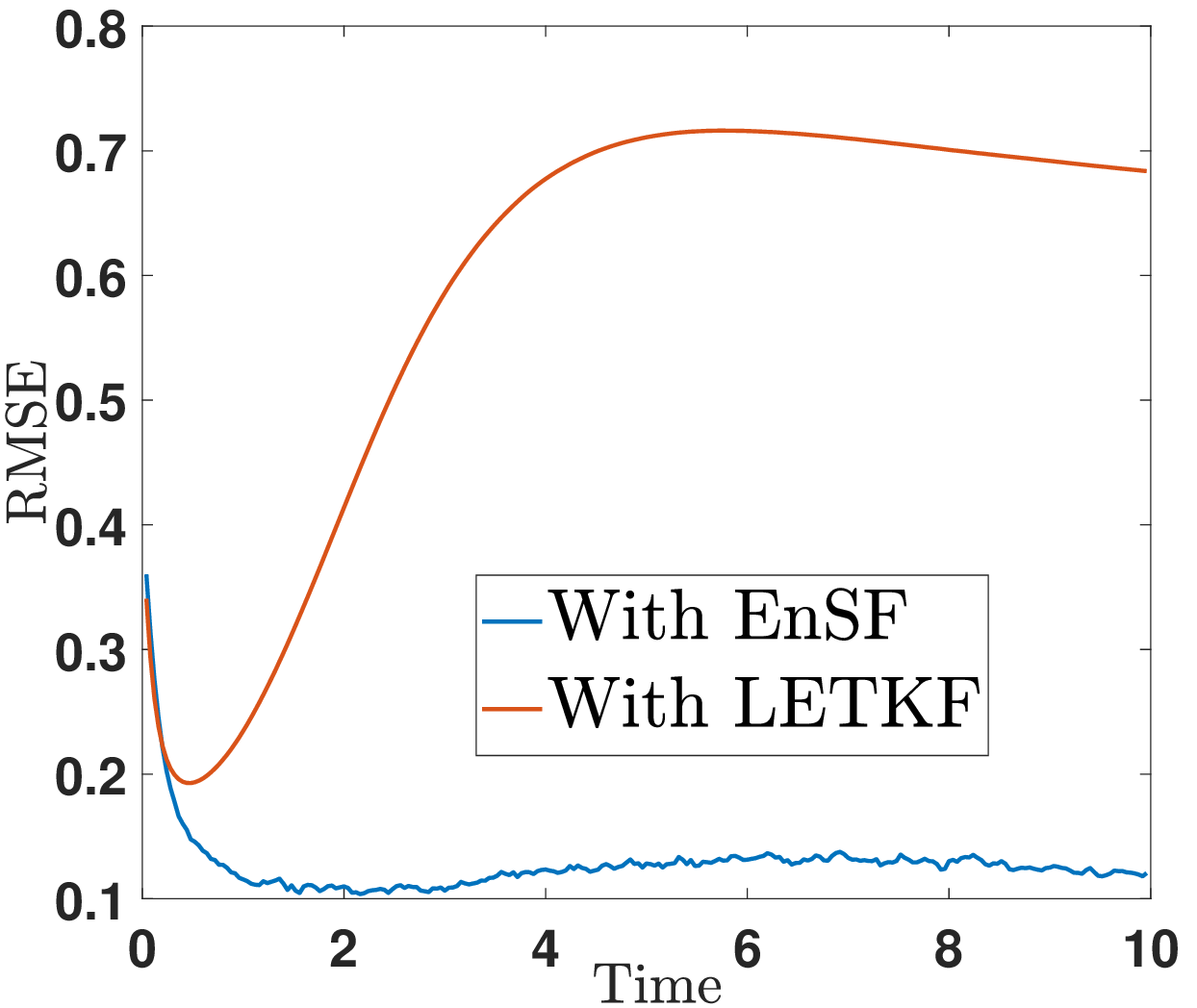}
\end{minipage}%
\begin{minipage}{0.33\textwidth}
    \includegraphics[scale = 0.22]{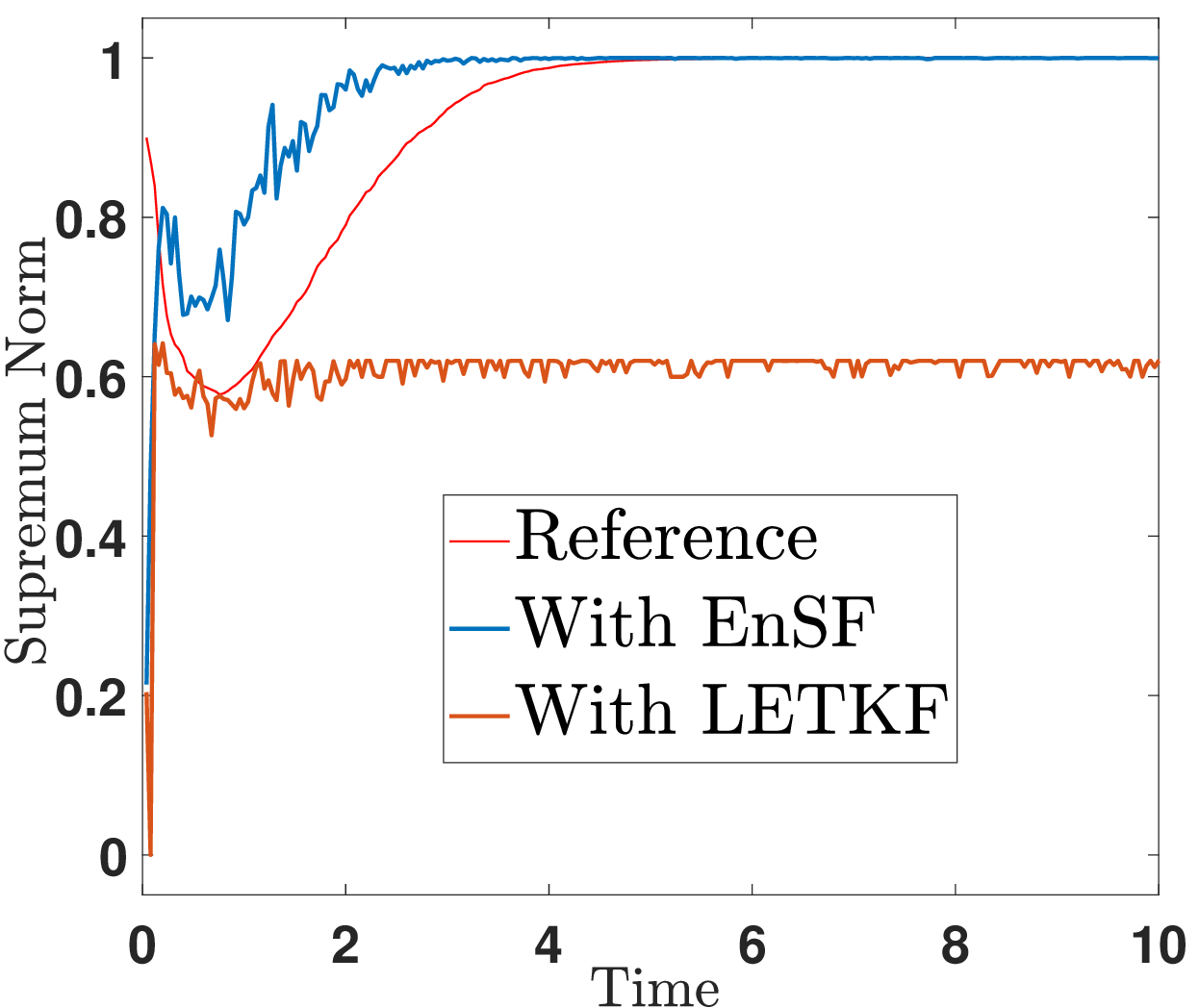}
\end{minipage}%
\begin{minipage}{0.33\textwidth}
    \includegraphics[scale = 0.22]{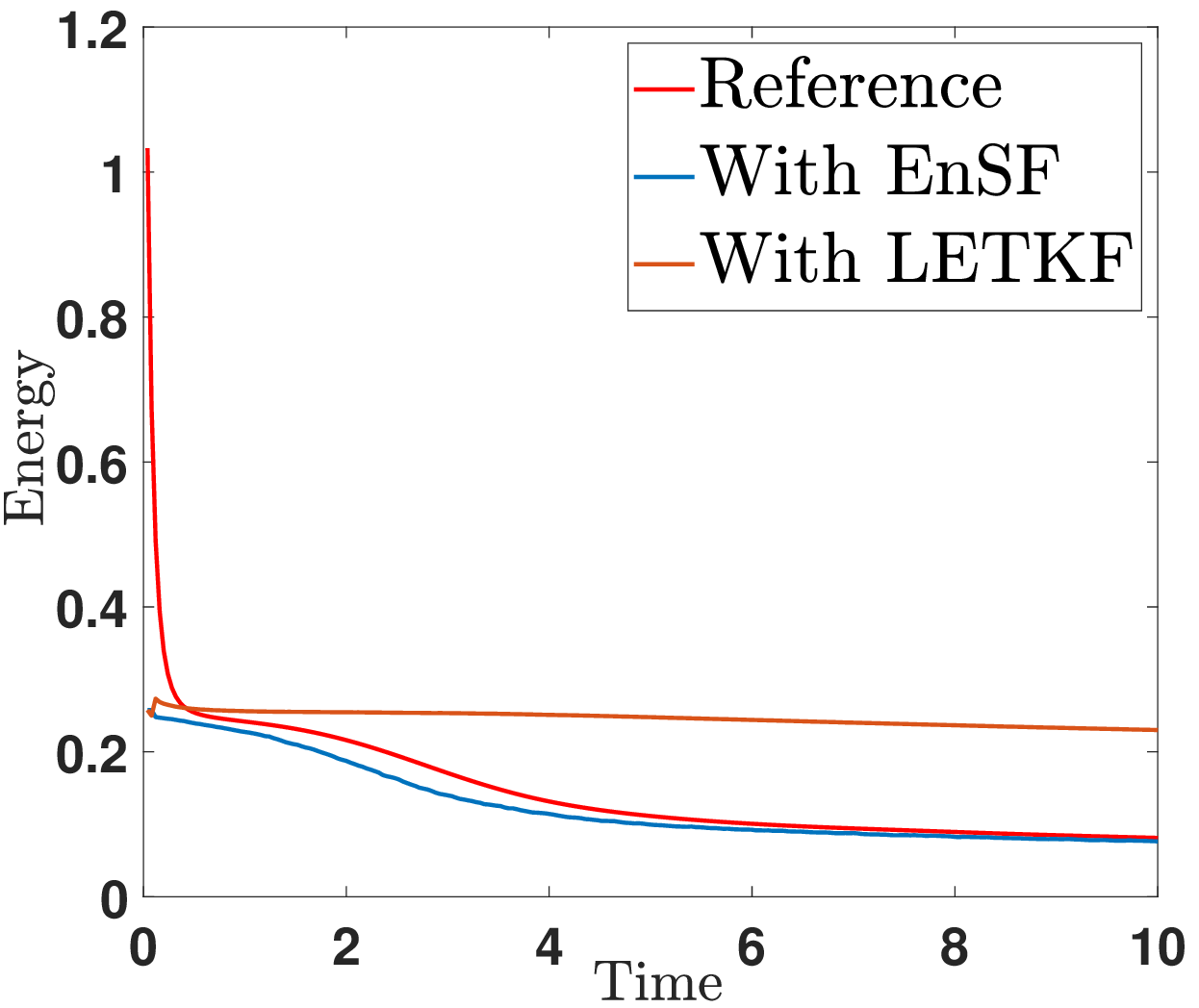}
\end{minipage}
\caption{\small Illustration of RMSEs, supremum norms, and discrete energies with $10\%$ observations for Case 3. (Left) RMSE. (Center) Supremum Norm. (Right) Energy.}
\label{DegCons_10Obs_RMSE_Mass_Energy}
\end{figure}

The results for the $100\%$ and $70\%$ observation scenarios are shown in Figures~\ref{DegMob_100_70Obs} and~\ref{DegCons_100_70Obs_RMSE_Mass_Energy}, while those for the $10\%$ scenario, including comparisons with the LETKF, are presented in Figures~\ref{DegMob_10Obs},~\ref{DegCons_10Obs_3D}, and~\ref{DegCons_10Obs_RMSE_Mass_Energy}. Consistent with the findings in Cases 1 and 2, the results for the solution-dependent mobility with uncertainty case demonstrate that our estimates converge to the reference solution over time, preserve the key properties of the Allen–Cahn equation, and consistently outperform the LETKF in terms of accuracy, supremum norm, and energy.

\section{Conclusion}\label{Conclusion}

In this work, we introduced a novel adaptive learning framework for stochastic partial differential equations (SPDEs) by leveraging score-based diffusion models within a recursive Bayesian inference setting. The proposed methodology encodes the underlying physics into the score function using simulation data and incorporates observational information via a likelihood-based correction in a reverse-time SDE. This allows the model to iteratively refine solution estimates as new data becomes available, thereby enhancing predictive accuracy and reducing epistemic uncertainty. Our diffusion model-based approach addresses key limitations of classical optimal filtering methods — such as the dimensionality constraints of particle filters and the linearity assumptions in Kalman-type filters — and enables efficient, physics-informed generative sampling for data assimilation.

We validated the effectiveness and robustness of our approach through comprehensive numerical experiments on benchmark SPDEs, including the Burgers’ equations, Navier–Stokes equations, and Allen–Cahn equations under uncertainty. These tests demonstrate the capability of the EnSF to recover hidden physical states and perform accurate solution updates even in the presence of sparse and noisy observations.

Future directions include extending the framework to multi-scale systems and exploring its integration with adaptive mesh refinement strategies for enhanced spatial resolution and computational efficiency. 

\newpage

\bibliographystyle{plain}
\bibliography{references}

\end{document}